\documentclass[apj,twocolumn,twocolappendix,numberedappendix]{openjournal}
\usepackage{newtxtext,newtxmath,enumitem,multirow,ctable}
\usepackage[T1]{fontenc}
\usepackage[breaklinks,colorlinks,citecolor=blue,urlcolor=blue]{hyperref}

\newcommand{\Alf}{{Alfv\'en}}

\newcommand{\paperone}{Paper {\small I}}
\newcommand{\papertwo}{Paper {\small II}}
\newcommand{\paperthree}{Paper {\small III}}

\newcommand{\orcidauthor}[3]{\author{\href{http://orcid.org/#1}{#2$^{#3}$}}}

\shorttitle{Super-Eddington, Flux-Frozen Disks}
\shortauthors{Hopkins et al.}

\begin{document}

\title{\vspace{-0.8cm}Zooming In On The Multi-Phase Structure of Magnetically-Dominated Quasar Disks:\\ Radiation From Torus to ISCO Across Accretion Rates\vspace{-1.5cm}}

\orcidauthor{0000-0003-3729-1684}{Philip F. Hopkins}{1,*}
\orcidauthor{0000-0003-1598-0083}{Kung-Yi Su}{2}
\orcidauthor{0000-0002-8659-3729}{Norman Murray}{3}
\orcidauthor{0000-0001-8867-5026}{Ulrich P. Steinwandel}{4}
\orcidauthor{0000-0002-5375-8232}{Nicholas Kaaz}{5}
\orcidauthor{0000-0002-7484-2695}{Sam B. Ponnada}{1}
\orcidauthor{0009-0002-8417-4480}{Jaeden Bardati}{1}
\orcidauthor{0000-0003-1661-2338}{Joanna M. Piotrowska}{1}
\orcidauthor{0000-0001-7167-6110}{Hai-Yang Wang}{1}
\orcidauthor{0000-0002-0087-3237}{Yanlong Shi}{3}
\orcidauthor{0000-0001-5769-4945}{Daniel Angl{\'e}s-Alc{\'a}zar}{6}
\orcidauthor{0000-0002-0491-1210}{Elias R. Most}{1}
\orcidauthor{0000-0002-4086-3180}{Kyle Kremer}{1}
\orcidauthor{0000-0002-4900-6628}{Claude-Andr\'{e} Faucher-Gigu\`{e}re}{6}
\orcidauthor{0000-0002-3977-2724}{Sarah Wellons}{7}

\affiliation{$^{1}$TAPIR, Mailcode 350-17, California Institute of Technology, Pasadena, CA 91125, USA}
\affiliation{$^{2}$Black Hole Initiative, Harvard University, 20 Garden St, Cambridge, MA 02138, USA}
\affiliation{$^{3}$Canadian Institute for Theoretical Astrophysics, University of Toronto, Toronto, ON M5S 3H8, Canada}
\affiliation{$^{4}$Center for Computational Astrophysics, Flatiron Institute, 162 5th Ave., New York, NY 10010 USA}
\affiliation{$^{5}$CIERA and Department of Physics and Astronomy, Northwestern University, 1800 Sherman Ave, Evanston, IL 60201, USA}
\affiliation{$^{6}$Department of Physics, University of Connecticut, 196 Auditorium Road, U-3046, Storrs, CT 06269, USA}
\affiliation{$^{7}$Department of Astronomy, Van Vleck Observatory, Wesleyan University, 96 Foss Hill Drive, Middletown, CT 06459, USA}

\thanks{$^*$E-mail: \href{mailto:phopkins@caltech.edu}{phopkins@caltech.edu}},

\begin{abstract}
Recent radiation-thermochemical-magnetohydrodynamic simulations resolved formation of quasar accretion disks from cosmological scales down to $\sim 300$ gravitational radii $R_{g}$, arguing they were ``hyper-magnetized'' (plasma $\beta\ll1$ supported by toroidal magnetic fields) and distinct from traditional $\alpha$-disks. We extend these, refining to $\approx 3\,R_{g}$ around a $10^{7}\,{\rm M_{\odot}}$ BH with multi-channel radiation and thermochemistry, and exploring a factor of $1000$ range of accretion rates ($\dot{m} \sim 0.01 - 20$). At smaller scales (depending on $\dot{m}$), we see the disks maintain steady accretion, thermalize and self-ionize, and radiation pressure grows in importance, but large deviations from local thermodynamic equilibrium and single-phase equations of state are always present. Trans-\Alf{ic} and highly-supersonic turbulence persists in all cases, and leads to efficient vertical mixing, so radiation pressure saturates at levels comparable to fluctuating magnetic and turbulent pressures even for $\dot{m} \gg 1$. The disks also become radiatively inefficient in the inner regions at high $\dot{m}$. The midplane magnetic field remains primarily toroidal at large radii, but at super-Eddington $\dot{m}$ we see occasional transitions to a poloidal-field dominated state associated with outflows and flares. Large-scale magnetocentrifugal and continuum radiation-pressure-driven outflows are weak at $\dot{m} < 1$, but can be strong at $\dot{m}\gtrsim1$. In all cases there is a scattering ``photosphere'' above the disk extending to $\gtrsim 1000\,R_{g}$ at large $\dot{m}$, and the disk is thick and flared owing to magnetic support (with $H/R$ nearly independent of $\dot{m}$), so the outer disk is strongly illuminated by the inner disk and most of the inner disk continuum scatters or is reprocessed at larger scales, giving apparent emission region sizes as large as $\gtrsim 10^{16}\,{\rm cm}$.
\end{abstract}


\maketitle

\section{Introduction}
\label{sec:intro}

Accretion disks around supermassive black holes (SMBHs) are the engines that power active galactic nuclei (AGN) and quasars, the most luminous sustained sources known \citep{schmidt:1963.qso.redshift,salpeter64}. Mass inflow rates exceeding $\gtrsim 1\,{\rm M_{\odot}\,yr^{-1}}$ are required to explain typical quasars at redshifts $z \gtrsim 2$ \citep{shen:bolometric.qlf.update}, and to provide most of the SMBH mass today \citep{soltan82}. The effects of ``AGN feedback'' in the form of radiation, outflows, and jets launched from such disks near the BH \citep{laor:warm.absorber,crenshaw:nlr,dunn:agn.fb.from.strong.outflows,sturm:2011.ulirg.herschel.outflows,cafg:2012.egy.cons.bal.winds,faucher-giguere:2012.felobal.model,Zakamska2016,Williams2017} can also have dramatic effects \citep{silkrees:msigma,king:msigma.superfb.1,dimatteo:msigma,murray:momentum.winds,hopkins:lifetimes.methods,hopkins:lifetimes.obscuration,torrey:2020.agn.wind.bal.gal.fx.fire} potentially explaining BH-host galaxy correlations and co-evolution \citep{magorrian,FM00,Gebhardt00,hopkins:bhfp.obs,aller:mbh.esph,kormendy:2011.bh.nodisk.corr} and shaping galaxy ``quenching'' and masses \citep{croton:sam,hopkins:qso.all,hopkins:groups.ell,wellons:2022.smbh.growth,mercedes.feliz:2023.agn.feedback.positive.negative,cochrane:2023.agn.winds.galaxy.size.effects}. 

\begin{figure*}
	\centering
	\includegraphics[width=0.97\textwidth]{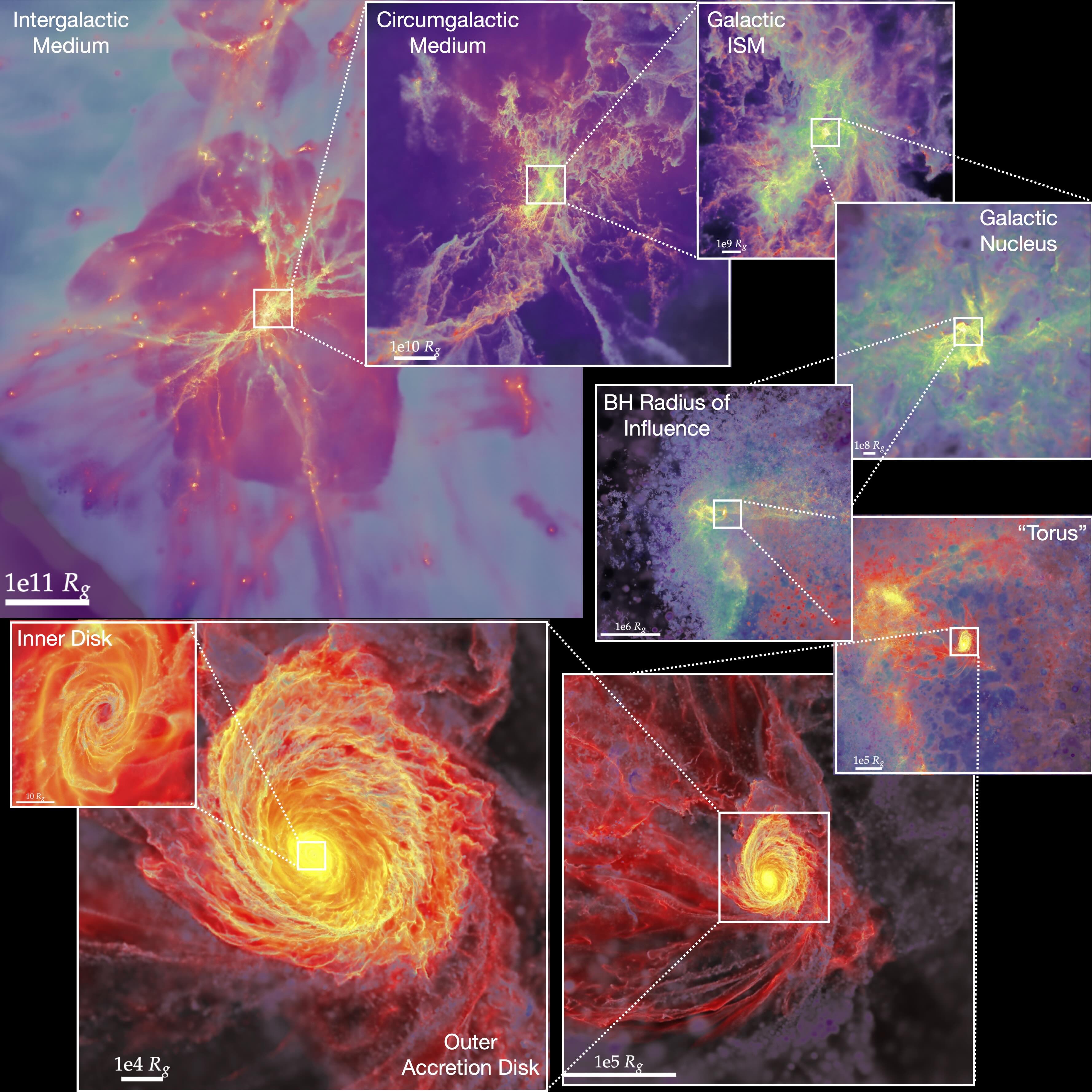} 
	\caption{Images of the simulation on various scales. Bar shows the scale in units of $R_{g} \equiv 2 G M_{\rm BH}/c^{2}$ for the $1.3 \times 10^{7}\,{\rm M_{\odot}}$ BH at the galaxy center. Intensity scales with log of projected surface density (factor of $\sim 10^{4}$ range in each) and colormaps reflect the dominant pressure source (thermal plus radiation in {\em purple-magenta}, turbulent/kinetic in {\em blue-green}, magnetic in {\em red-yellow}). The simulation from \paperone\ begins from a fully-cosmological box and trace $\sim 10^{-15}\,$G magnetic fields at redshift $z\approx 150$, zooming in to the formation of a massive starburst galaxy, which undergoes multiple mergers and gas-rich accretion at $z\sim 4.5$ (galaxy-integrated star formation rate $\sim 200\,{\rm M_{\odot}\,yr^{-1}}$), wherein a massive ($\sim 10^{8}\,{\rm M_{\odot}}$) molecular cloud-complex is disrupted around the BH radius of influence ($\sim 5\,$pc here) and forms the accretion disk. Here we continue to zoom in to the inner disk, seeing the accretion disk persist down to $\sim 3\,R_{g}$. 
	\label{fig:zoomies}}
\end{figure*}

The theoretical study and modeling of accretion disks is a rich field, and it has been recognized for decades that the structure and basic qualitative nature of the accretion disk, let alone structures surrounding or part of the outer disk such as the broad line region and dusty ``torus,'' are sensitive to the outer boundary conditions ``supplying'' the disk \citep{pringle:accretion.review,frank:2002.accretion.book,abramowicz:accretion.theory.review,jafari:2019.mhd.ppd.review}. These flows come from interstellar medium (ISM) scales $\gg$\,pc where star formation, non-equilibrium atomic+molecular radiation-thermochemistry, optically-thin cooling, and stellar feedback processes such as supernovae and stellar radiation must be modeled. In \citet{hopkins:superzoom.overview} (henceforth \paperone) we presented the first radiation-magnetohydrodynamic simulations to follow all of these physical effects in a single simulation beginning from a cosmological simulation with box size $\sim 100\,$Mpc and refining down to scales of $\sim 80\,$au or $\sim 300\,R_{g}$,\footnote{Here we define the gravitational radius equal to the Schwarzschild radius $R_{g} \equiv 2\,G\,M_{\rm BH}/c^{2}$, or $\approx 3.9 \times 10^{12}\,{\rm cm}$ here.} and showed that indeed a clearly-defined, steady-state accretion disk forms interior to the BH radius of influence ($R_{\rm BHROI} \sim 5\,$pc, interior to which the BH dominates the gravitational potential). As shown in Fig.~\ref{fig:zoomies}, this particular event occurs in a high-redshift ($z\sim 4.5$) massive starburst (star formation rate $\gtrsim 200\,{\rm M_{\odot}\,yr^{-1}}$) galaxy which is clumpy and merging, when a massive ($\gg 10^{8}\,{\rm M_{\odot}}$) star-forming cloud complex passes by and is tidally disrupted by the pre-existing SMBH (mass $\sim 1.3\times10^{7}\,{\rm M_{\odot}}$), fueling up to $\sim 20-30\,{\rm M_{\odot}\,yr^{-1}}$ steady-state inflows which fall into the BHROI and circularize forming the accretion disk at $\sim 0.1-1\,$pc. 

In \citet{hopkins:superzoom.disk} (\papertwo), we studied the detailed properties of these disks, showing that they were qualitatively unlike classic thermal-pressure-dominated $\alpha$-disk models like that of \citet{shakurasunyaev73} (hereafter SS73), nor did they resemble magnetically ``arrested'' or ``elevated'' disks \citep{bisnovatyi.kogan:1976.mad.disk,narayan:2003.mad.disk}. Instead, these disks are supported by strong magnetic fields with plasma $\beta \equiv \beta_{\rm thermal} \equiv c_{s}^{2}/v_{A}^{2} \sim P_{\rm gas,\,thermal}/P_{\rm B} \sim 10^{-6}-10^{-2} \ll 1$ (see e.g.\ \citealt{gaburov:2012.public.moving.mesh.code}),\footnote{We adopt $c_{s} \equiv \sqrt{k_{B} T_{\rm gas}/\mu m_{p}}$ throughout, i.e.\ the sound speed refers specifically the isothermal gas sound speed (as opposed to some ``effective'' wavespeed), for the reasons shown explicitly in \S~\ref{sec:lte} where we discuss the degree to which strict tight-coupling of radiation is inapplicable here.} where the field is dominated by the mean toroidal component and secondary radial and turbulent/fluctuating components (with weak, turbulently-sourced mean poloidal field), amplified primarily via flux-freezing from modest/typical ISM fields. They are also trans-\Alf{ically} turbulent, which necessarily means the turbulence is highly supersonic and that the medium has cooling times $t_{\rm cool}$ much smaller than dynamical times $t_{\rm dyn} \sim \Omega^{-1} \equiv (G\,M_{\rm BH}/R^{3})^{-1/2}$ (with cooling balancing heating). In \citet{hopkins:superzoom.imf}, we studied how star formation is completely suppressed in such disks, noting that they resolve the decades-old problem that thermal-pressure dominated disks should be gravitationally unstable (at high accretion rates) outside a few dozen gravitational radii. And in \citet{hopkins:superzoom.analytic} (\paperthree) we presented some simple analytic similarity models which describe the simulation outer disks and attempted to extrapolate them to smaller radii. Since then, similar disks have been seen under very different conditions, using different codes and numerical methods to simulate accretion from large-to-small scales in massive ellipticals (at much lower accretion rates; \citealt{guo:2024.fluxfrozen.disks.lowmdot.ellipticals}) and from star clusters onto intermediate-mass BHs \citep{shi:2024.imbh.growth.feedback.survey}. This suggests they may be ubiquitous in AGN.

However, these studies leave a number of behaviors at smaller radii uncertain. Most notably, the particular accretion episode studied in \paperone-\papertwo\ features a supercritical Eddington-scaled accretion rate $\dot{m} \equiv \dot{M} / (M_{\rm BH}/t_{S}) = 0.1\,\dot{M}\,c^{2}/L_{\rm Edd} \sim 100$ (with $t_{S} \approx 5\times10^{7}\,{\rm yr}$). But with the simulation only reaching down to $\sim 300\,R_{g}$, the actual accretion luminosity in the {\em resolved} portion of the disk is everywhere sub-Eddington. Based on the analytic extrapolations in \paperthree, much of the interesting behavior would be expected to appear at the next order of magnitude down in scale, $\sim 30-300\,R_{g}$. At these radii, for a simple radiatively-efficient model akin to SS73, we expect an order-unity fraction of the total luminosity to be produced. And somewhere in this range, if $\dot{m} \gg 1$ were maintained, models predict that the radiation pressure becomes comparable to (or larger than) the magnetic pressure. Moreover, if the accretion disk is effectively optically thick and radiation transport were strictly diffusive, so the midplane temperature $T_{\rm mid} \sim T_{\rm eff}\,\tau^{1/4}$, then simple analytic arguments suggest any disk would have radiation pressure internal to the disk sufficient to inflate the disk to $H \gtrsim R$ and transition to either (a) radiatively inefficient (slim disk-like) accretion or (b) very strong outflows (greatly decreasing $\dot{m}$), at radii $x_{g} \equiv R/R_{g} \gtrsim 3\,\dot{m}$. But this is just interior to the boundary in \papertwo. And as noted in \paperthree, the strong turbulence in these disks could {\em qualitatively} change this conclusion, with major implications for both the structure of magnetically-dominated disks (at high-$\dot{m}$) and the possibility of super-critical accretion. 

A number of other important physical transitions are predicted to occur, based on the extrapolations in \paperthree, at scales $\lesssim 1000\,R_{g}$. Most notably, within this range the radiation should become hard enough that the disk fully self-ionizes (it is primarily atomic at $\gtrsim 1000\,R_{g}$), and the scattering+absorption optical depth should become large enough that the ``effective'' optical depth $\sim \sqrt{\kappa_{s}\,(\kappa_{a}+\kappa_{s})}\,\Sigma_{\rm gas}$ exceeds unity and the disk could thermalize with optically-thick quasi-blackbody cooling. So in this range it may still be important to include physics from \paperone-\papertwo\ usually neglected in accretion disk simulations such as detailed non-equilibrium atomic/molecular chemistry, optically-thin plus thick multi-group radiation and coupled heating/cooling (including recombination, H$^{-}$, H, He and metal line, as well as continuum processes), and perhaps even self-gravity. And the study in \paperone-\papertwo\ only considered one accretion rate, so it is challenging to make extrapolations or test different analytic models for the disk structure.

In this paper, we therefore extend the simulations presented in \paperone-\papertwo\ by further refining down to $\sim 3\,R_{g}$ around the SMBH, while also modifying the outer boundaries to produce a range of accretion rates. This allows us to capture the key transitions in the disk described above as a function of both radius and accretion rate, before reaching radii where the physics limitations of the simulations -- we do not include general relativity [GR] nor explicit treatment of X-ray radiation/ionization nor weakly-collisional two-temperature plasmas -- become problematic. 
In \S~\ref{sec:methods} we summarize our numerical methods and different tests performed (reviewing the basic physics \S~\ref{sec:methods:overview}, initial conditions \S~\ref{sec:methods:ics}, and accretion rate survey \S~\ref{sec:methods:mdot}). 
In \S~\ref{sec:results} we present and discuss the results, including convergence and steady-state behaviors (\S~\ref{sec:results:steady}), results similar to those at larger radii (\S~\ref{sec:results:similar}) and unique to smaller radii (\S~\ref{sec:results:new}), luminosities and radiative efficiencies (\S~\ref{sec:results:radeff}), and variability/bursts/state changes (\S~\ref{sec:bursts}). 
In \S~\ref{sec:discussion} we discuss the physics of these results, attempting to understand the radiation pressure and its saturation (\S~\ref{sec:rad.pressure.saturation}), radiative efficiencies (\S~\ref{sec:rad.efficiency}), scaling with accretion rate $\dot{m}$ (\S~\ref{sec:mdot}), comparison to analytic similarity models (\S~\ref{sec:modified.radial}), physics of extended scattering/reprocessing regions (\S~\ref{sec:scattering}) and implications for the emergent spectrum (\S~\ref{sec:Trad}), properties of outflows (\S~\ref{sec:outflows}) and of the broad-line region (\S~\ref{sec:blr}), deviations from local thermodynamic equilibrium and simple equation-of-state models (\S~\ref{sec:lte}), and sensitivity to different physics (\S~\ref{sec:sensitivity}). 
We summarize and conclude in \S~\ref{sec:conclusions}. In the Appendices we include more details of the cooling physics and opacities (\S~\ref{sec:opacities}) and the full set of physical and numerical variations considered (\S~\ref{sec:methods:variants}), discuss the effective spatial resolution of the simulations (\S~\ref{sec:spatial.resolution}) and show tests where the thermal scale-length is always forced to be well-resolved (\S~\ref{sec:resolution.thermal.scale}).

\begin{figure*}
	\centering
	\includegraphics[width=0.97\textwidth]{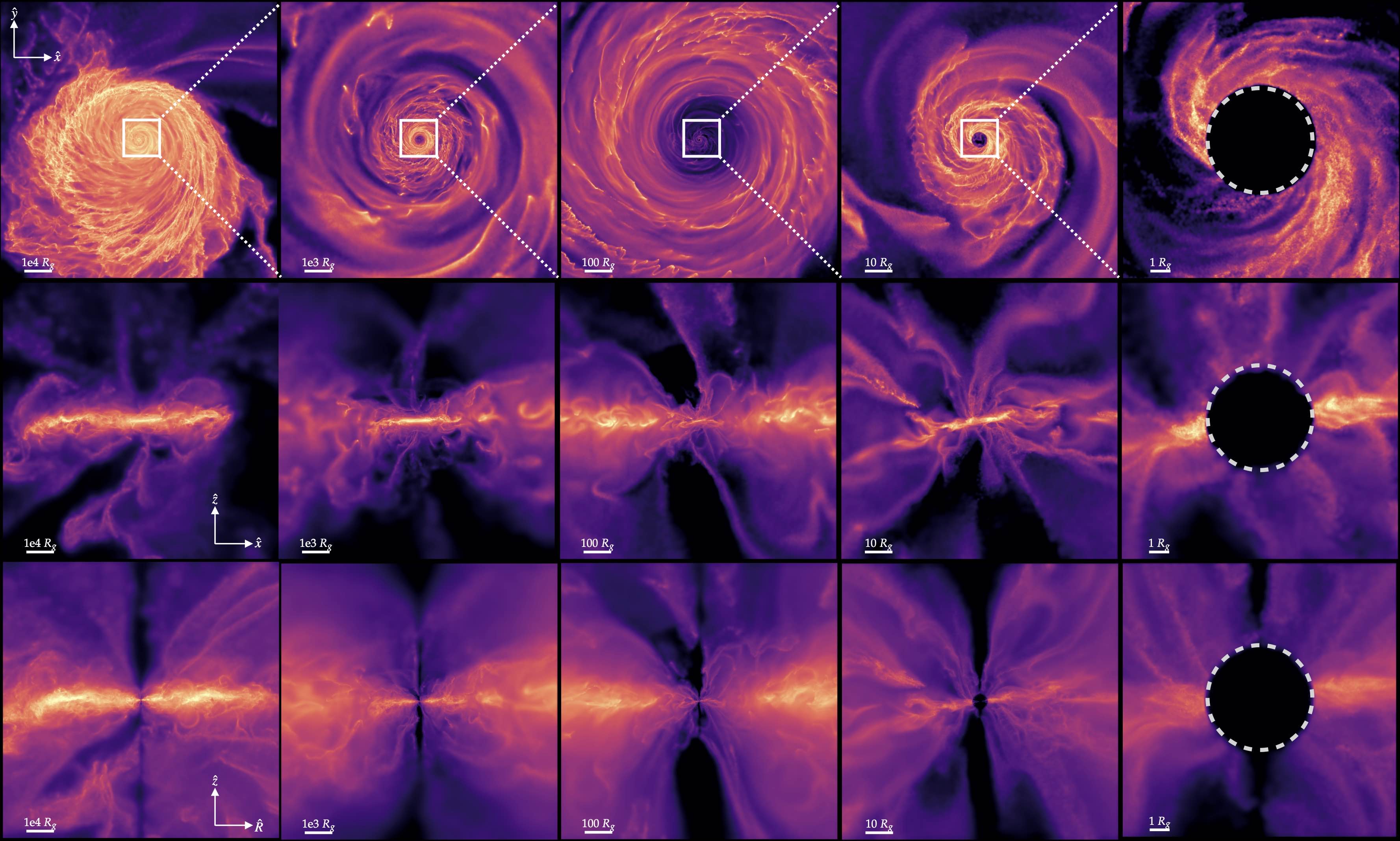} 
	\caption{Surface density projection (increasing black-to-yellow, scaled independently in each panel to show the $1-99\%$ range in each image) of the accretion disk at various scales from $\sim 10^{5}\,R_{g} \sim 0.1\,$pc (the outer circularization radius) to $\sim 3\,R_{g}$ (the ISCO and inner accretion boundary of the simulation; {\em dashed circle}), with scales labeled ({\em left-to-right}). Rows show different projections: face-on slice through the $x$-$y$ plane ({\em top}), edge-on slice through the $x$-$z$ plane ({\em middle}), and edge-on cylindrical $R$-$z$ projection (in a wedge of azimuthal opening angle $\sim 30^{\circ}$). The accretion flow is clearly disky, but with a thick, turbulent structure with material at large $H/R$ at all radii here.
	\label{fig:zoomies.cyl}}
\end{figure*} 

\begin{figure*}
	\centering
	\includegraphics[width=0.98\textwidth]{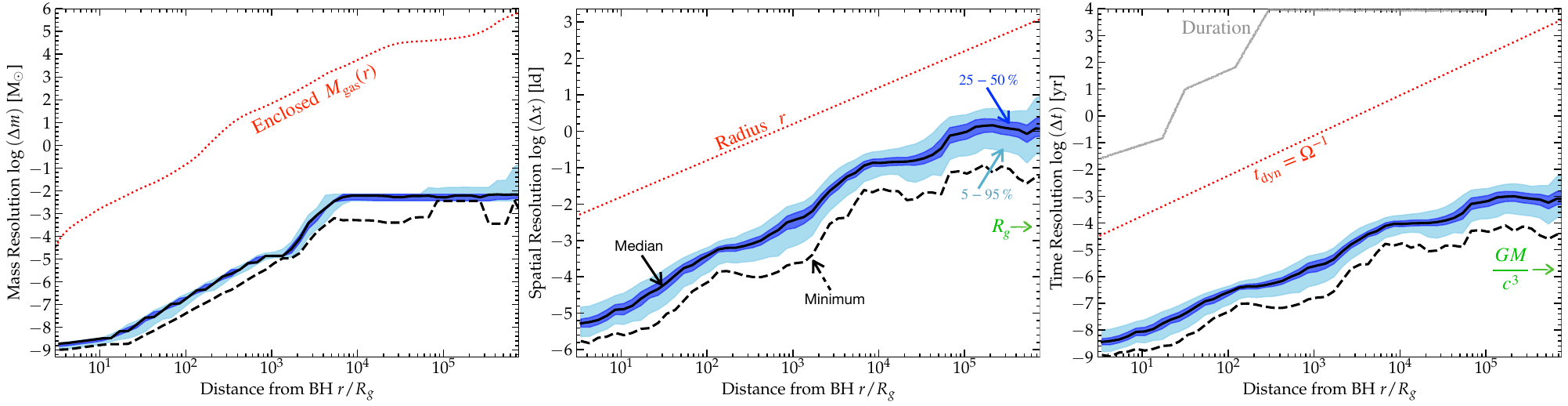} 
	\caption{Resolution of our simulations as a function of distance from the BH $r$ (in units of $R_{g}$), in the accretion disk from the ISCO at $\sim 3\,R_{g}$ to beyond the circularization radius at $\sim 1\,$pc ($\sim 10^{6}\,R_{g}$). In spherical shells, we plot the gas-mass-weighted median ({\em solid}) and minimum ({\em dashed}) cell mass $\Delta m$ ({\em left}), linear size $\Delta x \equiv (\Delta m/\rho)^{1/3}$ ({\em center}), and numerical timestep $\Delta t$ ({\em right}), with their interquartile and $90\%$ range. We compare the total enclosed disk mass $M_{\rm gas}(<r)$, size scale $r$, and dynamical time $t_{\rm dyn} \equiv 1/\Omega \approx \sqrt{r^{3}/G M_{\rm BH}}$, as well as the total duration/run-time of the simulations after reaching the maximum refinement level at each radius.  We also note the value of $R_{g} \equiv 2 G M_{\rm BH}/c^{2}$ and $G M_{\rm BH}/c^{3}$. In the midplane of the disk region we study here, $\Delta x \sim 0.0001-0.001\,R$ (akin to a $1000^{3}-10,000^{3}$ simulation at each log radius), with the smaller value representative of the resolution in dense substructures and the larger representative of the lower-density material in the disk.
	\label{fig:resolution}}
\end{figure*}

\section{Simulations \&\ Methods}
\label{sec:methods}

\subsection{Overview \&\ Basic Physics}
\label{sec:methods:overview}

The simulations here are direct continuations/extensions of those presented in \paperone, to which we refer for details. In brief, we initialized a $\sim (100\,{\rm cMpc})^{3}$ cosmological periodic box at redshift $z\sim 100$, with a trace primordial magnetic field ($\sim 10^{-15}\,$G), and follow it as a cosmological galaxy formation simulation using the full combined Feedback In Realistic Environments (FIRE) and STARFORGE physics \citep{hopkins:fire2.methods,hopkins:fire3.methods,grudic:starforge.methods,guszejnov:2020.starforge.jets}, with initial ``coarse'' adaptive resolution reaching $\sim 10\,$pc in the galactic nucleus. At redshift $z\sim 4.5$ the galaxy is a massive, high-redshift starburst (star formation rate $\gtrsim 200\,{\rm M}_{\odot}\,{\rm yr^{-1}}$) and a massive molecular cloud complex is perturbed to encounter the supermassive black hole of mass $\sim 1.3\times10^{7}\,{\rm M_{\odot}}$. We therefore hyper-refine at this time to follow the evolution in the central $\lesssim 10-100\,$pc of the galaxy. In \paperone\ we refine to a uniform target resolution of $\Delta m \sim 0.005\,{\rm M_{\odot}}$ in the central $\lesssim 5\,$pc (the BH radius of influence, BHROI, where it begins to dominate the total enclosed mass and gravitational potential), following the gas inflow from the tidally disrupted GMC complex. This material circularizes at $\sim 0.1-1\,$pc, forming a stable, non-star-forming, magnetically-dominated (\papertwo) accretion disk around the BH, which we evolved in \paperone-\paperthree\ down to an inner boundary at $80\,{\rm au}\approx 300\,R_{g}$ from the BH (where $R_{g} \equiv 2\,G\,M_{\rm BH}/c^{2}$ is the Schwarzschild radius). Here we extend this refinement; the range of scales is illustrated in Fig.~\ref{fig:zoomies}. 

The simulations include magnetic fields (using the hybrid \citealt{powell:1999.8wave.cleaning,dedner:2002.divb.cleaning.scheme} plus constrained-gradient method from \citealt{hopkins:mhd.gizmo,hopkins:cg.mhd.gizmo}), with anisotropic Braginskii viscosity and conduction \citep{su:2016.weak.mhd.cond.visc.turbdiff.fx,hopkins:gizmo.diffusion}; a variable local cosmic-ray background \citep{hopkins:m1.cr.closure,hopkins:cr.spectra.accurate.integration,hopkins:cr.multibin.mw.comparison,hopkins:2021.sc.et.models.incompatible.obs} here modeled using the simple sub-grid method from \citet{hopkins:2022.cr.subgrid.model}; self-gravity with adaptive self-consistent softenings scaling with the resolution and high-order Hermite integrators capable of accurately integrating $\gtrsim 10^{5}$ orbits in hard binaries \citep{grudic:2020.tidal.timestep.criterion,grudic:starforge.methods,grudic:2021.accelerating.hydro.with.adaptive.force.updates,hopkins:tidal.softening}; metal enrichment and dust destruction/sublimation \citep{ma:2016.disk.structure,gandhi:2022.metallicity.dependent.Ia.rates.statistics.fire,choban:2022.fire.dust.growth.destruction.chemistry}; super-massive black hole seed formation and growth via gravitational capture of gas \citep{hopkins:qso.stellar.fb.together,shi:2022.hyper.eddington.no.bhfb,wellons:2022.smbh.growth}; (proto)star formation and accretion and explicit feedback from stars in the form of protostellar jets, main-sequence stellar mass-loss, radiation, and supernovae \citep{grudic:2022.sf.fullstarforge.imf,guszejnov:2022.starforge.cluster.assembly,guszejnov:environment.feedback.starforge.imf,guszejnov:starforge.environment.multiplicity}. The inner boundary is an inflow/sink boundary around the BH. The simulations evolve explicit multi-group radiation-hydrodynamics using the M1 approximation \citep{levermore:1984.FLD.M1} with adaptive-wavelength bands \citep{hopkins:radiation.methods,hopkins:2019.grudic.photon.momentum.rad.pressure.coupling,grudic:starforge.methods} coupled explicitly to all the thermo-chemical processes, together with radiative cooling and thermo-chemistry incorporating cosmic backgrounds, radiation from local stars, re-radiated cooling radiation, dust, molecular, atomic, metal-line, and ionized species opacities, cosmic rays, and other processes. This allows us to self-consistently model the thermochemistry and opacities in gas with densities from densities $n \sim 10^{-8} - 10^{16}\,{\rm cm^{-3}}$ and temperatures $\sim 1-10^{10}\,$K in a range of radiation and cosmic ray environments (see \paperone). 

Our default RHD treatment includes several different evolved radiation groups: OIR, NUV, EUV, and ionizing (H, He) bands with fixed photon frequency boundaries at $\sim (0.41,\,3.4,\,8,\,13.6,\,24.6,\,100)$\,eV, plus an adaptive self-radiation band which separately evolves the re-emitted continuum radiation from gas and/or dust as a greybody-type spectral shape but with explicitly-evolved energy and photon number density, e.g.\ explicitly evolving its radiation temperature $T_{\rm rad}({\bf x},\,t)$ or effective frequency $\langle \nu_{\rm rad} \rangle({\bf x},\,t)$ independently in each cell, rather than assuming local thermodynamic equilibrium (LTE) or strictly blackbody/thermal radiation. The latter can contribute radiation to all of the fixed-wavelength narrow bands, if $T_{\rm rad}$ is hot enough (as it is here), and all the relevant opacities and thermochemical couplings are appropriately Rosseland or Planck integrated for the evolved $T_{\rm rad}({\bf x},\,t)$ (shape/frequency of the band, as well as gas properties like $T_{\rm gas}$, density, metallicity, ionization state, etc.; see Appendix~\ref{sec:opacities}). We also specifically evolve the coupled anisotropic photon-gas momentum equations with terms up to and including $\mathcal{O}(v_{\rm gas}^{2}/c^{2})$ \citep{mihalas:1984oup..book.....M}, rather than assuming the common ``trapped photon'' limit (which treats the photons as an isotropic ``pressure'' in the Riemann problem). Details of the opacities and radiation-thermochemical coupling relevant on these scales are given in Appendix~\ref{sec:opacities}. The narrow fixed bands are designed primarily for situations like ISM chemistry where photo-electric heating, Lyman-Werner radiation, H ionization, and other processes are primarily sensitive to photons in a narrow frequency range and dominated by light emitted by stars, not the cooling radiation of gas. So the latter (adaptive) band contains most of the radiation energy density at the radii of interest here, with $T_{\rm rad}({\bf x},\,t)$ as we show ranging from $\lesssim 30\,$K to $\gtrsim 10^{8}\,$K. This allows us to flexibly handle continuum opacities over an extremely broad range of conditions without making assumptions about a single grey opacity or even source of opacities, whether the system is in LTE or non-LTE, optically thick or thin, etc. However, we caution that our treatment does not include resonant line scattering, so we cannot self-consistently capture line-driven winds.
 
In this paper we extend these simulations by further refining in the inner regions, shrinking our inner sink boundary in a couple of stages, first from $\sim 300\rightarrow30\,R_{g}$ (with much higher gas resolution in the inner regions), then further to $\rightarrow 3\,R_{g}$ (with a Paczy{\'{n}}ski-Wiita [PW] BH potential). Because we are simulating even smaller radii, we also remove the ``reduced speed of light'' approximation used in \paperone\ for the radiation-hydrodynamics. There, a reduced $\tilde{c} \approx 0.1\,c$ was used; here given the large speeds reached in the center we simply adopt $\tilde{c}=c$, i.e.\ no reduction.\footnote{Note this has no immediate effect on evolved conserved quantities, such as radiation energy densities, only on the dynamical equations used to evolve the radiation.} We retain all of the other physics above. Fig.~\ref{fig:zoomies.cyl} illustrates this higher-resolution region.

Appendix~\ref{sec:methods:variants} describes a number of experiments with numerical and physical parameters which we ran for much shorter times, including different refinement schemes, resolution variations, varying the inner/outer boundary conditions, re-initializing some quantities in the initial conditions, removing star formation and stellar evolution physics, switching between PW and Keplerian potentials, and experimenting with variations to the detailed cooling-radiation-thermochemical coupling and approximations. These generally have no impact on our qualitative conclusions, except where we otherwise specify.

\subsection{Initial Conditions and Resolution}
\label{sec:methods:ics}

Specifically, for our fiducial simulation here, we restart from the final snapshot of the \paperone-\papertwo\ simulations described above (which had been evolved for $\sim 10^{5}$ dynamical times at the inner radius of $\sim 300\,R_{g}$), and refine further. We initially reduce the size of the inner accretion boundary to $30\,R_{g}$, and we turn off the speed of light reduction throughout. We apply an additional refinement layer, enforcing a target mass resolution of $\Delta m \propto (r/0.01\,{\rm pc})^{4}$ in a boundary layer between $\sim 0.001-0.02\,$pc (exterior to which the rest of the accretion disk has target resolution $\Delta m \approx 0.005\,{\rm M_{\odot}}$) to a minimum and approximately uniform mass resolution of $\Delta m \approx 3\times 10^{-7}\,{\rm M}_{\odot}$ interior to $<0.001$\,pc ($\sim 200\,$au or $\sim 800\,R_{g}$). This initial uniform-resolution high-resolution region contains approximately $\approx 10^{7}$ roughly equal-mass cells, which then increases to our target resolution after the simulation turns on (see Fig.~\ref{fig:resolution}). The ``outer boundary'' of our highest-resolution region of this simulation is therefore at $\sim 1000\,R_{g}$, and the outer boundary of the intermediate resolution region (the high-resolution region from \paperone) is at $\sim 4\times 10^{6}\,R_{g}$ ($\sim 5\,$pc), but our box extends to $\sim 100\,$Mpc ($10^{14}\,R_{g}$). After $\sim 10\,$yr physical time of evolution ($\sim 10^{4}\,\Omega^{-1}$ at this inner boundary of $\sim 30\,R_{g}$) we iterate, moving the inner boundary to $\sim 3\,R_{g}$ and applying one further refinement layer with the target resolution decreasing as $\propto r^{2}$ interior to $<0.001\,$pc to a uniform $\Delta m \sim 10^{-9}\,{\rm M}_{\odot}$ in the inner disk.\footnote{We also enforce a secondary, but less important in practice, spatial refinement criterion requiring the cell size be $<0.03\,r$ at all $r$ in the high-resolution region.} We evolve this highest-resolution setup for a time $\sim 10^{5}\,G\,M_{\rm BH}/c^{3}$, i.e.\ $\sim 0.2\,$yr (1800 hours) physical time or $\sim 7000\,\Omega^{-1}$ at the inner boundary ($3\,R_{g}$) and $\sim 200\,\Omega^{-1}$ at the previous inner boundary ($\sim 30\,R_{g}$). For the BH specifically, we replace its Keplerian acceleration in the gravity tree with that of a Paczy{\'{n}}ski-Wiita potential ($G\,M_{\rm BH}/(r-R_{g})$), though for the run with innermost radius of $\sim 30\,R_{g}$, this is never more than a few percent-level correction and our tests below suggest it is largely negligible, but upon going to $\sim 3\,R_{g}$ it will be more important. For the self-gravity of the gas (which must still obey the Poisson equation) we retain Newtonian gravity. 

The vast majority of the material at the large radii in the simulations will do nothing (at resolved scales) over the duration of our small-scale simulation here. So in this simulation, for numerical convenience, we effectively ``freeze'' all the matter at radii $r > 1$\,pc ($\sim 10^{6}\,R_{g}$), for memory/storage purposes saving it in a separate snapshot and simply adding the potential from this material analytically to the ``live'' material at small radii. The dynamical time $t_{\rm dyn} = \Omega^{-1}$ (with $\Omega\approx \sqrt{G\,M_{\rm BH}/r^{3}}$) at our innermost boundary is $\approx 3\times10^{4}\,{\rm s} \sim 8\,$hours at $30\,R_{g}$ and $\approx 1000\,{\rm s}$ at $3\,R_{g}$, but at this ``frozen'' boundary it is $\approx 1.3\times 10^{11}\,{\rm s}$ ($\sim 4000\,$yr), so this is a reasonable approximation. It is more useful to think of the material at $\sim 0.01-1\,$pc in our simulation (where the total time we evolve is $\lesssim 1$ orbital time) as forming our outer boundary condition.

Note that for a standard $\alpha$-disk (SS73) with the dissipated energy scaling as $\propto \dot{M}\,\Omega^{2}\,(1-\sqrt{3\,R_{g}/R})$ around a non-rotating BH, we would expect $\sim 25\%$ ($\sim 100\%$) of the total integrated bolometric luminosity to be released at $\gtrsim 30\,R_{g}$ ($\gtrsim 3\,R_{g}$). Appendix~\ref{sec:methods:variants} describes a variety of experiments detailed below, where we modify the refinement scheme or physics, and run for much shorter times, and tests where we re-run from a different ``initial'' snapshots (to see if a different inner disk begins to form). These do not change our conclusions except where we explicitly note below.

\subsection{Different Accretion Rates}
\label{sec:methods:mdot}

Note that for this BH mass, we define the critical/Eddington accretion rate for a reference radiative efficiency $=0.1$, $\dot{M}_{\rm crit} \equiv L_{\rm Edd}/0.1\,c^{2} \sim 0.3\,{\rm M_{\odot}\,yr^{-1}}$, so $\dot{m} \equiv \dot{M}/\dot{M}_{\rm crit}$. 

The ``default'' initial condition in \S~\ref{sec:methods:ics} begins from the supercritical accretion rate at the outer boundaries predicted from our original galaxy-scale simulation. Here we wish to explore the effects of different accretion rates, in a controlled manner. One possibility is to repeat the zoom-in exercise from \paperone\ at different cosmic times or in different galaxies, beginning again from large-volume cosmological simulations \citep[see][]{daa:20.hyperrefinement.bh.growth}. While such experiments are valuable and will be the subject of future work, they are extremely computationally expensive, and do not provide a controlled comparison, since everything else about the system varies as well. To instead enable a more straightforward comparison here, we adopt a simpler approach of rescaling the initial conditions in \S~\ref{sec:methods:mdot}. Specifically, since we are focused only on radii well interior to the BHROI, we can rescale the expected accretion rate by a factor $f_{\dot{M}}$, taking our initial condition (fiducial snapshot from \papertwo) and multiplying all gas cell masses by $f_{\dot{M}}$ (keeping BH and other masses fixed), $\Delta m \rightarrow f \Delta m$. We rescale our ``target'' mass resolution in \S~\ref{sec:methods:ics} by the same factor. Following the analytic similarity models in \paperthree, we rescale the initial magnetic fields ${\bf B} \rightarrow f^{1/2}\,{\bf B}$ and temperatures $T \rightarrow f\,T$ such that the \Alf\ and sonic speeds $v_{A}/v_{\rm K}$, $c_{s}/v_{\rm K}$ remain constant. In practice we find the temperature rescaling is unnecessary, since the disks have cooling times much shorter than dynamical times they almost immediately come into their new temperature equilibria. However, some ${\bf B}$ rescaling is necessary, since the default initial conditions already feature large $v_{A}/v_{\rm K} \sim H/R \sim 0.1-1$, so rescaling the gas mass by $f_{\dot{M}} \ll 1$ without rescaling ${\bf B}$ would produce initial conditions with the gas everywhere unbound in the outer disk. 

We emphasize that this rescaling is applied only once to initial conditions (snapshot of the outer disk with inner truncation radius $\sim 300\,R_{g}$ from \papertwo). We then repeat our refinement and simulation identically as in \S~\ref{sec:methods:ics}, just with the target mass resolution shifted by the same factor so the spatial resolution, or number of cells in the disk, is roughly equal across all simulations. So this is akin to standard accretion disk studies using different initial conditions to represent different $\dot{m}$. Below and in Appendix~\ref{sec:methods:variants} we describe a number of detailed variants to how we initialize these disks in detail, which do not change our results.

In addition to our default initial condition ($f_{\dot{M}}=1$), we simulate two values of $f_{\dot{M}}$: $f_{\dot{M}}=0.03$ and $=0.001$, which non-linearly give rise to accretion rates $\dot{m} \sim 0.3-0.4$ and $\sim 0.02-0.04$ ($\dot{M} \sim 0.1$ and $\sim 0.007$\,${\rm M_{\odot}\,yr^{-1}}$), respectively.

\begin{figure*}
	\centering
	\includegraphics[width=0.9\textwidth]{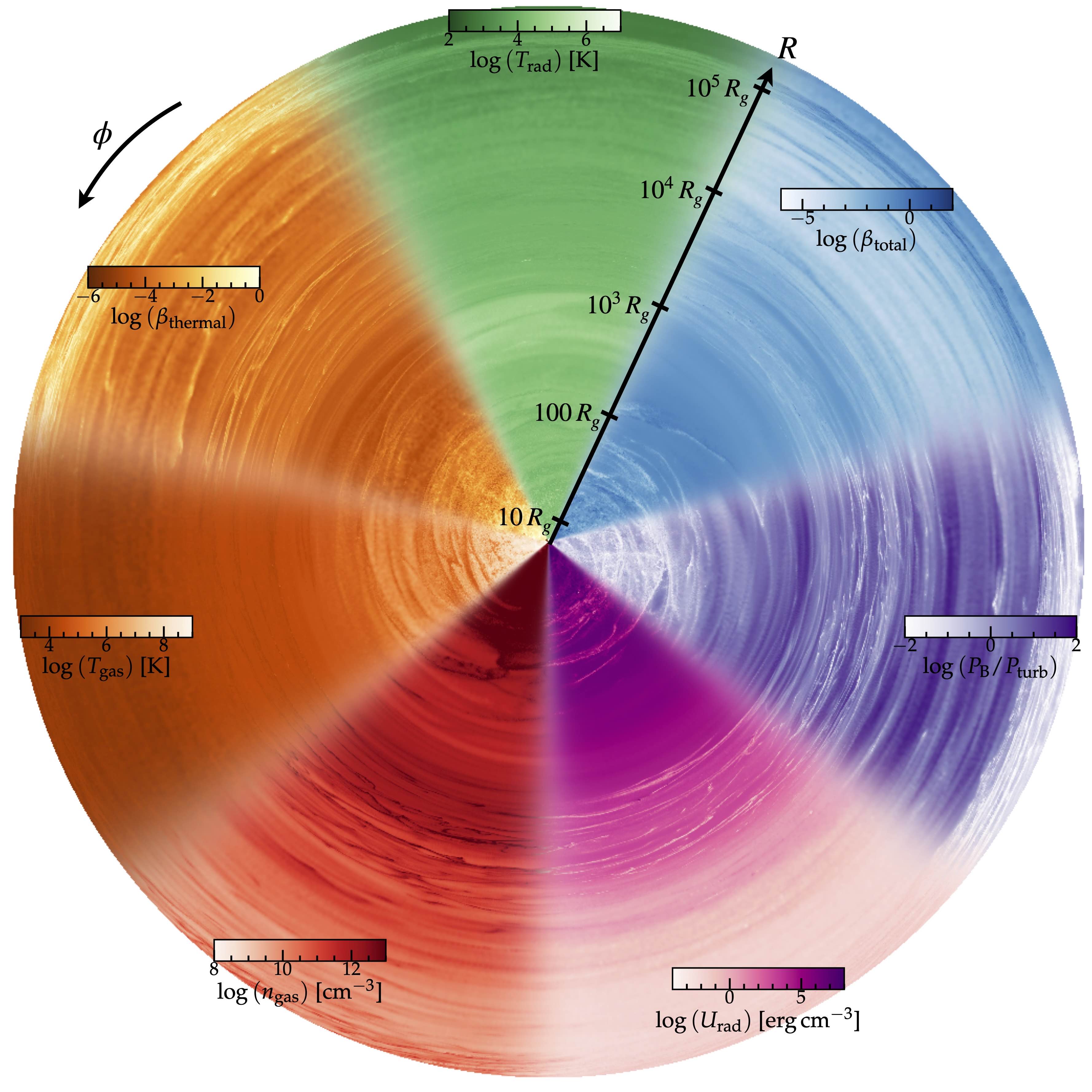} 
	\caption{Face-on slice through the midplane showing various quantities in the disk in different colormaps. The radial coordinate $R$ has a logarithmic stretch as labeled, from $\sim 3\,R_{g}$ to $\sim 1\,$pc. Different colormaps show: 
	{\bf (1)} thermal $\beta_{\rm thermal} \equiv P_{\rm therm}/P_{\rm mag}$ (ratio of thermal-to-magnetic pressure), 
	{\bf (2)} radiation temperature $T_{\rm rad}$, 
	{\bf (3)} total $\beta_{\rm tot} \equiv (P_{\rm therm}+P_{\rm rad})/P_{\rm mag}$ pressure ratio (assuming strong radiation coupling), 
	{\bf (4)} magnetic-to-turbulent pressure ratio $P_{\rm B}/P_{\rm turb}$ ($P_{\rm turb} \equiv (1/2)\rho\delta {\bf v}_{\rm turb}^{2}$), 
	{\bf (5)} radiation energy density $U_{\rm rad}$, 
	{\bf (6)} gas density $n_{\rm gas} \equiv \rho / m_{p}$, 
	{\bf (7)} gas kinetic temperature $T_{\rm gas}$. 
	As implied by Fig.~\ref{fig:zoomies}, the disks generally have $\beta_{\rm thermal} \ll 1$, with $\beta_{\rm total} \ll 1$ in the outer disk but more modest $\beta_{\rm total} \sim 0.1$ in the innermost disk, and power-law-like gradients in the mean $T_{\rm gas}$, $n_{\rm gas}$, $U_{\rm rad}$. Multi-phase structure is visible in most maps, with trans-\Alf{ic} turbulence (local $P_{\rm B}/P_{\rm turb}$ fluctuating below/above unity at all radii). 
	\label{fig:maps.pinwheel}}
\end{figure*}

\begin{figure*}
	\centering
	\includegraphics[width=0.32\textwidth]{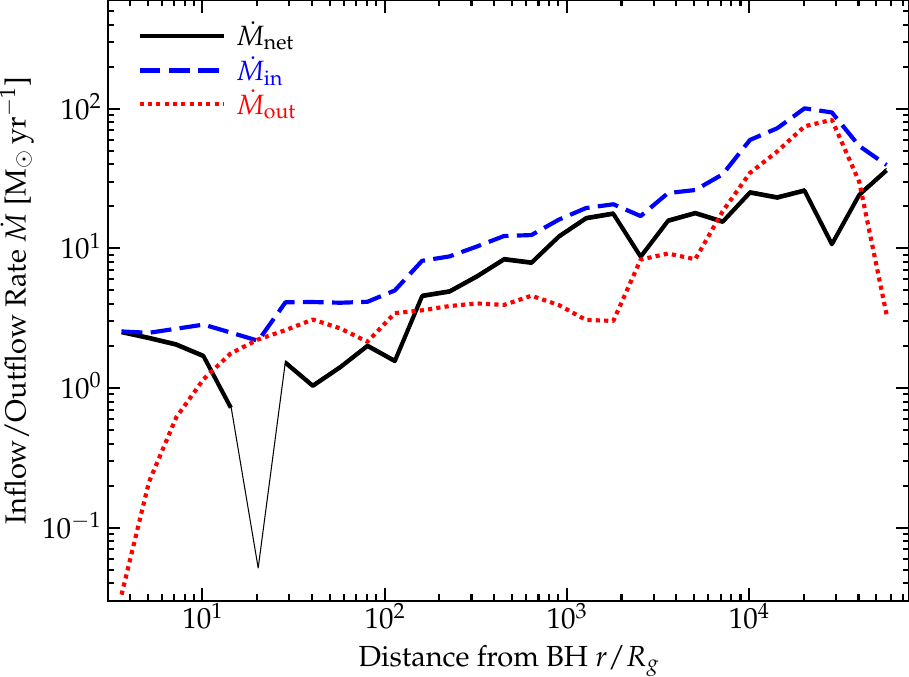} 
	\includegraphics[width=0.32\textwidth]{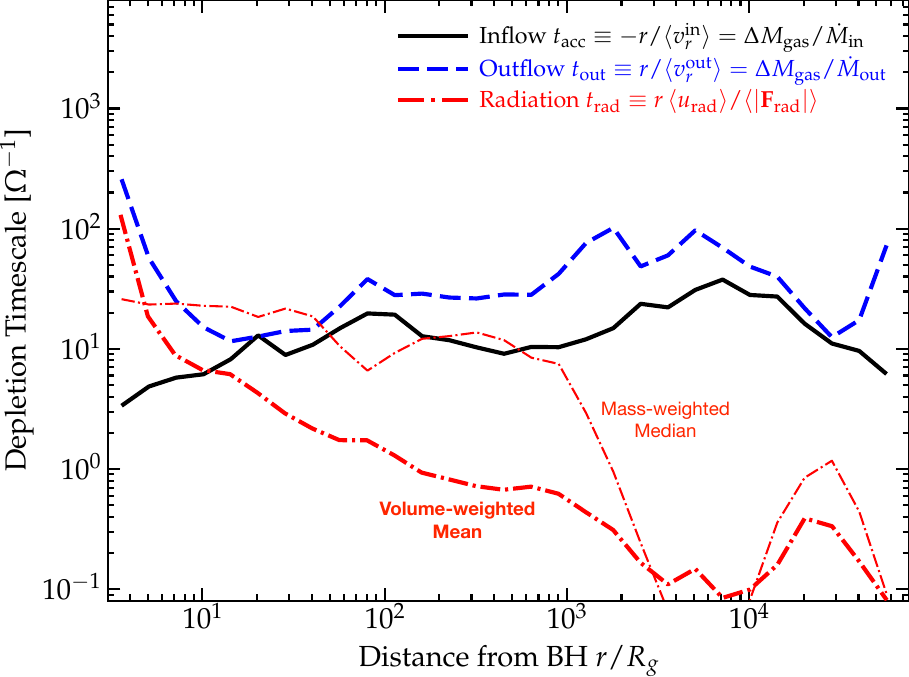} 
	\includegraphics[width=0.32\textwidth]{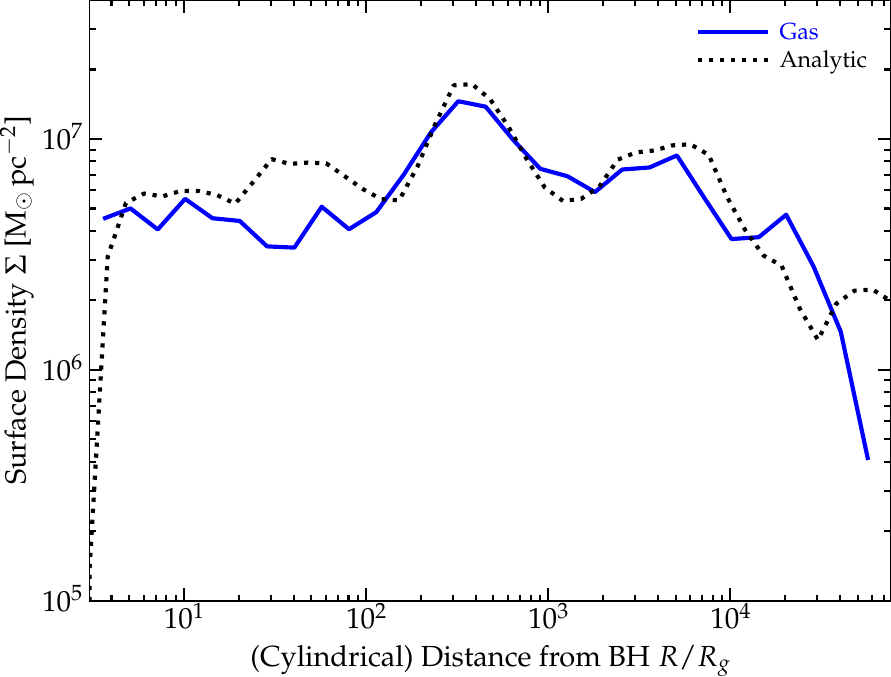} 
	\includegraphics[width=0.32\textwidth]{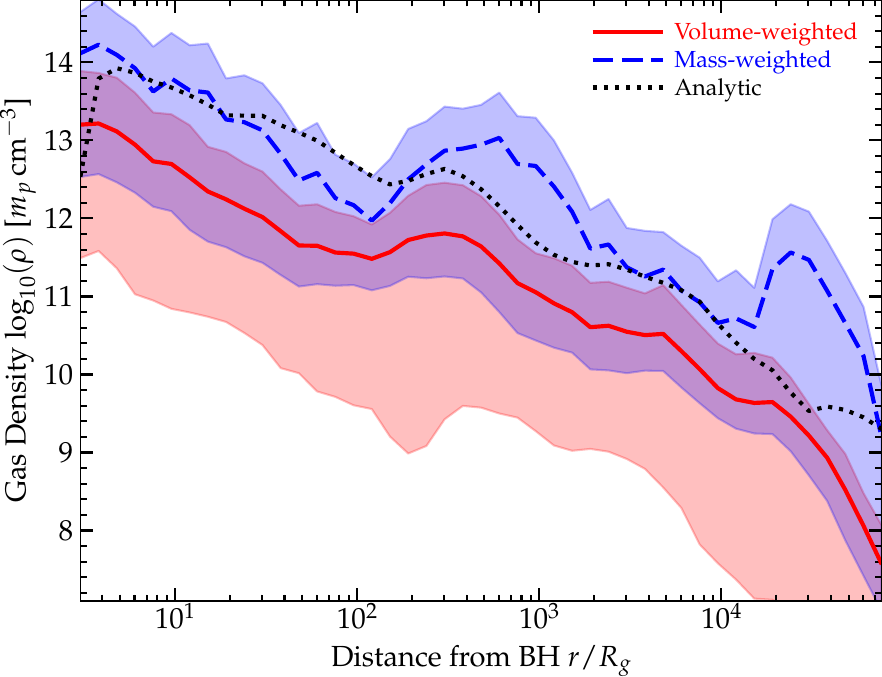} 
	\includegraphics[width=0.32\textwidth]{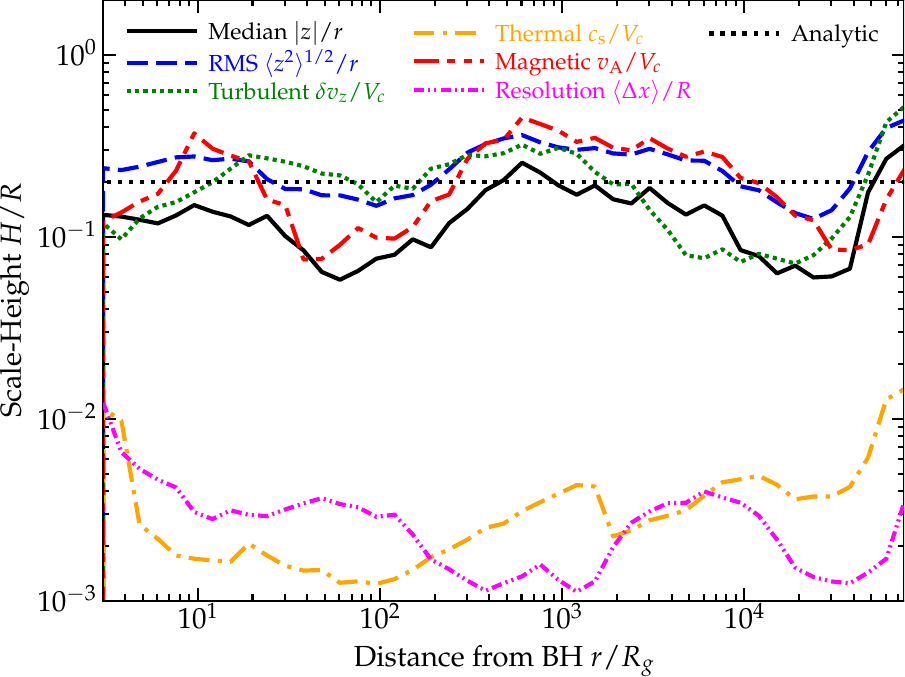} 
	\includegraphics[width=0.32\textwidth]{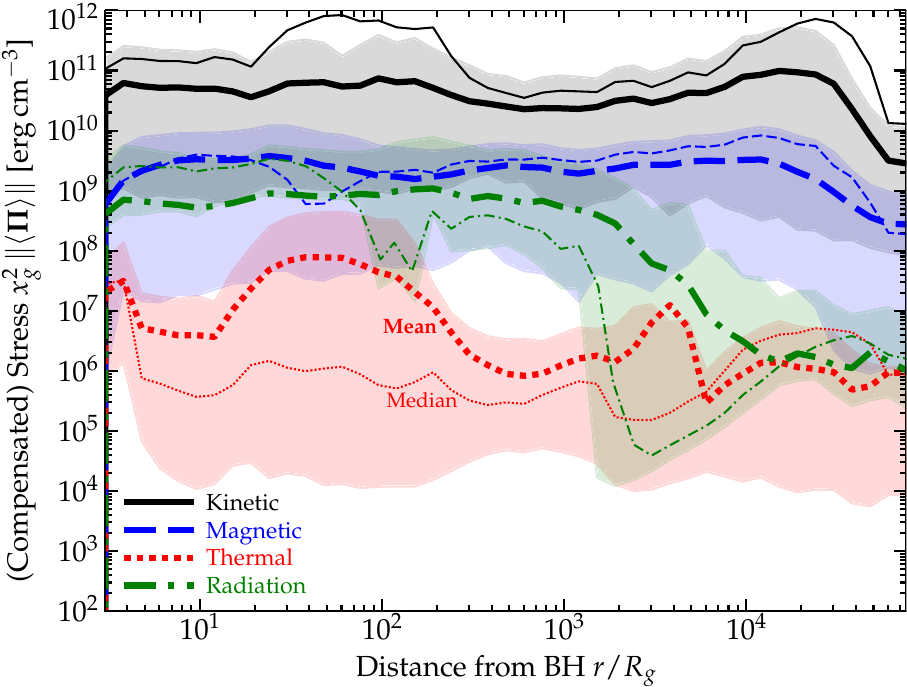} 
	\includegraphics[width=0.32\textwidth]{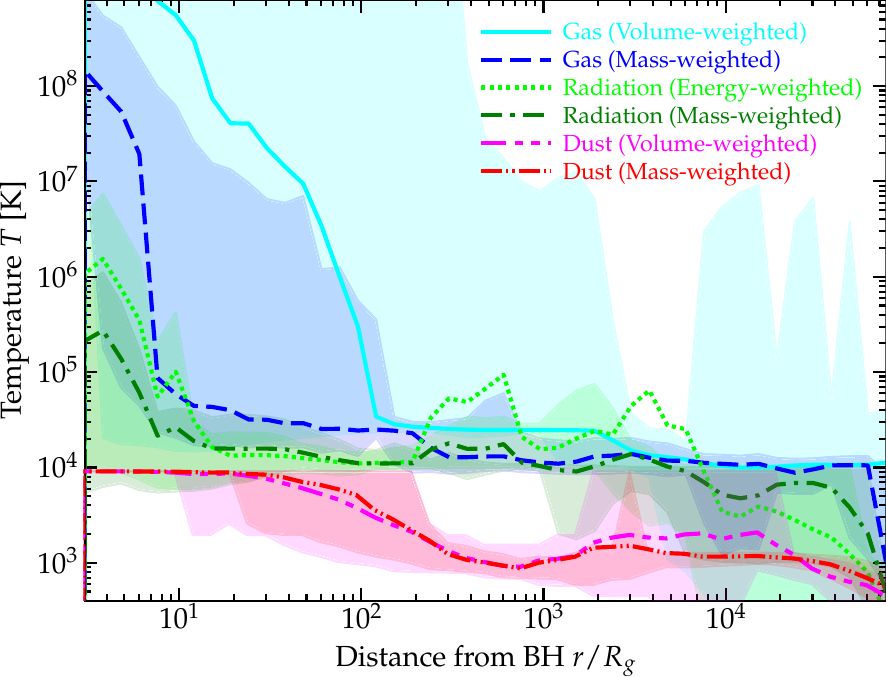} 
	\includegraphics[width=0.32\textwidth]{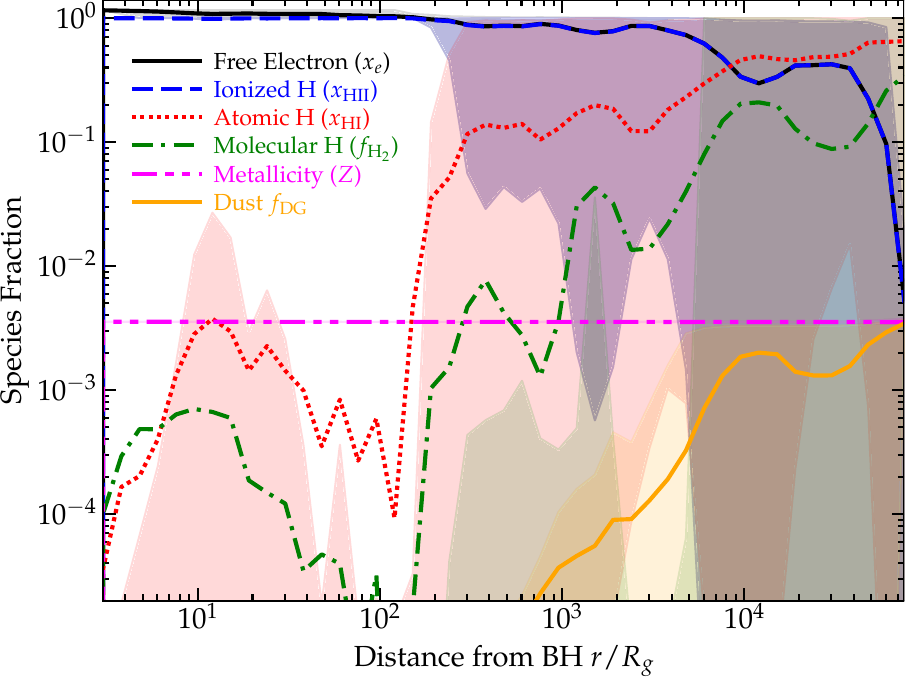} 
	\includegraphics[width=0.32\textwidth]{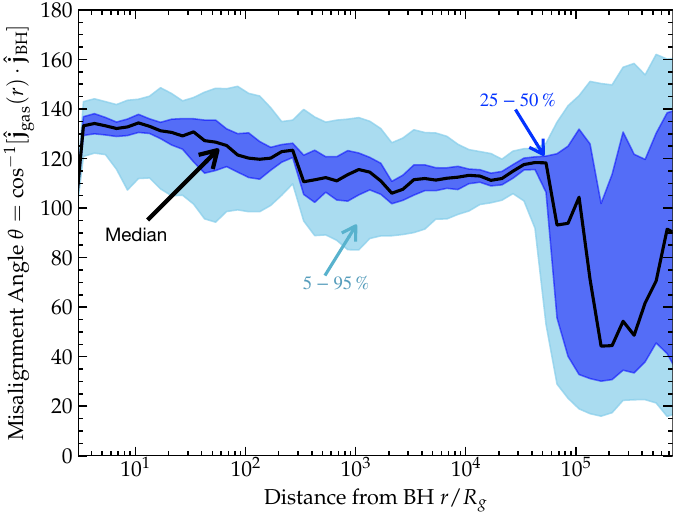} 
	\caption{Radial profiles of quantities with distance from the BH (versus cylindrical $R$ or spherical $r$), at the final time in the simulation.
	{\bf (1)} Inflow/outflow rate (summing all elements with $v_{r}<0$ or $>0$) and net inflow rate $\dot{M}_{\rm net}$ through each annulus. Accretion remains highly supercritical with $\dot{M} \sim 3-5\,{\rm M_{\odot}\,yr^{-1}} \sim 10-20\,L_{\rm Edd}/0.1\,c^{2}$ down to $\sim 3\,R_{g}$.
	{\bf (2)} Accretion/outflow (depletion) timescales $\sim \dot{M}/M_{\rm disk}$, in units of $t_{\rm dyn} \equiv \Omega^{-1} = \sqrt{r^{3}/G M_{\rm BH}}$. We also show the radiation escape timescale from the gas $t_{\rm rad} \sim r \langle u_{\rm rad}\rangle/\langle |{\bf F}_{\rm rad}|\rangle$, with both the volume-weighted mean (dominated by polar escape in low-density channels) and mass-weighted median (approximate escape times from the disk midplane). Accretion is dynamical ($t_{\rm acc} \sim 1-2\,t_{\rm orbit}$), and radiation transport advective in the inner disk.
	{\bf (3)} Mean gas surface density $\Sigma_{\rm gas} \equiv {\rm d} M_{\rm gas}/(2\pi\,R {\rm d}R)$ in cylindrical annuli, projected face-on. We compare the prediction from the flux-frozen disk similarity model in \paperthree\ (\S~\ref{sec:modified.radial}), giving the measured $\dot{M}$. The agreement suggests the disk similarity is valid.
	{\bf (4)} Gas 3D densities $n\equiv \rho/m_{p}$, volume-weighted and mass-weighted median (lines) and $90\%$ inclusion interval (shaded) shown. We again compare the flux-frozen disk analytic model. Structure arises from spiral arms (Fig.~\ref{fig:zoomies.cyl}), with highly supersonic turbulence generating a broad range of $\rho$.
	{\bf (5)} Disk scale-height $H/R$ (mass-weighted median $|z|$ or rms $\langle z^{2}\rangle^{1/2}$), and predicted mass-weighted scale-height from turbulence (vertical $\delta v_{z}/V_{\rm c}$), magnetic ($v_{A,\,\phi}/V_{\rm c}$), and gas thermal support ($c_{s}/V_{\rm c}$). We also compare the spatial resolution (Fig.~\ref{fig:resolution}), and analytic prediction $H/R \approx v_{{\rm turb}}/V_{\rm c} \approx v_{A}/V_{z} \approx \psi \approx 0.2$. The disk is supported by magnetic+turbulent pressure and the scale-heights are well-resolved.
	{\bf (6)} Kinetic ($\langle \rho |{\bf v}|^{2}/2 \rangle$), magnetic ($\langle |{\bf B}|^{2}/4\pi \rangle$), thermal ($\langle \rho u_{\rm th} \rangle \approx \langle3 n k_{\rm B} T_{\rm gas}/2\rangle$), and radiation ($\langle U_{\rm rad}\rangle$) energy densities ($90\%$ range shaded, volume-weighted mean and mass-weighted median in {\em thick} and {\em thin} lines respectively). Radiation pressure becomes more prominent at small $R$, but saturates at $\sim 10\%$ of the magnetic pressure (comparable to the turbulent pressure; \S~\ref{sec:rad.pressure.saturation}).
	{\bf (7)} Gas $T_{\rm gas}$, radiation $T_{\rm rad}$, and dust $T_{\rm dust}$ temperatures, volume and mass-weighted (median and $90\%$ range). Reprocessing maintains a relatively cool radiation temperature.	
	{\bf (8)} Abundances of free electrons ($x_{e}$), ionized (HII), atomic (HI), and molecular (H$_{2}$) gas, and metal ($Z$) and dust ($f_{\rm DG}$) mass-fraction. Dust is mostly sublimated at $T_{\rm dust} \gtrsim 1400\,$K (note $T_{\rm dust}$ is plotted even when dust is sublimated, but has no meaning/impact), but the disk remains highly multi-phase until $\lesssim 100\,R_{g}$.
	{\bf (9)} Angular momentum vector and distribution (relative to $\hat{\bf j}_{\rm BH}$, the pre-existing estimated spin direction of the BH) of the gas within each annulus. Outside the circularization radius, we see a very broad distribution and different mean, as expected, while interior there are still strong warps at the $\sim 10-40^{\circ}$ level. Lens-Thirring precession could strongly modify this in GR.
	\label{fig:profiles.general}}
\end{figure*}

\begin{figure*}
	\centering
	\includegraphics[width=0.32\textwidth]{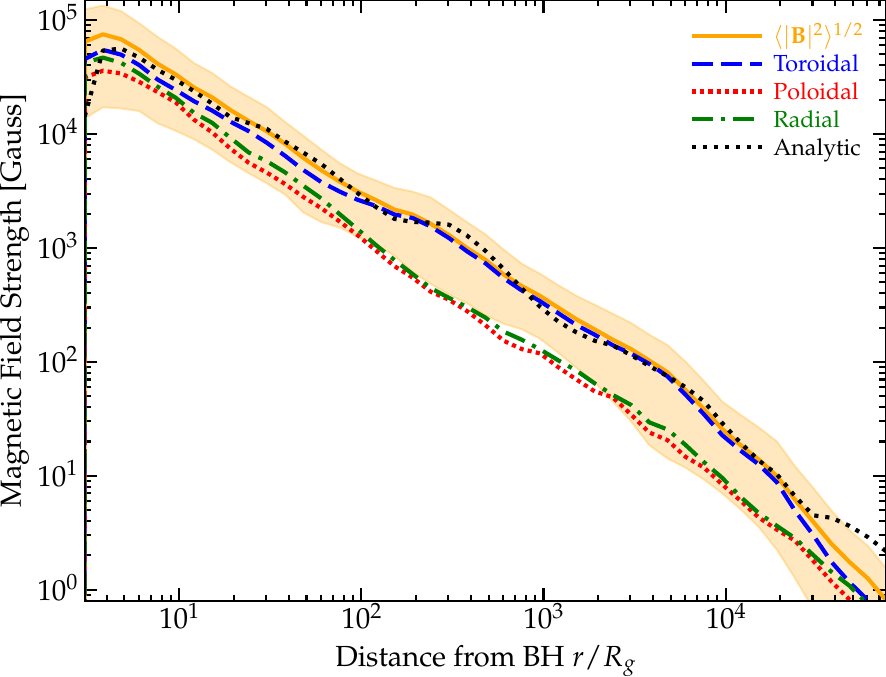} 
	\includegraphics[width=0.32\textwidth]{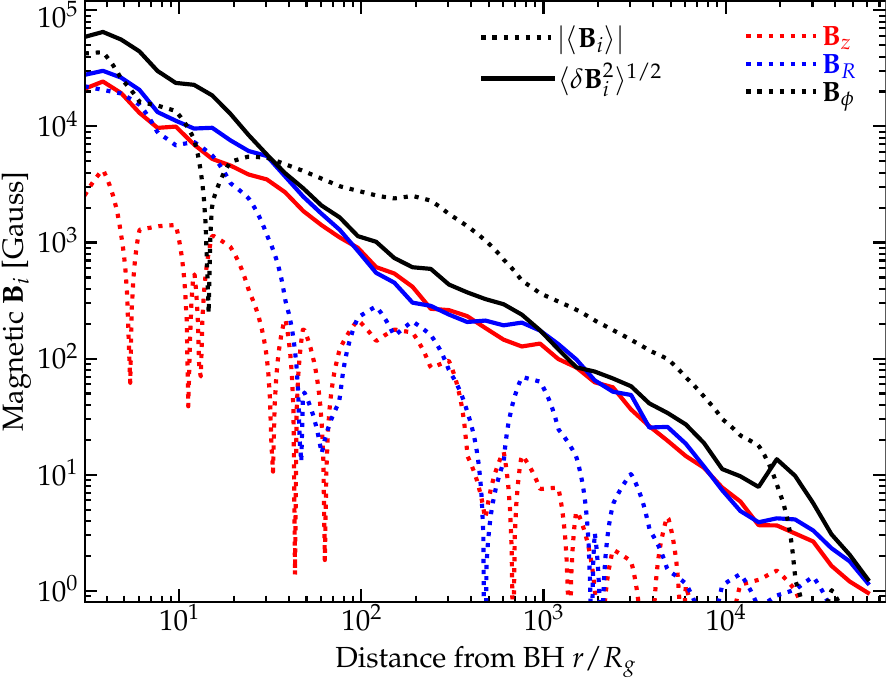} 
	\includegraphics[width=0.32\textwidth]{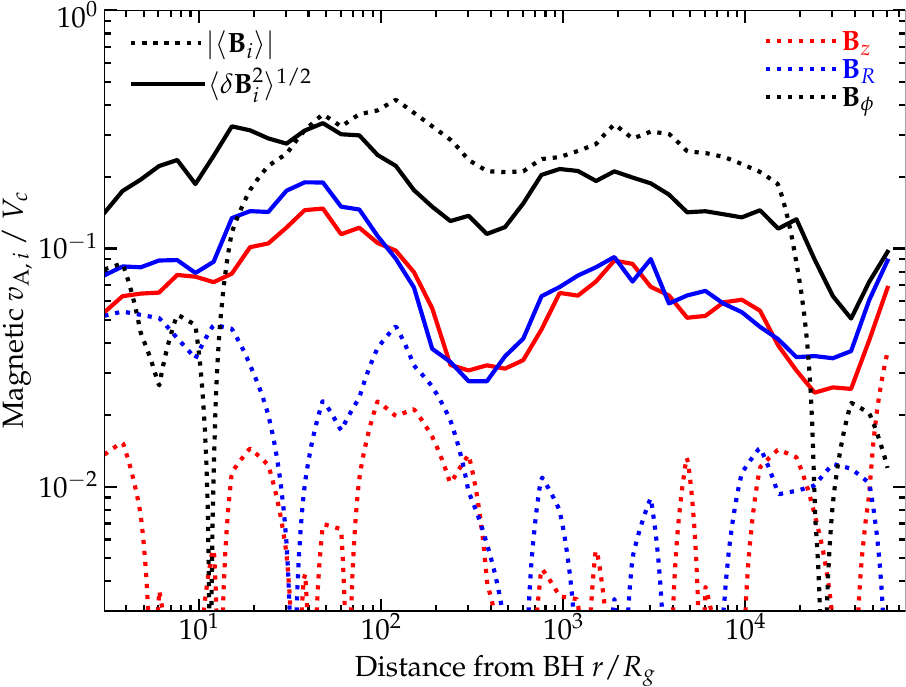} 
	\includegraphics[width=0.32\textwidth]{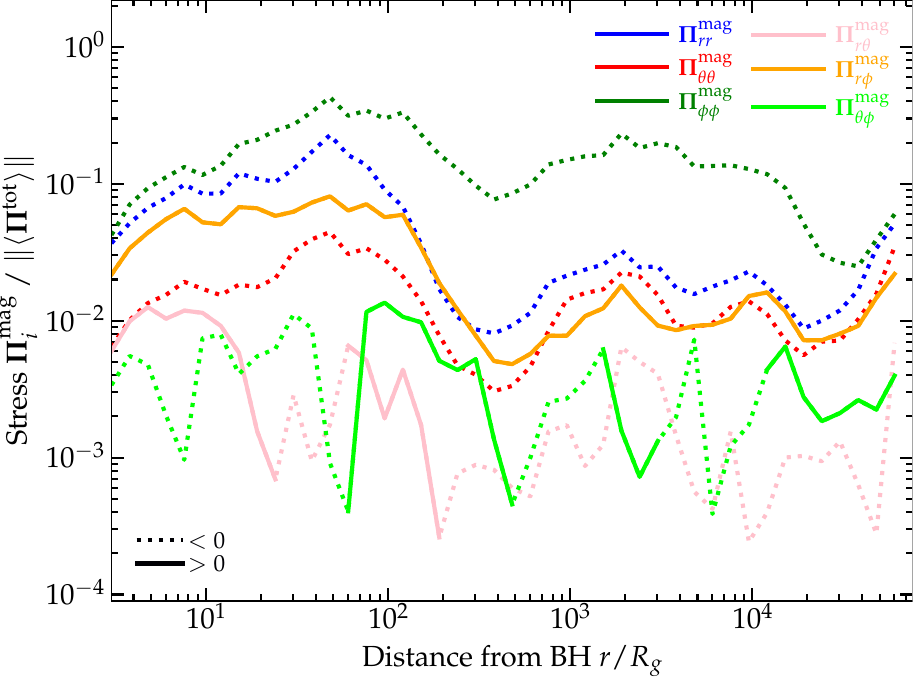} 
	\includegraphics[width=0.32\textwidth]{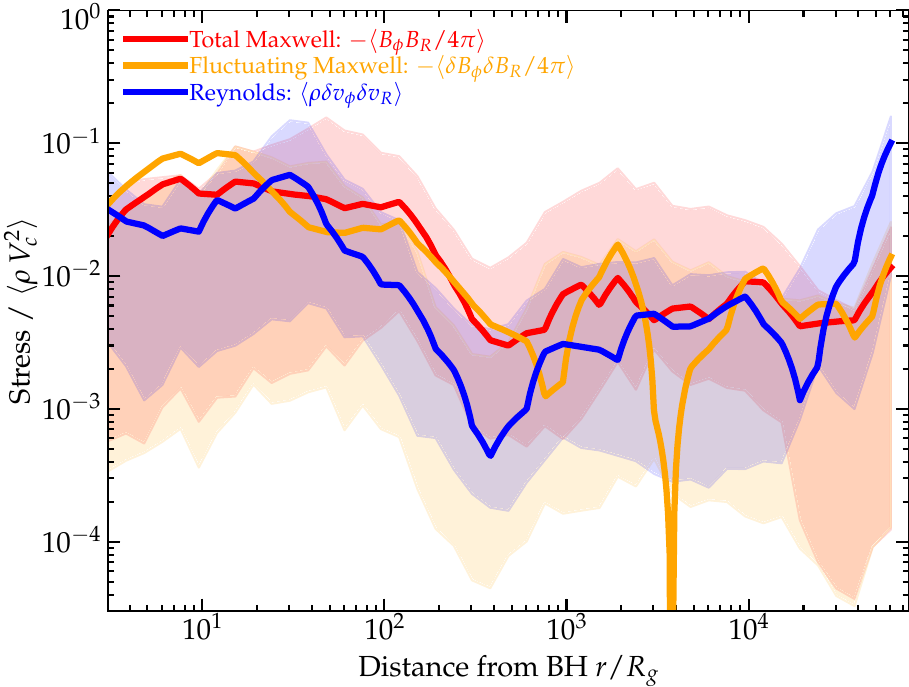} 
	\caption{Volume-weighted magnetic profiles, as Fig.~\ref{fig:profiles.general}. 
	{\bf (1)} Magnetic field strengths: volume-weighted rms $\langle B_{i}^{2} \rangle^{1/2}$ toroidal/poloidal/radial ($\langle B_{\phi,\,z,\,R} \rangle^{2} \rangle^{1/2}$) components and total $|{\bf B}|$, with $90\%$ range for $|{\bf B}|$. We compare the analytic prediction for $|{\bf B}|$ from the same similarity models as Fig.~\ref{fig:profiles.general}, which agree well again.
	{\bf (2)} Mean $|\langle {\bf B}_{i} \rangle|$ and standard deviation $\langle \delta {\bf B}_{i}^{2}\rangle^{1/2}$ of the toroidal $B_{\phi}$, poloidal $B_{z}$, and radial $B_{z}$ magnetic fields. The mean toroidal field dominates except in the innermost radii where it is more comparable to fluctuating fields.
	{\bf (3)} Mean and dispersion of the \Alf\ speeds ${\bf v}_{A,\,i} \equiv {\bf B}_{i}/\sqrt{4\pi\rho}$, relative to $V_{c}$. 
	{\bf (4)} Different components of the Maxwell stress tensor $\boldsymbol{\Pi}^{\rm mag} \equiv -{\bf B}{\bf B}/4\pi + \boldsymbol{I}|{\bf B}|^{2}/8\pi$, with positive ({\em solid}) and negative ({\em dotted}) values indicated, normalized to the norm of the total stress tensor from all sources. The toroidal field dominates the magnetic pressure but contributions from fluctuating fields are non-negligible. Angular momentum transport is dominated by the $r\phi$ component, with magnitude comparable to these.
	{\bf (5)} Mean and $90\%$ range of the standard total and fluctuating Maxwell stress and Reynolds stress (all comparable), consistent with the accretion rate through each annulus (Fig.~\ref{fig:profiles.general}).
	\label{fig:profiles.magnetic}}
\end{figure*} 

\begin{figure*}
	\centering
	\includegraphics[width=0.32\textwidth]{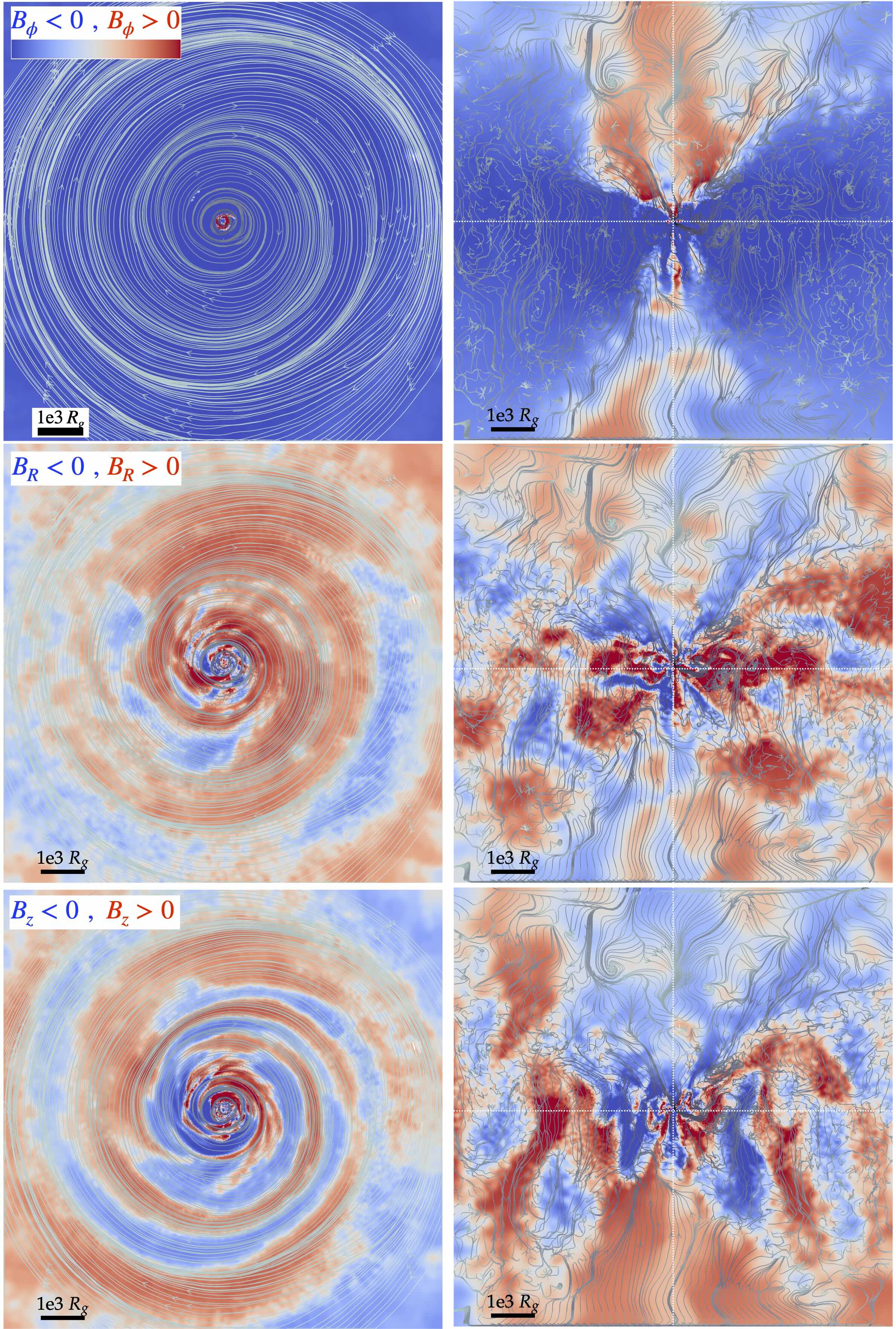}
	\includegraphics[width=0.32\textwidth]{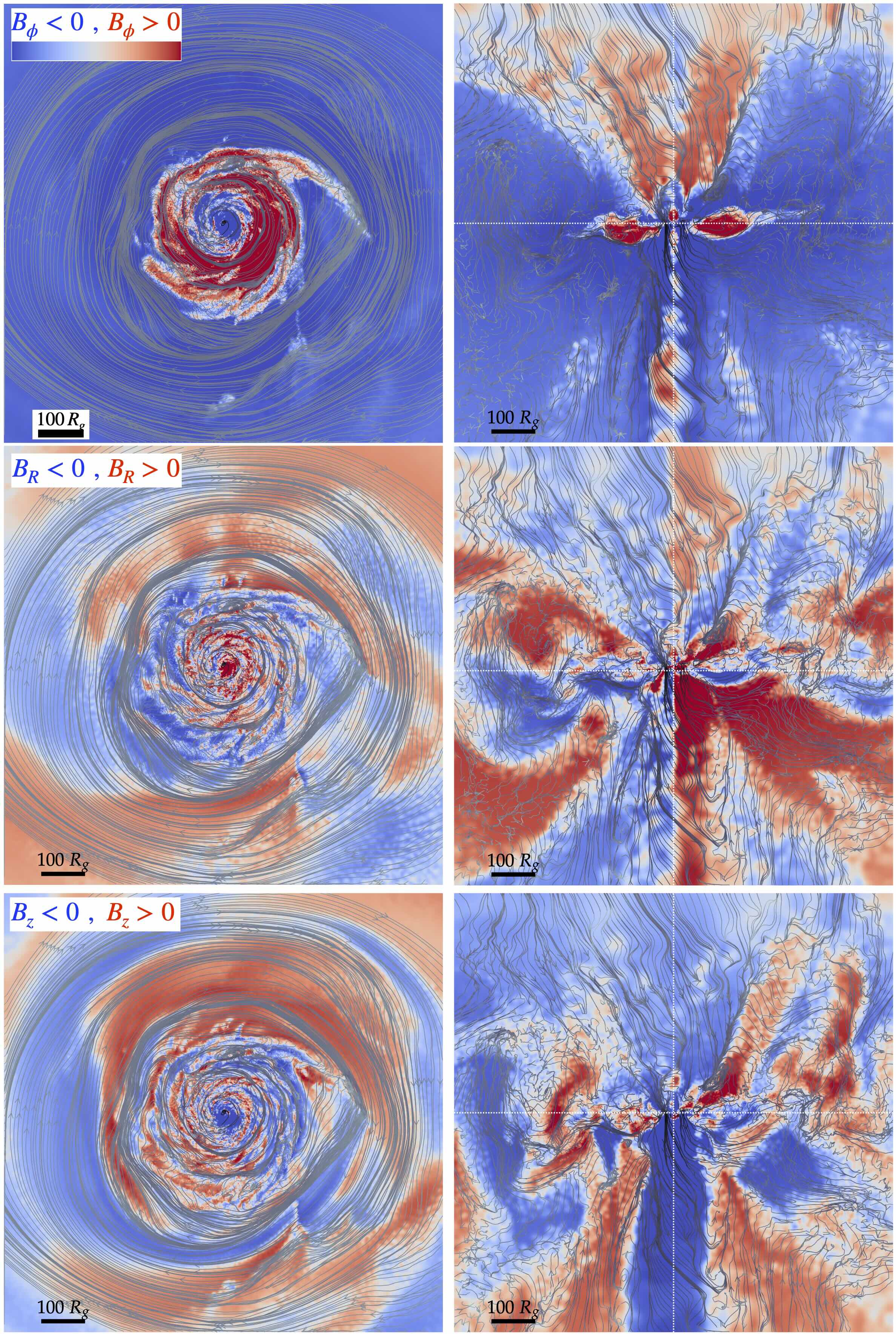}
	\includegraphics[width=0.32\textwidth]{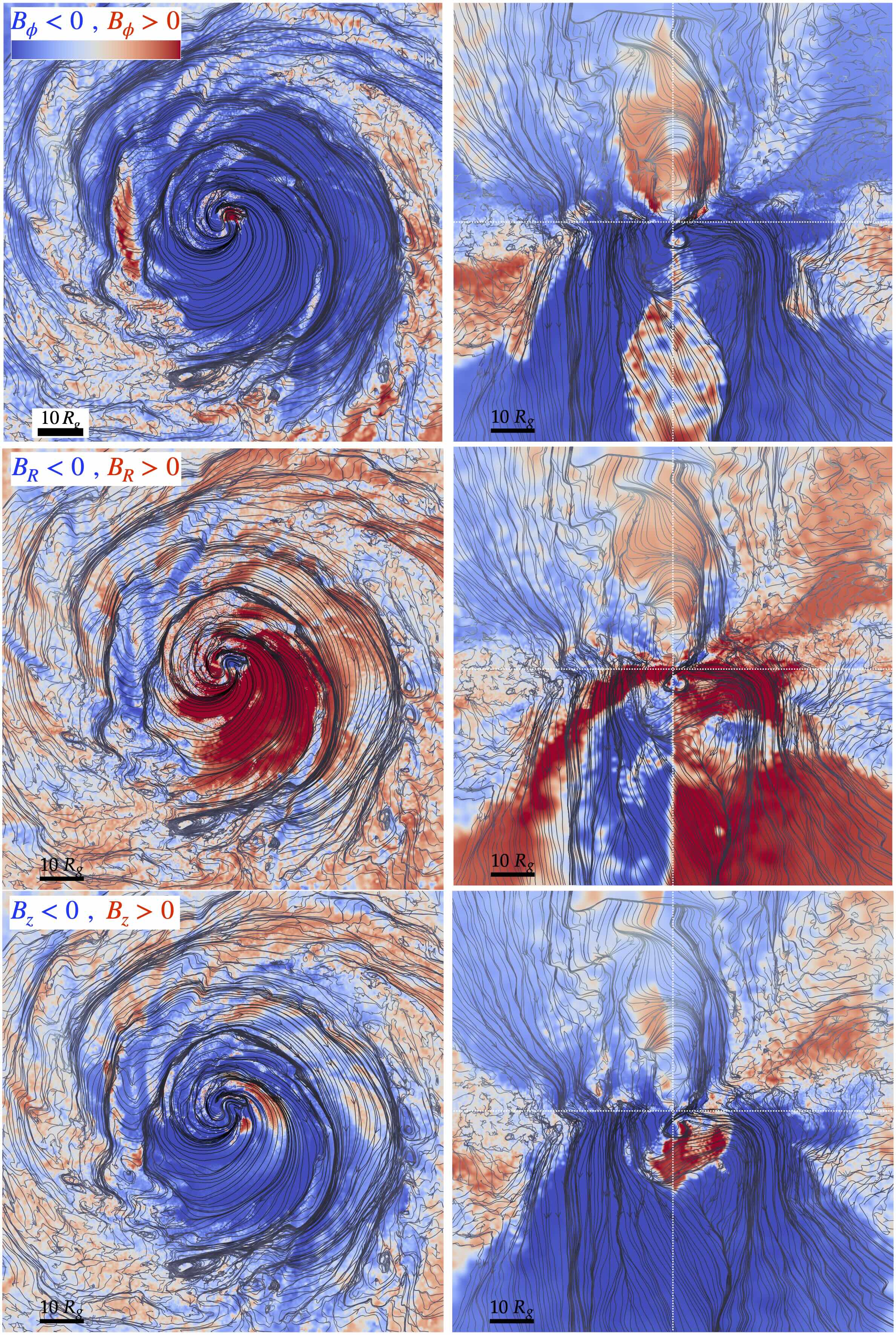}
	\caption{Face-on ($xy$ midplane slice) and edge-on (cylindrical $Rz$ wedge) maps of the magnetic field lines in the disk. 
	We show three different scales, from large scales in the disk ({\em left}) to intermediate ({\em center}) to inner disk ({\em right}), as labeled. 
	In each panel, lines show the projected field lines in the plane, while colors encode the sign of the field components labeled: azimuthal $B_{\phi}$ ({\em top}), radial $B_{R}$ ({\em middle}), vertical $B_{z}$ ({\em bottom}). There is a strong mean toroidal field on all scales. A $B_{R}-B_{\phi}$ anti-correlation continues to all scales, expected from mean-field amplification by induction with the accretion flow and powering strong Maxwell stresses and accretion (Fig.~\ref{fig:profiles.general}). At small radii coherent poloidal fields can thread the midplane from flux carried in from larger $R$, but they clearly fluctuate and the coherence length of the poloidal field is limited to $\sim H$, which partly explains the lack of large-scale coherent escaping magnetocentrifugal outflows (there is no field with coherence length $\gg R$, and such winds are highly mass-loaded, so will ``fail'' or ``fountain'' or ``recycle'' at heights $|z| \sim H$).
	\label{fig:maps.b}}
\end{figure*}

\begin{figure*}
	\centering
	\includegraphics[width=0.32\textwidth]{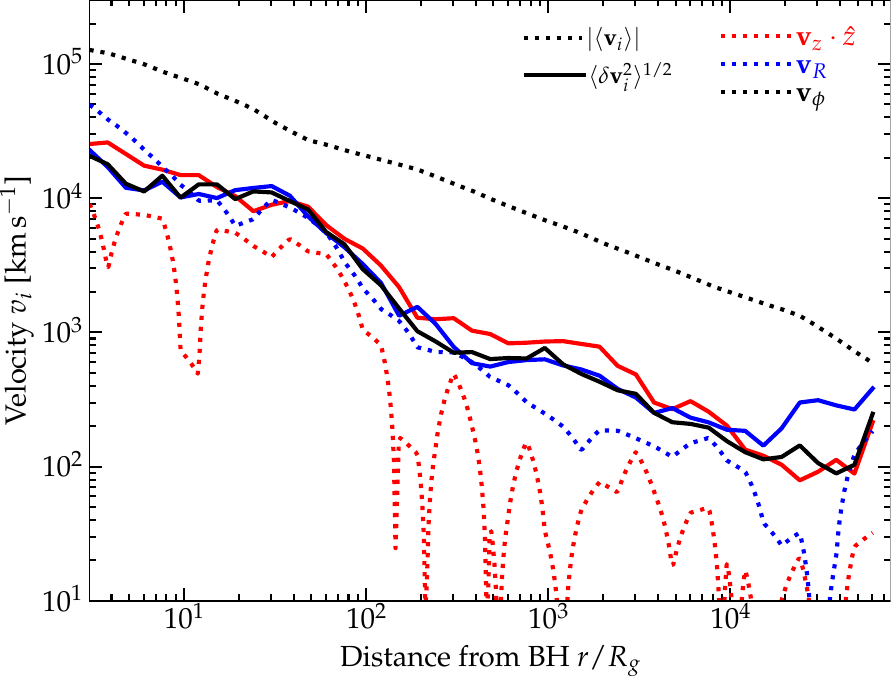} 
	\includegraphics[width=0.32\textwidth]{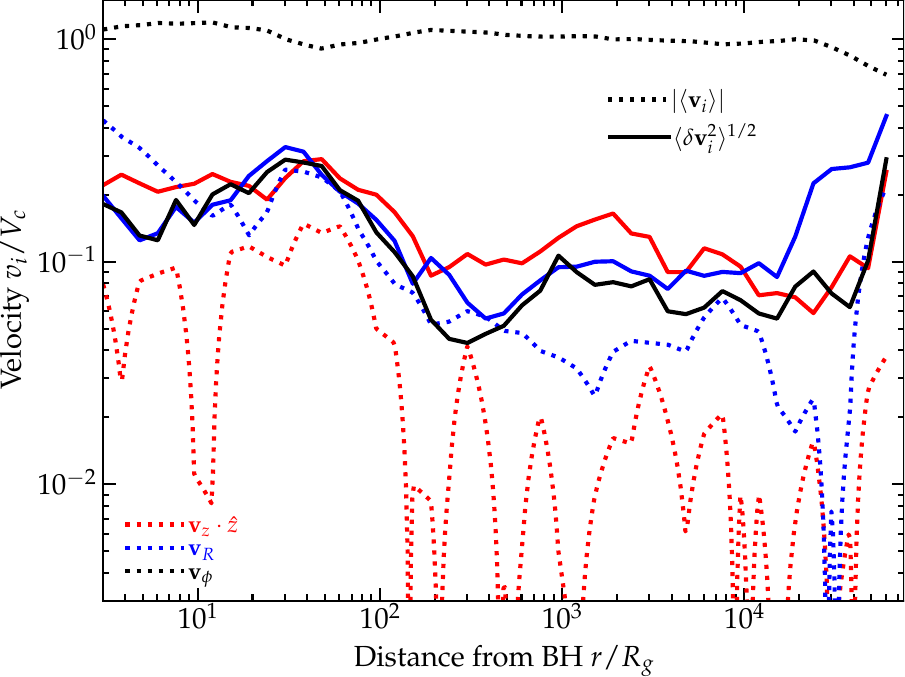} 
	\includegraphics[width=0.32\textwidth]{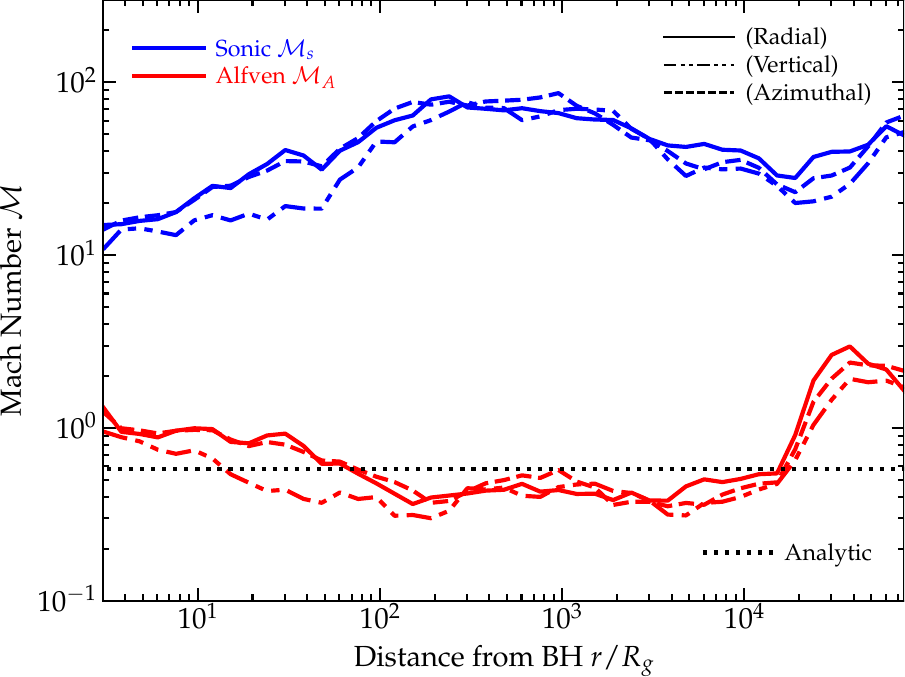} 
	\includegraphics[width=0.32\textwidth]{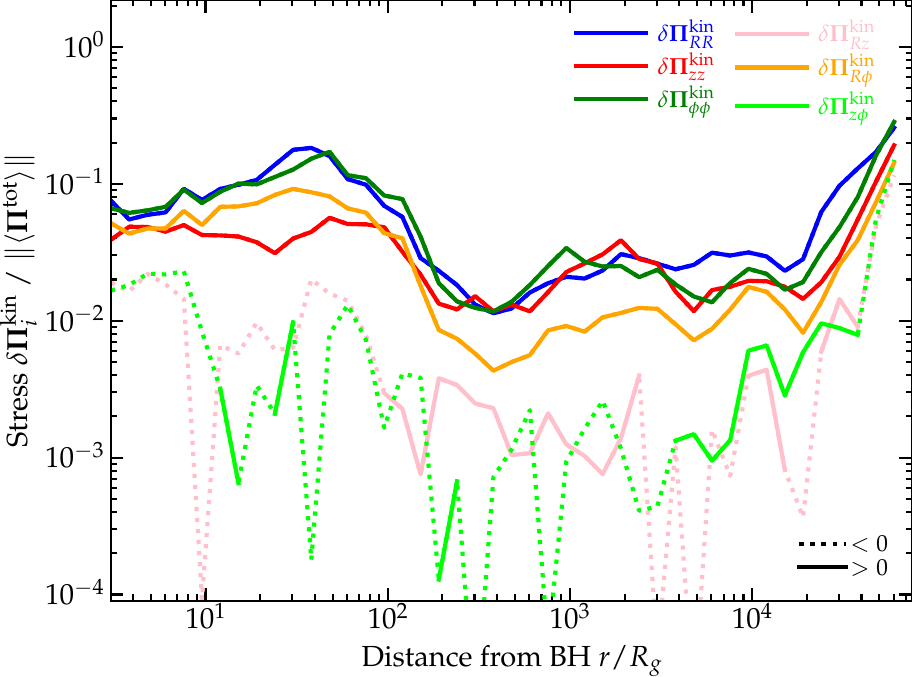} 
	\caption{Mass-weighted velocity/turbulence profiles, as Fig.~\ref{fig:profiles.general}. 
	{\bf (1)} Mean $|\langle {\bf v}_{i} \rangle|$ and standard deviation $\langle \delta {\bf v}_{i}^{2}\rangle^{1/2}$ of the azimuthal $B_{\phi}$, vertical/outflow $v_{z}\cdot \hat{z}$, and radial $v_{R}$ velocity fields. The vertical flow is primarily inflow onto the disk at all radii, while the mean radial advection speed is comparable to the turbulent speed (itself quasi-isotropic) in the inner disk.
	{\bf (2)} Same, normalized to the Keplerian velocity $v_{\rm K}$ at each radius.The thick disks with rapid transport (Fig.~\ref{fig:profiles.general}) follow from the large velocities relative to $V_{\rm c}$.
	{\bf (3)} Turbulent \Alf\ ($\mathcal{M}_{A} \equiv \delta v/v_{A}$) and sonic ($\mathcal{M}_{s} \equiv \delta v/c_{s}$) Mach numbers, divided into components ($\delta v_{R,\,\phi,\,z}$), with the analytic prediction for the 1D $\mathcal{M}_{A} \approx 1/\sqrt{2}$ from Fig.~\ref{fig:profiles.general}. Turbulence is trans-\Alf{ic} and highly super-sonic.
	{\bf (4)} Different components of the Reynolds stress tensor $\delta \boldsymbol{\Pi}^{\rm kin} \equiv \rho \delta {\bf v} \delta {\bf v}$, with positive ({\em solid}) and negative ({\em dotted}) values indicated, normalized to the norm of the total stress tensor from all sources. The $R\phi$ stress is the dominant source of angular momentum exchange and comparable to the level of the turbulent pressure.
	\label{fig:profiles.velocity}}
\end{figure*} 

\begin{figure*}
	\centering
	\includegraphics[width=0.32\textwidth]{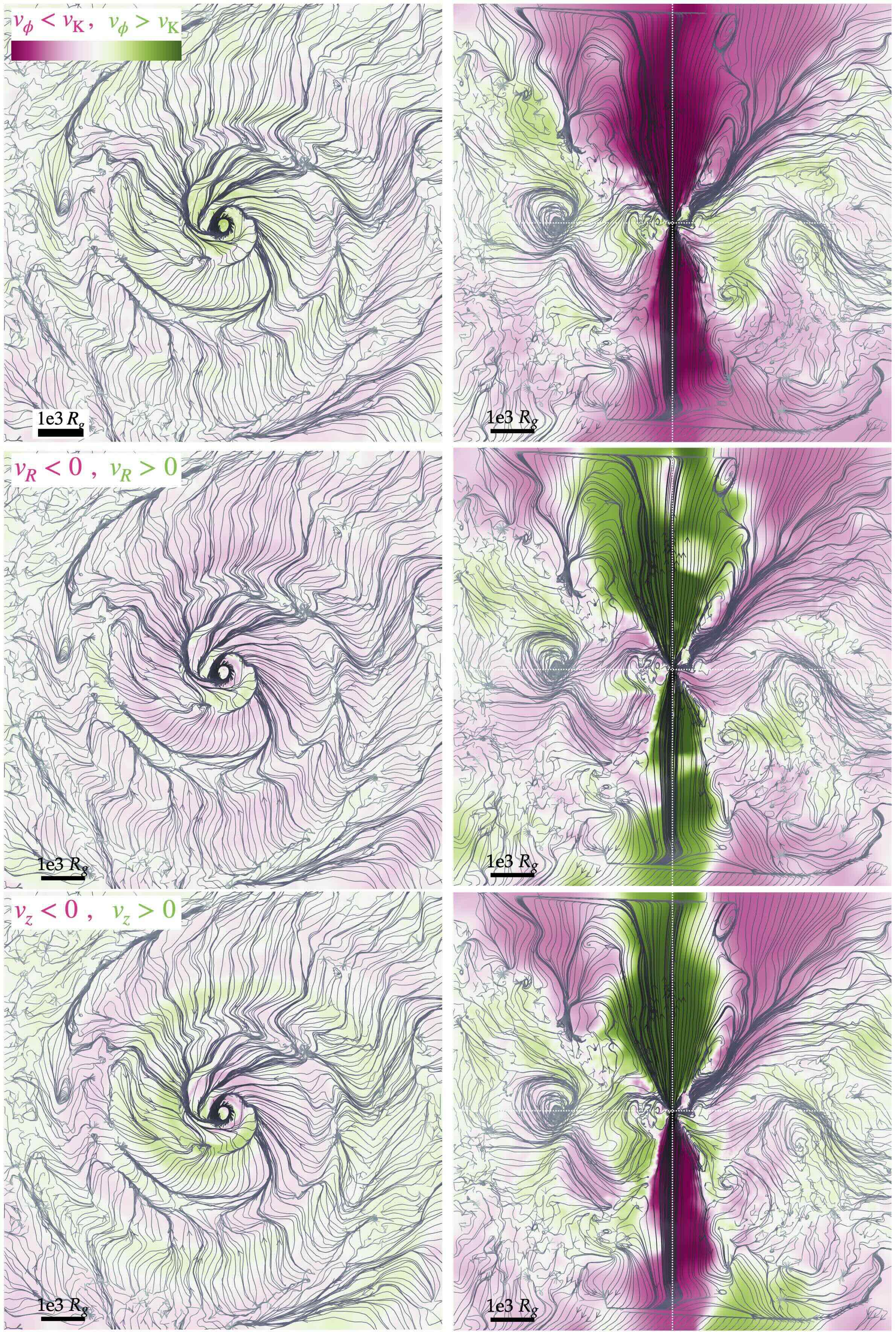}
	\includegraphics[width=0.32\textwidth]{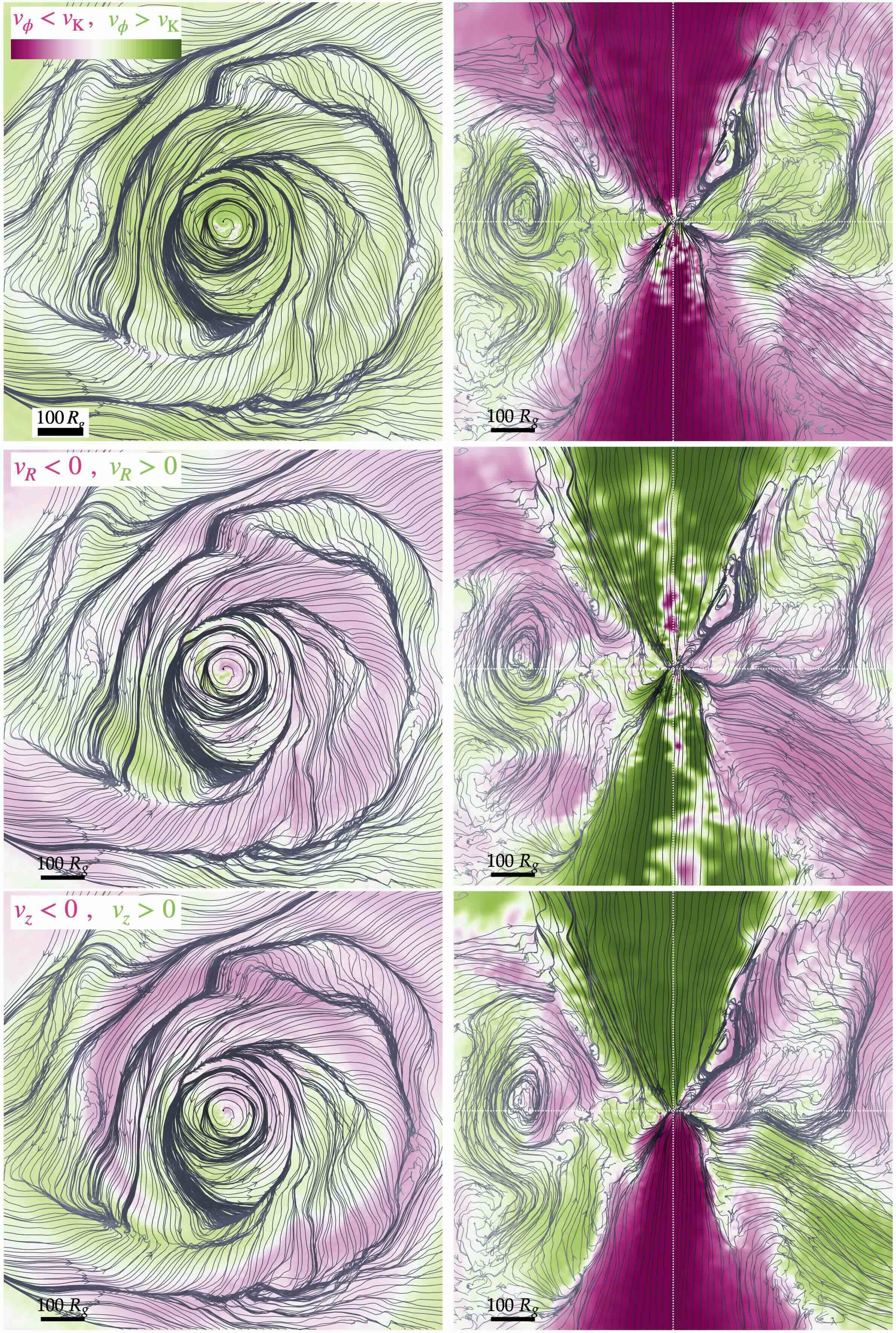}
	\includegraphics[width=0.32\textwidth]{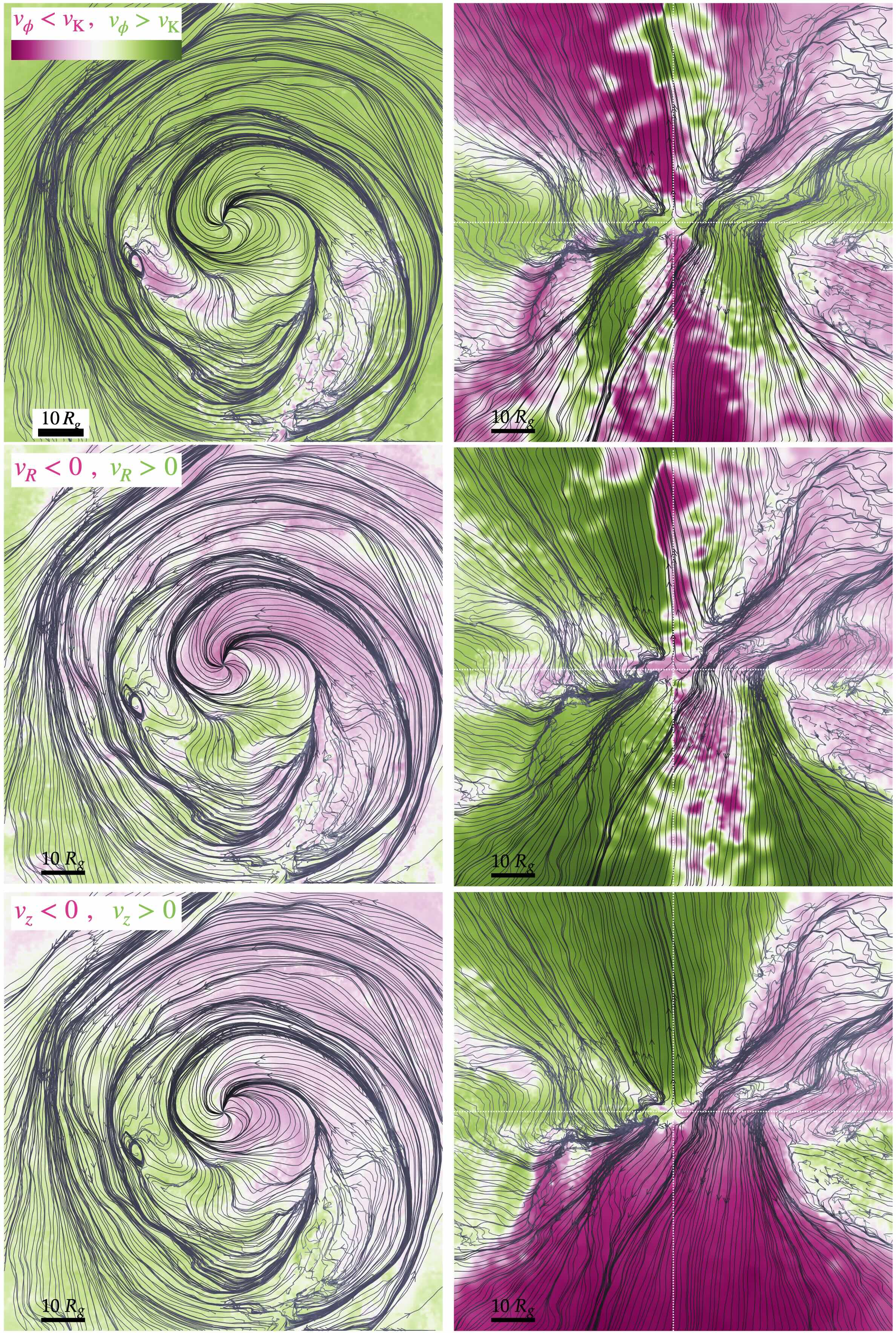}
	\caption{Face-on and edge-on maps of the velocity field lines in the disk, as Fig.~\ref{fig:maps.b}, on different scales. 
	For $v_{\phi}$, we subtract the expected Keplerian velocity for a circular orbit at the given $R$, $\sqrt{G M_{\rm BH}/R}\,\hat{\phi}$, in order to show the more interesting residuals from (primarily) rotational motion. 
	These coincide with the positions of spiral structures in the morphology (Fig.~\ref{fig:zoomies.cyl}). 	
	The rotation in the polar region is mostly sub-Keplerian while in the midplane it is mildly super-Keplerian (owing to the Paczy{\'{n}}ski-Wiita potential and rapid accretion meaning gas does not fully circularize at each radius). Radial and vertical flows are primarily inflow, though both show midplane turbulence and residuals associated with spiral structure.
	\label{fig:maps.vel}}
\end{figure*}

\begin{figure}
	\centering
	\includegraphics[width=0.99\columnwidth]{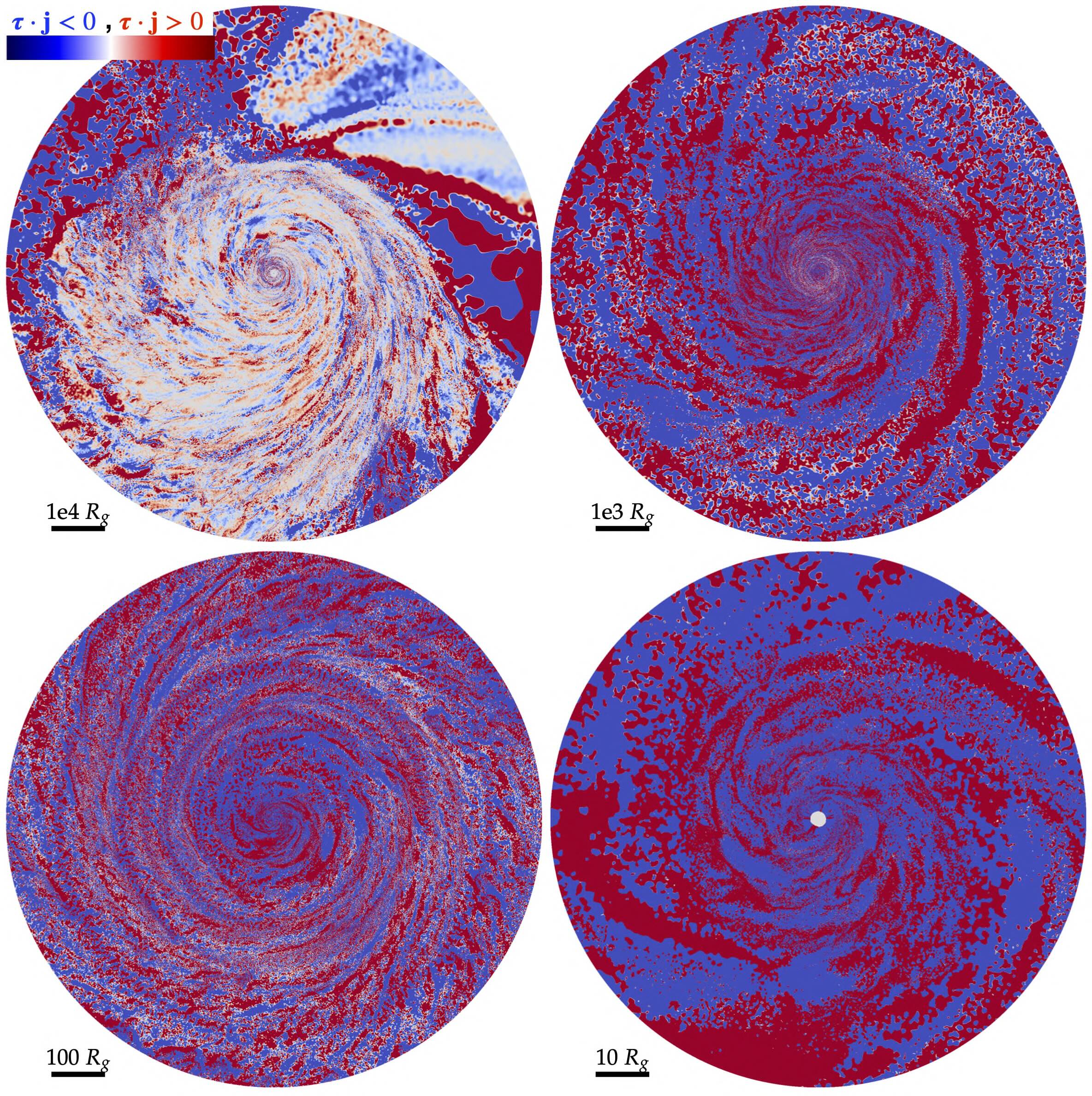} 
	\caption{Map of the instantaneous torques on each gas parcel, directly from the simulation using the MHD+radiation acceleration computed in that timestep (at the final timestep), projected face-on in an $R\phi$ slide through the midplane on different scales. There is both a small-scale (turbulent) component and a large-scale coherent (spiral) component visible.
	\label{fig:maps.torque}}
\end{figure}

\section{Results}
\label{sec:results}

Fig.~\ref{fig:zoomies}-\ref{fig:zoomies.cyl} illustrate the morphology and gas structure on a wide range of scales, though we will focus primarily on the smaller scales illustrated in Fig.~\ref{fig:zoomies.cyl} which are newly-resolved here. We first focus on a case study of our default ($f_{\dot{M}}=1$) super-critical simulation, before studying the effects of varying accretion rates below. Figs.~\ref{fig:profiles.general}, \ref{fig:profiles.magnetic}, \ref{fig:profiles.velocity} \&\ \ref{fig:profiles.flux} plot radial profiles of dynamical, magnetic, velocity, and radiation profiles (averaged within annuli, but at a given time). We choose the final time to which our simulation was run for properties at a given time, but highlight how these do (or do not) evolve below. Fig.~\ref{fig:profiles.vertical} shows the vertical profiles of different quantities at a given radius. Figs.~\ref{fig:maps.b}, \ref{fig:maps.vel},\ref{fig:maps.torque}, \&\ \ref{fig:maps.flux} show corresponding maps of the field of torques, magnetic fields, velocity fields, and radiation flux fields, while Figs.~\ref{fig:maps.pinwheel} \&\ \ref{fig:maps.pinwheel.edgeon} show colormaps of other quantities on a wide dynamic range of scales, face-on and edge-on. Figs.~\ref{fig:profiles.mdot}, \ref{fig:profiles.stress.vs.mdot} compare quantitative profiles versus $\dot{m}$ while Figs.~\ref{fig:maps.fields.mdot}, \ref{fig:maps.pinwheel.mdot}, \ref{fig:maps.edgeon.mdot} compare projected maps of fields and different scalar quantities versus $\dot{m}$.

\subsection{Convergence to Local Steady-State}
\label{sec:results:steady}

Initially in our simulation, there is (by construction) a ``cavity'' at $R \lesssim 300\,R_{g}$. Gas quickly begins to fill this in, with some inflow through each radius on its dynamical time, but for some time this still leads (as expected) to a surface density $\Sigma_{\rm gas}(R)$ and inflow rate $\dot{M}(R)$ which decrease with smaller $R \ll 300\,R_{g}$. This means gas mass is building up, and after a time of several order the inflow/accretion timescale in Fig.~\ref{fig:profiles.general}, we see that the central gas profile reaches quasi-steady-state behavior, with approximately constant $\dot{M}(R)$ as a function of radius, out to radii $\gtrsim 10^{4}\,R_{g}$ where the evolved time since refinement is more comparable to the accretion timescale. This means that if we continued to evolve our simulation even further in time, it is likely that the central accretion rate and gas surface densities would continue to rise until ``catching up to'' their values at larger radii. But those values are themselves evolving on even longer timescales (\paperone), as the massive gas clouds complex which is being tidally disrupted and fueling the accretion episode continues to accrete, driving evolution in disk properties (like mass and angular momentum direction) on timescales as large as $\sim 10^{7}-10^{8}$\,yr. 

We also see that sometimes episodes of strong bipolar outflows, or motions owing to spiral structure, occasionally produce net outflow at some ranges of radii. So it is unclear if the system can ever attain a true ``steady-state.'' If such a structure is present at radii $\sim 1000\,R_{g}$ at a time when we have refined to $\sim 3\,R_{g}$, and persists for tens of orbits at that radius, it would require evolving our simulations at the ISCO for a time $\gg 10^{7}\,G M_{\rm BH}/c^{3}$, which is computationally infeasible. And in this time, other structures with non-steady-state inflow/outflow will appear at other radii. It appears inevitable that at radii where we resolve many orbital times (e.g.\ the lower-resolution, less-refined simulations from \papertwo\ at $R \gtrsim 300\,R_{g}$, or the simulations here at $\ll 300\,R_{g}$), the accretion rate can systematically increase and decrease by factors of several (and occasionally order-of-magnitude) over many orbital times, as expected in dynamical accretion models. 

It is sufficient for our purposes here that the central regions of interest have reached {\em local} steady-state. For all our subsequent analysis, we focus on this quasi-steady-state regime after the first few tens of orbits at small radii, as the initial ``buildup'' is necessarily sensitive to how we initialize the refinement and resolution change. We see these local steady-state conditions maintained for hundreds to thousands of dynamical times at different radii.

\subsection{Behaviors Continuing from Larger Radii}
\label{sec:results:similar}

Several behaviors we see appear to be natural extensions of what was seen at larger radii in \paperone-\papertwo:

{\bf Accretion persists.} We see (Fig.~\ref{fig:profiles.general}) a quasi-steady $\dot{M}$ in both space and time, with accretion rates $\dot{M} \sim 3-6$, $0.1-0.2$, and $0.005-0.01$\,${\rm M_{\odot}\,yr^{-1}}$ in the $f_{\dot{M}}=1,\,0.03,\,0.001$ runs, respectively (Fig.~\ref{fig:profiles.mdot}), corresponding to Eddington-scaled $\dot{m} \sim 10-20$, $0.3-0.6$, $0.02-0.04$, given the BH mass of $\sim 1.3 \times 10^{7}\,{\rm M_{\odot}}$. We will discuss the luminosities this corresponds to below, but it for the default ($f_{\dot{M}}=1$) simulation corresponds to a BH growth/mass-doubling timescale of $\sim 0.5-2\times10^{6}\,{\rm yr}$, which is much shorter than the global duration of the accretion episode expected from the large-scale cloud-disruption dynamics (and seen in the low-resolution ``parent'' simulation from which this particular inflow time was chosen for our zoom-in study). The local gas accretion time is largely independent of accretion rate at $\sim 3-20\,\Omega^{-1}$ (i.e.\ a couple orbital times) through most of the disk, though it rises in these units at smaller radii, meaning that the accretion is fundamentally dynamical, rather than secular. In the general state the accretion is clearly not ``arrested'' (it continues, and is disk-like, per Fig.~\ref{fig:zoomies}, rather than quasi-spherical or dominated by clumps falling through field lines, though see \S~\ref{sec:bursts}). Instead it proceeds through the midplane (Fig.~\ref{fig:maps.vel}), with $\langle v_{r} \rangle$ (Fig.~\ref{fig:profiles.velocity}) as expected given the density profile and accretion rate (Fig.~\ref{fig:profiles.general}).

{\bf There is a Thick, Flared Accretion Disk.} The disk has a large covering factor/scale height, $H/R \sim 0.1-0.5$ (Figs.~\ref{fig:zoomies.cyl} \&\ \ref{fig:profiles.general}), weakly dependent on distance from the SMBH and $\dot{m}$ (Figs.~\ref{fig:maps.pinwheel.edgeon}, \ref{fig:maps.edgeon.mdot}). But nonetheless it is clearly a disk: the gas is primarily supported in the radial direction by centrifugal motion, on primarily close-to-circular/tangential Keplerian orbits ($\langle v_{\phi} \rangle \approx V_{c}(r)$; Fig.~\ref{fig:profiles.velocity}) with small-to-modest global/coherent eccentricity (Fig.~\ref{fig:zoomies}, \ref{fig:maps.pinwheel}, \ref{fig:maps.vel}). The vertical support of the thick disk comes from a combination of turbulence (with $\delta v_{z}$ being trans-\Alf{ic} and an order-unity fraction $\sim 0.3$ of $v_{A,\,\phi}$; Figs.~\ref{fig:profiles.general} \&\ \ref{fig:profiles.velocity}) and magnetic (with gas supported by the in-plane fields $v_{A,\,\phi}$; Figs.~\ref{fig:profiles.general} \&\ \ref{fig:profiles.magnetic}), again all weakly-dependent on $\dot{m}$ (Fig.~\ref{fig:profiles.mdot}). Interestingly, the systematically smaller median gas-mass-weighted $|z|/R$ appears to closely track the predicted turbulent scale-height $\delta v_{z}/\Omega$, while the larger rms mass-weighted $\langle z^{2} \rangle^{1/2}/R$, reflecting the contribution from the more extended tails of the vertical density distribution visible in Fig.~\ref{fig:profiles.vertical} (which also show slow vertical decay of $B_{\phi}$), appears to more closely track the \Alf\ scale height $v_{A\,\phi}/\Omega$.

{\bf The Disk Remains ``Hyper-Magnetized.''} Specifically the plasma $\beta_{\rm thermal} = P_{\rm gas,\,thermal}/P_{\rm B} \sim 10^{-5} - 10^{-2} \ll 1$ even in the midplane at all $\dot{m}$ (Figs.~\ref{fig:profiles.mdot}, \ref{fig:maps.pinwheel.mdot}, \ref{fig:maps.edgeon.mdot}), as at larger radii (though at the largest $\dot{m}$, radiation pressure can become more comparable to magnetic, as shown below). The values (see Fig.~\ref{fig:profiles.general} \&\ \ref{fig:profiles.magnetic} for the dispersion in $P_{B}/P_{\rm thermal}$) range widely depending on the multi-phase structure of the gas, more driven by large variations in the gas temperature in this structure (hence $P_{\rm thermal}$) rather than magnetic pressure (though that varies significantly as well, especially as the inner disk thermalizes and becomes warmer and hotter coronal gas becomes more prevalent above/below the disk at smaller radii). As we show below, removing magnetic fields from the inner disk (even while retaining radiation pressure) produces qualitatively different outcomes and disk structure.

{\bf Outflows are Modest.} We see (usually radiation pressure-driven) outflows which are not negligible at some radii, but are in general modest, usually comprising tens of percent to an order one fraction of the inflow rate (though see \S~\ref{sec:outflows}). As discussed further there are transient phases where the outflow rate at some radii exceeds the local inflow rate, though they do not generally exceed the inflow rate from the largest radii $\gtrsim 10^{5}\,R_{g}$ (around the BHROI) where the inflow rates are as large as $\sim 30\,{\rm M_{\odot}\,yr^{-1}}$. So they typically do not change the mass or accretion structure of the disks themselves at the orders-of-magnitude level, though they are certainly of interest for AGN ``feedback.'' Note that {\em some} of the apparently large outflow rates in Fig.~\ref{fig:profiles.general} (e.g.\ the apparent large outflow ``spike'' at $\sim 3\times10^{4}\,R_{g}$) can clearly be seen in Figs.~\ref{fig:profiles.velocity}, \ref{fig:maps.vel}, \ref{fig:maps.torque} to {\em not} actually be large-scale winds, but rather related to in-plane disk motions such as spiral/eccentric modes (plainly visible in Fig.~\ref{fig:zoomies}) in the midplane causing local inflow-outflow sloshing/epicyclic motions (e.g.\ non-zero oscillating $v_{r}$), while the mean sense of the radial velocities is largely inflow, especially on large scales. The true ``outflows'' in the sense usually meant will be explored more below but tend to be bipolar and composed of more tenuous gas. This is broadly consistent with previous simulations even of highly super-Eddington, radiation-pressure-dominated inner accretion disks, which generally see $\dot{M}_{\rm out} \sim \dot{M}_{\rm in}$ \citep{jiang:2019.superedd.sims.smbh.prad.pmag.modest.outflows}. We stress that we do not include line driving or GR/spin effects, nor evolve the simulations long enough to follow wind-launching from large scales (e.g.\ the torus or narrow-line region), so the only outflows we can follow are driven by magnetic fields and/or continuum radiation pressure and almost certainly under-estimate the true magnitude of outflows, as discussed further below (\S~\ref{sec:outflows}).

{\bf The Disk Features Radiative, Trans-\Alf{ic} Turbulence.} We see that both velocity (Fig.~\ref{fig:profiles.velocity}) and magnetic field fluctuations (Fig.~\ref{fig:profiles.magnetic}) are roughly isotropic with \Alf\ Mach number of order unity, remarkably constant as a function of radius and $\dot{m}$ (Fig.~\ref{fig:profiles.mdot}). Midplane fluctuations with coherence lengths $\sim H$ are clearly visible in all cases (Figs.~\ref{fig:maps.pinwheel}, \ref{fig:maps.b}, \ref{fig:maps.vel}, \ref{fig:maps.pinwheel.edgeon}, \ref{fig:maps.fields.mdot}). 
Since $\beta_{\rm thermal} \ll 1$, this necessarily means the sonic Mach number $\mathcal{M}_{s} = v_{\rm turb}/c_{s} \gg 1$, so the turbulence is highly supersonic, enabled by efficient cooling ($t_{\rm cool} \ll \Omega^{-1}$), leading to compressive shocks which feature large density fluctuations (up to $\sim \mathcal{M}_{s}^{2}$, producing the variance seen in midplane density at a given radius) and multi-phase structure (density and temperature spreads in Fig.~\ref{fig:profiles.general}) and visible as well in the extremely small-scale density structure in Figs.~\ref{fig:zoomies}-\ref{fig:zoomies.cyl} reaching down to scales of order the expected sonic length/shock width/Sobolev length (all dimensionally similar at size $\sim H/\mathcal{M}_{s}^{2} \sim c_{s}^{2}/v_{A}\Omega$). As shown in \S~\ref{sec:lte}, the scattering mean-free-paths of photons even in the inner disk at the highest $\dot{M}$ here are generally larger than the intrinsic shock width, so radiation pressure does not strongly alter our intiution.

{\bf Accretion Is Driven by In-Plane Maxwell and Reynolds Stresses.} The torques on the gas are clearly turbulent/local on scales $\sim H$ (Fig.~\ref{fig:maps.torque}), and the Maxwell and kinetic/Reynolds stresses which involve angular momentum transport are dominated by the usual in-plane $R\phi$ terms (Figs.~\ref{fig:profiles.magnetic} \&\ \ref{fig:profiles.velocity}) for all $\dot{m}$ (Fig.~\ref{fig:profiles.stress.vs.mdot}). As in \papertwo, the net accretion rates $\dot{M}$, inflow speeds $\langle v_{R} \rangle$, and torques $\langle \tau \rangle$ are all consistent with the magnitude of the Maxwell ($-\langle B_{\phi} B_{R}/4\pi\rangle$) and Reynolds ($\langle \rho \delta v_{\phi} \delta v_{R} \rangle$) stresses, with the Maxwell stresses generally larger by a factor of a few in the inner disk compared to the Reynolds stresses and, notably, the {\em mean} Maxwell stress ($-\langle B_{\phi} \rangle \langle B_{R} \rangle/4\pi$) comparable to or larger than the fluctuating Maxwell stress ($-\langle \delta B_{\phi}  \delta B_{R} \rangle/4\pi$). The similarity of these across values of $f_{\dot{M}}$ (Fig.~\ref{fig:profiles.stress.vs.mdot}) suggests that the basic similarity arguments in \paperthree\ provide a reasonable first-order description of the disk properties.

{\bf Spiral Structure Persists.} There are large-amplitude, clearly visible spiral ($m=1$) modes throughout in all cases, as noted above (Figs.~\ref{fig:zoomies}, \ref{fig:maps.pinwheel}, \ref{fig:maps.b}, \ref{fig:maps.vel}, \ref{fig:maps.torque}). These evolve in time but are always visually present and appear at all radii in the disk. Decomposing the disk into non-axisymmetric modes, the dominant mode is indeed $m=1$, as expected for a gravitationally stable (non-fragmenting, negligibly self-gravitating) disk in a close-to-Keplerian potential \citep{tremaine:slow.keplerian.modes,jacobs:longlived.lopsided.disk.modes,touma:keplerian.instabilities,bacon:m31.disk,hopkins:inflow.analytics,hopkins:slow.modes,hopkins:cusp.slopes}. And indeed these have also been seen in many other accretion disk simulations on these scales \citep{heinemann.papaloizou:2009.spiral.mode,hopkins:zoom.sims,hopkins:m31.disk,hopkins:qso.stellar.fb.together,gaburov:2012.public.moving.mesh.code}, especially of high-accretion rate or super-Eddington disks even without any self-gravity \citep{jiang:2019.superedd.sims.smbh.prad.pmag.modest.outflows,kudoh:2020.strong.b.field.agn.acc.disk.sims.compare,davis:2020.mhd.sim.acc.disk.review,kaaz:2022.grmhd.sims.misaligned.acc.disks.spin}. 

{\bf Star Formation Is Completely Shut Down.} At all $\dot{m}$ and all radii we analyze ($\ll$\,pc scales), the disk has Toomre $Q \gg1$, reaching $\gg 10^{6}$ if we include magnetic support (not explicitly shown, but easily inferred from the properties plotted in Fig.~\ref{fig:profiles.general}, and plotted at larger radii in \papertwo). The magnetic critical mass is also much larger than the entire disk mass. So we see star formation is strongly suppressed relative to inflow at all radii $\lesssim$\,pc, and zero stars form over the duration of the simulation at radii $r\lesssim 0.05\,$pc. Only a couple stars have orbits taking themselves into radii $\lesssim 0.01\,$pc, and these (as well as their winds/jets/radiation) have a negligible effect on the dynamics (see \citealt{hopkins:superzoom.imf}, for details). This is not trivial: if we assumed an SS73-like $\alpha$ disk with $\alpha=0.01-0.1$ and the same BH mass and accretion rate, we would predict $Q\ll 1$ at all radii $\gtrsim 200-500\,R_{g}$.

{\bf Magnetic Fields Continue to Scale with Radius.} We see growth of the magnetic fields in a Lagrangian sense as gas accretes inwards, giving a spatial scaling typical $|{\bf B}| \propto \dot{m}^{1/2} \,r^{-1}$ very approximately (Fig.~\ref{fig:profiles.general}, \ref{fig:profiles.mdot}), or $|{\bf B}| \propto H^{-1}$ or $\rho^{1/4}r^{-3/4}$ at a given $\dot{m}$ (Fig.~\ref{fig:profiles.magnetic}). This is consistent with the expectations from simple models of ${\bf B}$ supplied by flux-freezing or net advection of magnetic flux from larger radii, with a large-scale mean toroidal field (see \papertwo), although the actual physics governing the mean (let alone fluctuating/turbulent) field can be much more complicated (see \citealt{johansen.levin:2008.high.mdot.magnetized.disks,gaburov:2012.public.moving.mesh.code,bai:2012.mri.saturation.turbulence}, plus \citealt{squire:2024.mri.shearing.box.strongly.magnetized.different.beta.states}, Guo et al., Tomar et al., in prep.). Note at larger radii, these scalings are approximately the same as the simpler $B \propto \rho^{2/3}$ adopted in \paperthree; here we see the latter is not quite as accurate an approximation at small radii and large $\dot{m}$ where radiation pressure modifies the profiles, which we discuss below. The mean toroidal and radial fields are anti-correlated (Fig.~\ref{fig:profiles.magnetic} \&\ \ref{fig:maps.b}), again as expected.

{\bf The Magnetic Field Remains Predominantly Toroidal.} The field is ordered, with the clearly-dominant component being the midplane toroidal/azimuthal field at most radii (Fig.~\ref{fig:profiles.magnetic}) and $\dot{m}$. There is a non-negligible but smaller mean radial field. And there is a quasi-isotropic turbulent/fluctuating field, of comparable amplitude to the mean radial field, $\sim 10\%$ of the mean azimuthal field (Figs.~\ref{fig:profiles.magnetic}, \ref{fig:maps.b}). The mean/coherent vertical component through the midplane generally remains the weakest component. Again this fits the expectation from \papertwo\ for accretion of gas into the disk initially via streamers or other in-plane structures, which amplify the toroidal-radial mean field and weaken the vertical field. This places the mean vertical/poloidal field in the ``turbulence-dominated'' regime (so turbulent resistivity suppresses formation of a large coherent mean vertical field). But it is striking that this coherent toroidal field structure persists at all $\dot{m}$ even in the strongly turbulent, radiation-pressure-dominated zone. However, in \S~\ref{sec:bursts} we show note that the field can change in episodes associated with strong outflows.

All of these are qualitatively consistent with analytic extrapolations in the simple models \paperthree, from the behaviors in \paperone-\papertwo\ from larger scales.

\begin{figure*}
	\centering
	\includegraphics[width=0.45\textwidth]{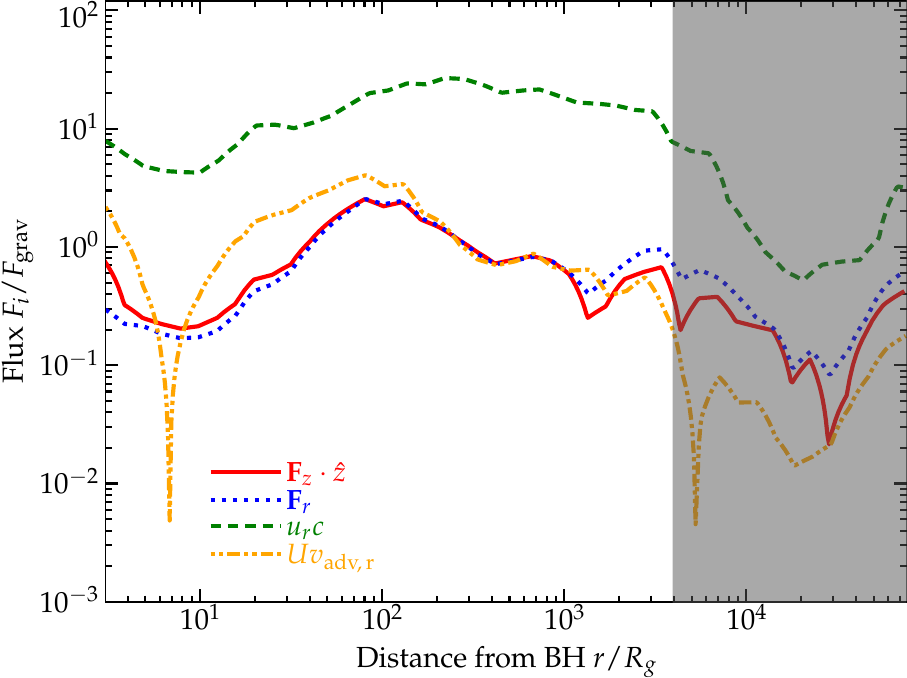} 
	\includegraphics[width=0.45\textwidth]{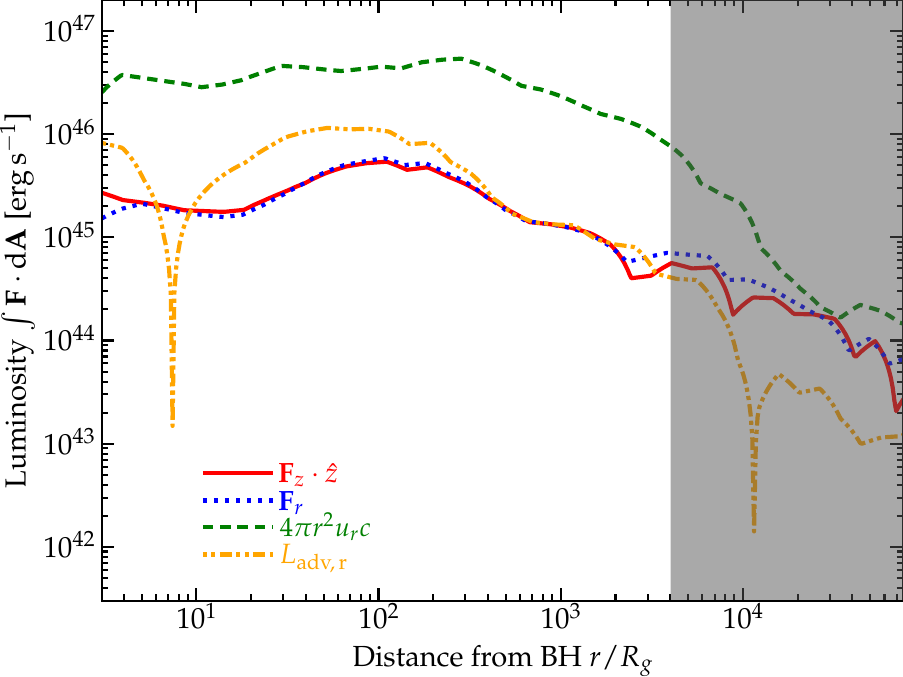} \\
	\hspace{0.8cm}
	\includegraphics[width=0.45\textwidth]{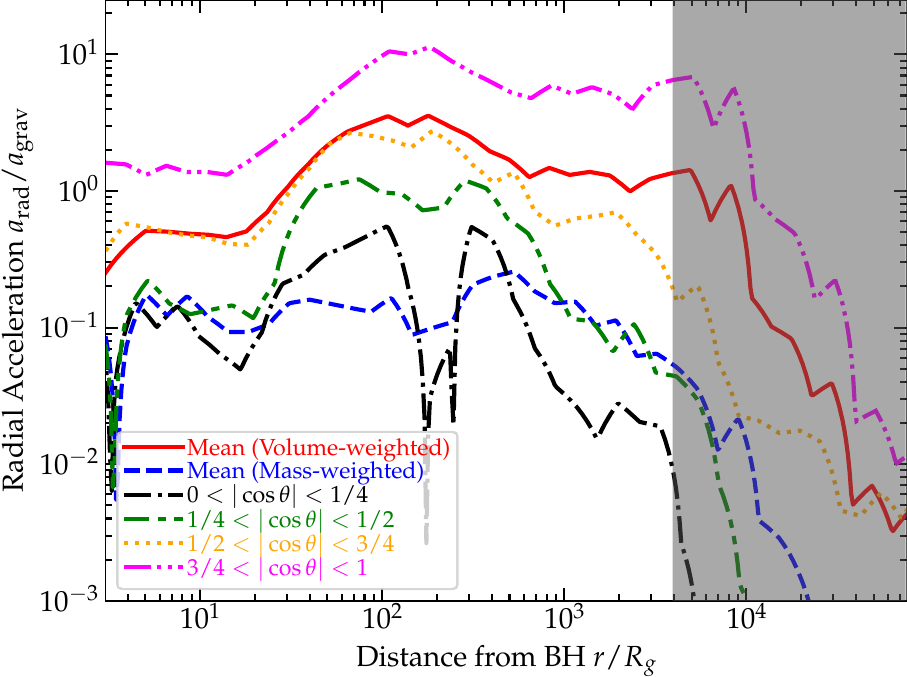} 
	\hspace{0.1cm}
	\includegraphics[width=0.49\textwidth]{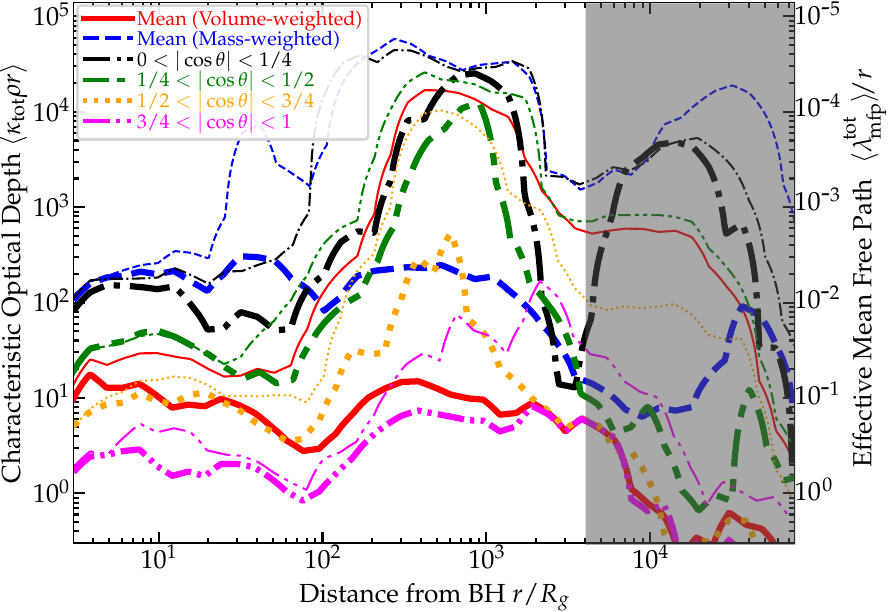} 
	\caption{Volume-weighted radiation profiles, as Fig.~\ref{fig:profiles.general}, for our high-$\dot{M}$ ($f_{\dot{M}}=1$) simulation.
	{\bf (1)} Mean flux ${\bf F}_{i}$ by vertical/radial components, normalized to the total flux expected for a radiatively efficient thin disk truncated at the ISCO: $F_{\rm grav}^{\alpha} \equiv 3/(4\pi) \dot{M} \Omega^{2} (1 - \sqrt{3 R_{g}/R})$. 
	We also compare the ``optically thin'' flux/radiation energy density $u_{\rm rad}\,c$, and the advective flux of total internal energy $v_{r} (e_{\rm rad} + e_{\rm therm} + e_{\rm B})$. The disk becomes radiatively inefficient and advection-dominated at small radii.
	{\bf (2)} Same, but the integral of each in thin shells to represent the total energy flow rate per unit time. The standard Eddington luminosity $L_{\rm Edd} = 3.2\times10^{4}\,L_{\odot} (M_{\rm BH}/M_{\odot}) = 1.6\times10^{45}\,{\rm erg\,s^{-1}}$ here. The luminosity saturates modestly super-Eddington (though we do not resolve the innermost near-horizon regime). 
	{\bf (3)} Radial acceleration ${|\bf a}_{\rm rad}\cdot \hat{\bf r}|$ relative to gravity $|{\bf a}_{\rm grav}\cdot \hat{\bf r}$ (computed in-code), weighted by gas volume or mass or volume-averaged in different wedges of polar angle $|\cos{\theta}|$ ($=1$ along the angular momentum axis). In the polar direction radiation exceeds gravity, especially at somewhat larger radii $\gtrsim 100\,R_{g}$, but in the midplane (where most of the gas mass resides) radiation pressure is weak.
	{\bf (4)} Characteristic optical depth or photon mean-free-path $\lambda^{\rm abs,\,eff}_{\rm mfp}$, estimated as $\langle \kappa_{\rm tot} \rho r \rangle \sim r/\langle \lambda^{\rm tot}_{\rm mfp} \rangle$ as a function of polar angle and radius (in terms of total absorption+scattering opacity). We show both mass-weighted (or volume-weighted where noted) median opacities ({\em thin}), typical of the disk midplane, and Rosseland mean opacities in each shell ({\em thick}), more representative of what is seen by escaping radiation.
	Shaded radii show where the radiation equilibration timescales $\sim \tau_{\rm tot} r / c$ tend to exceed ($\gg 10^{5}\,G M_{\rm BH}/c^{3}$)  our total runtime since reaching maximum refinement near the ISCO (where much of the radiation is produced). 
	\label{fig:profiles.flux}}
\end{figure*} 

\begin{figure*}
	\centering
	\includegraphics[width=0.9\textwidth]{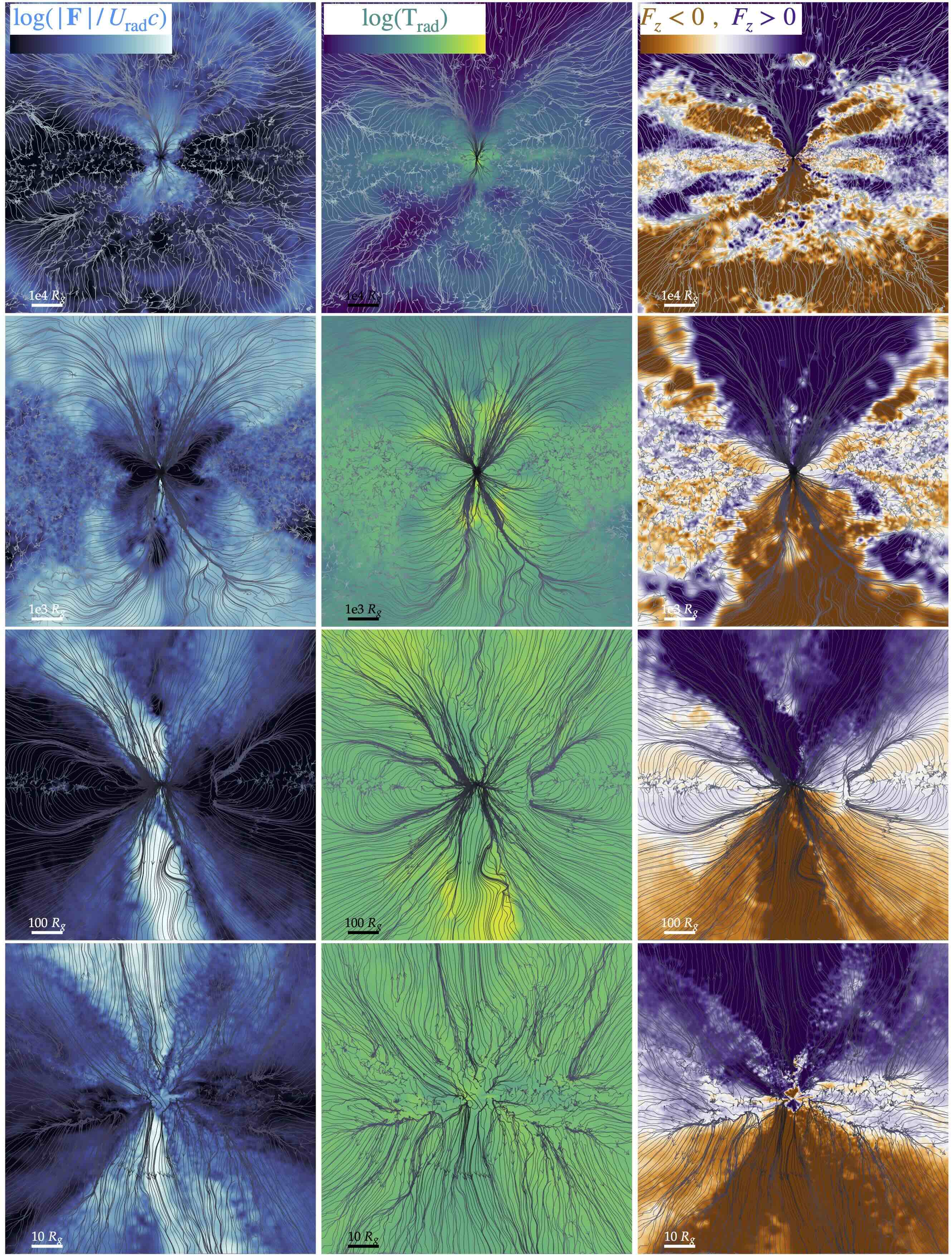}
	\caption{Radiation flux maps, as Fig.~\ref{fig:maps.b}. 
	Lines show radiation streamlines (following the flux vectors ${\bf F}_{\rm rad}$), edge-on ($Rz$) in a cylindrical wedge, with box side-length decreasing {\em top-to-bottom}. 
	Fluxes are summed over all bands followed, and defined in the Lagrangian frame (so does not include the advective flux). 
	{\em Left:} Colors show $|{\bf F}_{\rm rad}|/U_{\rm rad}c$, from $0$ (trapped; {\em black}) to $1$ (free-streaming; {\em white}) on a logarithmic scale. 
	{\em Center:} Colors show radiation temperature, as Fig.~\ref{fig:maps.pinwheel}.
	{\em Right:} Colors show sign(${\bf F}_{z}$), as Figs.~\ref{fig:maps.b} \&\ \ref{fig:maps.vel}. In the polar direction radiation escapes, but along the disk surface much of the radiation scatters into or is absorbed into the disk, which is externally heated/irrated at $R \gtrsim 100\,R_{g}$ so the outer disk flux vectors flow {\em into} the disk. 
	\label{fig:maps.flux}}
\end{figure*}

\begin{figure*}
	\centering
	\includegraphics[width=0.75\textwidth]{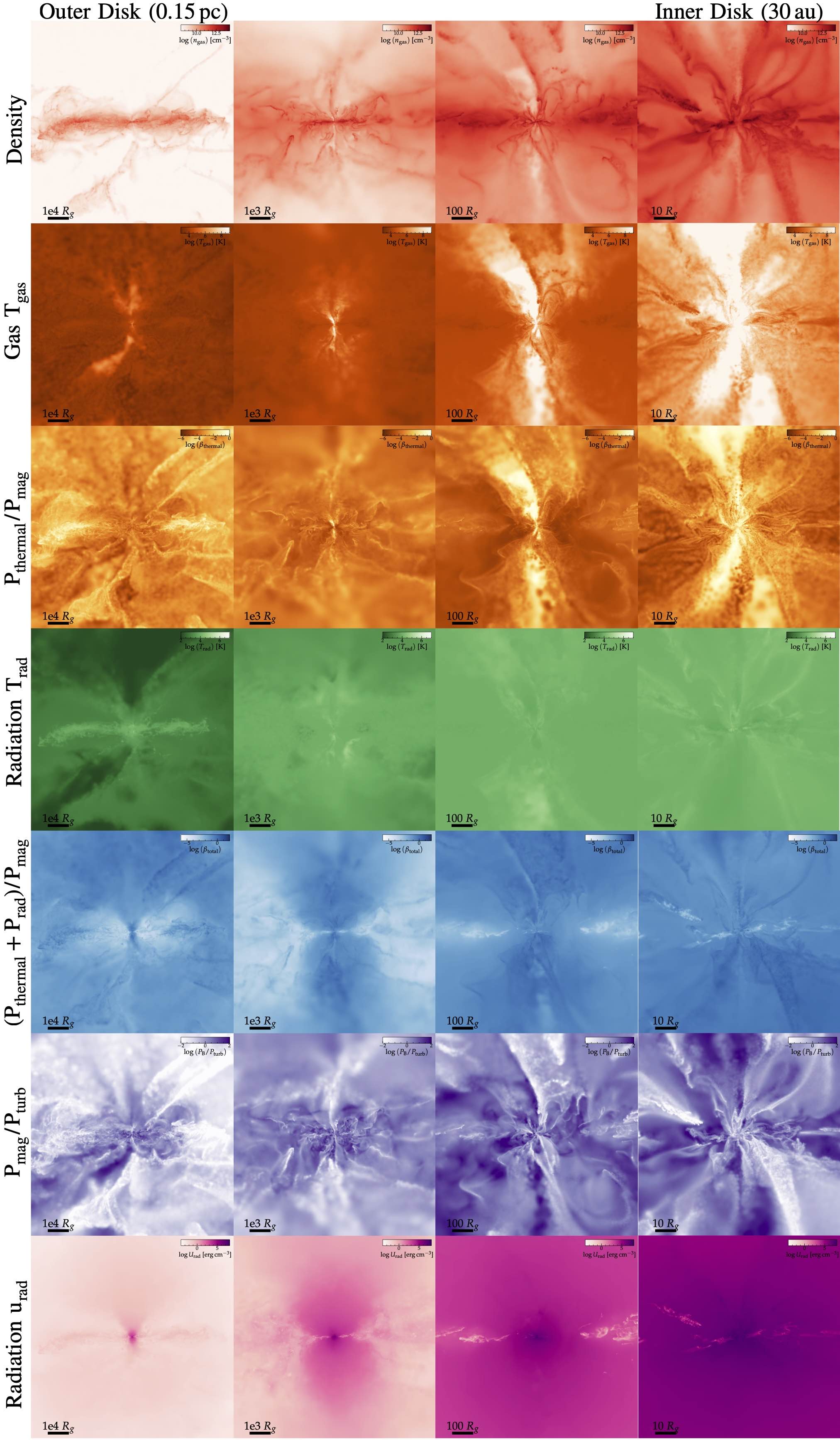} 
	\caption{Edge-on $Rz$ cylindrical wedge maps of quantities from Fig.~\ref{fig:maps.pinwheel}, in different box sizes (zooming in {\em left-to-right}). 
	\label{fig:maps.pinwheel.edgeon}}
\end{figure*}


\subsection{Behaviors That Change In the Inner Disk}
\label{sec:results:new}

Unlike \paperone-\papertwo, we now resolve the transition from a disk with negligible to important radiation fields, in all three of our different accretion-rate cases.

{\bf The Disk Self-Ionizes.} As temperatures rise towards the inner disk and the local radiation field becomes harder, the disk transitions from primarily atomic to fully ionized. The ionized gas mass fraction begins to rise significantly, but there is still a mix of ionized and atomic hydrogen, interior to $\lesssim 2 \times10^{16} {\rm cm}$ ($\lesssim 10\,$light-days or $\lesssim 6000\,R_{g}$). This rises steadily until at $\lesssim 10^{15}\,{\rm cm}$ or $\sim 0.5\,$light-days ($\lesssim 300\,R_{g}$) the atomic mass fraction drops steeply, in our default ($f_{\dot{M}}=1$) simulation, though the radius where this occurs appears to scale with accretion rate (Fig.~\ref{fig:profiles.mdot}). For each of our different $\dot{M}$ runs, these radii are within a factor of $\sim 2$ of the analytically-predicted radius for the same $M_{\rm BH}$ and $\dot{M}$ for the simplified magnetically-dominated disk models in \citet{hopkins:multiphase.mag.dom.disks}.

{\bf The Flow Becomes Increasingly Radial.} The inner regions at $\ll 300\,R_{g}$ exhibit an increasing $\langle v_{R} \rangle$ and radial Reynolds stress $\Pi_{RR}$ (Fig.~\ref{fig:profiles.velocity}) which also manifests in stretching the magnetic field lines to enhance $B_{R}$ (Fig.~\ref{fig:profiles.magnetic}, \ref{fig:profiles.stress.vs.mdot}). 

{\bf Radiation Pressure Saturates at Similar Levels to Turbulent and Magnetic Pressures.} At radii $\gtrsim 0.01\,{\rm pc} \sim 10^{4}\,R_{g}$, $P_{\rm rad} \sim P_{\rm therm} \ll P_{\rm turb} \sim P_{\rm mag}$ -- i.e.\ radiation pressure is order-of-magnitude similar to thermal pressure and both are negligible dynamically. Relative to other pressures/energy densities in the disk, we see $P_{\rm rad}$ and $U_{\rm rad}$ rise as expected (Figs.~\ref{fig:maps.pinwheel}, \ref{fig:profiles.general}, \ref{fig:maps.pinwheel.edgeon}), so $\beta_{\rm tot} \equiv (P_{\rm rad}+P_{\rm therm})/P_{\rm B}$ increases,\footnote{Here and throughout, we define $P_{\rm rad} \equiv u_{\rm rad}/3$, and $\beta_{\rm tot} \equiv (P_{\rm rad}+P_{\rm therm})/P_{\rm B}$, which follow standard literature conventions. However we caution that the interpretation of $\beta_{\rm tot}$ and $P_{\rm rad}$ depends on the degree of radiation-gas coupling: $P_{\rm rad}$ will only act like a ``pressure'' dynamically in the local tight-coupling (infinitely small radiation mean-free-path) limit, which we show in \S~\ref{sec:lte} is rarely true in the simulations. A more meaningful comparison of the actual dynamical importance of radiation and magnetic fields or gravity is manifest in the salient forces/accelerations, discussed in \S~\ref{sec:discussion}.} until saturating  at values $P_{\rm rad} \sim 0.1-1 \, P_{\rm B}$ or so, i.e.\ radiation energy density comparable to the turbulent or fluctuating magnetic energy density, but notably a factor of a few smaller than the mean magnetic energy density. There is substantial variation in $P_{\rm rad}$ in the multi-phase regions of the disk and between the midplane and polar regions (Figs.~\ref{fig:maps.pinwheel}, \ref{fig:maps.pinwheel.edgeon}). Because the simulations with different $\dot{m}$ have radiation pressure roughly $\propto \dot{m}$, and magnetic pressure $\propto \rho\,v_{A}^{2} \propto \dot{m}$, this ratio does not depend strongly on $\dot{m}$ (Fig.~\ref{fig:profiles.mdot},  \ref{fig:maps.pinwheel.mdot}, \ref{fig:maps.edgeon.mdot}).

{\bf Energy Transport Becomes Advective at the smallest radii.} Per Fig.~\ref{fig:profiles.flux}, in the outer portions of the disk, $R \gtrsim 10^{-3}\,{\rm pc} \sim 3\times10^{15}\,{\rm cm} \sim {\rm ld}\sim 800\,R_{g}$, the radiation flux escaping from the disk is the primary energy loss mechanism, and corresponds to the expected gravitational energy flux in a Keplerian thin disk, $F_{\rm rad} \sim F_{\rm grav} \sim (3/8\pi)\dot{M}\Omega^{2}\,(1-\sqrt{3 R_{g}/R})$ \citep{shakurasunyaev73}. This is comparable to but somewhat larger than the transport of energy through the disk, an order-unity fraction of which goes into compression of the toroidal field, since $B_{\phi}^{2} \propto R^{-(2-3)}$ rises rapidly towards smaller radii; see \paperthree\ and Figs.~\ref{fig:profiles.general}. And the radiation energy densities in the midplane are higher than $F/c$ by a factor of $\sim \tau$, more or less as expected (Figs.~\ref{fig:profiles.flux}, \ref{fig:profiles.mdot}). Note that most of the total flux (Fig.~\ref{fig:profiles.flux}) at large radii is the flux from the inner disk, which both illuminates the outer disk and represents escaping radiation in the polar regions. But interior to $R \ll 1\,$ld in our supercritical (highest accretion rate, $f_{\dot{M}}=1$) simulation, and within $\lesssim 10\,R_{g}$ in the sub-critical simulations (approaching the ISCO), we see the energy transport becomes primarily advective, and the escaping radiation flux falls increasingly below $F_{\rm grav}$ (Fig.~\ref{fig:profiles.mdot}). We also see this in the accretion timescales becoming comparable to or shorter than radiation escape timescales in Figs.~\ref{fig:profiles.general}, \ref{fig:profiles.mdot}. Note, per Fig.~\ref{fig:profiles.general}, that most of the advected energy density is not in radiation: summing the energies of advected turbulent motions, radiation, and magnetic fields, this is largely dominated by the advected magnetic mean-field ($B_{\phi}$) energies (Fig.~\ref{fig:profiles.general}), at these radii. 

\begin{figure*}
	\centering
	\includegraphics[width=0.32\textwidth]{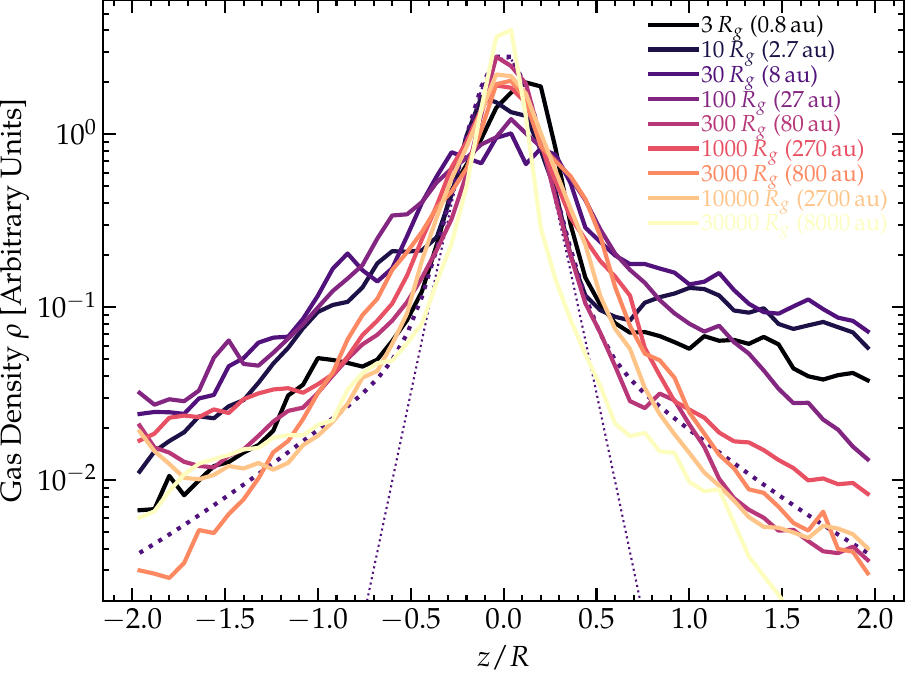} 
	\includegraphics[width=0.32\textwidth]{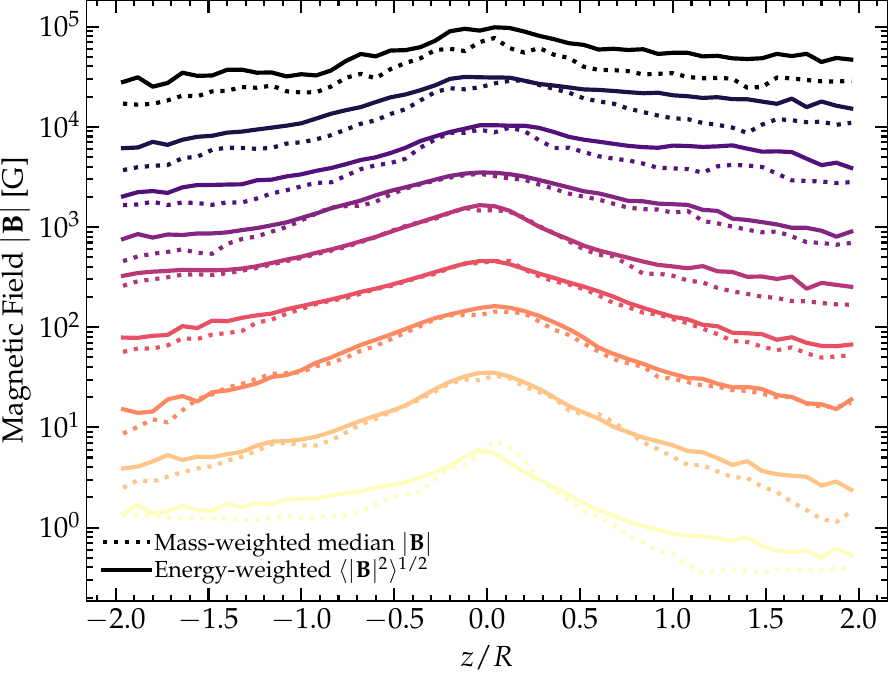} 
	\includegraphics[width=0.32\textwidth]{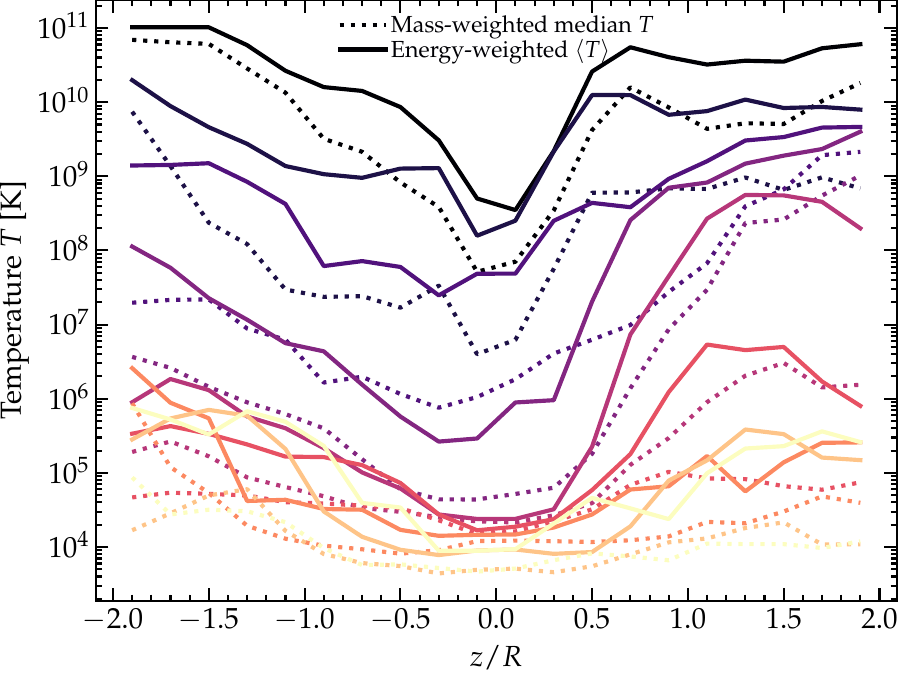} 
	\includegraphics[width=0.32\textwidth]{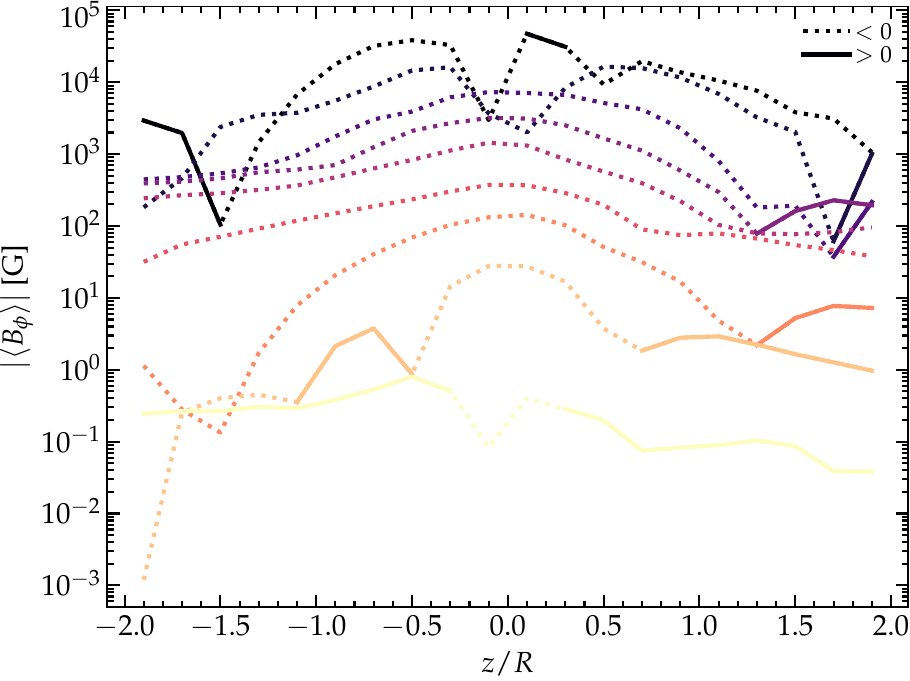} 
	\includegraphics[width=0.32\textwidth]{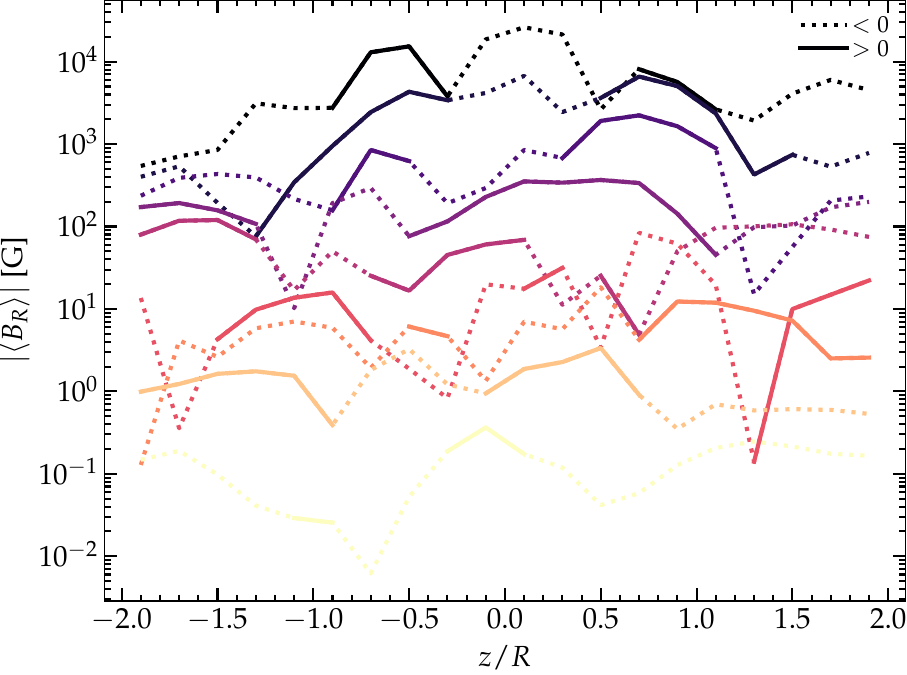} 
	\includegraphics[width=0.32\textwidth]{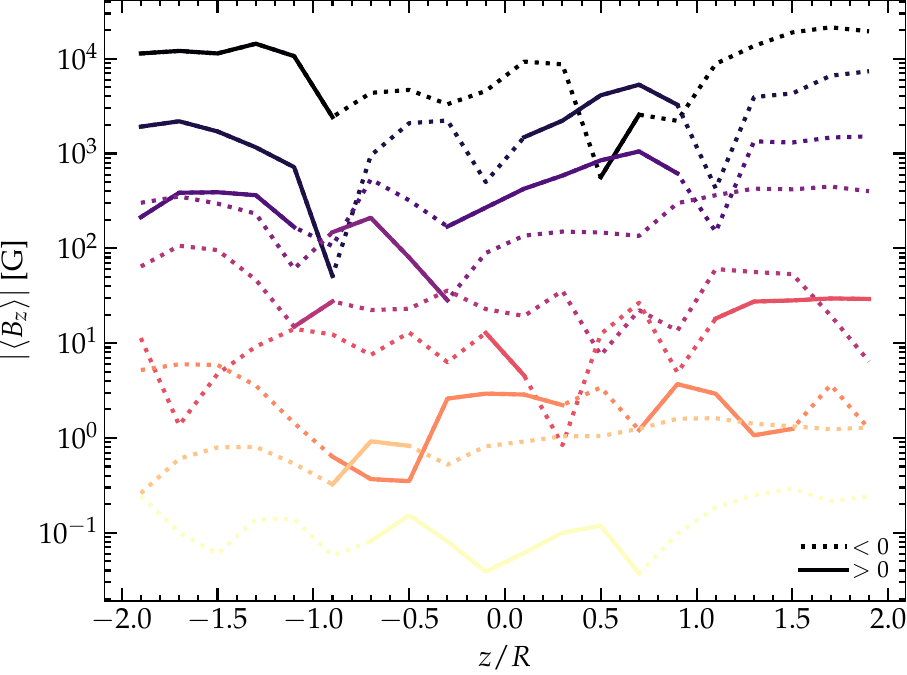} 
	\includegraphics[width=0.32\textwidth]{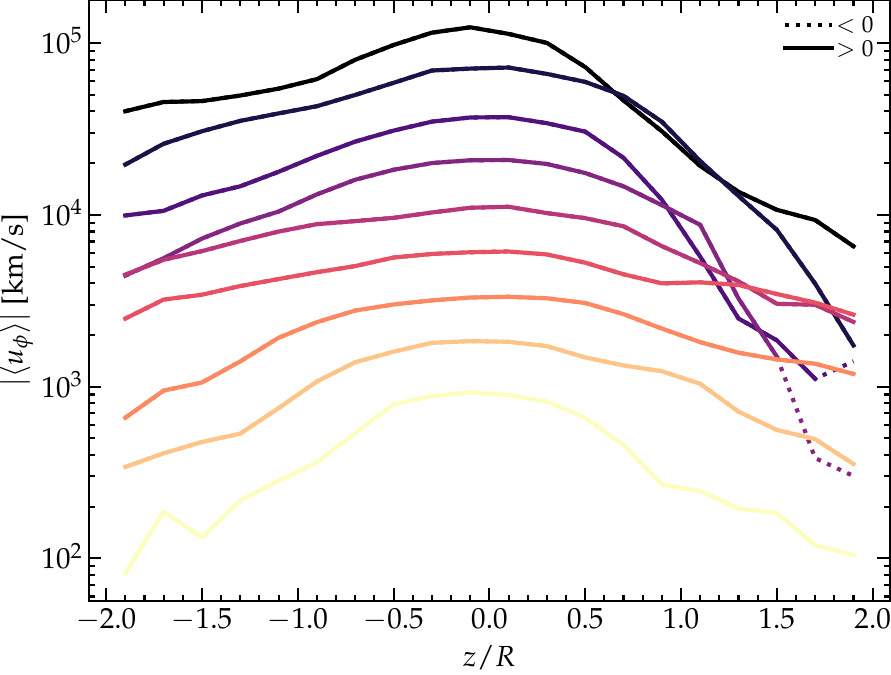} 
	\includegraphics[width=0.32\textwidth]{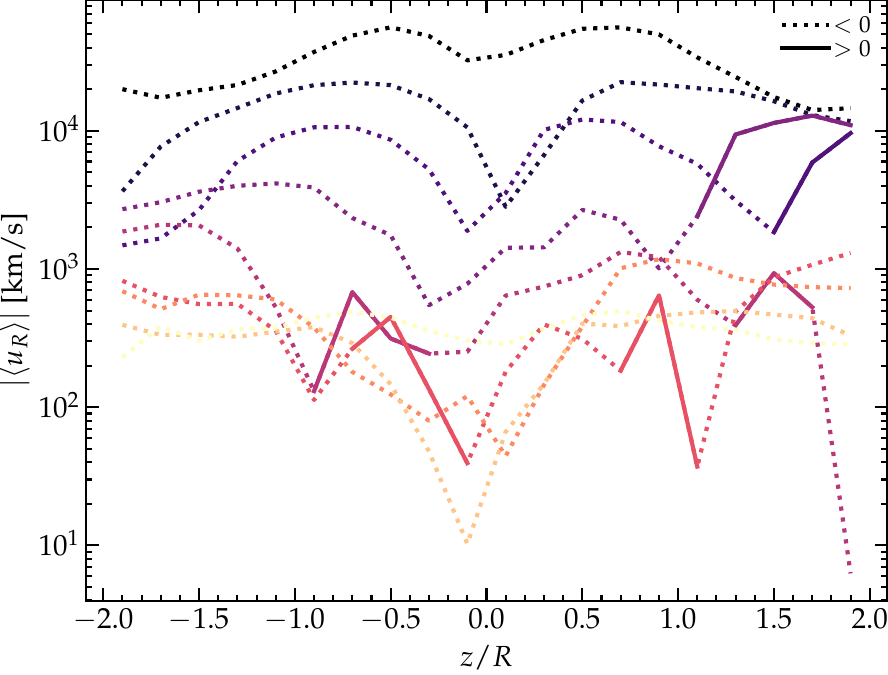} 
	\includegraphics[width=0.32\textwidth]{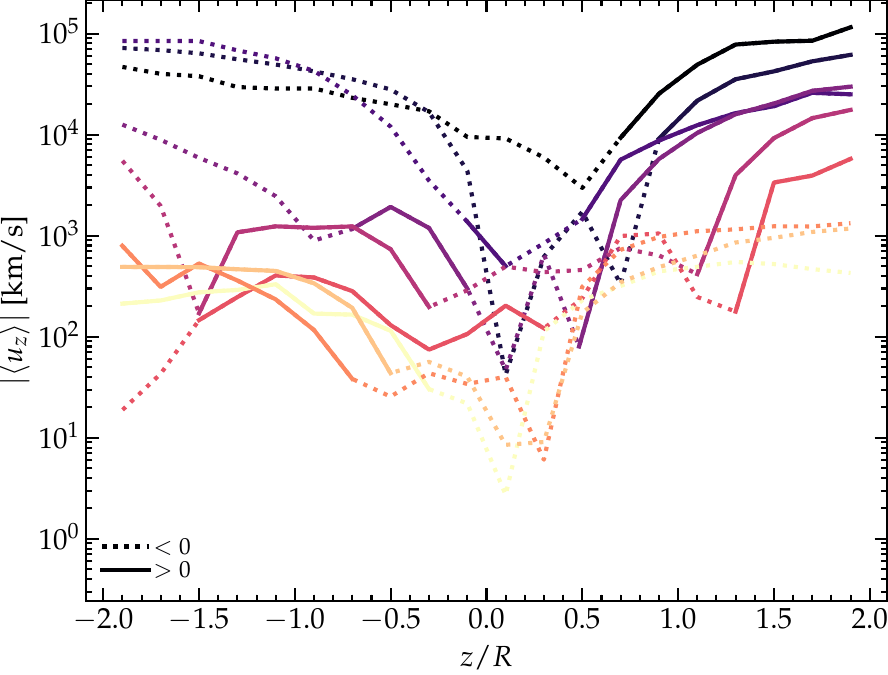} 
	\includegraphics[width=0.32\textwidth]{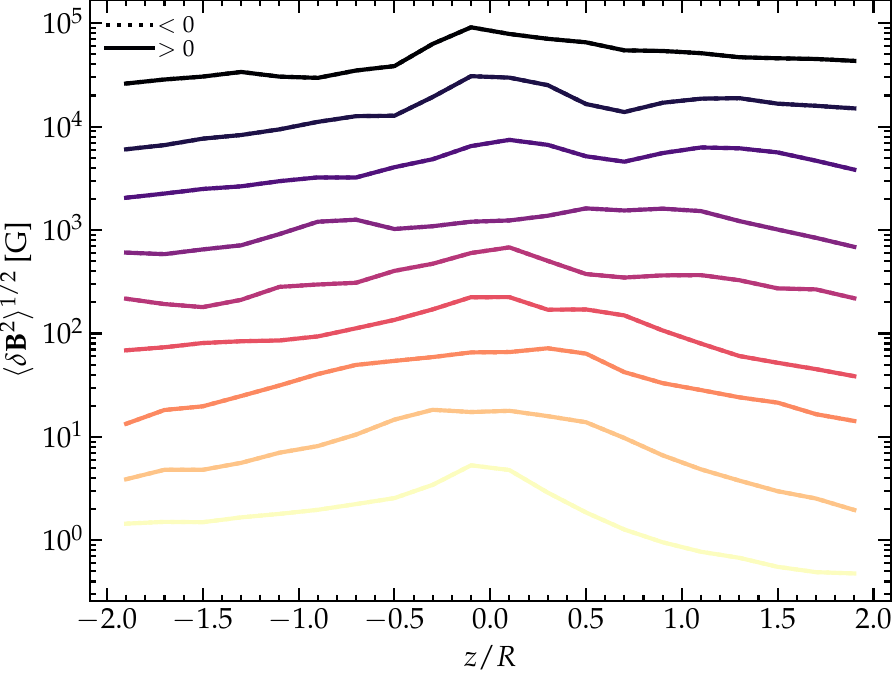} 
	\includegraphics[width=0.32\textwidth]{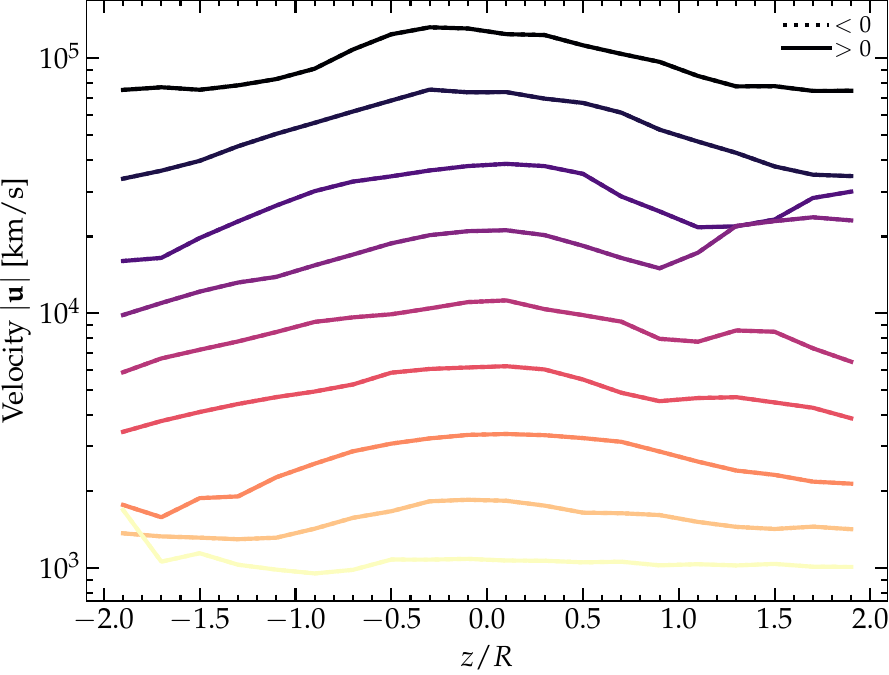} 
	\includegraphics[width=0.32\textwidth]{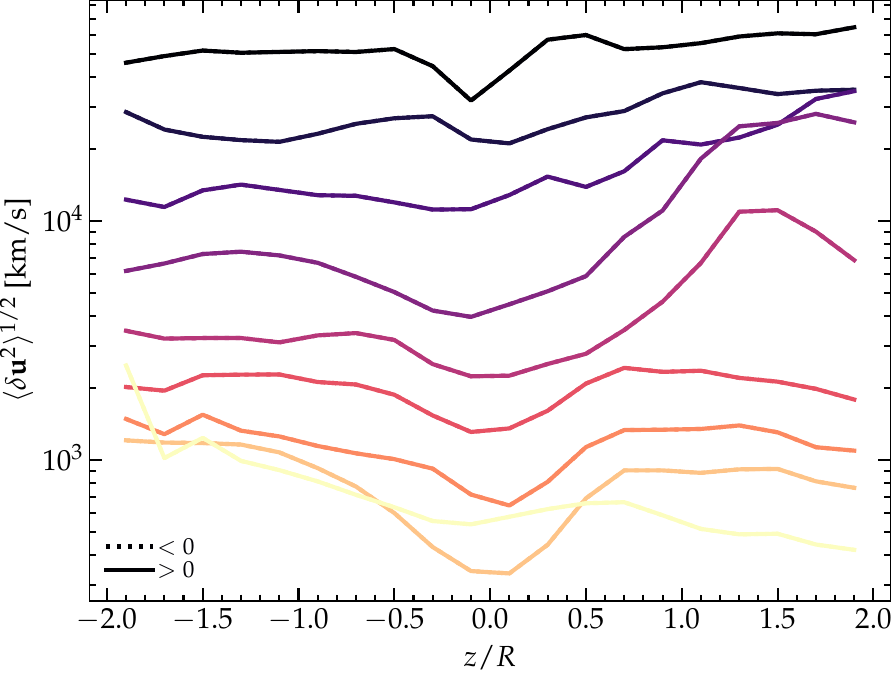} 
	\caption{Vertical profiles at different cylindrical radii. 
	{\bf (1)} Density $\rho$.
	{\bf (2)} Magnetic field $|{\bf B}|$ (primarily $B_{\phi}$).
	{\bf (3)} Gas temperature $T_{\rm gas}$.
	{\bf (4)} $\langle B_{\phi} \rangle$.
	{\bf (5)} $\langle B_{R} \rangle$.
	{\bf (6)} $\langle B_{z} \rangle$.
	{\bf (7)} $\langle v_{\phi} \rangle$.
	{\bf (8)} $\langle v_{R} \rangle$.
	{\bf (9)} $\langle v_{z} \rangle$.
	{\bf (10)} $\langle |\delta {\bf B}|^{2} \rangle^{1/2}$.
	{\bf (11)} $\beta$.
	{\bf (12)} $|{\bf v}|$.
	{\bf (13)} $\langle |\delta {\bf v}|^{2} \rangle^{1/2}$.
	\label{fig:profiles.vertical}}
\end{figure*}

{\bf Radiation Escapes in a Widening Bicone \&\ Illuminates the Disk.} Fig.~\ref{fig:maps.flux} and Figs.~\ref{fig:maps.fields.mdot}, \ref{fig:maps.edgeon.mdot} show that the radiation flux escapes in a broad-angle bicone, the effective opening angle of which appears to broaden at larger radii. The disk midplane is shielded, and features very little net flux in the Lagrangian frame (Fig.~\ref{fig:profiles.flux}), but the surface layers are strongly illuminated at all $\dot{m}$ (Fig.~\ref{fig:maps.fields.mdot}). This owes partly to the thick, flaring disk structure (Figs.~\ref{fig:zoomies.cyl}, \ref{fig:maps.pinwheel.edgeon}, \ref{fig:maps.edgeon.mdot}) but also to indirect scattering onto the disk. We see this in Fig.~\ref{fig:maps.flux} as flux streamlines at intermediate angles appear to ``curve down'' before hitting the disk, a behavior which is more dramatic at higher $\dot{m}$. This occurs because of scattering in the lower, but still significant gas densities above the disk (and those gas densities are higher at higher $\dot{m}$), where the gas is mostly ionized and so has a high scattering opacity but densities slowly decreasing with $|z|$ (Fig.~\ref{fig:profiles.vertical}). The effective size of the emitting region is therefore large: we can see from Fig.~\ref{fig:maps.flux} that a significant ($\mathcal{O}(1)$) fraction of the sightlines from the inner disk will have their flux reprocessed at radii as large as the outermost disk ($\gtrsim 10^{16}\,{\rm cm}$). This is confirmed in \citet{kaaz:2024.hamr.forged.fire.zoom.to.grmhd.magnetized.disks}, and will be modeled in more detail in future work (Bardati et al., in prep.).

{\bf There Is An Extended Scattering ``Photosphere.''} We see in Figs.~\ref{fig:profiles.flux} \&\ \ref{fig:maps.flux} that there is significant scattering, with deflection of photon trajectories and slowing the escape of photons (making $u_{\rm rad} \gg |{\bf F}_{\rm rad}|/c$) even in the broad regions above the disk in the innermost regions. Measuring the mean column densities in different polar angles $\theta$, we see that the scattering optical depth ($\tau$ is primarily from electron-scattering here) in our highest-$\dot{M}$ simulation remains $\sim 5-10$ even along a broadly polar set of angles $3/4<|\cos{\theta}|<1$, out to a thousands of $R_{g}$, i.e.\ $\sim 0.5-1 \times 10^{16}\,{\rm cm^{-3}}$. This helps explain the scattering ``down'' into the disk described above, and enhances the reprocessing of radiation by the outer disk as well as providing a sort of ``warm absorber''-like Comptonizing medium for the continuum radiation. In our intermediate $f_{\dot{M}}=0.03$ simulation with $\dot{m} \sim 0.3-0.5$, the optical depths along the polar direction or in a volume-weighted sense are systematically smaller (as expected, owing to the systematically lower gas mass supply and therefore densities at all radii), but they remain non-negligible (order-unity) out to similar radii (Fig.~\ref{fig:profiles.mdot}). Only in our lowest accretion rate case ($f_{\dot{M}}=0.001$) with $\dot{m} \sim 0.02$ (where the disk itself is only marginally optically-thick) do we see optical depths $\ll 1$ along the polar/volume-filling directions. Still, even in this case, a non-negligible fraction of the light from the inner disk ($\gtrsim 10\%$) will be either scattered down from polar regions (with optical depths $\sim 0.01-0.1$) or intercepted by the outer disk (which still features $H/R \sim 0.1-1$ at large radii; Fig.~\ref{fig:maps.edgeon.mdot}). 

{\bf Emergent Radiation Temperatures are Modest.} In the innermost regions, there are regions where the radiation temperature becomes quite large, $\sim 10^{6}$\,K (Figs.~\ref{fig:profiles.general}, \ref{fig:profiles.mdot}). However, this is strongly reprocessed and, because of the radiative efficiency being such that most of the emission comes from a very broad range of radii, means that the emergent luminosity (even along the vertical ``cone'' in Fig.~\ref{fig:maps.flux}), has radiation temperatures which are much more modest (Figs.~\ref{fig:maps.pinwheel}, \ref{fig:profiles.general}, \ref{fig:maps.pinwheel.edgeon}). Indeed, in the direct escaping ``cone'', the closest to a ``direct'' sightline to the inner disk (Fig.~\ref{fig:maps.flux}), the escaping radiation at the furthest distances photons have diffused here gives an emergent effective blackbody temperature of $\sim 1-3\times10^{4}$\,K (Fig.~\ref{fig:profiles.general}) where it is escaping vertically from the outer thermalized radii of the disk (see Fig.~\ref{fig:profiles.mdot}, \ref{fig:maps.edgeon.mdot} for different $\dot{m}$). Note that in the outer disk, the effective temperatures appear to drop below $< 10^{4}\,{\rm K}$. However, examination of the optical depths, gas temperatures, and chemical structure of the disk (e.g.\ Fig.~\ref{fig:profiles.general}) shows that this occurs at radii where the disk becomes primarily atomic, dust is no longer sublimated, and one can no longer assume local blackbody equilibrium $T_{\rm rad} \ne T_{\rm gas} \ne T_{\rm dust}$. Since our effective radiation temperature evolved in-code is determined by separately evolving the photon energy density and number density, it is a {\em photon number weighted} effective radiation temperature. That means that even a small fraction ($\sim 1-10\%$) of the optical/UV radiation being reprocessed by dust and re-emitted with temperatures $T_{\rm dust} \sim 100-1000\,$K will drop the weighted radiation temperature to $T_{\rm rad} \ll 10^{4}\,$K. We expect a more detailed frequency-resolved radiation treatment would give rise to a highly multi-component spectrum in this regime (with separate contributions to the emergent QSO spectrum in escaping X-ray/EUV, UV/optical, NIR, and mid-far IR), but modeling this requires more detailed post-processing Monte Carlo radiation transport (Bardati et al., in progress). Given that this is happening embedded in a dusty torus and massive, high-redshift starburst galaxy (with galaxy-integrated star formation rate of $\sim 200\,{\rm M}_{\odot}\,{\rm yr}^{-1}$), it is likely that on large ($\gg $\,pc) scales the system will remain isotropically obscured and the radiation will be further downgraded, but this still gives us a useful proxy for the ``intrinsic'' quasar SED.

Some of these behaviors are anticipated, or at least qualitatively similar to standard thick or ``slim'' disk models for supercritical accretion \citep{paczynsky.wiita:1980.slim.disk,abramowicz:1988.slim.disks}. This is despite the fact that slim-disk models generally neglect magnetic pressure in the dynamics and energy transport of the disks entirely (the fields only appear implicitly in some assumed stress), and instead begin from assumptions like those in standard thin thermal-pressure-dominated $\alpha$-disks \citep{shakurasunyaev73}. Comparison to analytic models is discussed further below.

\subsection{Luminosities \&\ Radiative Efficiencies}
\label{sec:results:radeff}

We see in Fig.~\ref{fig:profiles.mdot} that for $\dot{m} \lesssim 1$, the bolometric luminosities $L_{\rm bol}$ are dominated by the energy from the central $\sim 10\,R_{g}$,\footnote{Note this is {\em not} the same as saying the effective or apparent size of the disk or ``emitting region'' at a given wavelength is $\sim 10\,R_{g}$, because almost all of this luminosity is reprocessed or scattered at larger radii.} and the inflow/accretion rates $\dot{M}$ are roughly constant with $R$ within the inner disk (central $\sim 100-1000\,R_{g}$). From direct comparison of these, we can estimate the radiative efficiency, $\epsilon_{r} \equiv L_{\rm bol}/\dot{M}\,c^{2}$, which is $\sim (0.15,\,0.1,\,0.02)$ for $f_{\dot{M}}=(0.001,\,0.03,\,1)$ ($\dot{m} \sim 0.02,\,0.3,\,20$). For $\dot{m} \lesssim 1$ these are similar to the canonical expected efficiency of any radiatively-efficient rotating disk, though we caution that in a full GRMHD calculation these could change at the order-unity level (as we only adopt a simplified PW potential). 

For $\dot{m} \gg 1$, a corollary of the point in \S~\ref{sec:results:new} that the flux becomes advective at radii well outside the ISCO is that the system becomes less radiatively efficient. Fig.~\ref{fig:profiles.flux} shows that the radiation fluxes fall well below $F_{\rm grav}$ at the smallest radii, and correspondingly the emerging luminosity has roughly equal contributions per logarithmic radius $R$ (so the total emergent luminosity grows logarithmically only as gas migrates further in, rather than increasing $\propto R^{2} F_{\rm grav} \propto R^{-1}$). The net luminosity emerging from the thermalized disk is $\sim$\,a few $\times 10^{45}\,{\rm erg\,s^{-1}}$ (a couple to a few times Eddington), more weakly dependent on the cutoff/innermost radius suggesting (seen visually in how the luminosity rises with radius weakly in Fig.~\ref{fig:profiles.flux}). This is consistent with the radiation interior to a trapping radius further out being advected inwards, also consistent with the flow seen in Fig.~\ref{fig:maps.flux}.

\subsection{Variability, Bursts, \&\ Magnetic State Changes}
\label{sec:bursts}

All of our simulations exhibit some variability in accretion rates, luminosities, and time-and-space-dependent fluxes and radiation temperatures (Fig.~\ref{fig:maps.demo.bursts}). This will be the subject of future study, as it both informs the physical understanding and interpretation of the simulations and observed quasar variability, and presents potential new observational tests and constraints on these accretion disk models, from X-ray through infrared wavelengths. Given the extensive comparisons needed to quantify this, we have focused here on qualitative properties which are robust to short-timescale variations, though note where they are important.

More radical, qualitative ``state changes'' in the simulations are rare, but we do see them in some cases (illustrated alongside more general variability in the morphology and topology of the magnetic and velocity fields of the inner disk in Fig.~\ref{fig:maps.demo.bursts}). Specifically, we have only seen this occur in the inner disk ($\lesssim 200\,R_{g}$), in our highest-$\dot{M}$ (super-critical, $f_{\dot{M}}=1$, $\dot{m} \gg 1$) simulation, and only if we refine to the highest level with the inner boundary at $\lesssim 3\,R_{g}$ (as opposed to even at $\sim 15-30\,R_{g}$), including our full physics (most importantly radiation pressure forces and MHD; see \S~\ref{sec:sensitivity}). And seeing such a change typically require running for many dynamical times ($\sim 10^{4}-10^{5}\,\Omega^{-1}$) at the smallest radii. Given all of that, we semi-stochastically see the following series of events. 
(1) Something (for example, a fluctuation in the accretion rate and disk size/incoming angular momentum direction/warp of the inner disk) makes the outflows somewhat stronger and, more importantly, wider angle and/or able to intercept/entrain some of the inner disk. In our simulations, these outflows begin as radiation-pressure driven (see \S~\ref{sec:outflows}), potentially explaining the critical role of radiation pressure and zooming into the smallest radii (where the radiation is generated). 
(2) The outflows then ``blow out'' a cavity, depleting the central ($\sim 10\,R_{g}$) disk density. 
(3) The field lines are stretched with the outflow, weakening the mean toroidal and strengthening the coherent poloidal/vertical field, until there is a stronger (volume-averaged) net vertical (seeded by turbulent poloidal loops) flux threading the midplane. 
(4) The outflow continues propagating, and resembles a ``flare'' at larger radii. As the outflow propagates, it carries radiation with it from the central disk which had been trapped/advected with the accretion flow, as well as blowing-off some surface layers of the disk, lowering the opacity locally and allowing trapped photons to escape. This produces a propagating ``pulse'' of photons with higher radiation temperature (potentially observed in EUV or X-rays) and instantaneously high radiative efficiency. Naively, the initial duration of the flare if it escapes would be something like a few $\sim 10^{2}-10^{4}\,G\,M_{\rm BH}/c^{3}$ (hours to months), with a total flare energy typically in the $\sim 10^{51}-10^{52}\,{\rm erg}$ range, but the observed radiation frequencies and durations could be strongly modified by reprocessing further out. 
 (5) The inner accretion disk ($<100-1000\,R_{g}$) is left in a much more MAD-like state (see \citealt{kaaz:2024.hamr.forged.fire.zoom.to.grmhd.magnetized.disks}). There is a dominant net vertical field $\langle B_{z} \rangle$ which has $v_{A,\,z} \sim 0.2-1\,v_{\rm K}$ (can hold up the diffuse gas), with accretion proceeding through this magnetosphere via Rayleigh-Taylor-like interchange instabilities or non-axisymmetric structures (clumps, filaments, etc.). 
(6) Run long enough, we have a couple examples which eventually have enough mass pile-up and accrete ``through'' this magnetospheric radius that they rebuild a toroidal-dominated dense disk in the innermost radii, which then fills the magnetosphere from the inside-out, returning the system to a mean toroidal-flux state. But this can require a few orbital times at the radii {\em outside} of the magnetosphere, which translates to $\gtrsim 10^{6}\,G\,M_{\rm BH}/c^{3}$ (so a very long time relative to near-horizon timescales). 

These transitions are broadly similar to those shown in \citet{kaaz:2024.hamr.forged.fire.zoom.to.grmhd.magnetized.disks}, who extend our simulations by taking one of our early snapshots of our highest-$\dot{m}$ ($f_{\dot{M}}=1$) simulation, and map it to their fully general-relativistic GR-MHD code {\small HAMR} (albeit with a more simplified treatment of radiation and thermochemistry). Interestingly, while the transitions above take a long time to occur here, they appear in \citet{kaaz:2024.hamr.forged.fire.zoom.to.grmhd.magnetized.disks} more rapidly near-horizon, and occur there even in runs without radiation. This may be because the different GR horizon boundaries and other effects produce a strong jet/wind launched from near-horizon scales in \citet{kaaz:2024.hamr.forged.fire.zoom.to.grmhd.magnetized.disks}, even without radiation. This suggests that outflows may play an interesting causal, or at least self-reinforcing, role in such state changes.

Given the restricted conditions/simulations under which we see these changes occur, we focus throughout on the more steady-state (non-burst) behaviors in the simulations. Still, these outbursts are interestingly similar in their timescales, energetics, and radiation properties to observed flares (and potentially some spectral state changes) seen in luminous/high-$\dot{m}$ AGN \citep[see e.g.][]{boller:1997.extreme.xray.variability.nls1.two.dex,graham:2017.extreme.qso.optical.variability.flares.properties,graham:2020.changing.state.qsos.usually.dimmer.but.sometimes.very.bright,trakhtenbrot:2019.flares.with.high.velocity.outflows.associated.in.qsos,ni:2020.extreme.xray.brightening.order.of.magnitude.in.quasar,frederick:2021.optical.flares.nls1.with.high.velocities,zhu:2021.extreme.xray.variability.radio.loud.qsos}. So they will be studied and compared to observations in detail in future work.

\section{Discussion}
\label{sec:discussion}

\subsection{Physics of Radiation Pressure Saturation in a Trans/Super-sonically Turbulent Disk}
\label{sec:rad.pressure.saturation}

In \paperthree\ and \citet{hopkins:multiphase.mag.dom.disks}, we argued analytically that even at $\dot{m} \gg 1$, the inner disk should feature roughly $P_{\rm rad} \sim P_{\rm mag} \sim P_{\rm turb} \gg P_{\rm thermal}$, not $P_{\rm rad} \gg P_{\rm mag}$. As the inner disk is optically thick, the radiation pressure should be $P_{\rm rad} \approx (4/3)\,(\sigma_{B}/c)\,T_{\rm mid}^{4}$, in terms of the radiation temperature $T_{\rm mid}$, with effective temperature $T_{\rm eff}$ defined by the surface flux $F_{\rm rad} = \sigma_{B}\,T_{\rm eff}^{4}$ and related to $T_{\rm mid}$ in general by some function $T_{\rm mid}^{4}=f_{\tau}\,T_{\rm eff}^{4}$ ($f_{\tau}$ defined below). If we assume gravitational energy losses come out in radiation so $\sigma_{B}\,T_{\rm eff}^{4} \approx (3/8\pi)\,\dot{M}\,\Omega^{2}$, then $P_{\rm rad} \approx (4\,f_{\tau}/3\,c)\,(3/8\pi)\,\dot{M}\,\Omega^{2}$. Meanwhile these models assume (consistent with the simulations) trans-\Alf{ic} turbulence, so magnetic and turbulent pressure scale similarly with $P_{\rm mag+turb} \equiv P_{\rm mag} + P_{\rm turb} \approx \rho\,(v_{A}^{2} + v_{t}^{2}) \approx \Sigma_{\rm gas}\,v_{t}^{2}/H$. If Reynolds and/or Maxwell stresses dominate inflow, or equivalently gravitational energy goes into kinetic/magnetic energy before being dissipated radiatively after shocks/reconnection, then $\Sigma_{\rm gas}\,v_{t}^{2}\,\Omega \sim (3/4\pi)\,\dot{M}\,\Omega^{2}$, so $P_{\rm mag+turb} \sim (2/v_{t})\,(3/8\pi)\,\dot{M}\,\Omega^{2}$. Thus $P_{\rm rad}/P_{\rm mag+turb} \sim (4\,f_{\tau}/3\,c) / (2/v_{t}) \sim v_{t} / (c/f_{\tau})$. 

So $P_{\rm rad} \gg P_{\rm mag+turb}$ requires $c/f_{\tau} \ll v_{t}$. If radiation transport is dominated by diffusion, then $f_{\tau} \sim \tau$ and $c/f_{\tau} \sim c/\tau$ is the bulk radiation transport speed. But if $c/\tau \ll v_{t}$, then diffusion cannot dominate radiation transport -- vertical transport will be turbulent/advective/convective, and we must modify $f_{\tau}$. Consider a mixing length-type approximation: the radiation flux in steady state should be vertically constant so $F_{\rm rad} \sim D_{\rm rad}\, \nabla a\,T_{\rm mid}^{4} \sim \sigma_{B}\,T_{\rm eff}^{4}$ with $D_{\rm rad}$ the radiation transport coefficient $\approx c/(\kappa\,\rho)$ for pure diffusion and $\approx v_{t}\,\ell_{t} \approx v_{t}^{2}/\Omega$ for turbulent transport with largest eddy turnover time $\Omega^{-1}$. This gives $T_{\rm mid}^{4}/T_{\rm eff}^{4} \equiv f_{\tau} \approx c\,H/D_{\rm rad}$ which in turn gives the usual $f_{\tau} \approx \tau$ for radiative diffusion, and $f_{\tau} \approx c\,H\,\Omega/v_{t}^{2} \approx c/v_{t}$ for turbulent transport. So when $\tau \lesssim c/v_{t}$, the transport will be via radiative diffusion with $f_{\tau} \sim \tau$ (provided we still have $\tau \gtrsim 1$), with $P_{\rm rad} \lesssim P_{\rm mag+turb}$, and when $\tau \gtrsim c/v_{t}$, we transition to turbulent vertical transport with $f_{\tau} \sim c/v_{t}$ and therefore $P_{\rm rad} \sim P_{\rm mag+turb}$. 

Key to this is that the turbulence is trans or super-sonic. In highly sub-sonic turbulence ($v_{t} \ll c_{s}$) in a shearing disk, the maximum turbulent coherence length/driving scale limited by shear is $\sim v_{t}/\Omega \ll c_{s}/\Omega \sim H$. This would mean that the turbulence would be confined to a midplane layer with height $\ll H$, so it is not even clear if we could safely assume eddies actually could carry radiation all the way to the photosphere at $|z| \sim H$. But even if we assume such turbulence is volume-filling in the disk with said coherence length, in the derivation above in the turbulence-dominated limit $f_{\tau} \rightarrow c\,H\,\Omega/v_{t}^{2} \rightarrow (c_{s}/v_{t})\,(c/v_{t})$, and so (1) the turbulent-transport-dominated limit is much harder to achieve, in practice, and (2) even in the said limit, we would have $P_{\rm rad}/P_{\rm mag+turb} \sim v_{t} / (c/f_{\tau}) \sim c_{s}/v_{t} \gg 1$. Thus, for {\em subsonic} turbulence, disks can become radiation-pressure dominated in the manner envisioned in e.g.\ SS73. But for a {\em trans/supersonically} turbulent disk, as inevitably occurs when $\beta_{\rm thermal} \ll 1$ in the disk, then the disk self-regulates to $P_{\rm rad} \sim P_{\rm mag+turb}$, and never truly becomes radiation-pressure dominated. Though the radiation pressure is, by definition, non-negligible here.

Note that saturation with $P_{\rm rad} \sim P_{\rm mag,\,turb}$ has also been widely seen in more idealized simulations of supercritical accretion flows beginning from a thin, thermal-pressure dominated SS73-like initial condition \citep{jiang:2019.superedd.sims.smbh.prad.pmag.modest.outflows,davis:2020.mhd.sim.acc.disk.review,utsumi:2022.superedd.rmhd.sims.accretion.mag.comparable.rad,huang:2023.rmhd.sims.rad.similar.mag.pressure}. There, the ``direction'' in which this solution is approached is somewhat opposite that here, as the radiation pressure puffs up the disk and drives stronger turbulence which leads to a turbulent dynamo saturating the turbulent $P_{\rm mag,\,turb}$ near $P_{\rm rad}$. But it suggests this is a robust state. The major difference is that these systems also have a large midplane mean-field component, so can saturate with the mean/total magnetic pressure somewhat larger than $P_{\rm rad}$. 

\subsection{Physics of the Radiative Efficiencies}
\label{sec:rad.efficiency}

For similar reasons, we expect the inner disk to be radiatively efficient for $0.01 \lesssim \dot{m} \lesssim 1$, and to become increasingly radiatively inefficient at $\dot{m} \gg 1$. As discussed in \citet{hopkins:multiphase.mag.dom.disks}, even with the turbulent mixing of radiation in \S~\ref{sec:rad.pressure.saturation}, because the opacities do not discontinuously transition to $\tau \ll 1$ at $|z| \sim H$ (the coherence length of the turbulence) -- i.e.\ there is no sharp ``edge'' of the disk and $H$ is not dramatically smaller than $R$ --  the turbulence should not qualitatively change the ability of radiation to escape in the radiatively saturated regime of \S~\ref{sec:rad.pressure.saturation}. There, the diffusive radiation escape speed is $\sim c/\tau_{\rm eff}$ (where $\tau_{\rm eff}$ can reflect the minimum $\tau$ seen over the eddy lifetime in the turbulence-dominated regime), which should become smaller in our highest-accretion-rate run ($f_{\dot{M}}=1$, $\dot{m} \gtrsim 20$) than the advective speed $v_{R}$ at $R \lesssim 300-1000\,R_{g}$ according to Figs.~\ref{fig:profiles.general} \&\ \ref{fig:profiles.velocity}. This corresponds well to where we measure the advective flux becoming larger than the comoving radiation flux in Fig.~\ref{fig:profiles.flux}. This is also consistent with the radii predicted for this transition from the analytic disk models in \citet{hopkins:multiphase.mag.dom.disks}. 

The major difference between this and classic advection-dominated ``slim disk'' scenario \citep{paczynsky.wiita:1980.slim.disk,abramowicz:1988.slim.disks} is that those models and other like them assume (as did SS73) that $\beta \gg 1$ throughout the midplane, which means that they only account for advection of thermal energy. Here, we plainly see from Fig.~\ref{fig:profiles.general}, and can directly check in Fig.~\ref{fig:profiles.flux}, that most of the advected energy is magnetic, particularly in the form of the dominant mean toroidal field. This modifies the solutions as discussed below.

On the other hand, for our $\dot{m} \ll 1$ simulations we see in Fig.~\ref{fig:profiles.mdot} that the disk is radiatively efficient, with radiation escape times $\sim r /(c/\tau_{\rm eff})$ which are fast because of the modest $\tau \sim 1$ in the disk; specifically, faster than accretion timescales down to the ISCO ($\sim r / |\langle v_{r}\rangle| \sim 10\,\Omega^{-1}$). This is expected from the arguments above and models in \citep{hopkins:multiphase.mag.dom.disks}. 

\begin{figure*}
	\centering
	\includegraphics[width=0.9\textwidth]{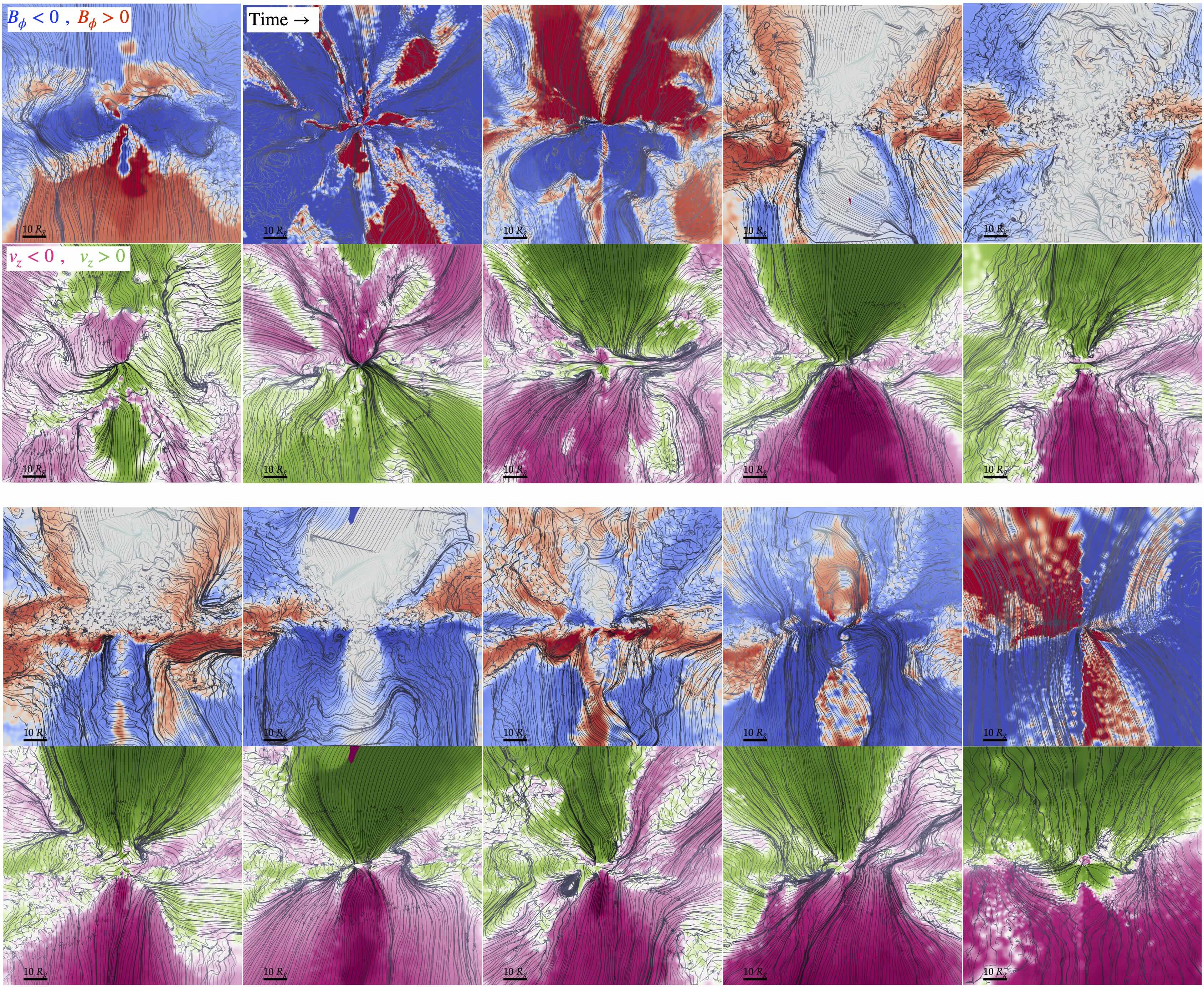} 
	\caption{Time series of edge-on maps of the magnetic field (showing sign of $B_{\phi}$, as Fig.~\ref{fig:maps.b}) and velocity (sign of $v_{z}$, as Fig.~\ref{fig:maps.vel}), in the inner disk ($<100\,R_{g}$) of our highest-$\dot{m}$ (supercritical; $f_{\dot{M}}=1$) simulation. Snapshots span a timescale $\sim 10^{6}\,G M_{\rm BH}/c^{3}\sim 2\,$years, but not uniformly (chosen to highlight different behaviors rather than quantitative time evolution). As discussed in \S~\ref{sec:bursts}, the detailed morphology can vary considerably, and there are even times of inflow and times where the symmetry properties of the toroidal field change from even about the midplane to odd, sometimes associated with flares and strong bursts of outflow which can disrupt the inner disk. These are discussed in \citet{kaaz:2024.hamr.forged.fire.zoom.to.grmhd.magnetized.disks}, and the time variability of these simulations will be studied in future work.
	\label{fig:maps.demo.bursts}}
\end{figure*}

\begin{figure*}
	\centering
	\includegraphics[width=0.32\textwidth]{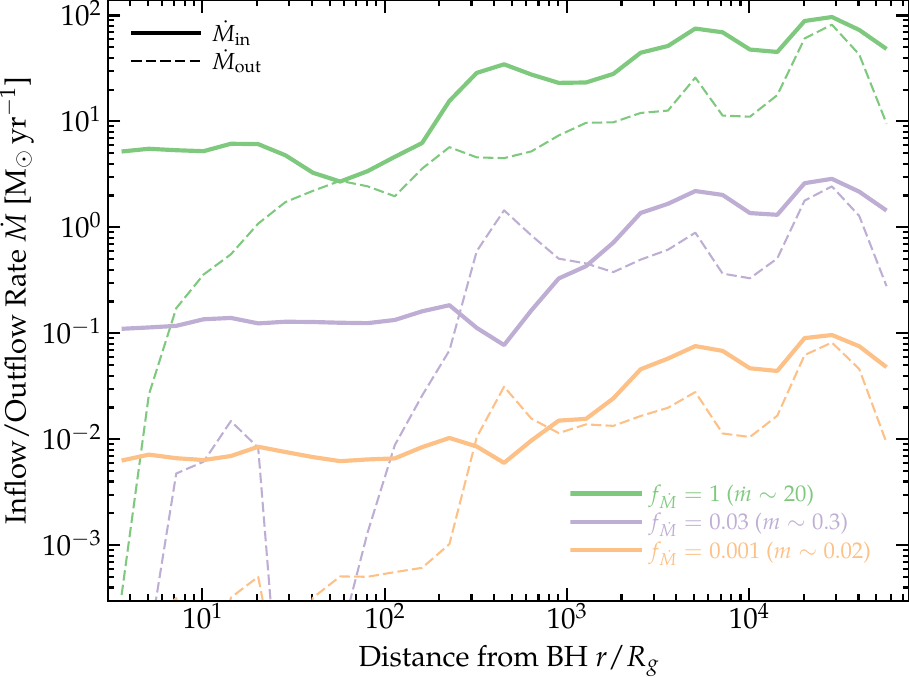} 
	\includegraphics[width=0.32\textwidth]{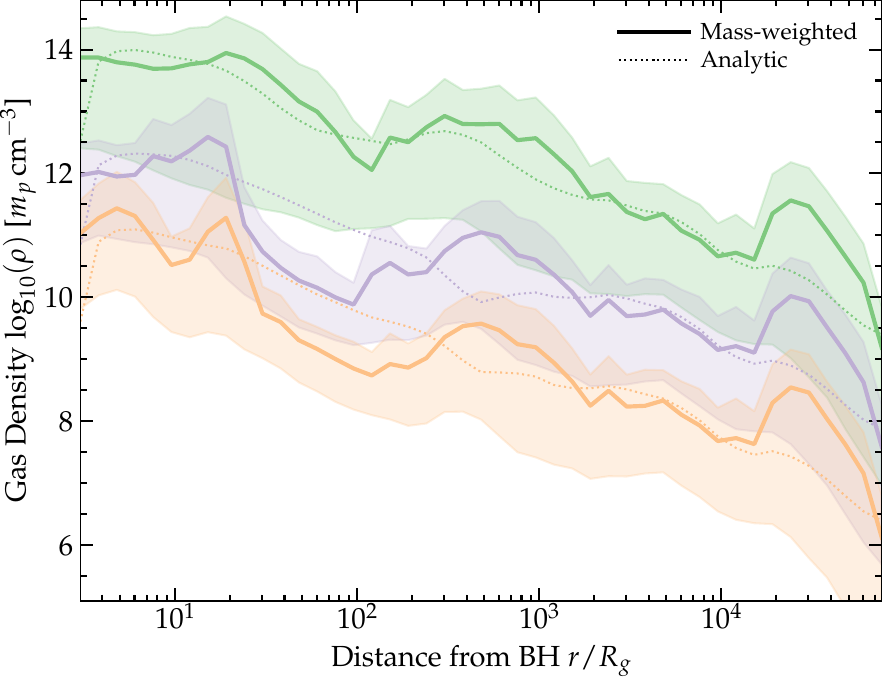} 
	\includegraphics[width=0.32\textwidth]{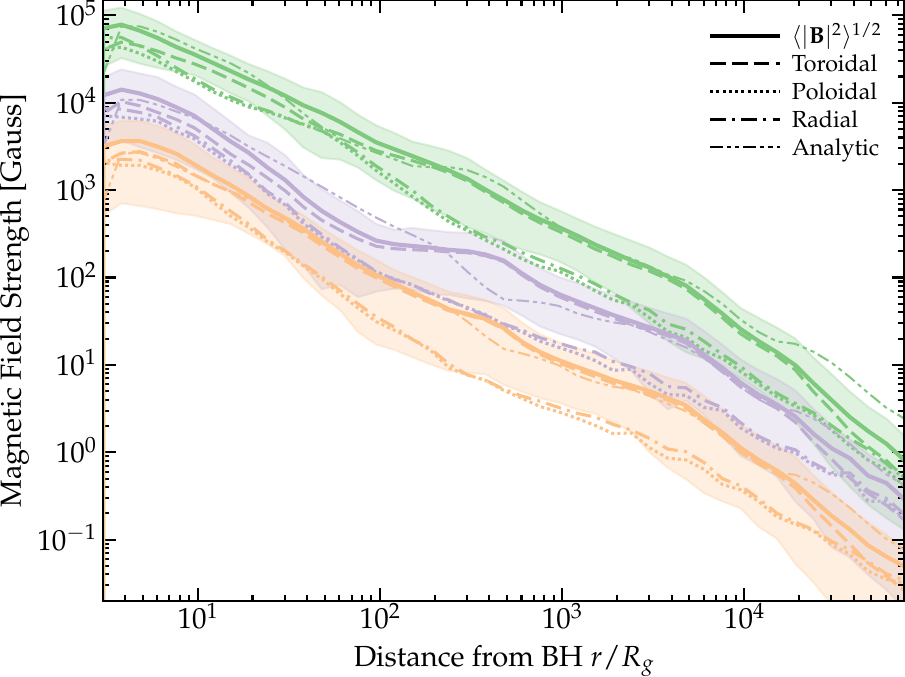} 
	\includegraphics[width=0.32\textwidth]{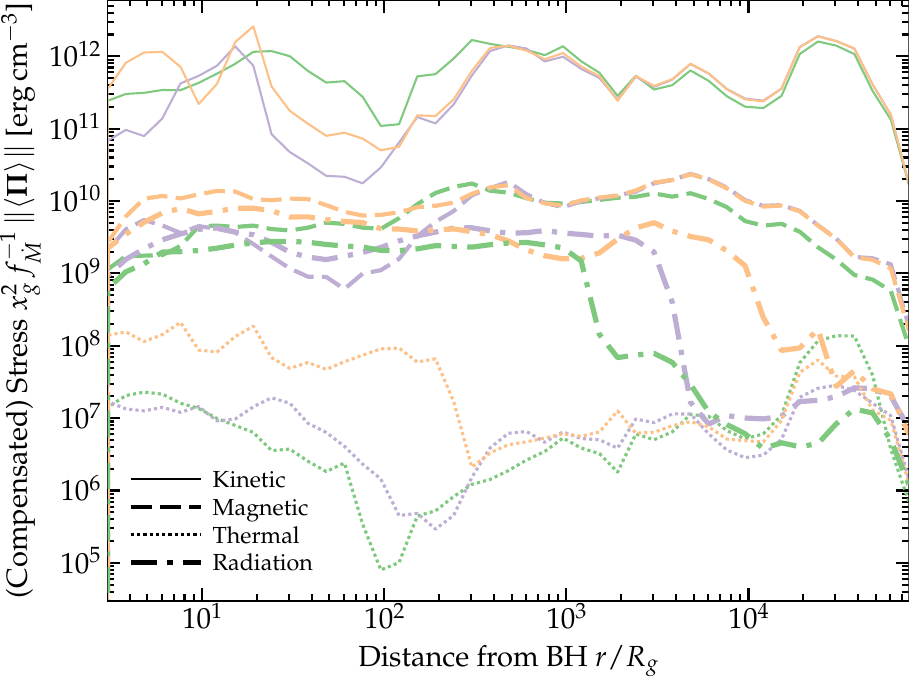} 
	\includegraphics[width=0.32\textwidth]{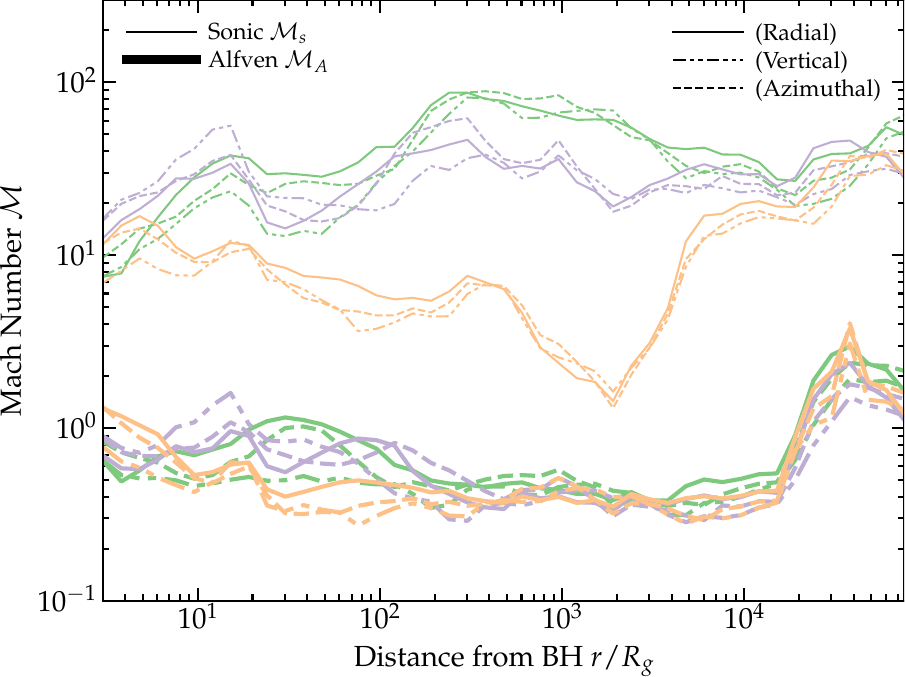} 
	\includegraphics[width=0.32\textwidth]{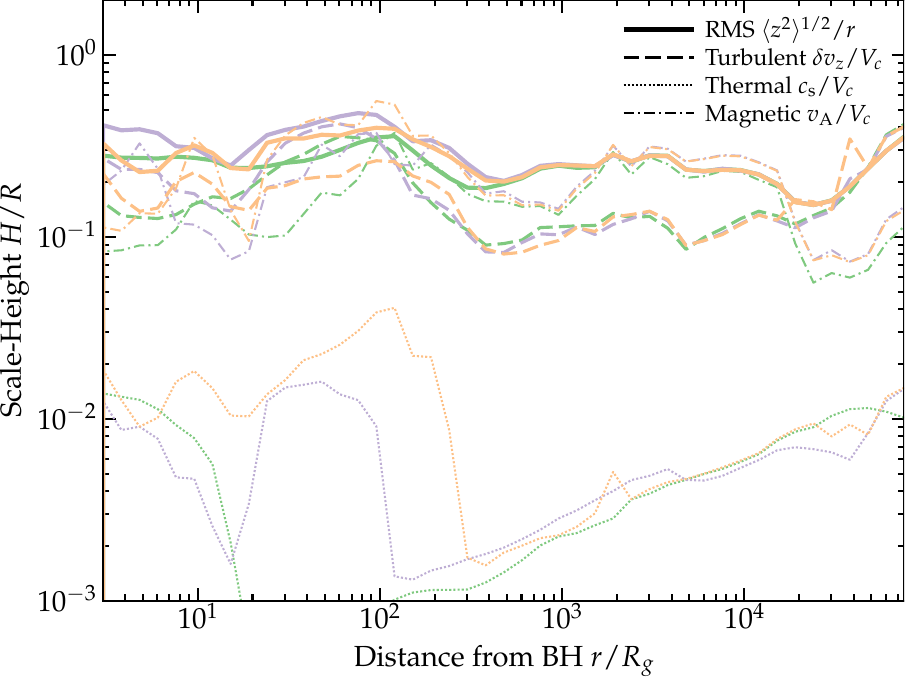} 
	\includegraphics[width=0.32\textwidth]{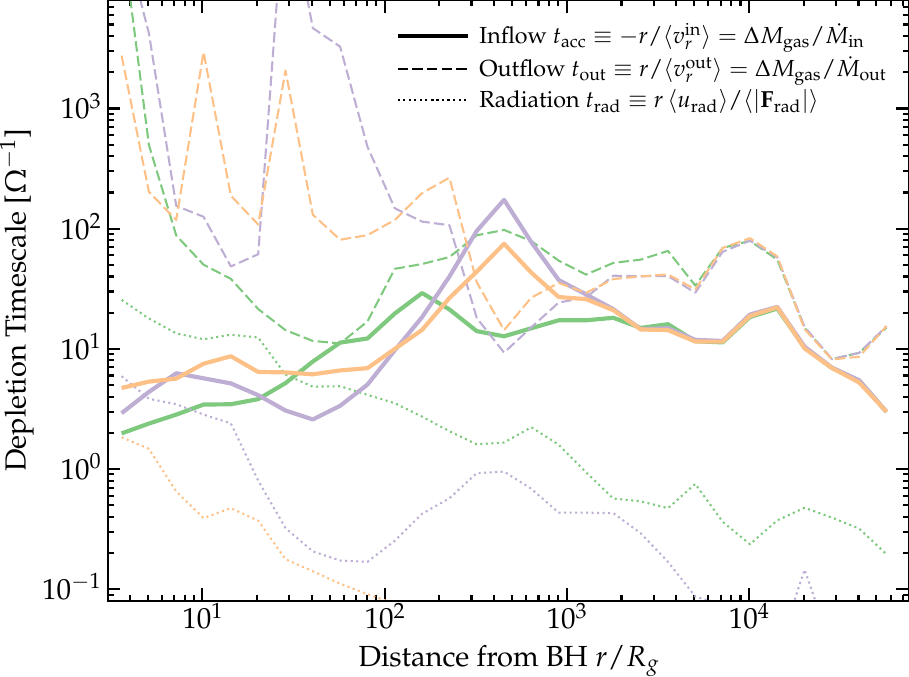} 
	\includegraphics[width=0.32\textwidth]{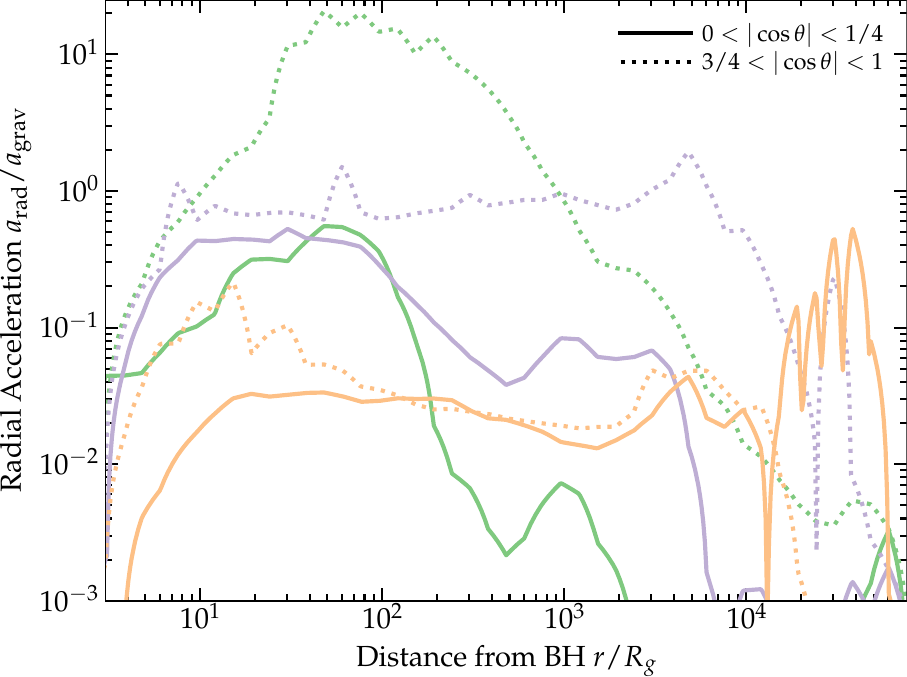} 
	\includegraphics[width=0.32\textwidth]{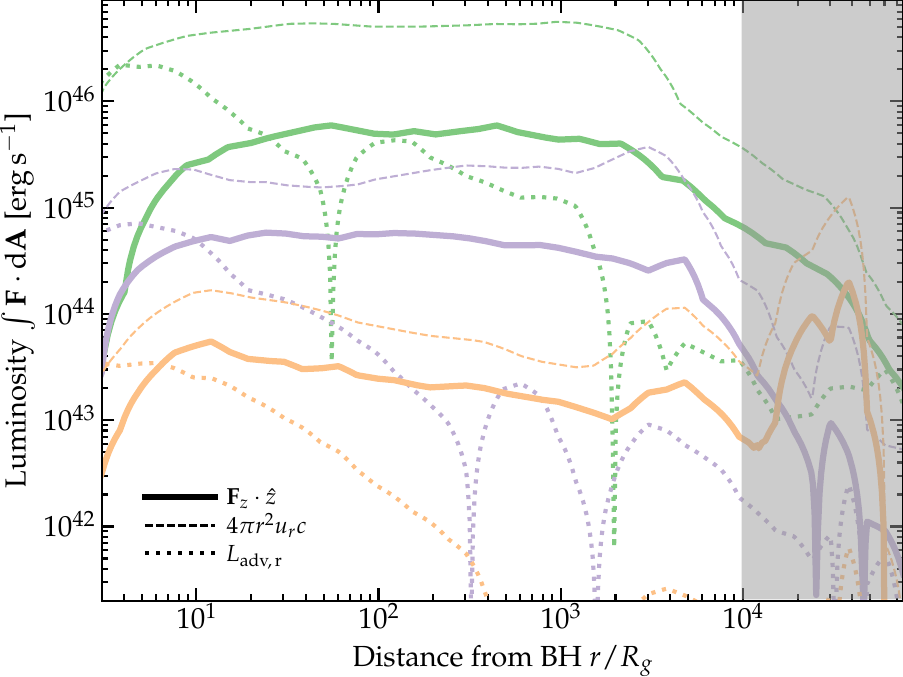} 
	\includegraphics[width=0.32\textwidth]{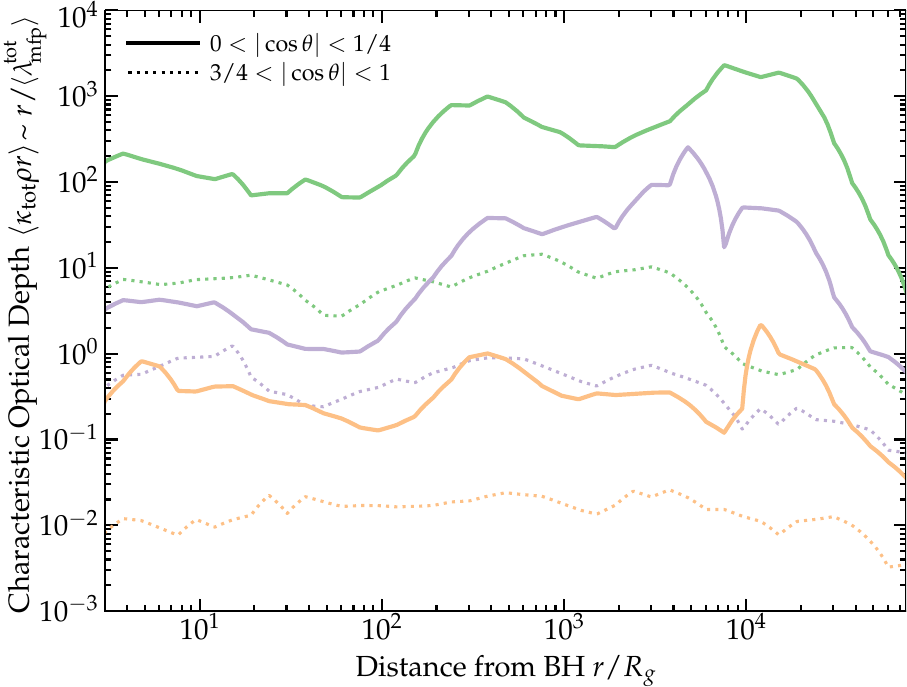} 
	\includegraphics[width=0.32\textwidth]{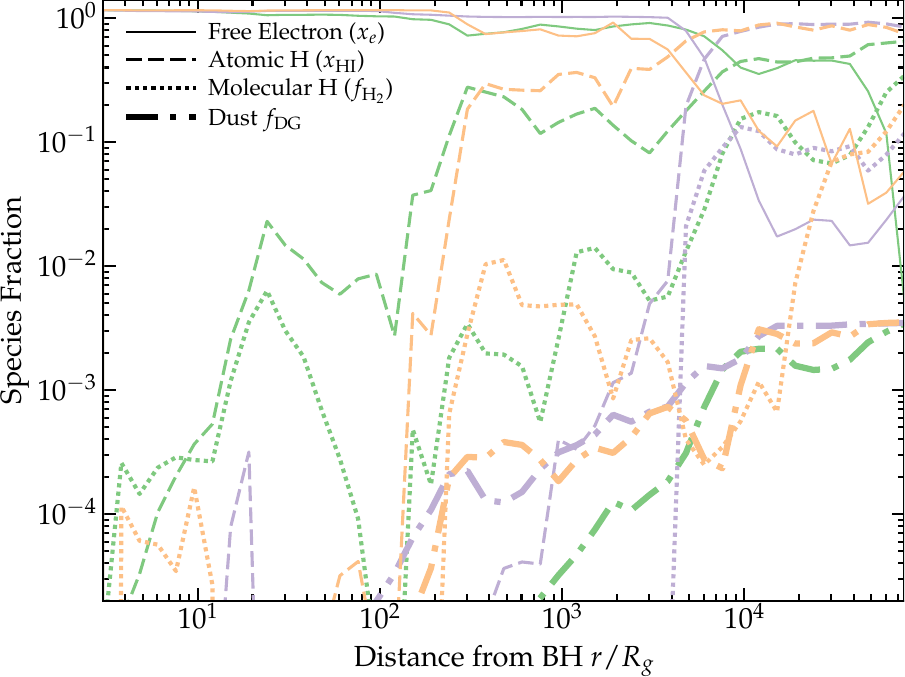} 
	\includegraphics[width=0.32\textwidth]{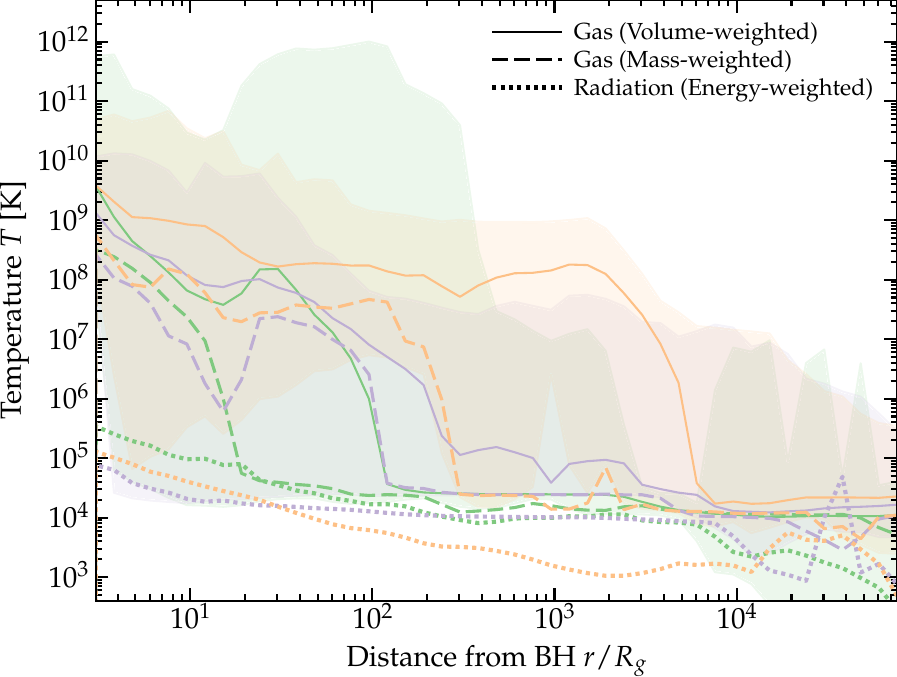} 
	\caption{Radial profiles as Figs.~\ref{fig:profiles.general}, \ref{fig:profiles.magnetic}, \ref{fig:profiles.velocity}, \ref{fig:profiles.flux}, but comparing our three simulations with different initial gas accretion rates (all else fixed): $f_{\dot{M}}=(1,\,0.03,\,0.001)$ or $\dot{m} \sim (20,\,0.3,\,0.02)$ ({\em green}, {\em purple}, {\em orange}, respectively). To leading order, properties like $\dot{M}$, $\rho$, $|{\bf B}|$, $\Sigma_{\rm gas}$ or $\tau$, and different energy densities scale with $\dot{m}$ while dimensionless quantities like $H/R$, $\mathcal{M}_{A}$, $t_{\rm acc}/t_{\rm orbit}$ are independent of $\dot{m}$, as predicted by simple magnetically-dominated disk similarity solutions (\paperthree). While thermochemistry is not scale-free, $\beta_{\rm therm} \ll 1$ at all radii, with $\beta_{\rm tot} \sim 0.1-1$ in the inner disk, in all cases. Below $\dot{m} < 0.02$, there ceases to be an optically-thick accretion disk. Continuum radiation pressure ($a_{\rm rad}/a_{\rm grav}$) becomes weaker at lower $\dot{m}$, and radiative efficiencies decrease with $\dot{m}$ as $\epsilon_{r} \sim (0.15,\,0.1,\,0.02)$ for $\dot{m}\sim(0.02,\,0.3,\,20)$. 
	\label{fig:profiles.mdot}}
\end{figure*} 

\begin{figure*}
	\centering
	\includegraphics[width=0.32\textwidth]{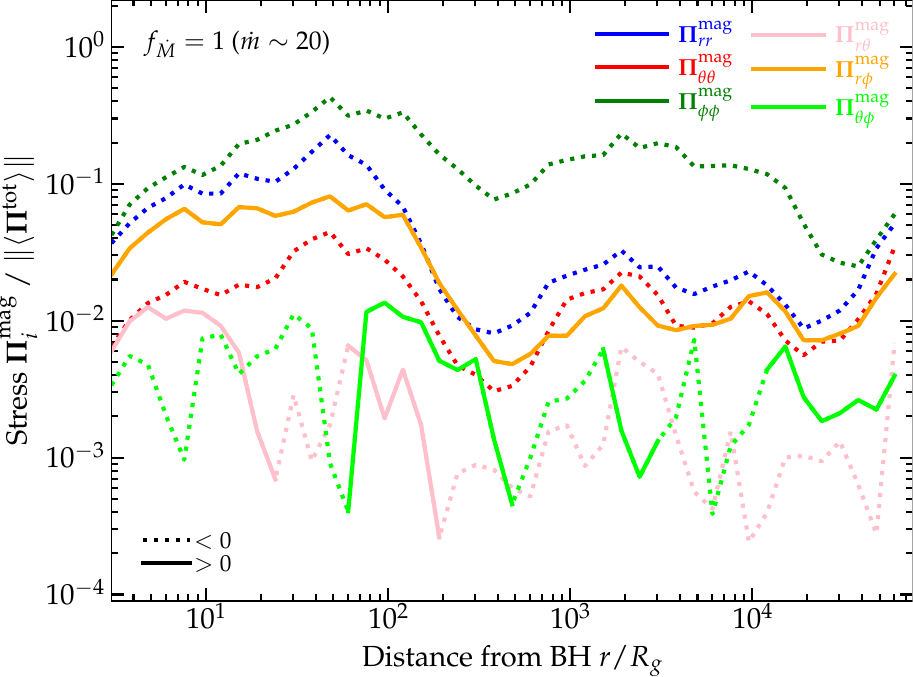} 
	\includegraphics[width=0.32\textwidth]{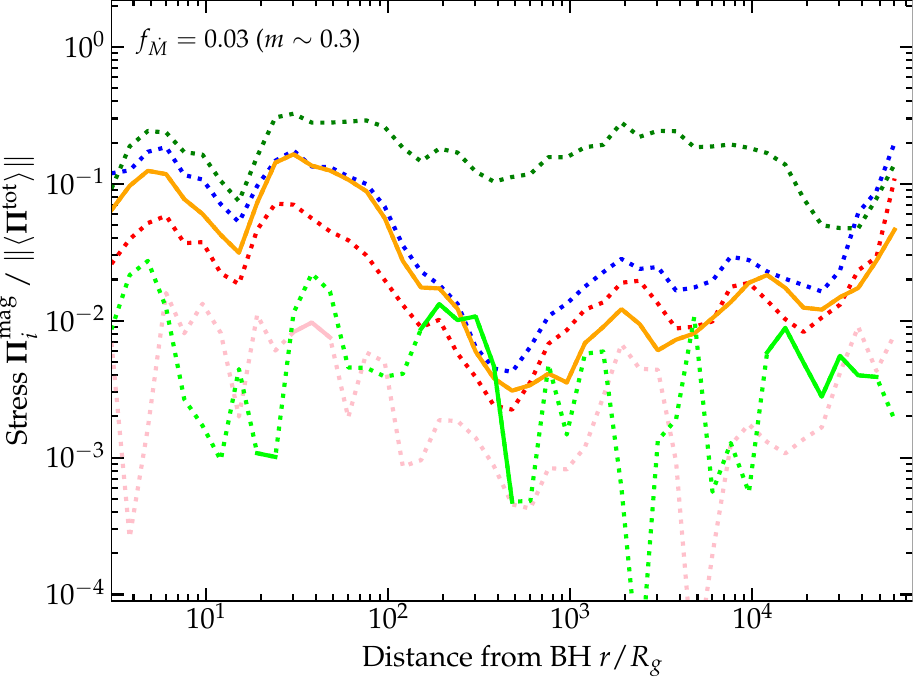} 
	\includegraphics[width=0.32\textwidth]{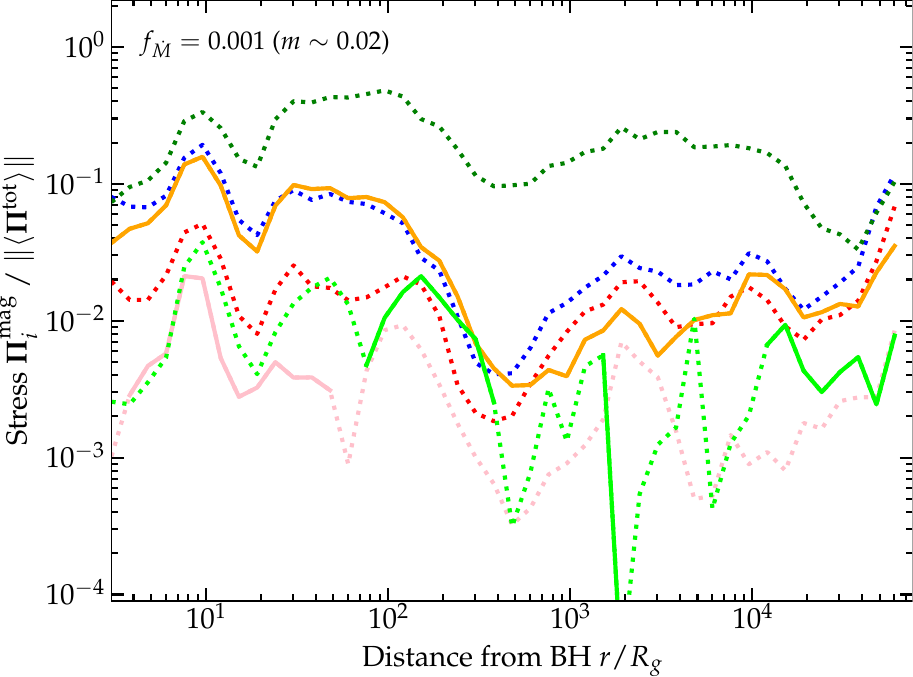} 
	\includegraphics[width=0.32\textwidth]{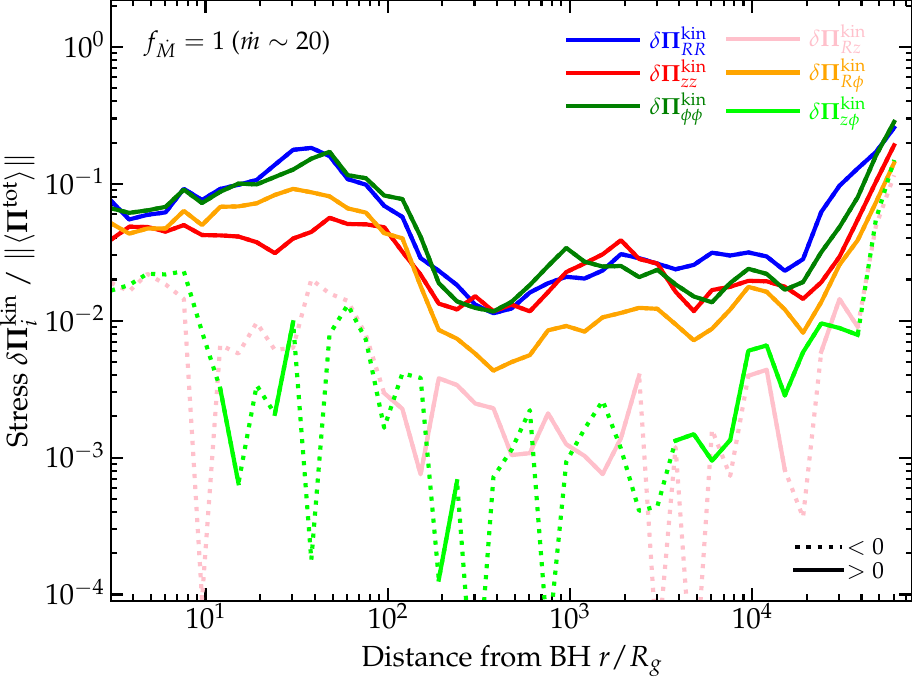} 
	\includegraphics[width=0.32\textwidth]{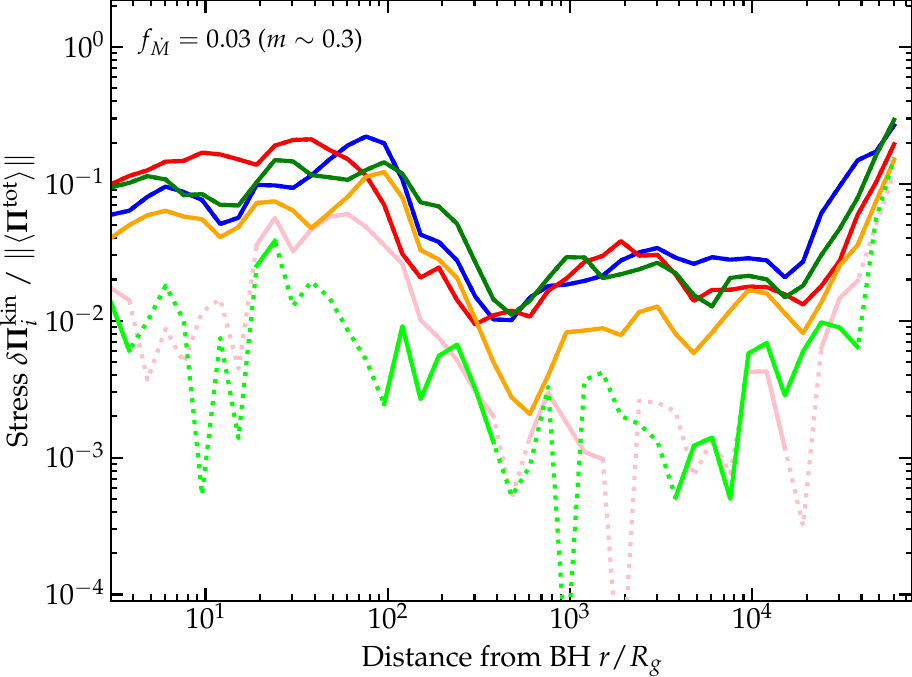} 
	\includegraphics[width=0.32\textwidth]{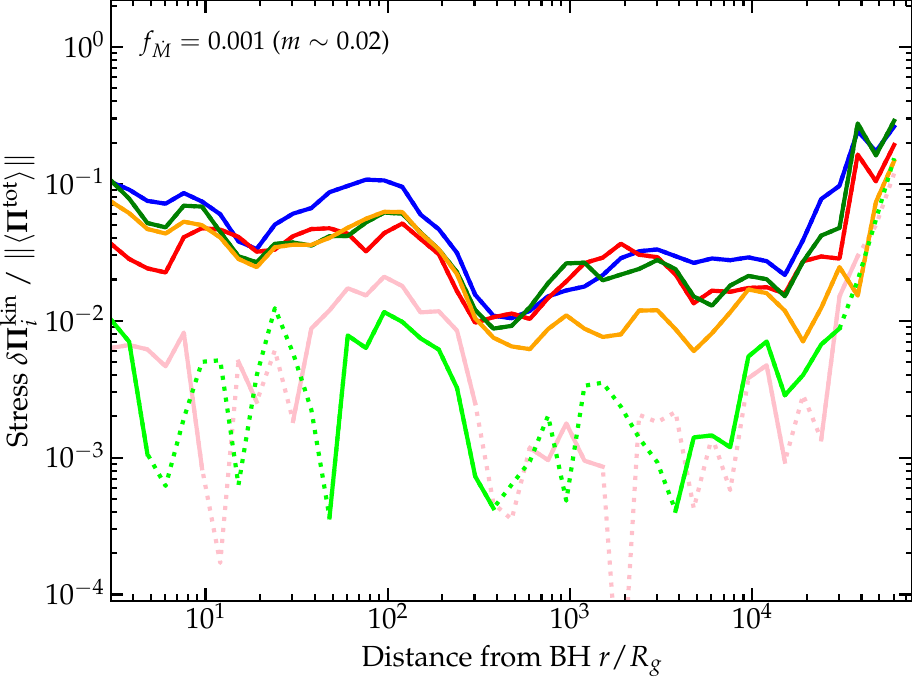} 
	\caption{Radial profiles as of Reynolds ({\em top}; as Fig.~\ref{fig:profiles.velocity}) \&\ Maxwell ({\em bottom}; as Fig.~\ref{fig:profiles.magnetic}) stresses, comparing our different accretion-rate simulations $f_{\dot{M}}=(1,\,0.03,\,0.001)$ or $\dot{m} \sim (20,\,0.3,\,0.02)$ ({\em left}, {\em middle}, {\em right}). The dimensionless scale of the stress is self-similar, and in all cases angular momentum transport is dominated by the $R\phi$ stress (comparable to the total stress) with mean Maxwell marginally larger than fluctuating Maxwell and/or Reynolds in the inner disk.
	\label{fig:profiles.stress.vs.mdot}}
\end{figure*}

\begin{figure*}
	\centering
	\includegraphics[width=0.95\textwidth]{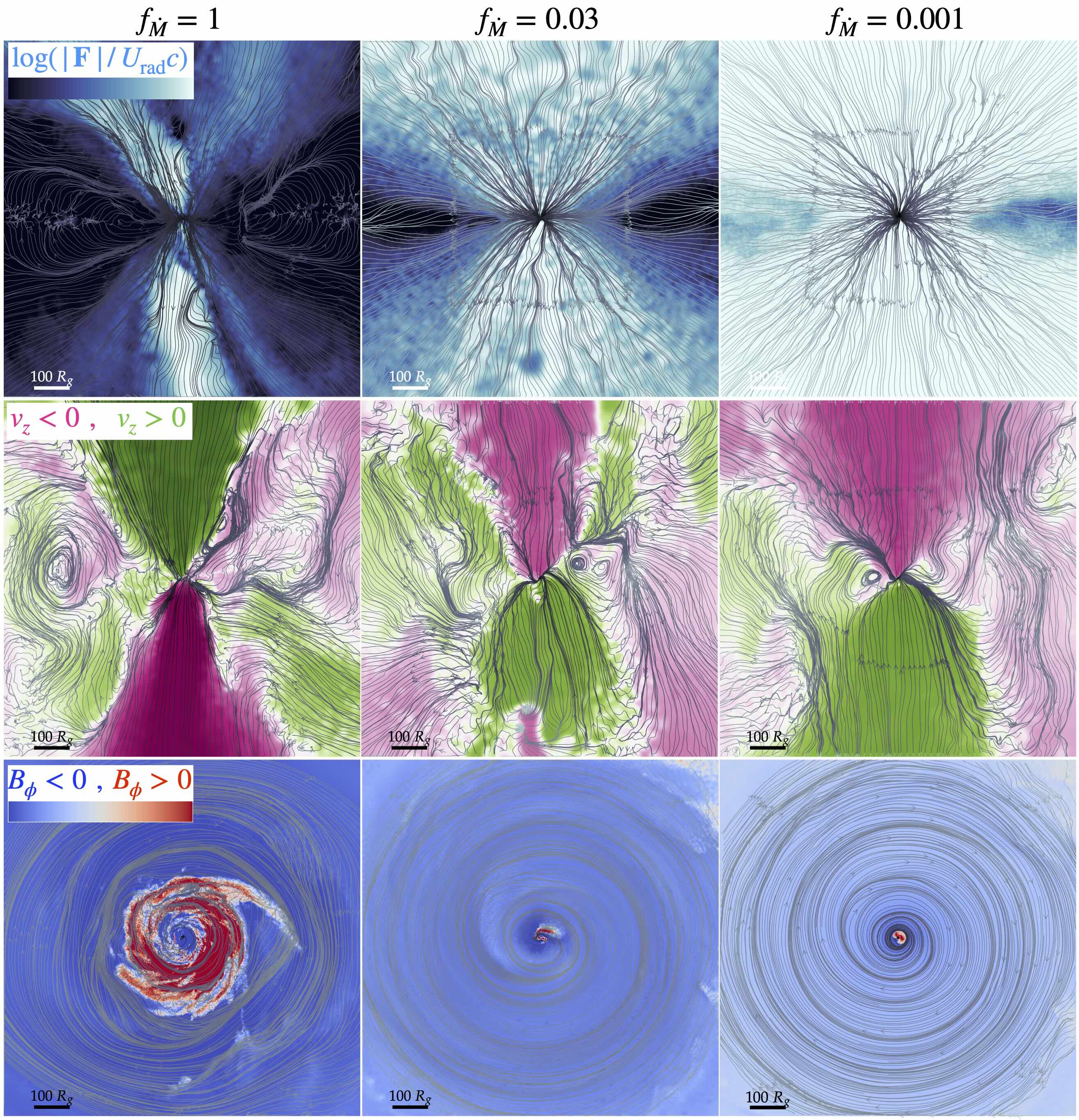} 
	\caption{Field-line maps comparing runs with different accretion rates $f_{\dot{M}}=(1,\,0.03,\,0.001)$ or $\dot{m} \sim (20,\,0.3,\,0.02)$ ({\em left}, {\em middle}, {\em right}) at fixed spatial scale. 
	{\em Top:} Radiation flux edge-on as Fig.~\ref{fig:maps.flux}. Radiation escapes in a widening bicone at lower $\dot{m}$, with less scattering so the flux approaches the free-streaming limit ($|{\bf F}| \rightarrow u_{\rm rad}\,c$). The reduced scattering at low $\dot{m}$ means we see more-radial flux streamlines without as much curvature ``into'' the disk (from scattering and absorption/reprocessing of inner disk photons by outer disk gas). 
	{\em Middle:} Velocity field edge-on as Fig.~\ref{fig:maps.vel}. All exhibit weak large-scale coherent outflows, but intermediate-scale semi-coherent outflows rising to $|z|/R \sim 0.1-1 \sim H$, i.e. potential ``failed winds'' or ``fountain flows'' become less prominent at lower $\dot{m}$.
	{\em Bottom:} Magnetic field face-on as Fig.~\ref{fig:maps.b}. The toroidal field is dominant in all cases with coherence lengths $\gtrsim R$.
	\label{fig:maps.fields.mdot}}
\end{figure*} 

\begin{figure*}
	\centering
	\includegraphics[width=0.95\textwidth]{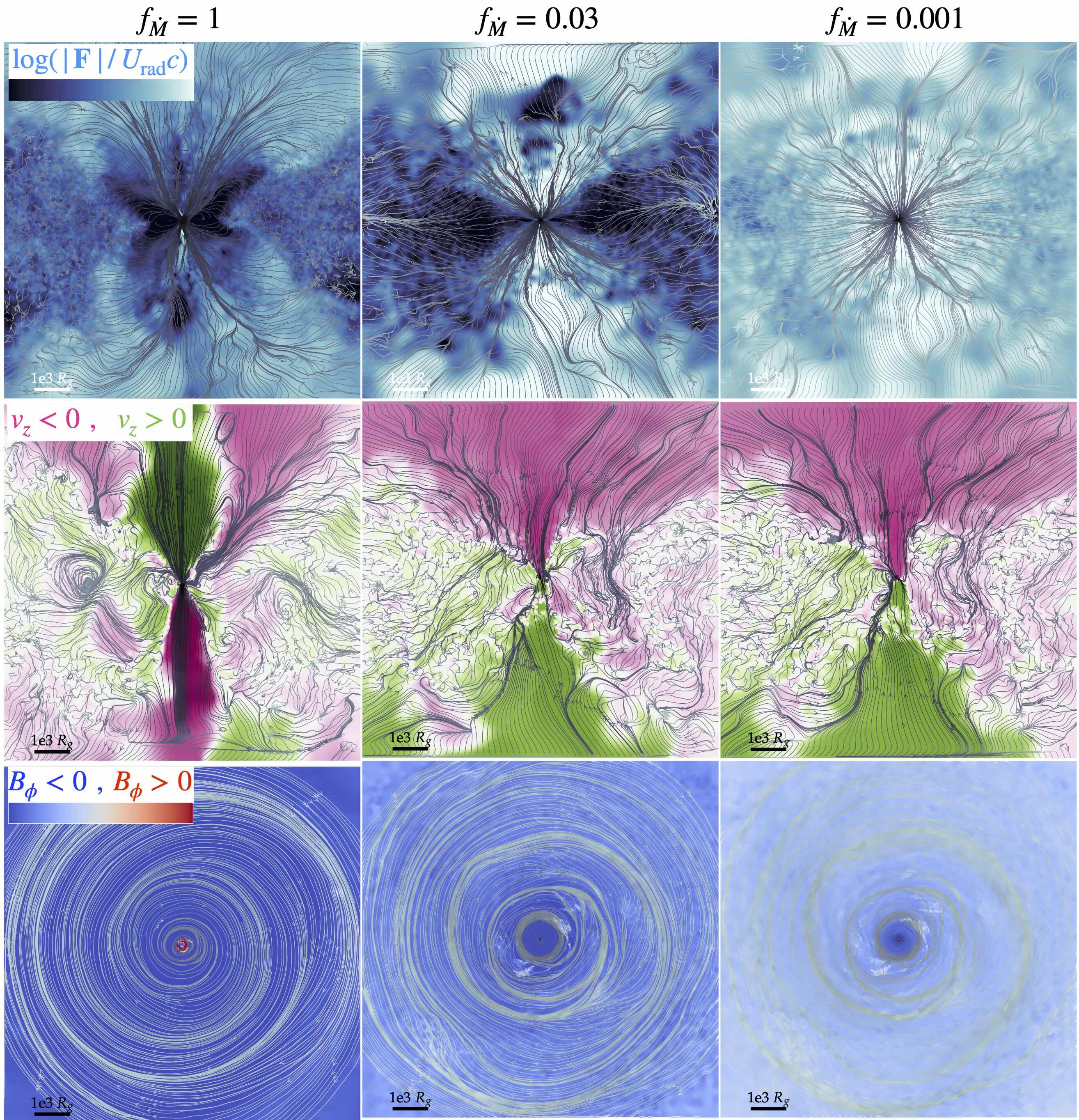} 
	\caption{As Fig.~\ref{fig:maps.fields.mdot}, at $10$ times larger scale. 
	\label{fig:maps.fields.mdot.alt}}
\end{figure*}

\begin{figure}
	\centering
	\includegraphics[width=0.98\columnwidth]{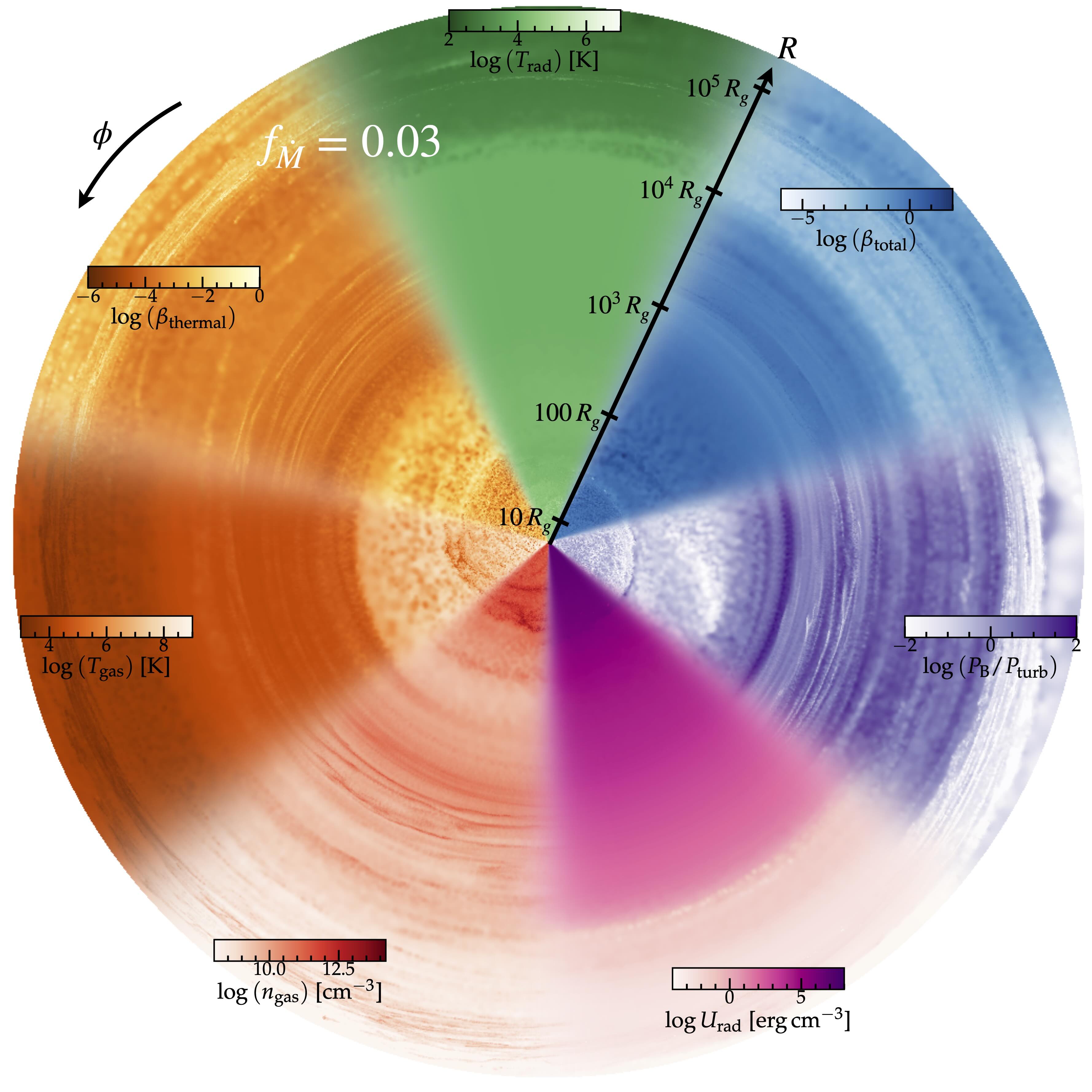} 
	\includegraphics[width=0.98\columnwidth]{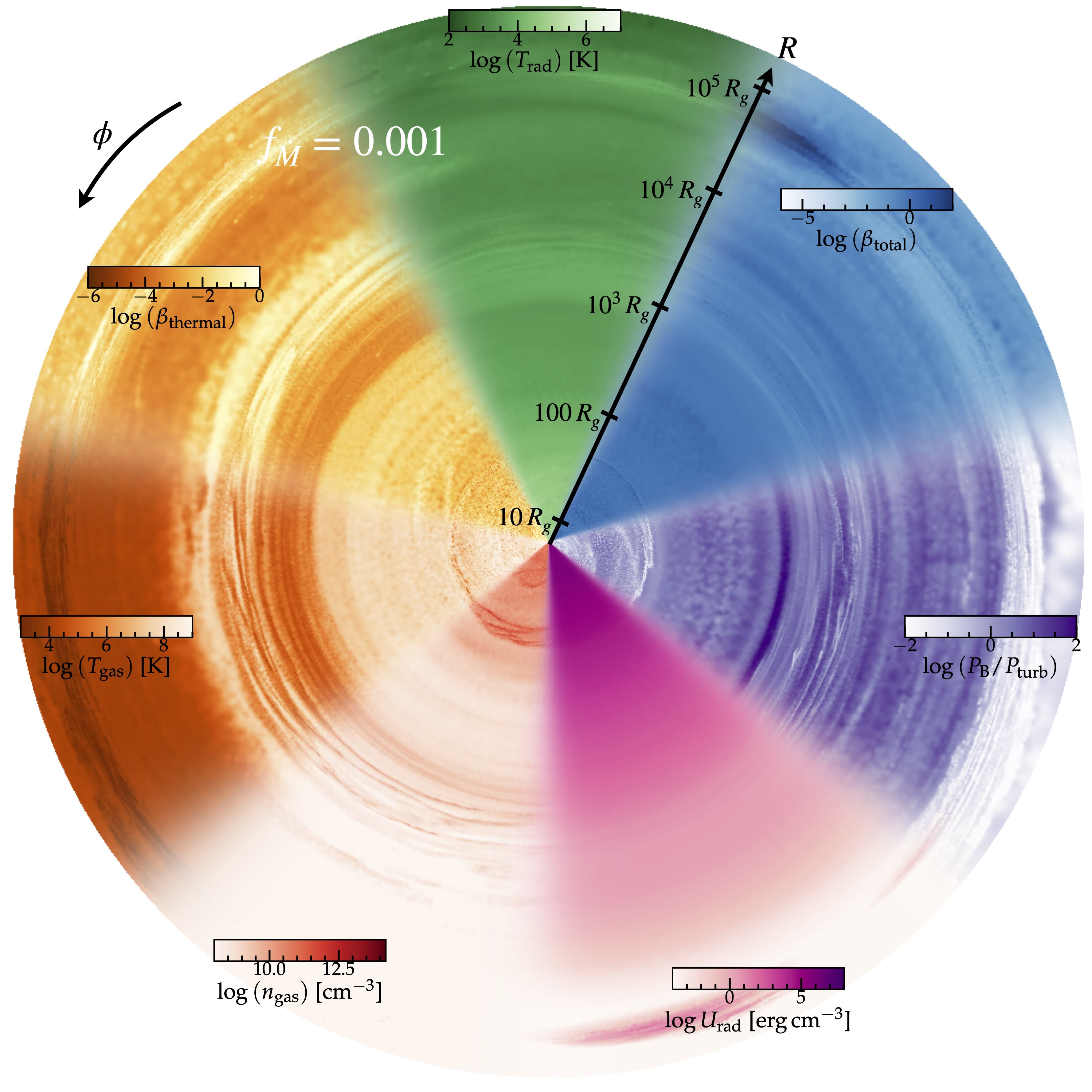} 
	\caption{Face-on projections showing the entire (log-scaled) dynamic range of the simulations with different accretion rates $f_{\dot{M}}=(0.03,\,0.001)$ or $\dot{m} \sim (0.3,\,0.02)$ ({\em top}, {\em bottom}), as Fig.~\ref{fig:maps.pinwheel}. Mass and radiation density decrease with lower $\dot{m}$ as expected, while $\beta$ increases weakly, and $\mathcal{M}_{A}$ remains similar.
	\label{fig:maps.pinwheel.mdot}}
\end{figure}

\begin{figure*}
	\centering
	\includegraphics[width=0.6\textwidth]{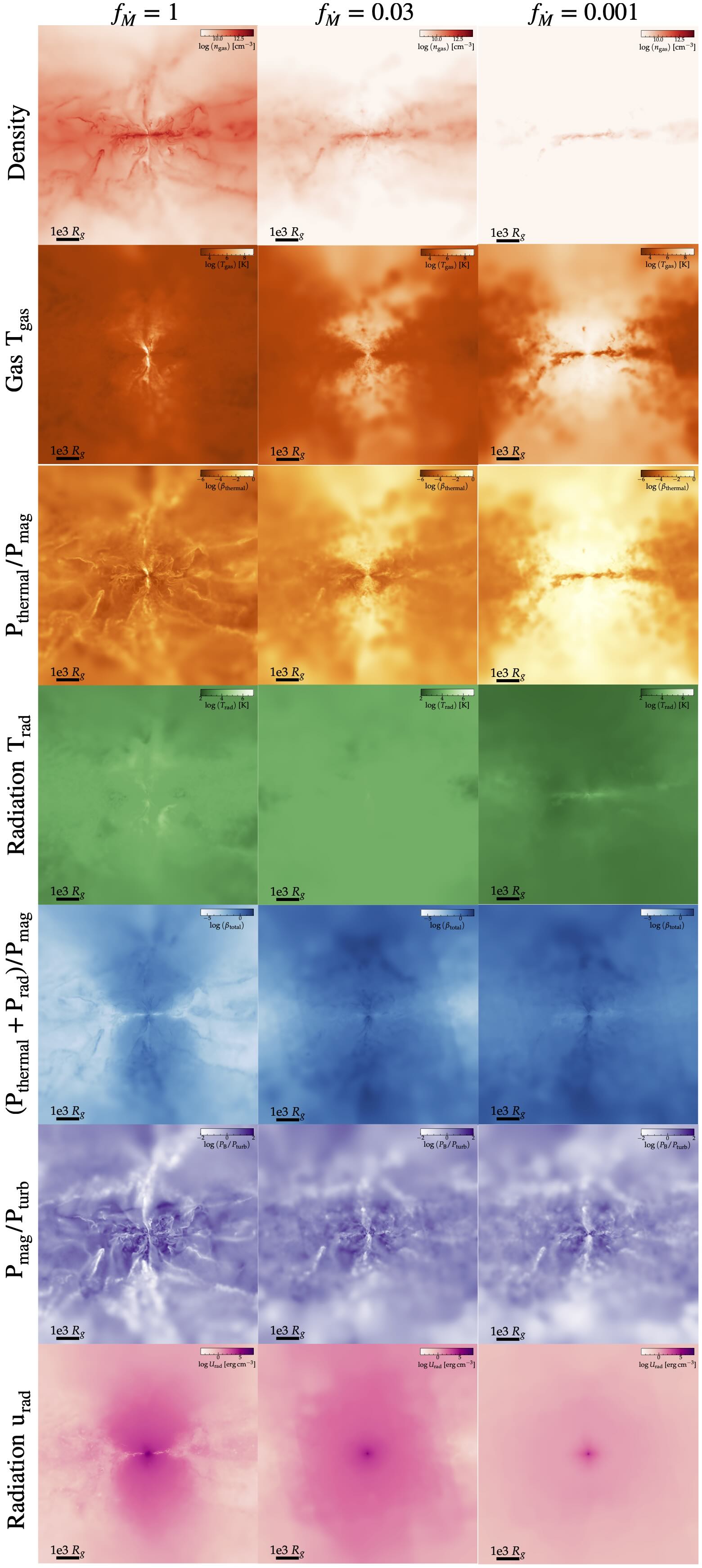} 
	\caption{Edge-on images as Fig.~\ref{fig:maps.pinwheel.edgeon}, but comparing runs with different accretion rates $f_{\dot{M}}=(1,\,0.03,\,0.001)$ or $\dot{m} \sim (20,\,0.3,\,0.02)$ ({\em left}, {\em middle}, {\em right}) at fixed spatial scale. Mass and radiation density decrease with lower $\dot{m}$ as expected, while $\beta$ increases weakly, and $\mathcal{M}_{A}$ remains similar (see Fig.~\ref{fig:profiles.mdot}). Vertical temperature structure in $T_{\rm gas}$ and $T_{\rm rad}$ becomes more prominent, with a more distinct ``sandwich'' corona in $T_{\rm gas}$ and vertical $T_{\rm rad}$ stratification at lower $\dot{m}$.
	\label{fig:maps.edgeon.mdot}}
\end{figure*}

\subsection{Dependence on Accretion Rate}
\label{sec:mdot}

We see in Figs.~\ref{fig:profiles.mdot} \&\ \ref{fig:profiles.stress.vs.mdot} that the disk {\em structural} properties scale with $\dot{m}$, to leading order, quite close to the similarity solution predictions in \paperthree. Recall the simulations have fixed $M_{\rm BH}$ and are compared at the same radii and times, but different (rescaled) initial gas mass. The initial conditions are rescaled versions of each other, but all begin with no gas interior to $<300\,R_{g}$, so predictions interior are entirely distinct and at larger radii we see as we evolve the runs they reach different local equilibria (only at the very largest $R$ to they reflect the initial conditions, because the simulation time is a small fraction of $\Omega^{-1}$ at e.g.\ $\gg 10^{5}\,R_{g}$). To leading order, we see the gas densities and kinetic, magnetic, and radiation energy densities/pressures, and absolute magnitude of the stresses, scale $\propto \dot{m}$ as predicted. This means dimensionless quantities like $\beta$, the accretion timescales $t_{\rm acc}/t_{\rm orbit}$, \Alf\ Mach numbers $\mathcal{M}_{A}$, and ratio of pressures/stresses in different components, and density/magnetic/turbulent scale heights $H/R$ are approximately independent of $\dot{m}$. This is already highly non-trivial: it implies that lower accretion rates are, to first order, simply a result of lower mass supply, not mass moving ``less quickly'' through the disk. Moreover it contrasts strongly with the predictions of standard $\alpha$ disk models like SS73, which predicts $\rho \propto \dot{m}^{-2}$ (or $\propto \dot{m}^{0.4}$) in the inner (outer) regions of these disks (their ``region (a)'' and ``region (b) or (c)'', respectively); or $H/R \propto \dot{m}^{0.2-1}$, $\delta v^{2}/v_{\rm K}^{2} \propto \dot{m}^{0-0.85}$, $t_{\rm acc}/t_{\rm orbit} \propto \dot{m}^{-2}$ to $\dot{m}^{0.4}$, etc. 

Per above (\S~\ref{sec:rad.efficiency}), the luminosity and radiative efficiency also scale with $\dot{m}$, roughly as expected though with the important aspect that advection makes the radiative efficiency not self-similar (as seen in other simulations, see e.g.\ \citealt{jiang.dai:2024.supereddington.accretion.simulations.review} for a review). Correspondingly weaker radiation-pressure driven outflows and $a_{\rm rad}/a_{\rm grav}$ at lower $\dot{m}$ (discussed below, \S~\ref{sec:outflows}) are expected.

The biggest systematic deviations from self-similarity are in the thermal properties (temperature, species abundances, opacities), which should not be scale-free. Still these roughly follow the predictions from more detailed analytic calculations in \citet{hopkins:multiphase.mag.dom.disks} of their properties in a disk whose structural properties follow the similarity solutions of \paperthree, and their variation is relatively weak with $\dot{m}$ (recall, $\dot{m}$ varies by a factor of $\sim 1000$ between the different runs). The dust sublimation radius moves out at higher luminosity or $\dot{m}$, but atomic and molecular gas can actually persist in the midplane further interior (at smaller $R$) at higher $\dot{m}$ owing to the higher densities of the midplane leading to both much stronger self-shielding, but also much higher recombination/molecule formation rates. This in turn non-linearly modifies the opacities (though the total optical depth, $\propto \kappa \Sigma_{\rm gas}$, varies more with $\Sigma_{\rm gas}$ which varies roughly as $\propto \dot{m}$ here, while the opacities $\kappa$ vary relatively weakly). As predicted in \citet[][see their Figs.~6, 12, \&\ 13]{hopkins:multiphase.mag.dom.disks}, the lower-$\dot{m}$ disks are hotter in terms of their volume-filling $T_{\rm gas}$ (note this is opposite the thermal-pressure dominated disk prediction, for which $T \propto \dot{m}^{0}$ to $\dot{m}^{0.4}$), and have more very hot (quasi-virialized) atmosphere/coronal gas, out to larger radii, even though the dust sublimation radius moves inwards. The behavior of the radiation temperature in Fig.~\ref{fig:profiles.mdot} needs a bit more interpretation, however: we discuss this in more detail in \S~\ref{sec:lte}, where we show it owes in part to highly non-LTE effects, where the role of optically-thin cooling and radiation becomes more important at lower $\dot{m}$.

\subsection{Analytic Models \&\ Modified Radial Scalings of Structural Properties}
\label{sec:modified.radial}

\citet{hopkins:multiphase.mag.dom.disks} argued that at $\dot{m} \gg 1$, the flow should become advection-dominated with most of the advected energy in the form of the mean toroidal field (as we see in Figs.~\ref{fig:profiles.general} \&\ \ref{fig:profiles.flux}). They noted that, akin to \citet{narayan.yi.95:adaf.self.similarity.outflows,narayan:bh.review.1998}, there is a simple similarity solution for this regime which was already discussed in \paperthree, namely that interior to the radius where advection dominates, the flow would be modified to:
\begin{align}
\frac{v_{\rm turb}}{v_{\rm c}}& \sim \frac{v_{A}}{v_{\rm c}} \sim \frac{H}{R} \rightarrow \psi \sim 0.2 \ , \\ 
\Sigma_{\rm gas} &\rightarrow \frac{\dot{M} \tilde{g}(x_{g})}{2\pi \psi^{2} \sqrt{G M_{\rm BH} R}} \\ 
\nonumber & \sim  4\times10^{7} \frac{\rm M_{\odot}}{\rm pc^{-2}} \left( \frac{\dot{M}}{3 {\rm M_{\odot}\,yr^{-1}}} \right) \sqrt\frac{1-x_{g}^{-1}}{x_{g}/100} \,\left(1-\sqrt\frac{3}{x_{g}} \right) \ .
\end{align}
The similarity parameter $\psi$ is set by continuity between the radii $R\lesssim R_{\rm adv}$ where the solution becomes advective, and the solutions at larger scales to be $\psi \approx (R_{\rm adv}/r_{\rm ff})^{1/6} \sim 0.2\,(\dot{m}/100)^{1/6}$ for the parameters here.  we have incorporated the correction factor $\tilde{g}(x) \equiv \sqrt{1-1/x}\,(1-\sqrt{x_{\rm in}/x})$ which accounts for both the shape of the PW potential (the first term) and assumption that the angular momentum vanishes at some $x_{\rm in}$ as in SS73 (who took $x_{\rm in}=3$ for the ISCO of a non-rotating BH). 

If we account for the fact that $\dot{M}$ is not actually exactly constant with radius at the time studied here (Fig.~\ref{fig:profiles.general}), we see in Figs.~\ref{fig:profiles.general}-\ref{fig:profiles.flux} that these equations, with constant $\psi \sim 0.2$, provide a remarkably accurate prediction of the disk dynamical and mass and density structure interior to the radius where the system becomes advective. Furthermore we see in Fig.~\ref{fig:profiles.mdot} that these analytic solutions explain the zeroth-order qualitative scalings with both $\dot{m}$ and $r$ in the simulations.

The similarity models do not perfectly reproduce the chaotic, non-linear complexity of the simulations, and the deviations can be significant (e.g.\ up to $\sim 1\,$dex in midplane $n$, Fig.~\ref{fig:profiles.mdot}). As discussed above, these deviations are often ``quasi-steady-state'' in the sense that they owe to global structures/modes (e.g.\ spiral arms, gaps, rings, interactions between outflows and inflows) that break the self-similarity and can persist for many orbital times at a given radius (as opposed to local turbulent fluctuations which have coherence times $\lesssim \Omega^{-1}$). So they cannot be trivially ``averaged out'' (especially since any time average here at large $R$ would average over many different local steady-state configurations at small $R$). But it is worth contrasting these magnetically-dominated disk models with a standard (assumed $\beta_{\rm thermal} \gg 1$) $\alpha$-disk model as an alternative similarity solution, as presented in detail in \paperthree\ and \citet{hopkins:multiphase.mag.dom.disks}. Per \S~\ref{sec:mdot}, the $\alpha$-disk model predicts a {\em qualitatively} different (sometimes opposite) dependence on $\dot{m}$ of key quantities like $n$, $H/R$, $\delta v_{\rm turb}$, $t_{\rm acc}$, $T_{\rm gas}$; likewise as shown in \citet{hopkins:multiphase.mag.dom.disks} the qualitative dependence on $r$ can be different for quantities like $H/R$ and $\tau$. But the most notable difference is in the characteristic values: $\alpha$-disk models (with the same $\dot{M}$, $M_{\rm BH}$, and $R$ simulated here) predict $H/R$ systematically smaller by $\sim 2-3\,$dex, $\tau$ and $\Sigma_{\rm gas}$ larger by $\sim 4-6$\,dex, $|{\bf B}|$ larger by $\sim 1-2$\,dex, and midplane densities $n$ larger by $\sim 6-9$\,dex \citep{hopkins:multiphase.mag.dom.disks}.

\subsection{Extended Scattering Structures \&\ Effective Size of the Emission Regions}
\label{sec:scattering}

The extended scattering structures visible in Figs.~\ref{fig:maps.flux}-\ref{fig:maps.pinwheel.edgeon} are a natural consequence of the large scale-height of the disk (Figs.~\ref{fig:zoomies.cyl}, \ref{fig:profiles.general}, \ref{fig:maps.pinwheel.edgeon}, \ref{fig:profiles.vertical}), and correspondingly relatively slow falloff of $\rho(z)$ in Fig.~\ref{fig:profiles.vertical}. This is discussed in more detail for the analytic models in \citet{hopkins:multiphase.mag.dom.disks}, but basically since $H/R \sim 0.1-1$ at the radii of interest here, (1) a large fraction of the radiation from the inner disk directly intercepts the outer disk, and (2) the density at $|z|\sim R$ cannot be vastly smaller than the midplane density (since even for an exponential vertical profile, this is just a few $e$-foldings $|z|/H \sim 1-10$). The latter means that an extended scattering halo/atmosphere in more polar directions is automatically expected, even without any ``additional'' component like outflows or jets. That directly leads to the curved flux lines seen in Fig.~\ref{fig:maps.flux}. 

Correspondingly, from Fig.~\ref{fig:profiles.flux}, we see that the radiation from the inner disk has an $\mathcal{O}(1)$ scattering component (but primarily scattering at these larger radii rather than thermal reprocessing), out to radii of of a few thousand $R_{g}$ or $\sim 10^{16}\,{\rm cm}$. This is intriguingly similar to the observed sizes of the optical/NUV emission observed in microlensing \citep{dai:2010.agn.microlensing.xray.optical.larger.than.expected,blackburne:2011.sizes.qso.acc.disks.microlensing.too.big,jimenez:2014.qso.disk.temp.profile.size.from.microlensing.large.flat,cornachione:2020.accretion.disk.microlensing.profiles.shallow} and reverberation mapping \citep{cai:2023.qso.sed.universal.w.flat.intrinsic.spectrum.into.uv.larger.than.expected.need.reprocessing.in.blr,ren:2024.reverb.mapping.sizes.larger.than.expected} studies. The fact that these studies also see sizes independently of wavelength has led to widespread speculation that this must owe a process like Thompson scattering (being achromatic) from an extended surface -- but the historical challenge has been that an $\mathcal{O}(1)$ fraction of the light must be reprocessed, which is not possible in most physical models of SS73-like $\alpha$ disks (with $H/R \sim 0.001-0.01$ at these radii) and/or winds from the disk surface \citep{czerny:2003.soft.excess.and.uv.profile.in.agn.from.reprocessing.warm.skin.warm.absorber,dai:2010.agn.microlensing.xray.optical.larger.than.expected,kamraj:2022.hard.xray.agn.corona.properties,tortosa:2022.hard.coronal.constraints.in.hyper.eddington.qsos}. In future work, it therefore seems important to forward-model these observations directly.

\subsection{Effective Temperature of the Emerging Radiation Field}
\label{sec:Trad}

In a disk which is radiatively efficient at all radii and very thin ($H\ll R$), so radiation from the surface escapes to infinity, we expect the effective radiation temperature of the radiation {\em emitted} at each radius/annulus $R$ to scale as $T_{\rm rad,\,eff}^{\rm em} \sim (3\,\dot{M}\,\Omega^{2}/8\pi\sigma_{B})^{1/4} \sim 15000\,{\rm K}\,(\dot{M}/10\,{\rm M_{\odot}\,yr^{-1}})^{1/4}\,(R/1000\,R_{g})^{-3/4}$ (scaling to the BH mass here). At the opposite extreme however, if the system is radiatively inefficient with most of the radiation coming from small $R$, and the radiation is reprocessed quasi-spherically (i.e. $H\sim R$ and the system remains optically thick at large heights), then $T_{\rm rad,\,eff}^{\rm em} \sim (L/4\pi r^{2}\sigma_{B})^{1/4} \sim 15000\,{\rm K}\,(L/6\times10^{44}\,{\rm erg\,s^{-1}})^{1/4}\,(R/1000\,R_{g})^{-1/2}$. For purposes of observational probes like the polarization studies in \citet{kishimoto:qso.spectrum.ir.bluer}, even these opposite extremes give very similar predictions for the temperature initially emitted at different radii, and both are  consistent with the ``un-processed'' spectral shapes seen therein. 

However, what can differ more dramatically between models is not the temperature initially emitted at each $R$, but the effective temperature of the integrated spectrum (including reprocessing) escaping to infinity, $T_{\rm rad,\,eff}^{\infty}$. In the thin, radiatively-efficient disk model with no reprocessing, the integrated spectrum is dominated by light from the innermost radii (around the ISCO), with $T_{\rm rad,\,eff}^{\infty} \sim 10^{6}\,{\rm K}$ predicted for the BH mass and accretion rate here. On the other hand, quasi-spherical reprocessing (from a disk/photosphere with appreciable $H/R$) acts like a photosphere, so the escaping spectrum has $T_{\rm rad,\,eff}^{\infty}$ determined to first order by the radiation temperature at the radii where the absorption optical depth drops below unity -- roughly $R \sim 500\,R_{g}$ in our simulations here (Fig.~\ref{fig:profiles.flux}). The latter predicts $T_{\rm rad,\,eff}^{\infty} \sim 10,000-30,000\,{\rm K}$ for our simulations (in good agreement with the analytic predictions in \citealt{hopkins:multiphase.mag.dom.disks}, who quote $T_{\rm rad,\,eff}^{\infty} \sim 4\times10^{4}\,{\rm K}\,\epsilon_{r,\,0.1}^{11/32}\,m_{7}^{-3/64}\,\dot{m}^{-7/32}\,\tilde{Z}^{-3/16} \sim 2 \times 10^{4}\,{\rm K}$ for parameters simulated here).\footnote{Again note as discussed in \S~\ref{sec:results:new} that the apparent sudden drop of $T_{\rm rad}$ at the dust sublimation radius to temperatures $\ll 10^{4}$\,K owes to the fact that in-code we evolve a photon-number-weighted effective temperature, so as soon as the dust absorbs and re-radiates even a fraction of the bolometric luminosity, this occurs. But really this simply indicates where the spectrum becomes multi-component with escaping UV/optical and dust emission in the IR.} This is much closer (compared to the thin-disk $T_{\rm rad,\,eff}^{\infty} \sim 10^{6}$\,K prediction) to what we see actually emerging in the optically-thin polar cone in the simulations, once the radiation is escaping (Figs.~ \ref{fig:maps.pinwheel}, \ref{fig:profiles.general}, \ref{fig:maps.flux}, \ref{fig:maps.pinwheel.edgeon}, \ref{fig:profiles.mdot}). It is also much closer to the actual effective temperatures observed (in total light) in quasar SEDs \citep{peterson:1997.agn.book,vandenberk01:composite.qso.seds,richards:seds,shen:bolometric.qlf.update,cai:2023.qso.sed.universal.w.flat.intrinsic.spectrum.into.uv.larger.than.expected.need.reprocessing.in.blr}. More detailed spectral predictions will be the subject of future work utilizing extensive radiative-transfer post-processing.

\subsection{On the Properties of Outflows (Absent Jets and Line Driving)}
\label{sec:outflows}

In all our runs, we can plainly see there there {\em is} a non-negligible amount of gas with large radial and/or vertical velocities sitting at large displacements from the midplane, i.e.\ with $v_{r} \gtrsim 0.1-1\, v_{\rm K}$ and $|z| \gtrsim 0.1-1\,R$, at a wide range of radii $R$ from $\sim 10-10^{5}\,R_{g}$ (see Figs.~\ref{fig:profiles.velocity}, \ref{fig:maps.vel}, \ref{fig:profiles.vertical}, \ref{fig:maps.fields.mdot}, \ref{fig:maps.fields.mdot.alt}). These may even appear the same as large-scale winds in the observable spectrum (this will be the subject of future work). However, given the highly supersonic turbulent motions with $\langle \delta v_{z}^{2} \rangle^{1/2} \sim 0.1\,v_{\rm K}$ and corresponding thick disk with $H/R \sim 0.1-0.3$ (Fig.~\ref{fig:profiles.general}), it is sometimes difficult to tell if these are ``failed winds'' or ``fountain flows'' or simply part of the turbulence. 

What we do not always see is coherent, stable, steady-state outflows filling most of the volume vertically ``above'' and ``below'' the disk (``rising'' directly from the disk vertically even in the outer disk at $\gg 100\,R_{g}$), as one might expect in simple analytic models for magnetocentrifugal quasar winds. But we do see some narrower opening-angle ``broad jet-like'' outflows driven by radiation pressure in our highest-$\dot{m}$ simulations. 

Note that we do {\em not} resolve the true jet-launching region nor include the physics necessary to properly model relativistic-jet formation,\footnote{Our Newtonian+PW method does not capture spin or other GR effects, but also treats the inner boundary as a sink domain (into which quantities like magnetic flux can simply advect and be removed from the active grid) in standard fashion, as opposed to a true horizon (which can be important the buildup of near-horizon poloidal flux; see \citealt{kaaz:2024.hamr.forged.fire.zoom.to.grmhd.magnetized.disks}).} so work with GRMHD simulations extending these to true horizon scales is needed (presented in \citealt{kaaz:2024.hamr.forged.fire.zoom.to.grmhd.magnetized.disks}). We also do not model radiation line-driving, so cannot properly represent line driven outflows from the broad-line region (one of the most popular candidates for quasar wind acceleration; see e.g.\ \citealt{proga:disk.winds.2000b}). But simpler, continuum radiation pressure or magnetocentrifugal pressure-driven outflows (at least in the inner, dust-free disk which we evolve for many dynamical times) are captured in our simulations.

\subsubsection{Magnetocentrifugal Outflows: Failed Winds, But Potential Drivers of Thick Disks and Trans-\Alf{ic} Turbulence} 
\label{sec:outflows:mhd}

In the outer disk, as in \papertwo, we see only weak outflow along occasional flow lines coming directly off the disk surface at large radii. Most of the apparent outflow in Figs.~\ref{fig:profiles.general}, \ref{fig:profiles.mdot} owes to radial non-circular motions, and the polar outflow at largest $\dot{m}$ is launched from relatively small radii. This implies magnetocentrifugal or buoyancy-driven outflows are weak through most of the disk. That may at first seem surprising in both cases given the magnetically-dominated disk, but should not be. Vertical buoyancy (e.g.\ Parker-like) modes can drive convective vertical turbulence (and may indeed play an important role in the turbulence here and even in sustaining the toroidal fields; see \citealt{johansen.levin:2008.high.mdot.magnetized.disks,gaburov:2012.public.moving.mesh.code,squire:2024.mri.shearing.box.strongly.magnetized.different.beta.states}), but induce relatively little outflow/overshoot above the density scale-height. Magnetocentrifugal outflows require a strong mean vertical field threading the disk to wind up into a (toroidal-vertical) magnetic tower \citep{spruit:1996.mhd.disk.winds.review,bai:2016.mhd.winds.ppds}. Indeed wind escape in classic \citet{blandfordpayne:mhd.jets}-type models assumes a vertical coherence length of field lines $\ell_{B_{z}}\rightarrow \infty$, while here we see the vertical coherence scale is $\sim H$. And by the criterion of \citet{spruit:1996.mhd.disk.winds.review}, the wind ``mass loading'' parameter $\mu_{\rm wind} = \Omega_{0} R_{0} v_{p,0} 4\pi \rho_{0}/B_{p}^{2} \sim v_{\rm K} / | \langle v_{A,\,z}\rangle | \sim 100-1000$  from Fig.~\ref{fig:profiles.magnetic} is extremely large, which implies a ``failed wind'' is likely as the wind terminal velocity is predicted to be $\ll v_{A} \ll (H/R)\,v_{\rm K}$. 

So almost any turbulence or exterior ram pressure from gas above/below the disk, or other instabilities would stall or disrupt the escape of said winds to $|z| \gg H$. Indeed as discussed in \citet{spruit:1996.mhd.disk.winds.review,ouyed:1999.numerical.sims.mhd.winds.vs.massloading,anderson:2005.mass.loading.mhd.winds}, when $\mu_{\rm wind} \gg 1$, especially if $|\langle B_{\phi} \rangle| \gg |\langle B_{z} \rangle |$ and/or $c_{s} \ll v_{A,\,\phi}$ (all of which are true here), the ``wind'' stalls or becomes confined within or just above the disk (at $|z| \sim H$). In-plane, in those studies, it then behaves more like one of many sources of weak oscillatory in-plane spiral motions. Out of plane, in the same studies, it continues to drive material up to $|z| \sim H$ then recycles in a failed wind or fountain flow. So although this will not lead to a true wind, it suggests that ``failed'' magnetocentrifugal or Blandford-Payne type flows could be an important source of the observed trans-\Alf{ic} vertical turbulence which is critical to maintain the disk scale height at large $H/R \sim 0.1-1$.

\subsubsection{(Continuum) Radiation Pressure-Driven Outflows}
\label{sec:outflows:rad}

Regarding radiation pressure, recall that we include continuum opacities but not line-driving, so cannot capture line-driven winds which may be extremely relevant given the large opening-angle, irradiated disk surface BLR-like regions (Fig.~\ref{fig:maps.flux}). Note also that we only evolve the simulations long enough for the radiation front to reach equilibrium out to $\sim (0.5-5) \times 10^{4}\,R_{g}$ (depending on whether it is free-streaming or slowed by scattering), once we reach our full refinement around the ISCO. But since most of the radiation is emitted at these small radii, this means that the full effects of radiation-pressure driven outflows at {\em larger} radii, most notably outside the dust sublimation radius (as large as $\sim 10^{5}-10^{6}\,R_{g}$ for the most optically-thin polar sightlines) cannot be modeled because we cannot run these simulations long enough. Instead such outflows are better studied in simulations which do not explicitly simulate the gas flows near the horizon (as we do here, imposing a steep timestep penalty), but inject radiation with a sub-grid model for the un-resolved accretion disk at some much larger inner boundary ($\gtrsim 1000\,R_{g}$). So we stress that the absence of strong radiation-pressure driven outflows from either the BLR or torus is an artifact of our numerical limitations here, and not a prediction. In future work, we will model these regions with detailed radiative transfer post-processing in order to better understand wind-launching conditions therein (Bardati et al., in prep.).

With that in mind, consider continuum radiation pressure {\em within} the accretion disk at radii $\lesssim 10^{4}\,R_{g}$, where our simulations are more predictive. It is not surprising that our lowest $\dot{m} \sim 0.02$ ($f_{\dot{M}}=0.001$) simulation does not exhibit strong radiation-pressure driven outflows. It is a factor of $\sim 50$ below the electron-scattering Eddington limit in luminosity and we confirm the local radiative acceleration $a_{\rm rad}$ (Fig.~\ref{fig:profiles.mdot}) is similarly below the strength of gravity $a_{\rm grav}$, as expected. Likewise the intermediate $\dot{m} \sim 0.3$ ($f_{\dot{M}}=0.03$) case is more marginal, but still has $a_{\rm rad}/a_{\rm grav} \sim 0.3$ as expected through the volume-filling region, with $a_{\rm rad} \ll a_{\rm grav}$ in most of the disk except in the very innermost disk ($\lesssim 100\,R_{g}$) where this ratio still remains sub-unity. In this latter case there could easily be appreciable radiation pressure driven outflows beginning to appear at $\gtrsim 10^{4}\,R_{g}$ from radiation pressure on dust, but again the evolution time of the simulation is only just long enough for the body of the radiation (emitted near the ISCO, so requiring our full refinement) to reach these radii. Just like with magnetocentrifugal outflows (\S~\ref{sec:outflows:mhd}), radiation pressure driving in Fig.~\ref{fig:profiles.flux} appears sufficient to accelerate material into non-trivial ``fountain'' or ``failed wind'' motions with $|\delta v| \sim 0.1-0.3\,v_{\rm K}$, comparable to the turbulent speeds, so some radiation-pressure-driven flows are present and could contribute to the observed velocity structure of lines and disk thickness in this ($\dot{m}=0.3$) simulation, but they are not creating large-scale coherent, escaping flows.

\begin{figure}
	\centering
	\includegraphics[width=0.97\columnwidth]{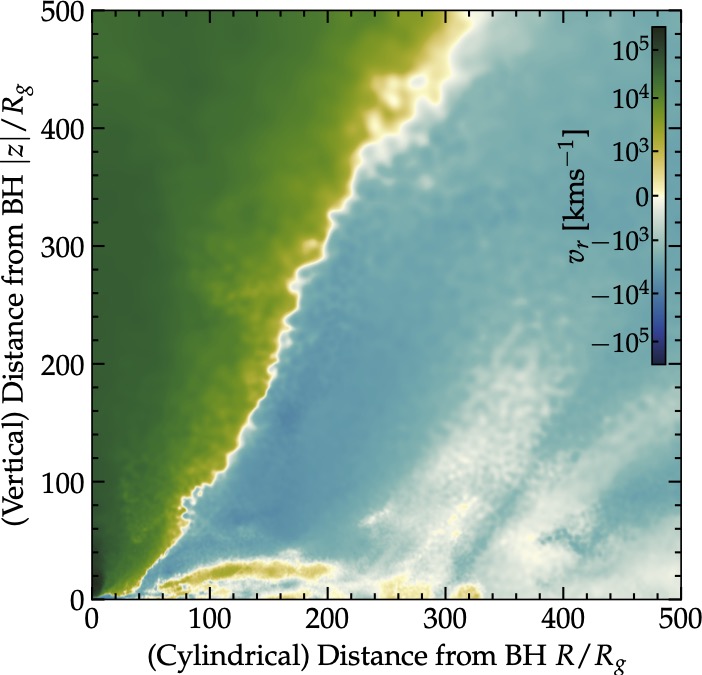} 
	\caption{Illustration of the bipolar radiation-pressure driven outflows in the inner disk for our super-critical ($f_{\dot{M}}=1$) case. We plot $|z|$ versus $R$ (out to $500\,R_{g}$) to collapse to a single quadrant and show colors indicating the radial velocity (labeled). Most of the volume, especially beyond this radius, shows inflow onto and through the disk. But there is a clear outflow bicone as seen in Fig.~\ref{fig:maps.vel}, narrowing at larger $r$. The outflow velocities range from relativistic in the inner tens of $R_{g}$ to $\gtrsim 10^{4}\,{\rm km\,s^{-1}}$ at $\sim 500\,R_{g}$ along the polar axis or $\sim 1000\,{\rm km\,s^{-1}}$ along the edge of the cone/disk surface.
	\label{fig:outflow.example}}
\end{figure}

The case of greatest interest here is our simulation with $\dot{m} \gg 1$ ($f_{\dot{M}}=1$), which does exhibit large-scale, coherent continuum radiation pressure driven outflows at the times shown (e.g.\ Figs.~\ref{fig:maps.vel}, \ref{fig:maps.fields.mdot}, \ref{fig:outflow.example}). In Fig.~\ref{fig:profiles.flux} we see that radiative acceleration is significantly weaker than gravity in the disk midplane. This is expected -- the disk itself is optically thick so is externally self-shielding (Figs.~\ref{fig:maps.pinwheel}, \ref{fig:maps.flux}, \ref{fig:profiles.mdot}), and is too heavy to be accelerated into radiation-pressure-driven outflow (Fig.~\ref{fig:maps.vel}) given the saturation of the radiation pressure from within the disk (\S~\ref{sec:rad.pressure.saturation}) producing  locally sub-Eddington fluxes from each annulus (Fig.~\ref{fig:profiles.flux}) in the outer disk. As seen in other supercritical simulations \citep[][and references therein]{jiang.dai:2024.supereddington.accretion.simulations.review}, the radiative acceleration is able to act most strongly in the less-dense, less Thompson-thick material in the polar direction, so the outflows driven tend to be strongly bipolar. The outflows feature velocities broadly comparable to the circular velocity at the corresponding radii, from $\sim 10^{3}$ to $\gtrsim 10^{5}\,{\rm km\,s^{-1}}$ at $\lesssim 100\,R_{g}$ (Fig.~\ref{fig:outflow.example}), with a  strong internal shear (leading the large radial Reynolds term $\langle \delta v_{r} \delta v_{r} \rangle$ in Figs.~\ref{fig:profiles.velocity} \&\ \ref{fig:profiles.stress.vs.mdot}), as expected \citep[e.g.][]{murray:1995.acc.disk.rad.winds}. And the total mass outflow rate (Figs.~\ref{fig:profiles.general}, \ref{fig:profiles.mdot}) is generally order-of-magnitude comparable to, but a factor of a couple to a few smaller than, the inflow rate \citep[similar to other recent simulations with very different numerical methods; see][]{kurosawa:disk.wind.feedback.efficiency,kuiper:2012.rad.pressure.outflow.vs.rt.method,jiang:2019.superedd.sims.smbh.prad.pmag.modest.outflows,toyouchi:2021.super.eddington.imbh.growth.dusty.nuclear.disks.sims,shi:2024.imbh.growth.feedback.survey,shi:2024.seed.to.smbh.case.study.subcluster.merging.pairing.fluxfrozen.disk,jiang.dai:2024.supereddington.accretion.simulations.review}. However the outflow rates can become larger in transient flare or outburst events (\S~\ref{sec:bursts}).

This means such outflows could be very significant in an ``absolute'' sense: carrying $\dot{M}_{\rm w} > 1\,{\rm M_{\odot}\,yr^{-1}}$ in mass, a momentum-flux $\dot{p}_{\rm w} \sim \dot{M}_{\rm w}\,v_{\rm w}$ comparable to $\sim L_{\rm bol}/c$ (given the numbers above), and an at-launch energy flux $\dot{E}_{\rm w} \sim 0.5\,\dot{M}_{\rm w}\,v_{\rm w}^{2} \sim 0.01-0.05\,L_{\rm bol}$, comparable to values often invoked in AGN feedback models for self-regulation of both SMBH masses and galaxy properties \citep{silkrees:msigma,murray:momentum.winds,begelman:msigma.feedback.model,croton:sam,hopkins:red.galaxies,hopkins:qso.all,ciotti:recycling.with.feedback.rad.vs.mech}. But they do not radically alter the structure of the disk itself, except perhaps in the most extreme burst/flare events (see \S~\ref{sec:bursts} and Fig.~\ref{fig:maps.demo.bursts}).

\subsection{Properties of the Broad-Line Region}
\label{sec:blr}

In \citet{hopkins:multiphase.mag.dom.disks}, we predicted that the basic properties of the BLR should emerge naturally from the surface layers of geometrically-thick, self-illuminated magnetically dominated disks. We will study this in more detail in the future, but it is worth briefly commenting on the global properties of the gas in our simulations at the radii expected for the BLR at the different luminosities simulated ($\sim 1-100\,$light-days, increasing with $\dot{m}$, e.g.\ \citealt{kaspi:2005.blr.size.reverb.mapping}). We see the gas at these radii (1) has typical velocities $\sim 10^{3}-10^{4}\,{\rm km\,s^{-1}}$; (2) exhibits covering factors $H/R \sim 0.1-0.3$; (3) is partially-ionized/partially-atomic; (4) has temperatures from few thousand to $\sim 10^{5}\,$K; and (5) densities $\sim 10^{9}-10^{13}\,{\rm cm^{-3}}$, with gas-mass-weighted mean $\sim 10^{10}-10^{11}\,{\rm cm^{-3}}$; (6) exhibits clear clumping/substructure at least down to scales $< 3\times10^{12}\,{\rm cm}$ (limited by our spatial resolution; the minimum predicted Sobolev length given the resolved disk properties from the expressions in \citealt{hopkins:multiphase.mag.dom.disks} ranges from $\sim 10^{10}-10^{13}\,{\rm cm}$); (7) has $|{\bf B}| \ll $G at the largest radii $\gtrsim 0.1\,$pc (well below maser Zeeman upper limits where they exist; see \citealt{modjaz:2005.agn.maser.modeling.Bfield.constraints.favor.fluxfrozen.disks,henkel:agn.masers,vlemmings:2007.maser.b.limits}); (8) has explicitly-calculated absorbed/reprocessed light fractions $\gtrsim 10\%$; (9) show modest optical depths from disk surface, implying line re-emission can escape from illuminated ``layer''; (10) has typical ionization parameters (combining the plotted densities and flux values) of $\sim 0.001-1$ (mostly varying with the large local density fluctuations in Fig.~\ref{fig:profiles.general}); (11) has a total gas mass $\sim 10-40\,{\rm M_{\odot}}$; and naturally (12) exhibits a ``thick disk'' geometry and kinematics. All of these properties agree remarkably well with what is observationally measured for the BLR \citep{krolik:1981.twophase.model.quasar.emission.lines,arav:1998.blr.not.discrete.clouds.but.inhomogeneous.system,krolik:1999.agn.book,kaspi:2005.blr.size.reverb.mapping,laor:2006.blr.could.be.smooth.disk.not.clumpy.but.must.be.turb,Peterson2006:BLR.review,gravity:2018.sturm.blr.rotating.thick.disk,gravity:2020.resolved.blr.size.disk.inside.dust.sub,devereux:2021.blr.mass.low.40msun.or.less.and.interior.to.sublimation.and.transition.to.xray.interior.favors.blr.as.disk,gravity:2021.resolved.blr.disk.hot.dust.coronal.regions,gravity:2024.blr.infrared.size.luminosity.relation.agn}. Clearly, a more quantitative analysis and comparison to observations is merited: but this requires post-processing the simulations with radiative line transfer and detailed metal ionization state calculations, and therefore is deferred to future work.

\subsection{Radiation-Thermal States \&\ Deviations from Photon-Trapped LTE}
\label{sec:lte}

\begin{figure}
	\centering
	\includegraphics[width=1.0\columnwidth]{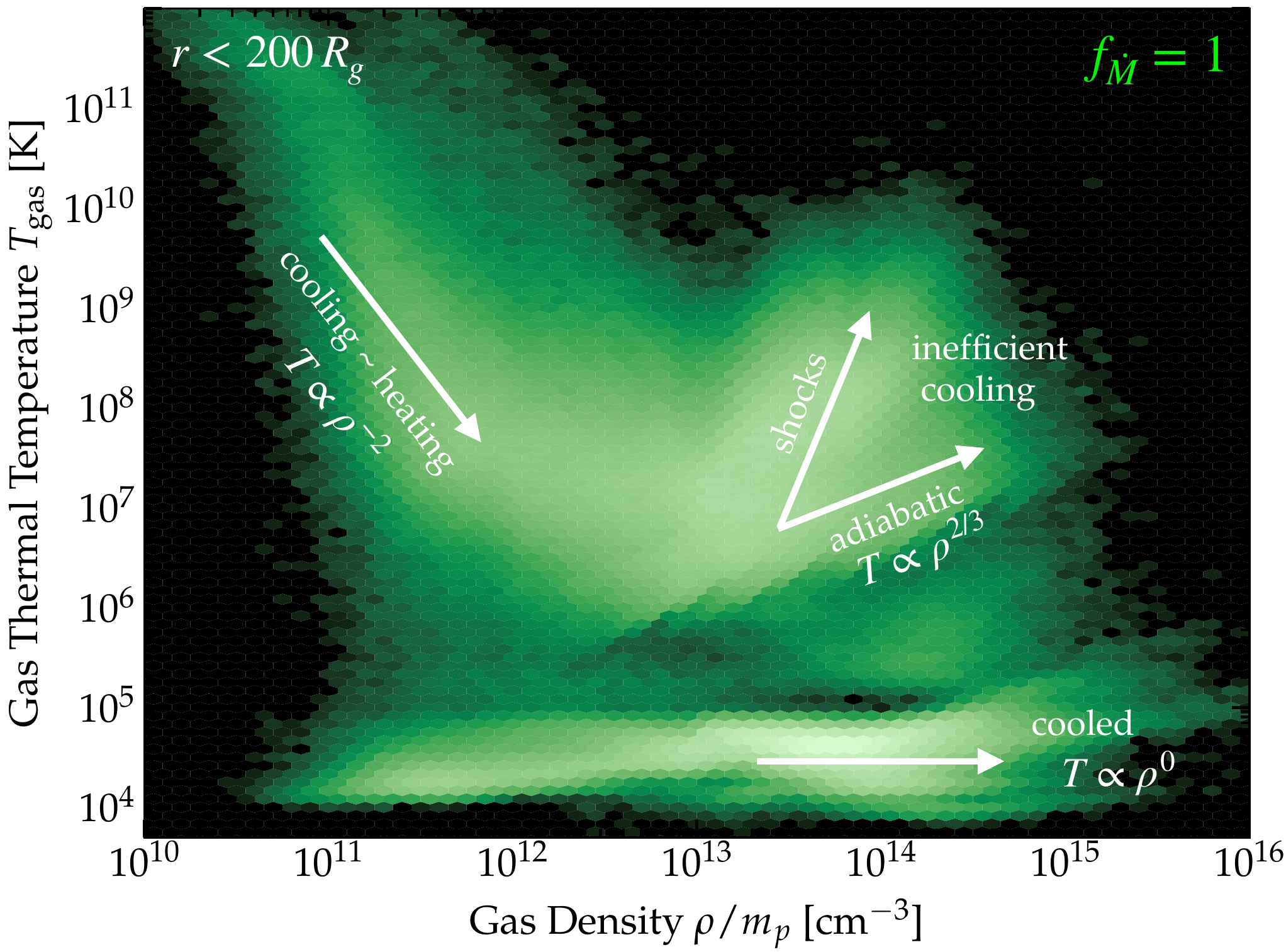} 
	\includegraphics[width=1.005\columnwidth]{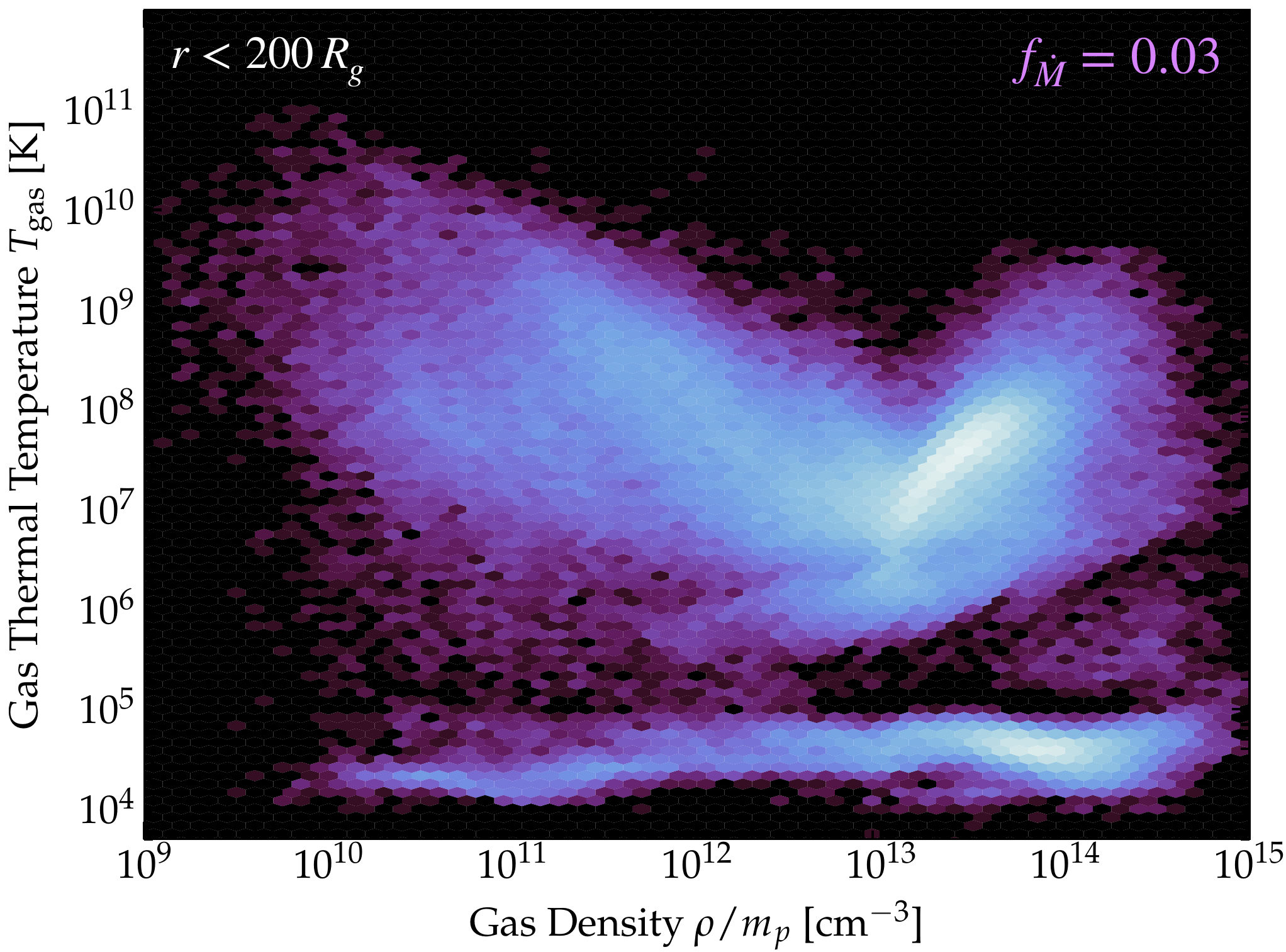} 
	\includegraphics[width=1.005\columnwidth]{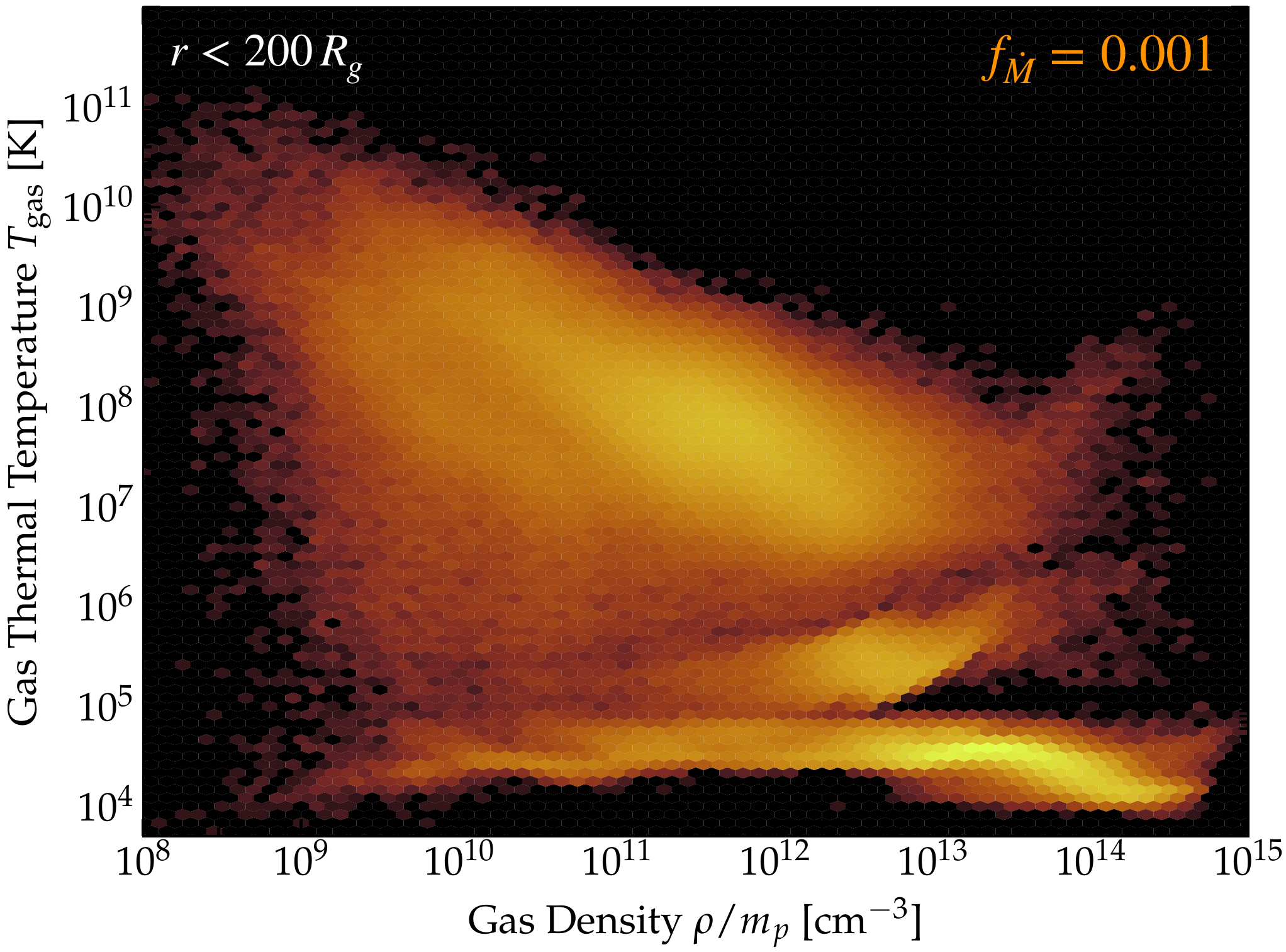} 
	\caption{Gas temperature $T_{\rm gas}$-density $\rho$ phase diagram in the inner disk ($r<200\,R_{g}$), for our three simulations with different $\dot{m}$ at the same time (labeled). Even in this region, complicated phase structure appears. We broadly label different behaviors. 
	Hot and low density gas is locally optically-thin with optically-thin radiative cooling ($\propto n^{2} \Lambda$) balancing external heating from e.g.\ turbulence, shocks, and reconnection ($\propto \rho \delta v_{\rm t}^{2}$), giving an inverse $T_{\rm gas} \propto \rho^{-2}$-like scaling.
	Efficiently-cooled gas sits near temperatures regulated by local collisional plus photo-ionization equilibrium, giving near isothermal $T_{\rm gas} \propto \rho^{0}$ behavior.
	Sufficiently hot, high-density ($n\gtrsim 10^{13}\,{\rm cm^{-3}}$) cannot cool rapidly and so is heated adiabatically ($T_{\rm gas} \propto \rho^{2/3}$) or more rapidly still in shocks.
	All runs exhibit similar behavior despite different $\dot{m}$, though the density scales shifts. At larger radii, multi-phase structure becomes even more pronounced, and atomic/molecular phases eventually appear at $T_{\rm gas} \lesssim 10^{4}\,$K.
	\label{fig:T.vs.rho}}
\end{figure}

\begin{figure}
	\centering
	\includegraphics[width=1.0\columnwidth]{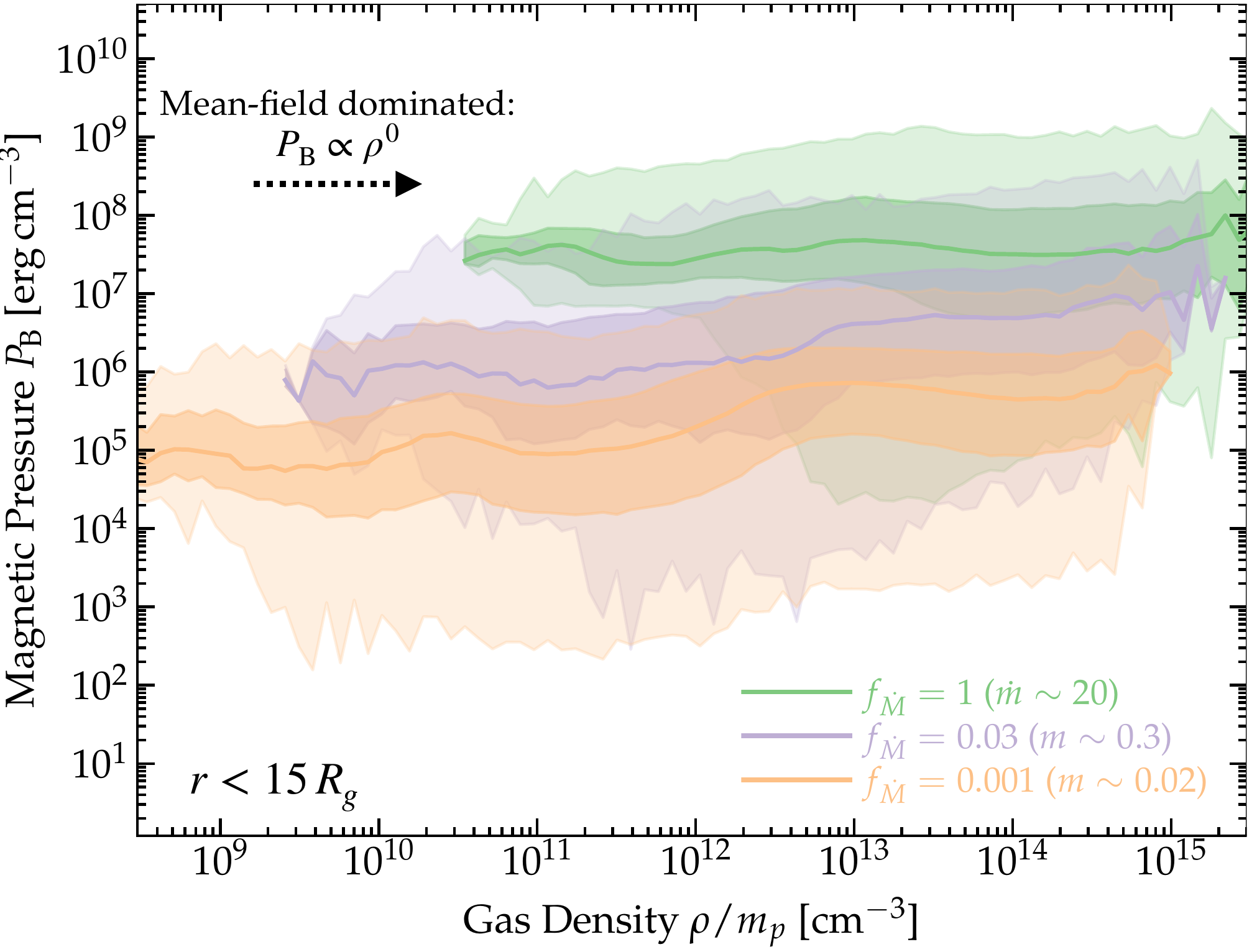} 
	\includegraphics[width=1.005\columnwidth]{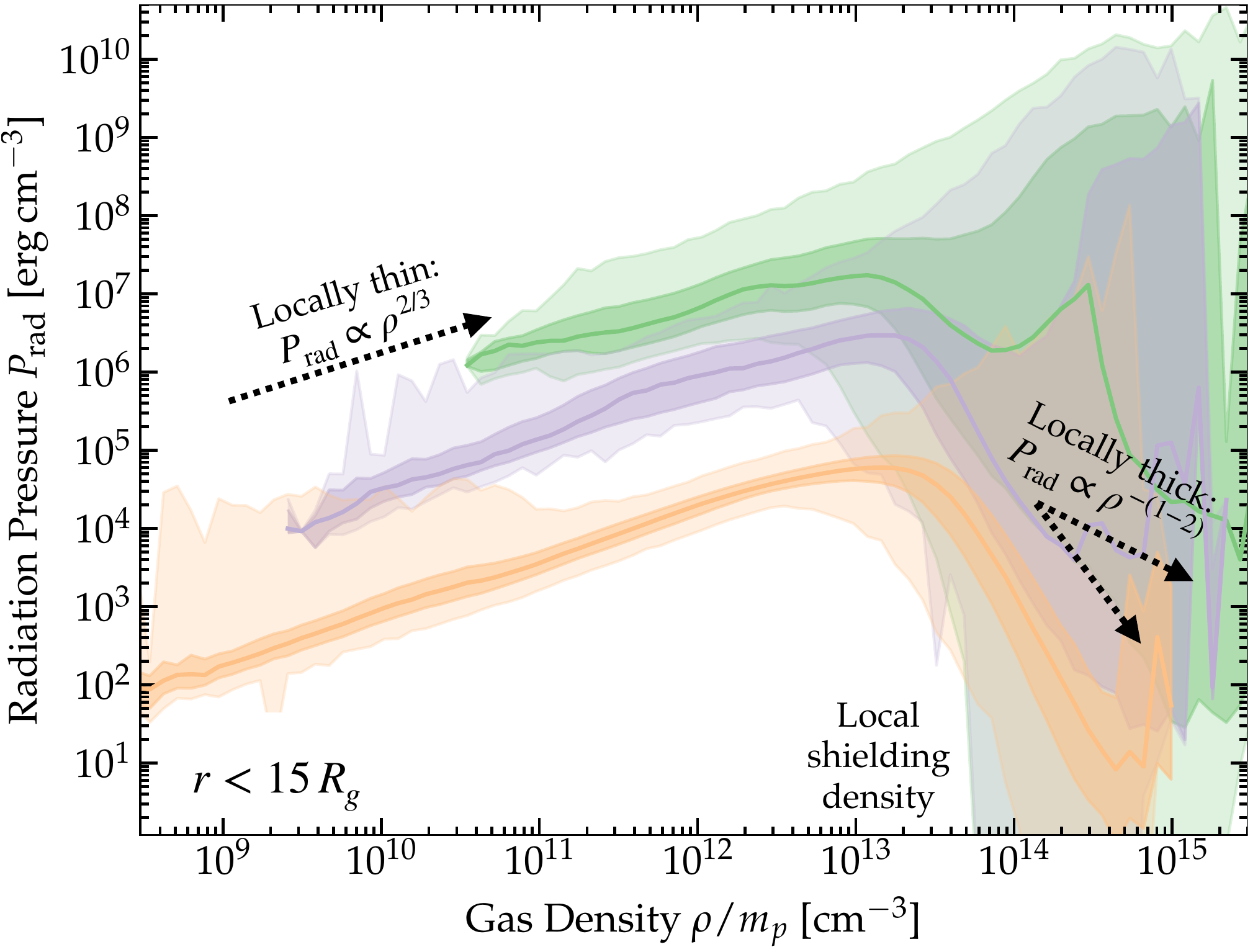} 
	\includegraphics[width=1.005\columnwidth]{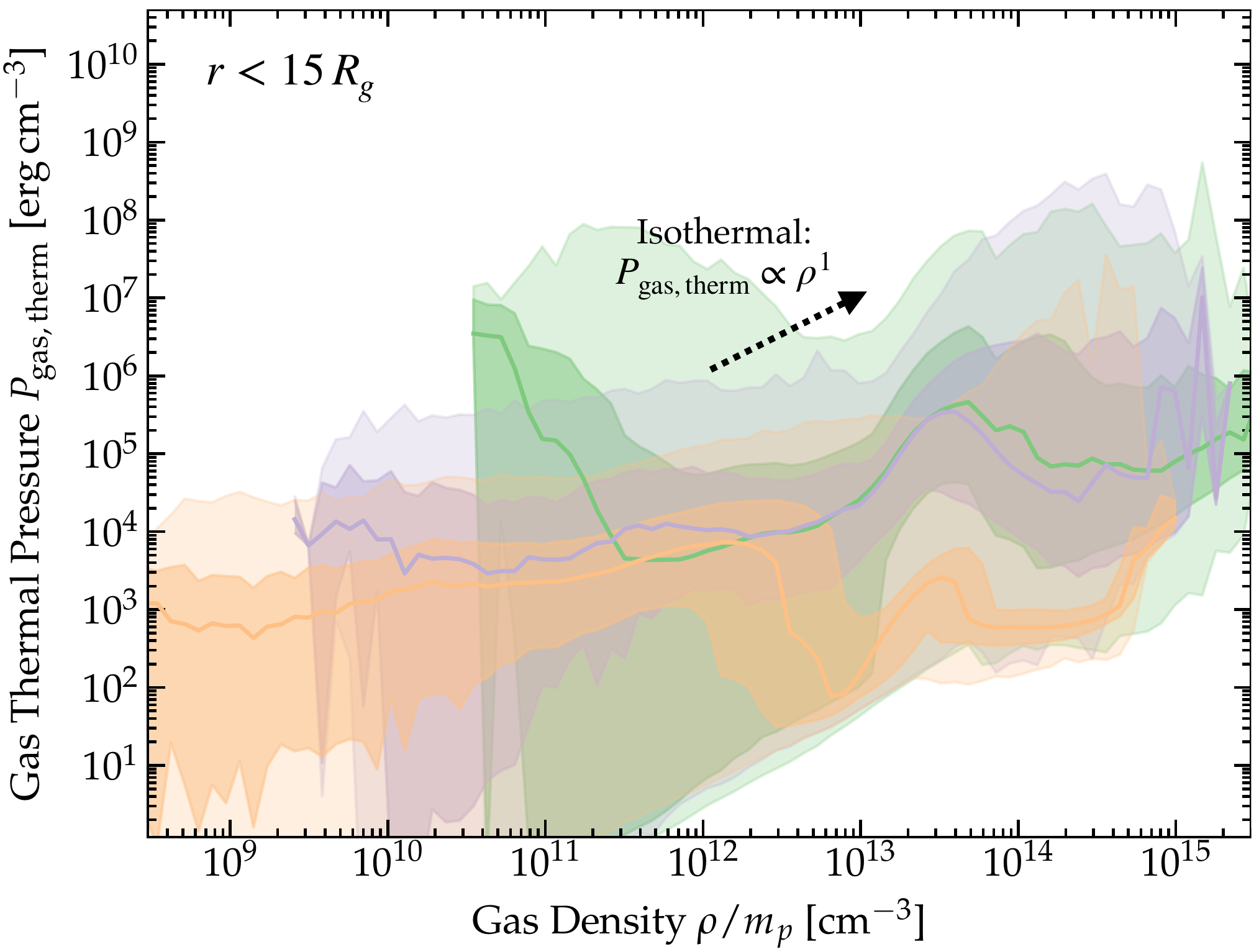} 
	\caption{Pressures $P$ versus gas density $\rho$ in the innermost disk ($r<15\,R_{g}$) for our three simulations with different $\dot{m}$ at the same time. We show the volume-weighted median ({\em line}), $\pm1\sigma$ ({\em dark shaded}), and $\pm 3\sigma$ ({\em light shaded}) $P$ at each $\rho$.
	{\em Top:} Magnetic pressure $P_{\rm B}$. The strong mean toroidal field dominates so this depends weakly on $\rho$ at a given $R$, though trans-\Alf{ic} turbulence generates significant fluctuations.
	{\em Middle:} Effective radiation pressure $P_{\rm rad,\,eff}$ (\S~\ref{sec:lte}). At a given $R$, we see a soft $P_{\rm rad} \propto \rho^{2/3}$ approximately at lower densities, tracing locally-optically-thin gas where not all of $u_{\rm rad}$ can couple strongly, which turns at a critical density and then falls rapidly as $\propto \rho^{-1}-\rho^{-2}$, tracing dense structures at the thermal scale-length which are strongly shielded from the external radiation.
	{\em Bottom:} Gas thermal pressure $P_{\rm gas,\,therm}$. Very crudely the median follows an isothermal-like track but with large deviations from multi-phase structure (\S~\ref{fig:T.vs.rho}).
	\label{fig:pressure.rinner}}
\end{figure}

\begin{figure*}
	\centering
	\includegraphics[width=1.0\textwidth]{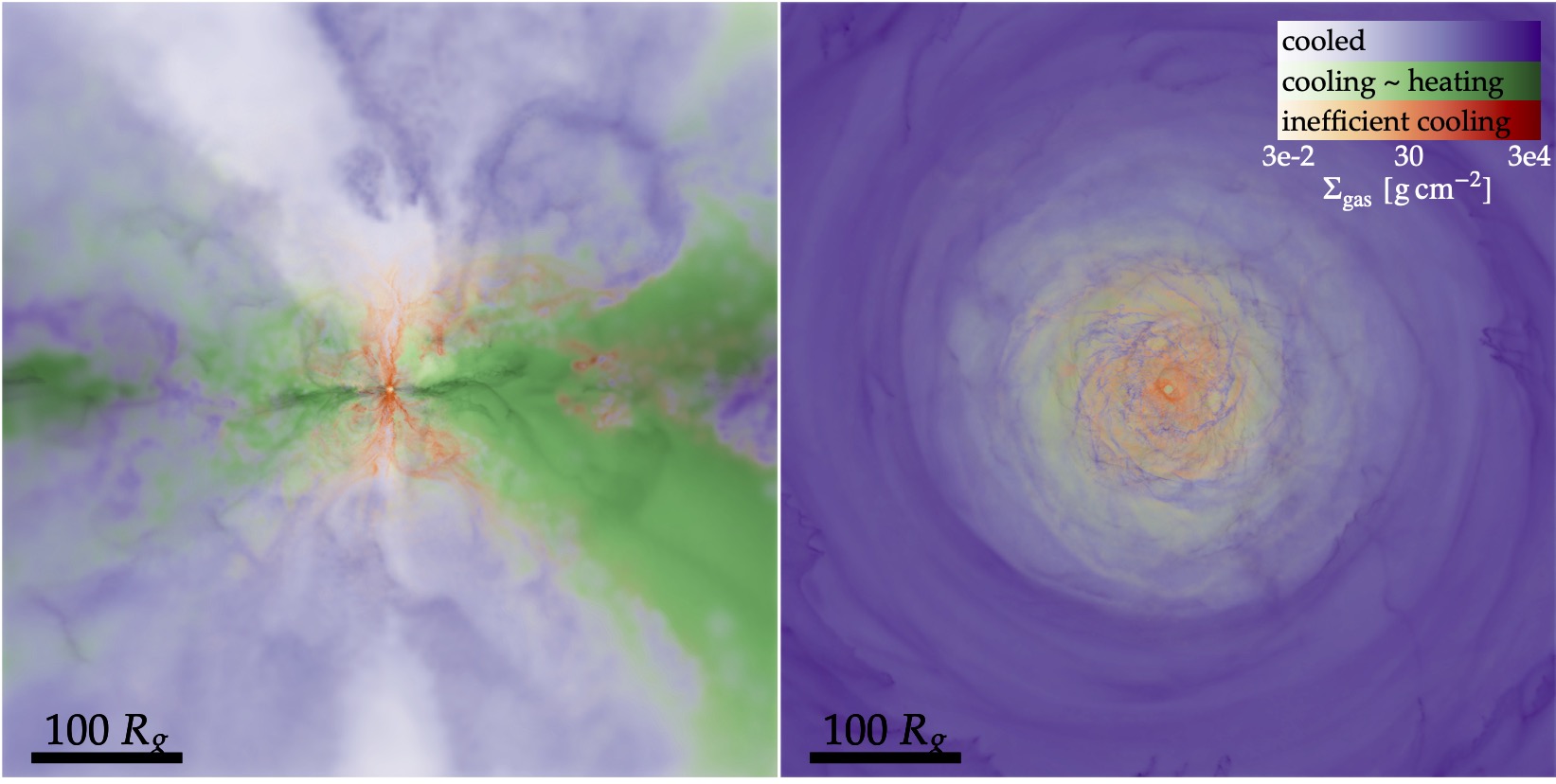}
	\caption{Edge-on and face-on projections (as Fig.~\ref{fig:maps.pinwheel.edgeon}) of gas surface density $\Sigma_{\rm gas}$ of the high-$\dot{m}$ ($f_{\dot{M}}=1$) snapshot from Fig.~\ref{fig:T.vs.rho}, within the same radii, highlighting the spatial distribution of gas in the different phases labeled in Fig.~\ref{fig:T.vs.rho}. Specifically gas which is quasi-isothermal at $T< 10^{6}\,$K (``cooled''), or warmer gas at either low densities ($\lesssim 10^{12.5}\,{\rm cm^{-3}}$, ``cooling $\sim$ heating'') or higher densities ($\gtrsim 10^{12.5}\,{\rm cm^{-3}}$, ``inefficient cooling''). At larger radii more of the gas is in cooler phases (as expected), but all three co-exist at similar radii, with some cold/cooled dense structures in the midplane, others cooling but hotter, and the highest surface-density (hence highest optical depth) structures in the jet/outflow shock regions showing less efficient cooling. 
	\label{fig:cooling.vs.heating.phase.image}}
\end{figure*}


Given our treatment of radiation-thermochemistry which allows for non-steady-state, non-LTE, multi-group, independent $T_{\rm gas}$, $T_{\rm rad}$, $T_{\rm dust}$ temperatures and non-thermal processes (\S~\ref{sec:opacities}), as well as incomplete gas-radiation coupling, it is worth examining the different pressures and deviations from LTE or tight coupling in the inner disk. This is a particularly interesting region, because historical radiation-MHD treatments (often motivated by SS73-type models, with orders-of-magnitude larger disk optical depths and densities) have often assumed single-radiation-group, thermal-only/blackbody, strictly LTE, perfect gas-radiation coupling (in addition to neglecting atomic/molecular/dust processes). 

We will see large deviations from these behaviors, as predicted in e.g.\ \citet{hopkins:multiphase.mag.dom.disks}, because these disks are only {\em globally} modestly optically thick to self-absorption (Fig.~\ref{fig:profiles.flux}). This means large-scale modified blackbody emission is expected, but that {\em local} deviations from equilibrium and tight coupling, e.g.\ in density structures which are much smaller than the disk (which we see are ubiquitous, \S~\ref{sec:results}) and therefore have much lower local/self optical depths, are also expected. Note that we focus on the more steady-state disk behaviors here, but the deviations from LTE and thermal radiation can become even larger in transient flare/outburst-type events (\S~\ref{sec:bursts}).

\subsubsection{Gas Thermal Phase Structure}
\label{sec:lte:gas.T}

Fig.~\ref{fig:T.vs.rho} shows the thermal gas phase structure in the inner disk (the multi-phase structure in the outer disk at $>300\,R_{g}$ is discussed extensively in \paperone, \papertwo, and \paperthree). The gas is clearly not single-phase, with at least three qualitatively different behaviors of a significant fraction of the mass/volume. At low densities and high ($\gtrsim 10^{6}\,$K) temperatures, the gas is locally optically-thin, with temperature set by approximate equilibrium between volumetric optically thin cooling (mostly free-free, so $\dot{u}_{\rm therm} \sim n^{2} \Lambda \propto \rho^{2} T_{\rm gas}^{1/2}$) balancing heating by shocks/turbulence/magnetic reconnection (all similar, given the trans-\Alf{ic} turbulence, with $\dot{u} \sim \rho \delta v_{\rm turb}^{3}/L_{\rm turb} \sim \rho v_{A}^{2} \Omega$), so we anticipate something like $T_{\rm gas} \propto \rho^{-2}$, broadly similar to what is seen. Gas which has cooled or is cooling even more efficiently sits at the collisional plus photo-ionization equilibrium at $T_{\rm gas} \sim 10^{4}-10^{5}\,$K as expected, and is nearly isothermal. This validates our expectation from analytic models \citep{hopkins:multiphase.mag.dom.disks} that $t_{\rm cool} \ll \Omega^{-1}$, even in the optically thin gas, throughout almost all of the disk (except some hot coronal gas). We have also verified this directly by checking the cooling rates in-code. Hotter, denser gas becomes optically-thick to its own cooling radiation {\em locally}, so has cooling times longer than its local dynamical/crossing times and so can be heated adiabatically ($T_{\rm gas} \propto \rho^{2/3}$, as it is effectively monatomic here being fully-ionized) or in non-radiative shocks (which give density enhancement of a fixed factor of a few, with large temperature enhancement $\sim \mathcal{M}_{s}^{2}$, so rise more steeply). 

We can obtain a simple estimate for the critical density $n_{\rm crit} \sim 10^{13}m_{p}\,{\rm cm^{-3}}$ of the transition in Fig.~\ref{fig:T.vs.rho}, if we ask when dense structures or shocks at the thermal scale length $H_{\rm therm} \sim c_{s}/\Omega$ will become opaque, $\tau_{\rm therm} \sim \kappa_{\rm es} H_{\rm therm} n_{\rm crit} \gtrsim 1$, giving $n_{\rm crit} \sim 10^{13}\,{\rm cm^{-3}}\,(R/10\,R_{g})^{-3/2}\,(x_{e}\,M_{\rm BH}/10^{7}\,M_{\odot})^{-1}\,(T_{\rm gas}/10^{7}\,{\rm K})^{-1/2}$.

\subsubsection{Pressures versus Density}
\label{sec:lte:P.n}

\begin{figure}
	\centering
	\includegraphics[width=1.0\columnwidth]{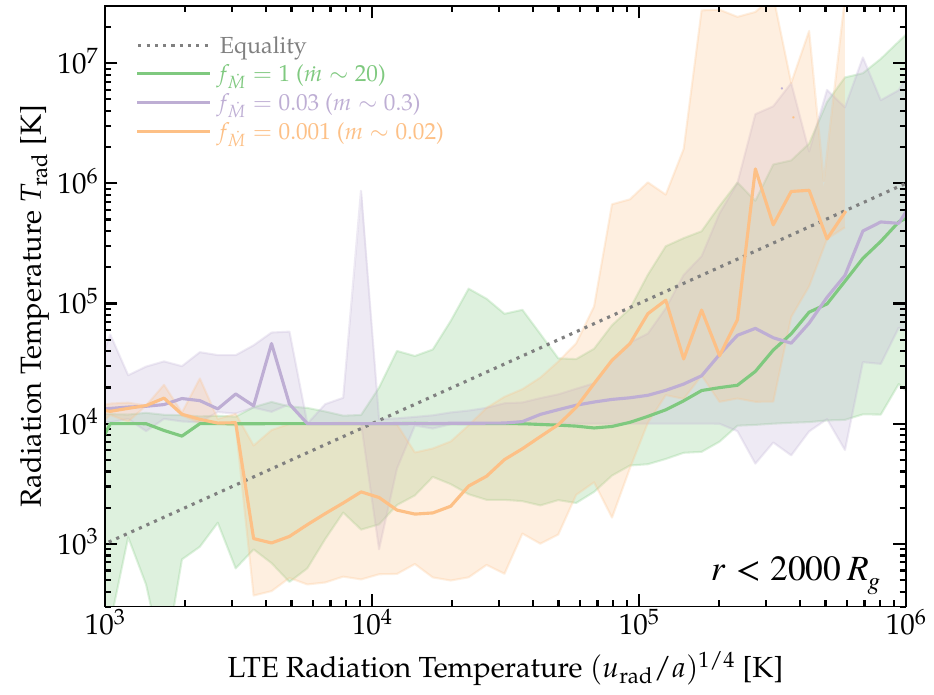} 
	\caption{Comparison of the explicitly-evolved simulation radiation temperature $T_{\rm rad}$ in each cell in the inner disk at $r < 2000\,R_{g}$ versus the expected radiation temperature in LTE $T_{\rm rad,\,LTE} \equiv (u_{\rm rad}/a)^{1/4}$, for our three different $\dot{m}$ simulations with volume-weighted median ({\em line}) and $\pm3\,\sigma$ range ({\em shaded}) at each $T_{\rm rad,\,LTE}$. While order-of-magnitude similar, the values of $T_{\rm rad}$ at a given $u_{\rm rad}$ can vary by $\sim 2\,$dex and the systematic offset can be as large as $\sim1\,$dex. Even in the inner disk, it therefore important to evolve $T_{\rm rad}$ explicitly. 
	\label{fig:T.vs.T.alt}}
\end{figure}

\begin{figure}
	\centering
	\includegraphics[width=1.005\columnwidth]{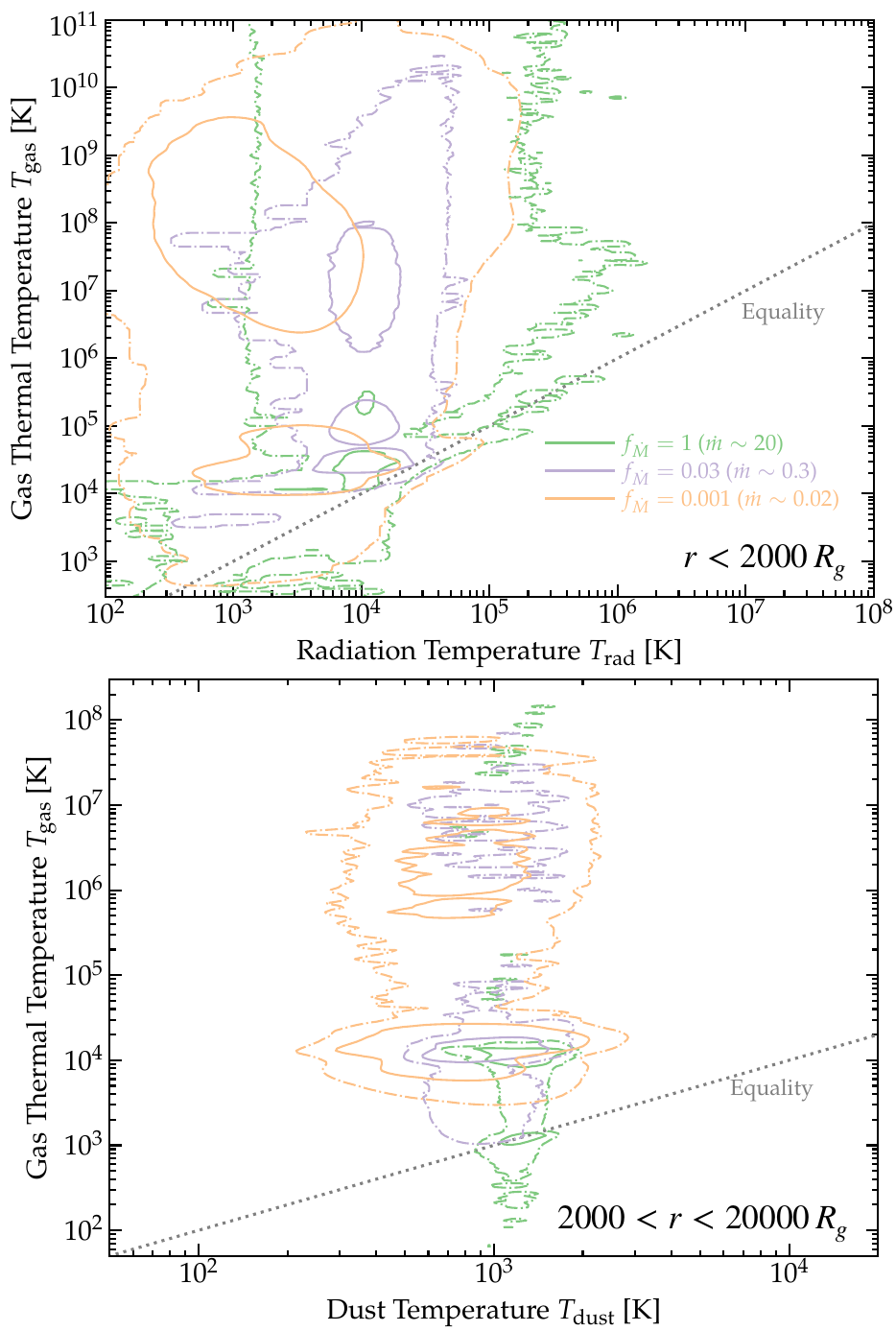} 
	\caption{Comparison of different evolved temperatures in the inner disk. For each, we show contours corresponding to the volume-weighted 2D histogram at iso-density contours near-peak ({\em solid}) and enclosing $>99\%$ of the volume ({\em broken}).  
	{\em Top:} Gas thermal temperature $T_{\rm gas}$ versus $T_{\rm rad}$. 
	{\em Bottom:} $T_{\rm gas}$ versus dust temperature $T_{\rm dust}$, only for cells with $>1\%$ of their potential dust mass (not sublimated; this requires we look at larger radii $>2000\,R_{g}$ to see). 
	The radiation, gas, and dust temperatures are highly NLTE, even in the central/inner disk at the highest $\dot{m}$, where the optical depths are maximized.
	\label{fig:T.vs.T}}
\end{figure}

Fig.~\ref{fig:pressure.rinner} compares different pressures versus density in the inner disk more broadly-defined. As expected and seen in \papertwo, given that we see the mean field generally dominates $|{\bf B}|$, the magnetic pressure as usually defined $P_{\rm B} \equiv |{\bf B}|^{2}/8\pi$ depends only weakly on $\rho$ at a given radius (almost all the weak trend in $P_{\rm B}$ versus $\rho$ in Fig.~\ref{fig:pressure.rinner} comes from the dependence of both on $R$ within the range plotted). But trans-\Alf{ic} turbulence does produce significant fluctuations. Magnetic pressure is not isotropic: this means as discussed in the main text, \papertwo, and \citet{hopkins:superzoom.imf}, that gas will generally ``feel'' a stiff magnetic equation-of-state in compression in the vertical or radial (perpendicular) directions, but soft in the azimuthal (parallel) direction, though the latter compressions are strongly sheared-out.

The thermal pressure in Fig.~\ref{fig:pressure.rinner}, $P_{\rm gas,\,thermal} \equiv (\rho/\mu m_{p})\,k_{B} T_{\rm gas}$, has a vaguely isothermal ``floor'' as expected from Fig.~\ref{fig:T.vs.rho} but reflects the same complicated multiphase structure. 

Less trivial is the radiation pressure. Here we define $P_{{\rm rad,\,eff},\,i} \equiv [\tau_{{\rm local},\,i}/(1+\tau_{{\rm local},\,i})]\,u_{{\rm rad},\,i}/3$ for each cell, where $\tau_{{\rm local},\,i} = \langle \kappa_{i} \rangle_{\nu}\,\rho_{i}/|\nabla \rho_{i}|$ reflects the local flux-weighted mean opacity and density gradient scale-length. The usual convention in the accretion disk literature is to simply take $P_{\rm rad} \equiv u_{\rm rad}/3$, and this used in e.g.\ the standard definition of $\beta_{\rm tot}$ in the text. But as noted there, this can be very misleading in some limits, like when the gas is locally optically-thin, and there can be arbitrarily small radiation forces/accelerations on the gas at any given $P_{\rm rad}$. More rigorously, we can compare the forces directly as in Fig.~\ref{fig:profiles.flux}, where we see $P_{\rm rad}$ only dominates the actual {\em acceleration} for gas in the inner disk in the highly polar regions for the highest $\dot{m}$. But the definition in Fig.~\ref{fig:pressure.rinner} gives us a better proxy for the ``effective'' radiation pressure which can actually act on a parcel of gas. Notably, we see a transition at a similar $n_{\rm crit}$ discussed in \S~\ref{sec:lte:gas.T}. Below this density, we expect local structures to be {\em locally} optically thin, so the full radiation energy density cannot effectively act as a pressure. Even if $u_{\rm rad}$ is homogeneous, we expect lower $P_{\rm rad,\,eff}$ at lower densities owing to the weaker coupling, and see this, with effective $P_{\rm rad,\,eff} \propto \rho^{2/3}$. Above the critical density, we see a huge spread in $P_{\rm rad}$. Structures here are self-shielded against the ambient radiation field. Thus at the highest $P_{\rm rad}$ we see hot, non-cooling structures from Fig.~\ref{fig:T.vs.rho}, while at the lowest $P_{\rm rad}$ we see structures at the disk midplane/dense clumps which are cooling (often $\sim 10^{4}-10^{5}\,$K dense gas from Fig.~\ref{fig:T.vs.rho}) but shielded from the collective radiation field. This is the natural result of the disk being {\em externally} illuminated over much of its size, with lower-$P_{\rm rad}$ center seen in Figs.~\ref{fig:maps.pinwheel}, \ref{fig:maps.flux}, \ref{fig:maps.pinwheel.mdot}. 

There appears, therefore, to be no regime in which one can ``safely'' take the tight-coupling limit, which would predict $P_{\rm rad,\,eff} \propto \rho^{4/3}$. This also means that one cannot with any meaning define an ``effective single-fluid sound-speed'' $c_{s,\,{\rm eff}}$ (commonly taken to scale as $c_{s,\,{\rm eff}} \approx \sqrt{\gamma_{\rm eff}[P_{\rm rad}/P_{\rm therm}]\,(P_{\rm rad}+P_{\rm therm})/\rho}$ such that $\gamma_{\rm eff} \rightarrow 5/3$ for $P_{\rm rad} \ll P_{\rm therm}$ and $\gamma_{\rm eff} \rightarrow 4/3$ for $P_{\rm rad}\gg P_{\rm therm}$; see \citealt{mihalas:1984oup..book.....M}), as these are only really valid in the infinite-optical-depth LTE and strong tight-coupling regime.

We can also verify that the qualitative behavior in Fig.~\ref{fig:pressure.rinner}  is similar, albeit with systematically lower pressures and densities, at different radii (making the same plot with the gas at e.g.\ $r=(15-30,\,50-150,\,150-300)\,R_{g}$). The predicted $n_{\rm crit}$ does shift weakly with radius as expected from our simple argument in \S~\ref{sec:lte:gas.T}. Outside this radius the radiation pressures are even weaker and gas more strongly multiphase (\papertwo).

\subsubsection{Temperatures versus LTE}
\label{sec:lte.temps}

Fig.~\ref{fig:T.vs.T.alt} plots one diagnostic of LTE in the inner disk, specifically comparing the explicitly-evolved simulation radiation temperature $T_{\rm rad}$, to the LTE expectation $T_{\rm rad,\,LTE} \approx (u_{\rm rad}/a)^{1/4}$. If the radiation is either not perfectly thermal/blackbody, or the system is not in strictly LTE (i.e.\ does not have absorption mean-free-path much smaller than other scale-lengths), these will not necessarily agree. Consistent the behaviors above and the disks being only modestly optically-thick, we see that while $T_{\rm rad}$, on average, is not wildly different from the LTE expectation, the local deviations can be large (up to $\gtrsim 1\,$dex in either direction). 

Fig.~\ref{fig:T.vs.T} further compares $T_{\rm gas}$ to $T_{\rm rad}$, as well as $T_{\rm dust}$ (where dust is not completely sublimated) to $T_{\rm gas}$. There is little or no correlation between the three, though $T_{\rm dust}$ correlates better with $T_{\rm rad}$ at much larger radii outside the sublimation radius (where dust reprocesses most of the radiation; \paperone) and $T_{\rm gas}$ does exhibit a lower envelope around $T_{\rm rad}$ (it is rare to have $T_{\rm gas} \ll T_{\rm rad}$ in the inner disk, though this can be true in the outer disk; see \papertwo).

\begin{figure}
	\centering
	\includegraphics[width=1.005\columnwidth]{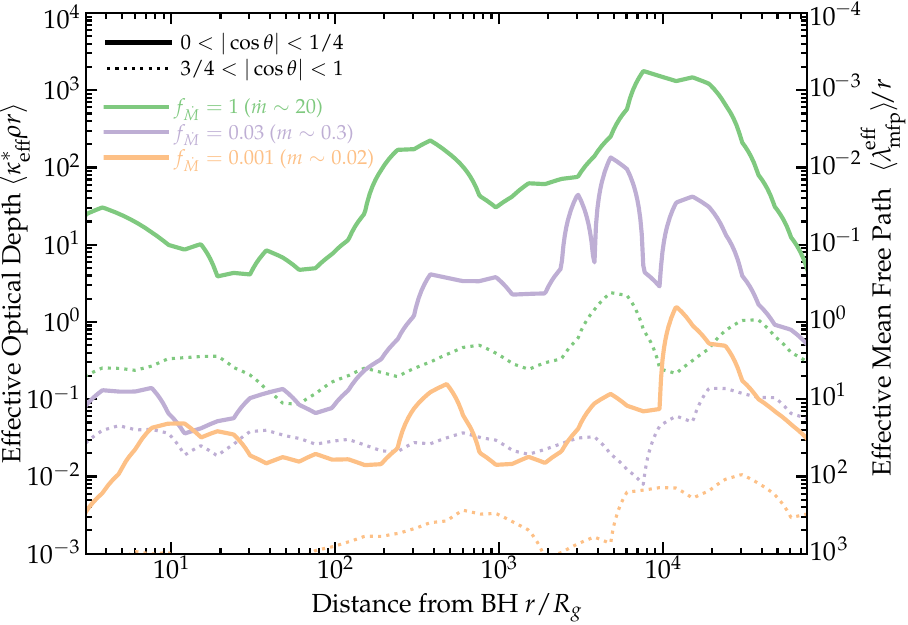} 
	\caption{Characteristic local effective optical depth, $\kappa_{\rm eff}^{\ast} \equiv \sqrt{\kappa_{\rm abs}\,(\kappa_{\rm abs} + \kappa_{\rm s})}$ in terms of absorption $\kappa_{\rm abs}$ and scattering $\kappa_{\rm s}$ opacities. As Fig.~\ref{fig:profiles.flux}, we plot the median (volume-weighted) $\kappa_{\rm eff}^{\ast}\rho r \sim r / \lambda^{\rm abs,\,eff}_{\rm mfp}$ in each annulus for cells within the given polar angle range (labeled). Note that calculating the integrated $\tau_{\rm eff}^{\ast} \equiv \int \kappa_{\rm eff}^{\ast}\rho d \ell$ for photon trajectories from e.g.\ the midplane at $r$ to escape can give factor of $\gtrsim 10-30$ larger $\tau_{\rm eff}^{\ast}$ in the inner disk; the plotted number is better thought of as the ratio $r$ to the median effective absorption mean-free path $\lambda^{\rm abs,\,eff}_{\rm mfp}$ seen by photons instantaneously at some $r$ and $\theta$. This is always smaller than the total optical depth ($\kappa_{\rm abs} < \kappa_{\rm s}$), and smaller than the global optical depth to escape, but shows that much of the volume in the inner disk is {\em locally} optically-thin to absorption in local density structures with scale-lengths $\lesssim H \ll r$.
	\label{fig:kappa.eff}}
\end{figure}

\subsubsection{Opacities and Criteria for LTE \&\ Tight-Coupling}

Fig.~\ref{fig:kappa.eff} plots a simple proxy for the characteristic local optical depths along different polar angles within the simulation, as Figs.~\ref{fig:profiles.general} \&\ \ref{fig:profiles.mdot}, but instead of the total opacity $\kappa_{\rm tot} \equiv \kappa_{\rm s} + \kappa_{\rm abs}$ (the sum of the scattering and absorption opacities), we plot the effective absorption opacity $\kappa_{\rm eff}^{\ast} \equiv \sqrt{\kappa_{\rm abs}\,\kappa_{\rm tot}}$. Almost everywhere in the inner disk, $\kappa_{\rm abs} \ll \kappa_{\rm s} \sim \kappa_{\rm tot}$, typically by factors of $\sim 100-1000$, so $\kappa_{\rm eff} \sim (0.03-0.1)\,\kappa_{\rm tot}$ (compare Fig.~\ref{fig:profiles.mdot}). In the outer disk (especially in the midplane at $\gtrsim 3000\,R_{g}$), dust survives, with (given the warm radiation temperatures) $\kappa_{\rm abs} \sim \kappa_{\rm s}$, so the difference is minimized. Note that we plot a median local $\kappa_{\rm eff}^{\ast}\rho r \sim r/\lambda^{\rm abs,\,eff}_{\rm mfp}$ (roughly, the ratio of $r$ to the effective absorption mean-free-path $\lambda^{\rm abs,\,eff}_{\rm mfp}$) in different annuli -- this can be smaller by as much as an order of magnitude compared to what one often refers to as the ``optical depth of the disk'' obtained by integrating $\int \kappa_{\rm eff}^{\ast}\rho d\ell$ along a hypothetical photon trajectory from the midplane to infinity, because we are not interested here in the {\em global} optical depth and reprocessing in the disk and atmosphere, but rather in the {\em local} mean free-paths $\lambda^{\rm abs,\,eff}_{\rm mfp}$ of the photons. Essentially, this gives $r/\lambda^{\rm abs,\,eff}_{\rm mfp}$ the inverse of the median absorption mean-free-path length in units of $r$ at the given $r$ and polar angle. Even in the midplane at our highest accretion rates/surface densities ($f_{\dot{M}}=1$), $\langle \kappa_{\rm eff}^{\ast} \rho r \rangle \sim r / \langle \lambda^{\rm abs,\,eff}_{\rm mfp} \rangle$ is modest ($\sim 10$), i.e.\ the typical effective absorption mean-free-path $\langle \lambda^{\rm abs,\,eff}_{\rm mfp} \rangle \sim 1/ \langle \kappa_{\rm eff}^{\ast} \rho \rangle \sim 0.1\,r \sim (0.3-1)\,H$. For lower accretion rates $\dot{m}\sim 0.02-0.3$ ($f_{\dot{M}}=0.001-0.03$) $\langle \kappa_{\rm eff}^{\ast} \rho r \rangle$ can be below unity, i.e.\ $\langle \lambda^{\rm abs,\,eff}_{\rm mfp} \rangle > r$ interior to $\lesssim 200\,R_{g}$, which is equivalent to saying that most of the gas is {\em locally} optically-thin to absorption over the typical local density gradient scale lengths.

For comparison, consider the requirements for local photon-trapping/tight-coupling and strict LTE. ``Full'' local trapping, as required for ``effective equation of state'' or ``effective sound speed'' or single-fluid gas-radiation models (as opposed to the much less stringent global trapping required for advection of radiation into the BH), requires the total/scattering mean-free-path $\lambda_{\rm s} \sim 1/\kappa_{\rm s} \rho$ be much smaller than the smallest density gradient length scales $\ell$, but also that the photon escape/diffusion time $\sim \ell/(c/\tau(\ell)) \sim \ell^{2}\kappa_{\rm s} \rho/c$ be much longer than the characteristic evolution times on $\ell$, given by e.g.\ turbulent crossing times $\ell/v_{\rm turb}(\ell)$. As noted in the text and derived in \citet{hopkins:multiphase.mag.dom.disks}, for a supersonically-turbulent disk the $\ell$ of greatest interest is the sonic/shock/Sobolev length (all similar) $\ell_{\rm sonic} \sim (c_{s}/v_{\rm turb})\,H_{\rm therm} \sim H/\mathcal{M}_{s}^{2}$ (where $c_{s}$ and $\mathcal{M}_{s}$ are defined with respect to the gas thermal sound speed). So the timescale criterion becomes $\kappa_{\rm s}\,\rho\,R \gg (c/v_{\rm turb}[\ell])\,(H/R)^{-1}\,\mathcal{M}_{s}^{2} \gtrsim 10^{5}-10^{8}$ (i.e.\ $\lambda^{\rm abs,\,eff}_{\rm mfp} \lesssim 10^{-8}-10^{-5}$), much larger than we see (Fig.~\ref{fig:profiles.flux}). The criterion for strict LTE is dimensionally similar but even more demanding because it requires reprocessing, so we must exceed the same threshold with the effective opacity $\kappa_{\rm eff}^{\ast}$ instead of $\kappa_{\rm s}$. So unless the disks were orders-of-magnitude more dense, we should expect to find (as we do) that local tight-trapping and LTE are not good local approximations.

\subsection{Sensitivity to Physics}
\label{sec:sensitivity}

In the inner disks ($\ll 1000\,R_{g}$), there are many physical processes formally included in our simulations which can be safely neglected, and a few which are of key importance. We review these briefly here.

\subsubsection{Physics Which Can Be Safely Neglected}
\label{sec:physics.can.be.neglected}

There is zero star formation at $\lesssim 0.05\,$pc in our longer-duration \papertwo\ simulations and these simulations, and negligible stellar mass (from migration of stellar orbits) in the inner disk, so simply removing all star formation and stellar evolution and stellar ``feedback'' physics yields identical results at the radii of interest here. Dynamical self-enrichment/pollution of metals is therefore also negligible. Related, local self-gravity of the disk is weak as the inner disk mass is $\ll 10^{-6}\,M_{\rm BH}$ and Toomre $Q$ ranges from $Q \sim 10^{5}-10^{10}$ (though it remains possible that inward-propagating, swing-amplified self-gravity modes from large radii could be important; see \citealt{hopkins:slow.modes,lin:2015.one.armed.spiral.propagation,duffell:2024.santa.barbara.binary.disk.code.comparison.gizmo.tests.comparable.ideal.performance}). 

Also per \papertwo\ \&\ \paperthree, even the atomic disk is sufficiently well-ionized that non-ideal (Ohmic resistivity, Hall drift, ambipolar drift) effects are smaller than turbulent or numerical resistivities and reconnection rates by factors of $\sim 10^{15}$. Likewise the fluid is extremely well into the collisional regime with electron collision frequencies $\sim 10^{-10}-10^{-6}\,\Omega$, so collisionless or kinetic effects and micro-physical Spitzer/Braginskii/atomic/molecular conduction and viscosity are negligible. Similarly in the {\em disk}, the temperatures are generally well below those where two-temperature effects (treated in a simplified manner in the simulations where relevant) are significant, though this is not true for all the gas in the diffuse disk atmosphere/corona (Fig.~\ref{fig:profiles.general}). And though we include a detailed non-equilibrium molecular chemical network, there is relatively little molecular gas at $<300\,R_{g}$ except at the lowest $\dot{m}$ (though atomic and molecular gas are certainly important in the outer disk). The external cosmic ray background is also weak on these scales (for both bulk disk support, and direct cosmic ray heating/cooling processes, though it may help non-trivially in supporting the relatively large ionization fractions), with CR energy densities falling below the plotted range in Fig.~\ref{fig:profiles.general}. Dust is largely sublimated at the interior radii $\sim 30-300\,R_{g}$, even in the self-shielded midplane of the disk, though it remains possible that with a more sophisticated model for dust re-formation some of the colder-phase gas we see could experience grain re-growth. So cosmic rays, dust, and detailed molecular thermo-chemistry are not directly dynamically important if one wishes to {\em only} focus on small scales $\ll 300\,R_{g}$.

On most of the scales we resolve, relativistic effects are generally small. For example (\S~\ref{sec:methods:variants}) it makes little difference if we treat the SMBH with a Keplerian or Paczy{\'{n}}ski-Wiita potential at most radii, though it does become significant around the ISCO. But such effects could be important indirectly, e.g.\ Lens-Thirring effects tilting the inner disk, and will grow rapidly in importance at smaller radii. And they could be non-linearly important, as e.g.\ GR effects could help launch jets, which could in turn carve a channel for radiation to more efficiently escape from the inner disk (see \citealt{kaaz:2024.hamr.forged.fire.zoom.to.grmhd.magnetized.disks}).

Numerically, we refer to \S~\ref{sec:methods:variants} to illustrate that numerical choices such as the details of our numerical refinement scheme, resolution, exact treatment of different radiation bands and opacities/cooling rates, and when precisely we select to ``zoom in'' from the parent simulation, do not qualitatively change our conclusions.

\begin{figure}
	\centering
	\includegraphics[width=0.97\columnwidth]{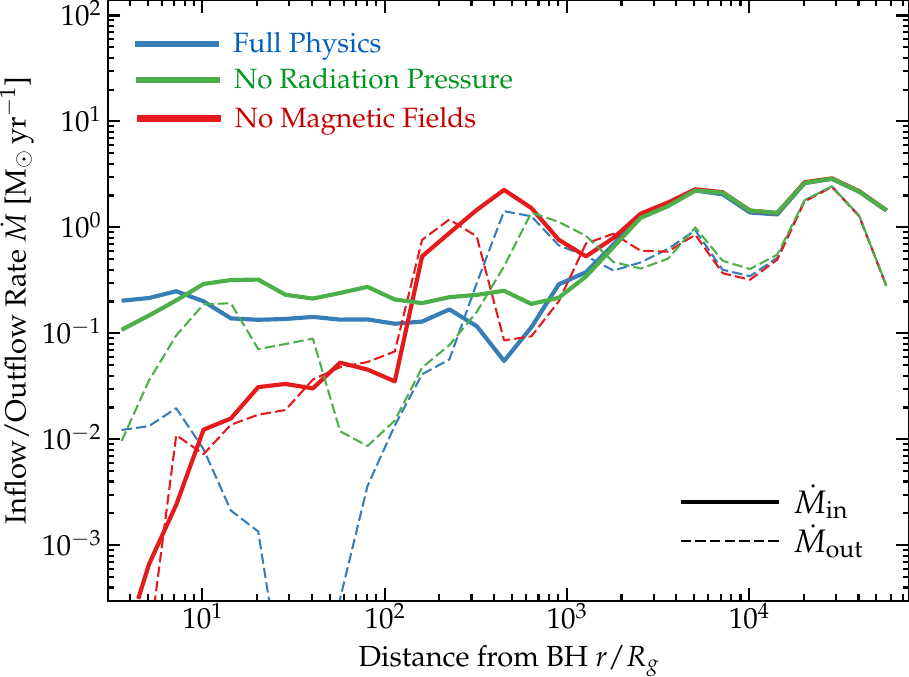} 
	\includegraphics[width=0.95\columnwidth]{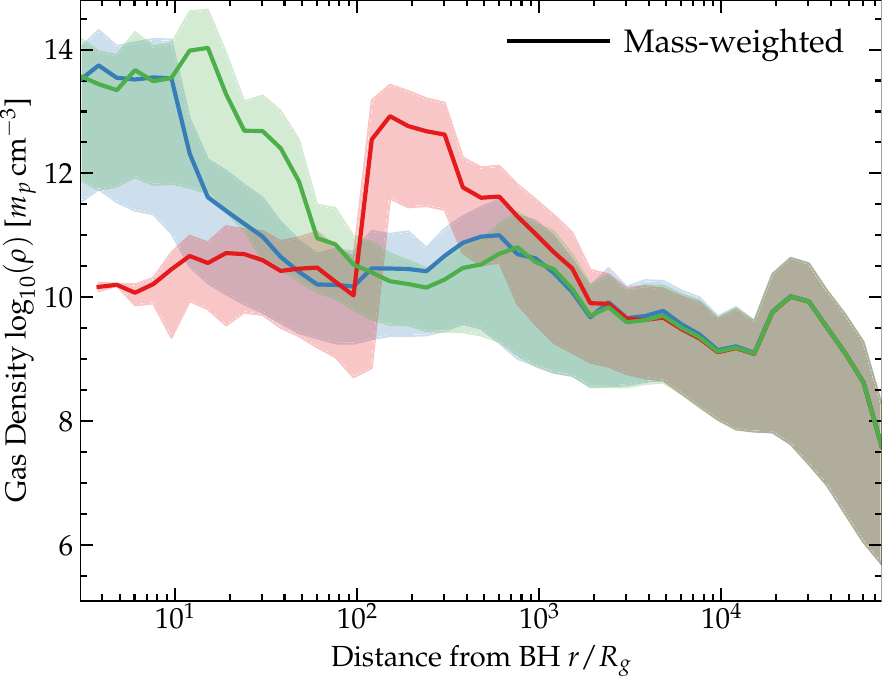} 
	\includegraphics[width=0.97\columnwidth]{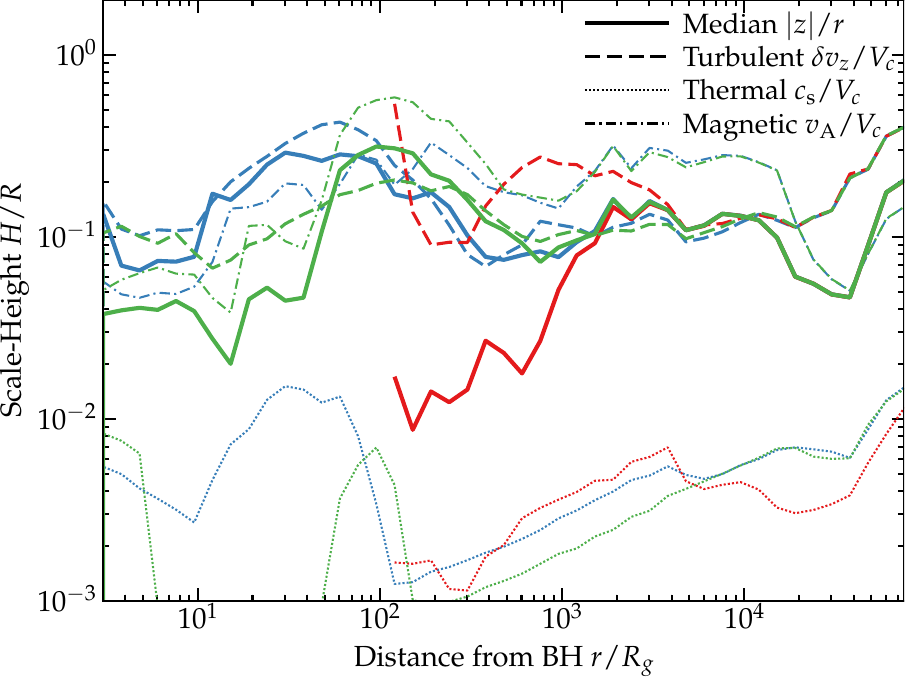} 
	\caption{Radial profiles of inflow/outflow rates $\dot{M}$ ({\em top}), densities ({\em middle}), and scale-heights ({\em bottom}) as Figs.~\ref{fig:profiles.general} \&\ \ref{fig:profiles.mdot}, but comparing our default/``full-physics'' $\dot{m}\sim 0.3$ ($f_{\dot{M}}=0.03$) simulation to two restarts (all at the same time since restart). In one, we remove radiation-hydrodynamics and assume optically-thin cooling (all cooling radiation is simply lost), so there is no radiation pressure. In another, we retain full radiation but remove magnetic fields. Both have significant effects (\S~\ref{sec:sensitivity:important}) in the inner disk (the larger radii being identical is simply an artifact of the short run-time of the simulations, showing the effects propagate inside-out with time, as expected). Removing radiation pressure leads to a somewhat thinner, less turbulent, cooler inner disk, with longer accretion times, larger in-plane outflow/radial sloshing motions, and higher densities at the same $\dot{M}$. Removing magnetic fields leads to collapse of the inner disk: the inflow rates drop by factors $\sim 100$ and are equal to outflows from weak eccentric/spiral/sloshing modes, an expending cavity forms, and the scale-height drops to nearly the resolution limit (we do not plot the ``no magnetic field'' case in $H/R$ interior to the cavity, because the density scale-height becomes unresolved).
	\label{fig:remove.B.rad}}
\end{figure} 

\begin{figure}
	\centering
	\includegraphics[width=0.97\columnwidth]{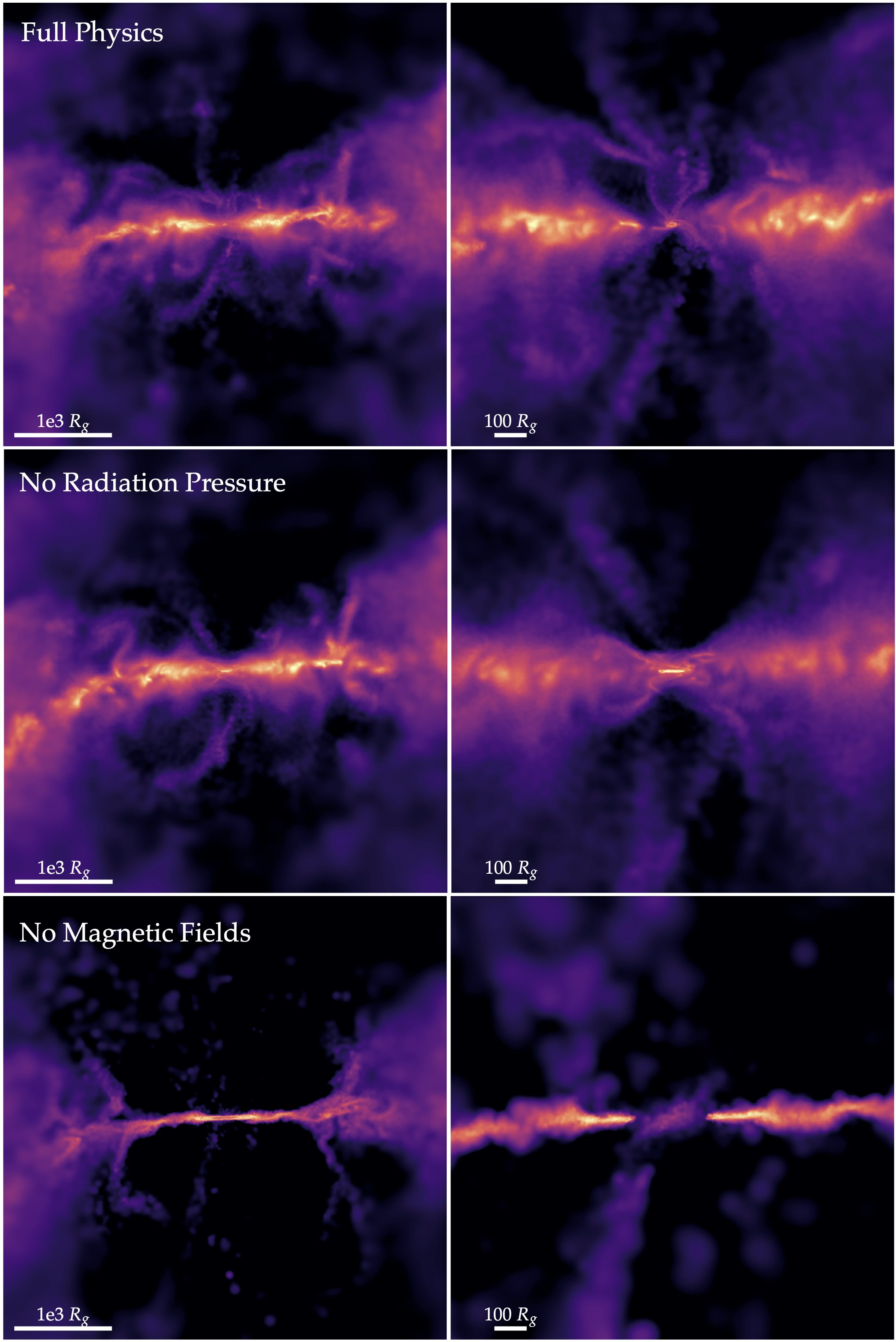} 
	\caption{Images of the simulations from Fig.~\ref{fig:remove.B.rad}, comparing our default/``full-physics'' $\dot{m}\sim 0.3$ run ({\em top}) to a restart at the same time with no radiation pressure ({\em middle}) or no magnetic fields ({\em bottom}), at two different scales ({\em left} and {\em right}). Without radiation pressure we can see a slightly thinner very inner disk at $\ll 100\,R_{g}$, the lack of the vertical (radiation-pressure driven) outflows from the very central disk, and a slightly smoother (i.e.\ less-turbulent/clumpy) disk on scales $\lesssim 1000\,R_{g}$. Without magnetic fields, the collapse to a much thinner disk and presence of a central hole/cavity (out to $\sim 100\,R_{g}$) owing to the decretion disk behavior are obvious.	
	\label{fig:remove.B.rad.images}}
\end{figure}

\subsubsection{Most Important Physics}
\label{sec:sensitivity:important}

The most important physics ``ingredients'' then (given initial and boundary conditions) for our focus here are MHD and radiation-thermochemistry in a BH potential. The MHD (being well-approximated by ideal MHD) and gravity (from the BH) are straightforward. But the radiation-thermochemistry undergoes several transitions over the range of radii we evolve: between optically thin and thick, radiation pressure-dominated and not, thermalized and non-thermalized radiation, atomic and ionized gas, scattering and absorption opacity-dominated, IR and optical/UV emission-dominated. It is therefore important to utilize radiation-themochemistry methods which can flexibly treat these different limits. 

Removing any radiation flux (in the Lagrangian frame), i.e.\ ``locking'' radiation to gas leads to the disk heating adiabatically and then ``exploding'' or ``pulsing'' (launching a radial ``bounce'') on of order one dynamical time, as expected -- this simply confirms that escaping radiation can be an important energy ``sink.'' Conversely, simply deleting all cooling radiation (i.e.\ assuming completely optically-thin escape of photons) so that cooling remains but there is zero radiation pressure, as shown in Figs.~\ref{fig:remove.B.rad} \&\ \ref{fig:remove.B.rad.images} leads to a colder, neutral atomic+molecular disk at large radii, which can become somewhat thinner/denser and less strongly-turbulent in the inner regions, and naturally never launches radiation-pressure driven outflows even for $\dot{m} \gg 1$. So radiation plays some significant roles here, but the effects are somewhat modest, even in regions where a naive comparison of $P_{\rm rad}$, $P_{\rm thermal}$, $P_{\rm B}$ would imply radiation pressure is dominant. This is discussed above in \S~\ref{sec:lte}, where we show that despite the disk being globally optically-thick and advecting photons into the BH, ``tight'' or ``local'' photon trapping (i.e.\ assuming that photon diffusion/escape times are much shorter than any other timescales in the problem even in small density sub-structures within the disk), as required for modeling the disk with some simply $\gamma=4/3$ ``effective'' equation of state or ``effective'' gas+radiation single-fluid sound speed, is not a good approximation. This cautions against such approximations, as well as showing that even more demanding approximations (like assuming the radiation and opacities/excitation states of different species all obey LTE) are not accurate in these simulations.

Figs.~\ref{fig:remove.B.rad} \&\ \ref{fig:remove.B.rad.images} show that removing magnetic fields has dramatic, leading-order effects, similar to those seen on larger scales in \papertwo. In order to start our simulation from a self-consistent zoom-in lacking magnetic fields, we could take the simulation without magnetic fields (zooming in from Mpc to $\sim 300\,R_{g}$ scales)  presented in \papertwo, and simply move the inner boundary inwards to $\sim 30\,R_{g}$ (keeping its resolution $\Delta m \approx 5\times10^{-8}\,{\rm M_{\odot}}$). As shown in \papertwo, if we do this, we find it leads to catastrophic fragmentation of the outer accretion disk, with SFRs higher by a factor of $\sim 1000$ at $\sim 0.1-10\,$pc and a factor of $\sim 10^{5}-10^{6}$ at $\sim 0.01-0.1$\,pc. The combination of this and orders-of-magnitude weaker stresses (weaker by a factor of $\sim 10^{2}-10^{4}$) means that outflow and SFRs are then much larger than inflow rates and the disk is starved, producing an inflow rate at least a factor of $\sim 1000$ times lower into $<300\,R_{g}$ (and indeed for most of the time, as shown in \papertwo, the inner region has a net Reynolds stress producing outflow, hence forms an expanding ``hole'' in the inner region or {\em decretion} disk). So a ``further zoom in'' would simply be uninteresting as there is no gas at $<300\,R_{g}$. So instead in Figs.~\ref{fig:remove.B.rad} \&\ \ref{fig:remove.B.rad.images} we take one of our simulations here in which strong magnetic fields developed self-consistently, and restart after turning off MHD (i.e.\ ``deleting the fields''). This is not self-consistent, but is a useful counterfactual: since turbulent, magnetic, and radiation pressures are order-of-magnitude comparable in the inner disk, i.e.\ $\beta_{\rm tot} \sim 1$, one might imagine turbulence and radiation somehow sustaining the disk without fields. But we find that removing magnetic fields, two dramatic changes occur even in the inner disk: (1) the disk loses vertical support and collapses to a vertical density scale-height approximately equal to the resolution limit, and (2) the stresses driving accretion not only become much weaker, but even reverse sign, with $\dot{M}_{\rm out} \approx \dot{M}_{\rm in}$ at all radii where we evolve for much longer than the dynamical time. These comparable inflow-outflow rates are dominated by random turbulent and sloshing/eccentric motions, with a small net $\dot{M}_{\rm out}$ driving overall loss of mass inside of the inner disk while the time-averaged accretion rate actually reaching the ISCO/SMBH decreases by at least a factor of $\sim 100$. Again, this is similar to the effects seen in our tests in \papertwo, and we find again similar results with additional resolution tests without magnetic fields in Appendix~\ref{sec:resolution.thermal.scale}. So despite $P_{\rm rad} \sim P_{\rm B}$, the effects of removing radiation pressure and magnetic fields are very different.

\subsubsection{Potentially Important Physics Neglected Here}
\label{sec:neglected.physics}

While the list of physics included in these simulations is extensive, it is by no means exhaustive. As argued above, there are a couple pieces of physics that could be important in certain regimes. The most obvious is GR. This should only become important on near-horizon scales, but our lack of relativistic physics means (as discussed in \S~\ref{sec:outflows}) that we cannot capture spin effects (important both for jets but also potentially mis-aligned disks and Lens-Thirring precession; see \citealt{kaaz:2022.grmhd.sims.misaligned.acc.disks.spin}), nor horizon-scale effects on the accretion of magnetic flux which could also influence jet-launching. 

Also as noted in \S~\ref{sec:outflows}, our multi-group radiation transport still lacks fine enough binning to properly treat line-driven winds. This could be critical for strong wind-launching from the surface of the disk in regions like the BLR. How important we expect this to be will be investigated in future work using more detailed Monte-Carlo radiation post-processing. We also adopt a simple model for dust sublimation, rather than explicitly following the evolution of dust grain populations (size evolution, destruction, condensation and coagulation), which could be important for the details of the location of the sublimation radii and for dust-radiation-pressure driven outflows from the inner edge of the torus \citep{soliman:agn.variability.from.dust.instabilities}.

And while we argued above that collisionless and two-temperature effects should be weak for the midplane accreting gas in the simulations here, it could definitely become important for the hottest ($T_{\rm gas} \gg 10^{9}\,$K), most tenuous coronal gas around the ISCO. Here more detailed plasma physics treatments in idealized simulations (e.g.\ \citealt{bambic:2024.coronal.simulations}) are needed.

\section{Conclusions}
\label{sec:conclusions}

We extend multi-group radiation-magnetohydrodynamics simulations with self-gravity and detailed non-equilibrium and NLTE thermochemistry, originally run from $\sim$\,Mpc scales to follow gas forming a quasar accretion disk, down to ISCO scales. We also extend these to explore different inflow rates. The accretion disks studied involves accretion at $\sim 0.005 - 10\,{\rm M_{\odot}\,yr^{-1}}$ around a $\sim 10^{7}\,{\rm M}_{\odot}$ SMBH, which for a reference radiative efficiency of $\epsilon_{r}^{0} = 0.1$ translates to sub-to-supercritical accretion rates spanning $\dot{m} \sim 0.02-20$.

Most of the dynamical properties of the disks follow a reasonable extrapolation from larger radii: they maintain (thermal) $\beta \sim 10^{-5}-10^{-2} \ll 1$, i.e.\ are strongly magnetically dominated, with a predominant toroidal mean magnetic field driven by flux freezing/advection (except in the innermost disk), with trans-\Alf{ic} and hence highly super-sonic and quasi-isotropic turbulent motions, and strong Maxwell and Reynolds stresses, in a thick, flared, multi-phase disk. All of these properties agree qualitatively with the analytic models presented in \paperthree, designed to extrapolate from the BHROI at a few pc to near-horizon scales. The most important extension of \papertwo\ is that we now resolve the transition radii where disks self-ionize, where most of the radiation is actually emitted, where radiation pressure can become important in the disks, and where (for the higher $\dot{m}$) a radiatively-efficient disk could become locally super-Eddington.

Here we show that even at $\dot{m} \gg 1$, supersonic turbulence makes the radiation transport globally advective, leading to a quasi-isothermal vertical structure with midplane temperature similar to the effective temperature (but with multiphase thermal structure within the disk), and $P_{\rm mag} \gtrsim P_{\rm mag,\,turb} \sim P_{\rm kin,\,turb} \sim P_{\rm rad}  \gg P_{\rm thermal}$, i.e.\ saturation of radiation and turbulent (kinetic+magnetic) pressures in equilibrium comparable to or a factor of a few smaller than the mean magnetic pressure (so the radiation+thermal $\beta_{\rm tot} \equiv (P_{\rm rad}+P_{\rm thermal})/ P_{\rm mag} \sim 0.1-1$ in the inner disk regions). Closely related, at high $\dot{m}$ the disk becomes radiatively inefficient at radii well interior to this transition, with a total emergent luminosity just a few times larger than Eddington. However we caution that any prediction for the ``total'' radiative efficiency at better than factor of a few accuracy may require GRMHD simulations that can more accurately extend to horizon scales, as much of the luminosity will be released on these scales or be carried back out by jets or other outflows \citep{kaaz:2024.hamr.forged.fire.zoom.to.grmhd.magnetized.disks}, though it is also clear here that modeling the thick-to-thin transition and NLTE, non-tight-coupling limit of radiation is important for the radiation temperature, pressure, and escape.  We also verify that these transitions towards advective transport modify the similarity solutions from \paperthree\ for disk properties, as predicted in \citet{hopkins:multiphase.mag.dom.disks}. We also show that despite $P_{\rm rad} \sim P_{\rm B}$ in the inner disk, radiation and magnetic fields act very differently on the gas, and removing magnetic fields causes much more dramatic effects on disk structure compared to removing radiation pressure. Large-scale coherent magneto-centrifugal winds do not emerge at any radii or accretion rates, but we show that this is expected given the large densities. However, ``failed'' magnetocentrifugal outflows or fountain flows could still be playing an important role driving turbulence and ``puffing up'' the disk.

With the ability to resolve an order-unity fraction of the expected emergent bolometric luminosity, our simulations show that radiative properties of the disk are strongly influenced by these transitions as well as the geometrically thick, low-density, modest optical-depth structure generic to flux-frozen disks. The radiation flux escapes in a wide-angle bicone, but strongly self-irradiates the disk itself. All of the disk ``surface layers'' from radii $\sim 10-10^{6}\,R_{g}$ are strongly illuminated, and an order-unity fraction (weakly decreasing at lower $\dot{m}$) of the entirely bolometric flux from the innermost disk is intercepted by and reprocessed or scattered in the outer disk. Even sightlines that naively would not intercept the disk (e.g.\ along more polar angles with $\cos{\theta} \gtrsim H/R$) can be scattered by the extended, accreting ``atmosphere'' of warm coronal gas at higher $\dot{m}$ (which is expected, given that $H/R$ is order-unity, there must be appreciable gas at $|z| \sim R$, and we see this in the slowly-declining vertical gas density profiles supported by even more slowly-declining magnetic pressure) back onto the disk. The thermal and radiation structure is complicated and multi-phase, and is not captured by LTE or single-fluid (tight gas-photon coupling) prescriptions. Continuum radiation pressure drives outflows in a relatively narrow, low-gas-density bicone from launching radii $\sim 10-100\,R_{g}$ at the highest $\dot{m} \gtrsim 1$, without outflow rates order-of-magnitude comparable to (though usually somewhat below) the accretion rates in the inner disk.

These radiation behaviors will have important implications for observed quasar variability, reverberation mapping, microlensing sizes and their wavelength dependence, and the spectral shape of the quasar emission -- as discussed in \citet{hopkins:multiphase.mag.dom.disks} it means the disk {\em itself} can easily provide the gas needed for historical ``components'' usually imagined to float above the (razor-thin) SS73-like accretion disk, like warm Comptonizing ``skins'', the corona, the warm absorber, the broad line region, and dusty torus. Detailed predictions for the spectra at different wavelengths and their variability will require radiative transfer post-processing calculations using these simulations, as our simple few broad-band on-the-fly radiation-MHD algorithm is insufficient to make such predictions. But we can directly illustrate the salient scattering and reprocessing, and see, for example, its critical effect on the emergent quasar radiation field. For example, the innermost disk features a radiation temperature of almost $\sim 10^{6}$\,K (roughly following $T_{\rm rad} \propto R^{-3/4}$), but by the time we follow this radiation to radii $\gtrsim 2000\,R_{g}$ where it is escaping in said bicone, the emergent effective temperature is $\sim 4\times10^{4}\,$K, in excellent agreement with observed SEDs of quasars at similar luminosities. And we show that properties of the gas in the surface layers of the disk at radii corresponding to the BLR agree remarkably well with what is needed to explain the more detailed spectral properties observed.

The physics in our code, while designed to follow a large range of scales from $\sim 10^{14}-10^{27}\,{\rm cm}$, have two key limitations that must be addressed at smaller radii. First, GR effects will become important, not just at the ISCO, but since (see \papertwo) the disk here is likely strongly misaligned with the pre-existing BH spin, where Lense-Thirring precession becomes important. Second, our multi-group radiation methods, while including sophisticated optically thin and thick treatments for a wide range of chemical processes, do not include either the effects of hard X-ray over-ionization (of e.g.\ Fe) or line transfer, critical for understanding both the opacities and potential line-driven winds launched by the inner disk. So the behavior of line-driven winds, the alignment of the central disk, and the launching of jets (presumably also near-horizon) remain important subjects for future work.

\begin{acknowledgements}
Support for PFH was provided by NSF Research Grants 1911233, 20009234, 2108318, NSF CAREER grant 1455342, NASA grants 80NSSC18K0562, HST-AR-15800. CAFG was supported by NSF through grants AST-2108230 and AST-2307327; by NASA through grants 21-ATP21-0036 and 23-ATP23-0008; and by STScI through grant JWST-AR-03252.001-A.
\end{acknowledgements}

\bibliographystyle{mn2e}
\bibliography{ms_extracted}

\begin{appendix}

\section{Cooling Physics and Opacities}
\label{sec:opacities}

Our default thermochemistry and cooling physics follows \citet{grudic:starforge.methods,hopkins:fire3.methods} and is described therein, with updates specified here. Briefly, the cooling rates $\Lambda({\bf x},\,t,\,T,\,\rho,\,Z,\,x_{e},\,...)$ include tabulations and fitting functions for a wide range of processes including detailed pre-tabulated rates for free-free and bound-free transitions (including spontaneous, collisional, Compton, cyclotron/synchrotron, Bremsstrahlung, fine-structure, ionization and recombination) of 11 different separately followed elements (H, He, C, N, O, Ne, Mg, Si, S, Ca, Fe), dust grains (modeling a full size spectrum from nm through $\sim 0.1\,\mu{\rm m}$ sizes following \citealt{weingartner:2001.dust.size.distrib,semenov:2003.dust.opacities}), free electrons, charged grains, ions (including important negative ions like H$^{-}$ as well as partially-ionized H, He, C, etc.), atomic and molecular (H$_{2}$, HD, CO, and heavier) species. For all of these the salient abundances of each species are evolved implicitly from the non-equilibrium network and the rates depend on the relative abundances, metallicities, temperatures, etc. 

A variety of non-radiative heating/cooling processes are explicitly accounted for as well, including for example cosmic-ray heating (via streaming and hadronic losses and ionization) and ionization; dust collisional and photo-electric heating/cooling;  charge exchange reactions; and the ``MHD work'' from adiabatic gas evolution, shocks, and reconnection (given by the Riemann problem in each cell). The radiative absorption/ionization and cooling mechanisms all couple to the evolved radiation fields but also account for additional radiation sources such as the (appropriately redshift-dependent; \citealt{cafg:2020.uv.background}) meta-galactic UV background and CMB (accounting for self-shielding at higher column densities). 

The dust-to-metals mass ratio is approximated by $f_{\rm dm} = 0.5\,\mathcal{F}_{\rm sub}(T_{\rm dust}/T_{\rm sub})$ with $T_{\rm sub} \equiv 1500\,$K and $\mathcal{F}_{\rm sub}(x) \equiv 0.5\,(1+9\,(1-x)/\sqrt{1+81\,(1-x)^{2}})\,\exp{[-(x/3)^{2}]}$. 
We approximate two-temperature (different proton and electron temperatures) plasma effects as detailed in \citep{cafg:2012.egy.cons.bal.winds,hopkins:qso.stellar.fb.together,torrey:2020.agn.wind.bal.gal.fx.fire}, evolving ion temperatures explicitly limited by Coulomb exchange at sufficiently high temperature. We account for self-absorption of ionizing photons within a cell explicitly (treating each as a slab). For self-absorption of continuum radiative transitions within a cell we implement a simple correction by multiplying the optically-thin emission $\Lambda_{{\rm rad},\,i}$ of that cell $i$ by $1/(1+ \tau_{{\rm self},\,i}^{2})$ where $\tau_{{\rm self},\,i} \equiv \kappa_{{\rm rad},\,i}(T_{\rm rad,\,em} = T_{{\rm gas},\,i},\,\rho_{i},\,T_{{\rm dust},\,i},\,Z_{i},\,..{\bf x}_{i},\,t_{i},\,...)\,\rho_{i}\,\Delta x_{i}/2$, i.e.\ some average continuum self-opacity. But we find that correction is generally small.

Also in our default simulations, we calculate the opacities for each cell and each evolved radiation band based on the combination of the explicitly-integrated chemical network results plus $T_{\rm rad}$ and other evolved radiation properties, including molecular (scaling with CO and H$_{2}$ abundances as $\kappa_{\rm mol} \sim 0.1\,(f_{\rm CO}+3\times10^{-9}\,f_{{\rm H}_{2}})$, where $f_{x}$ is a mass fraction and units are all in CGS), dust (scaling with the sublimation-dependent dust-to-metals ratio, dust temperature, $T_{\rm rad}$, and metallicities according to the detailed tables in \citealt{semenov:2003.dust.opacities}); free electron ($\kappa_{e} \sim 0.4\,x_{e}\,(1 + 2.7 \times 10^{11} \rho\,T_{\rm gas}^{-2})^{-1} [1 + (T_{\rm rad}/4.5\times 10^{8})^{0.86} ]^{-1}$ where $x_{e}$ is the free $e^{-}$ fraction per nucleon, $\rho$ and $T_{\rm gas}$ are the gas density and temperature and $T_{\rm rad}$ the radiation temperature); Kramers continuum ($\kappa_{\rm Kramers} \sim 4 \times 10^{25}\,(1+f_{\rm H})\,(f_{Z} e^{-1.5 \times 10^{5}/T_{\rm rad}} + 0.001\,x_{e})\,\rho\,T_{\rm rad}^{-3} T_{\rm gas}^{-1/2}$ where $f_{Z}$ is a total metal mass fraction {\em not} locked into dust); Rayleigh (atomic scattering $\kappa_{\rm Rayleigh} \sim f_{\rm HI}\,{\rm MIN}[5 \times 10^{-19}\,T_{\rm rad}^{4},\,0.2\,(1+f_{\rm H})]$); iron line-blanketing opacities ($\kappa_{\rm Fe} \sim 1.5 \times 10^{20}\,f_{Z}\,\rho\,T_{\rm rad}^{-2} e^{-(8000/T_{\rm rad})^{4} -(T_{\rm rad}/7 \times 10^{5})^{2} }$, an approximate fit to the tabulated scalings in \citealt{jiang:2015.rhd.star.sims.metal.opacities.for.agn.disks.as.well} where they are important); and H$^{-}$ ($\kappa_{\rm H^{-}} \sim x_{\rm H^{-}}\,(k_{\rm H^{-}}^{\rm bf}+k_{\rm H^{-}}^{\rm ff})$, $k_{\rm H^{-}}^{\rm bf} \sim 4.2\times 10^{7}\,(8760/T_{\rm rad})^{3/2} e^{-8760/T_{\rm rad}}$, $k_{\rm H^{-}}^{\rm ff} \sim 1.9 \times 10^{6} (8760/T_{\rm rad})^{2} e^{-8760/T_{\rm rad}} (0.6-2.5 \phi_{-}^{1/2}+2.5\phi_{-}+2.7 \phi_{-}^{3/2})$ with $\phi_{-}={\rm MIN}(T_{\rm gas}/5040,\,2)$, with the calculation of these and $x_{\rm H^{-}}$ following \citealt{john:1988.hminus.opacity,lenzuni:1991.opacities,glover:2007.low.metallicity.cooling.h2.tables}), as well as the salient absorption opacities in each narrow band \citep{hopkins:fire3.methods}. These all contribute to absorption opacities except free electron (scattering only) and dust (which contributes to both scattering and absorption per the tables given). 

Total energy is manifestly conserved in all radiative absorption/emission/cooling/heating operations. Radiative cooling is added directly into the photon energy budget, and absorption to the gas energy \citep{grudic:starforge.methods}. Note that these approaches mean that the gas, dust, radiation, and (implicitly) salient excitation temperatures (determining the relative species/ionization/excitation abundances) are all evolved {\em separately}, i.e.\ we do not impose the common simplifying assumption that gas and radiation temperatures must be in equilibrium nor that the system is blackbody-like. Similarly note the appearance of terms like $x_{e}$, $f_{\rm H}$, $x_{\rm H^{-}}$, $f_{\rm H_{2}}$ in the equations above, which are integrated explicitly in our thermochemical network; this means there is no assumption of local thermodynamic equilibrium (LTE), nor of any equilibrium for the slowest-evolving species, in our opacities. Related, our treatment is able to handle both optically thin or highly non-LTE or non-equilibrium limits, and thick or photon-trapped or blackbody limits, without prior assumptions.

However, the above split treatment of opacities and radiative cooling is not always strictly self-consistent, in the sense that the optically-thin radiative cooling emission per particle $n\,\Lambda_{{\rm cool},\,{\rm rad},\,i}$ calculated in the cooling operation will not always be exactly equal to $4\pi\,\langle \kappa_{{\rm abs},\,i}\rangle\,\mu_{i} m_{p}\,\sigma_{B} T_{{\rm gas},\,i}^{4}$, the emission rate one would estimate using the band-averaged opacities $\kappa_{{\rm abs},\,i}$ defined above. This is intentional: since the thermo-chemistry is strictly local for a given incident radiation field we can integrate emission rates over an arbitrarily fine-grained set of frequencies and species and include non-radiative and non-equilibrium effects per above, but for computational expense the explicitly-evolved radiation fields must be coarse-grained/binned, breaking this exact symmetry. But we can test the approximate consistency of the two with a couple of simple tests. First, for the opacities in our adaptive grey-band (focusing on that as the narrow-band cases are less complex), we could take $\kappa_{{\rm abs,\,eff},\,i}(T_{{\rm rad},\,i}) \rightarrow n\,\Lambda_{{\rm cool},\,{\rm rad},\,i}(T_{\rm gas}^{\prime},\,T_{{\rm rad},\,i},\,{\bf x}_{i},\,t_{i},\,n_{i},\,Z_{i},...)/(4\pi\,\mu_{i} m_{p}\sigma_{B}\,T_{\rm gas}^{4})$ where $T_{\rm gas}^{\prime} = T_{{\rm rad},\,i}$, i.e.\ forcing the mean absorption opacities to match the emission from the thermo-chemical network at the radiation temperature. We stress that this is not the default because it is often less accurate: it ignores the differences between e.g.\ Planck or Rosseland or flux mean opacities, and can significantly overestimate absorption opacities in certain limits especially relevant in the more diffuse outer disk radii (more ISM-like conditions), for example when the radiation bands are broad-band continuum but the cooling is dominated by recombination transitions or narrow lines, or when the radiation is otherwise highly non-thermal dominated or optically-thin (so the effective radiation temperature of the emission is not, in fact, closely related to the gas kinetic temperature). Second, we can conversely replace our detailed cooling rates with $\Lambda_{{\rm cool},\,{\rm rad},\,i} \rightarrow 4\pi\,n_{i}^{-1}\,\mu_{i} m_{p}\sigma_{B}\,T_{\rm gas}^{4}\,\kappa_{{\rm abs,\,eff},\,i}(T_{\rm rad}^{\prime},\,T_{{\rm gas},\,i}\,{\bf x}_{i},\,t_{i},\,n_{i},\,Z_{i},...)$ where $T_{\rm rad}^{\prime} = T_{{\rm gas},\,i}$, i.e.\ the value given by this our absorption opacity tables at a radiation temperature equal to the gas temperature. Again, this is less accurate for the reasons above. But in the limit where the radiation and gas kinetic temperatures are broadly similar, or the radiation is broad-band continuum-dominated, or the system is sufficiently optically-thick, these should all agree reasonably well (note these experiments still allow for arbitrarily non-LTE behavior, and separately evolve the various abundances and temperatures above). Indeed we find this is the case: in the inner disk of interest here (within a few hundred $R_{g}$), these experiments introduce quantitative differences, but do not qualitatively change any of our conclusions or key thermochemical or radiation properties of the system.

\section{Physical and Numerical Variations Considered}
\label{sec:methods:variants}

We have experimented with a number of alternative physics or numerics experiments and setups, to ensure the behaviors here are robust and to check different potential uncertainties. For brevity, we summarize here the variant runs, restarted at different times from snapshots of our fiducial simulation or parent simulation, and generally run for a much short time (though always still many dynamical times at the inner boundary), to test if they strongly influence the results.

{\bf Refinement variations:} We considered several different functions to define the ``target mass'' $\Delta m$ versus $R$, including smoother/shallower refinement beginning at larger radii, steeper refinement, or even refining entirely in the initial conditions (akin to simply running a higher-resolution IC statistically sampling the density field as interpolated from the low-resolution snapshot, and hoping there would be no contamination by low-resolution cells migrating inwards). Per \S~\ref{sec:resolution.thermal.scale} we also considered pure spatial-resolution refinement criteria (enforcing $\Delta x_{i} \equiv (m_{i}/\rho_{i})^{1/3} < 0.01\,r$). We disabled de-refinement of outflows (any material with $v_{r} > 1\,v_{\rm K}$ or $0.5\,v_{\rm K}$), or any gas with local thermal-scale lengths $c_{s}/\Omega$ that would become un-resolved upon de-refinement. We also shifted the ``pivot radius'' of refinement from $\sim 0.001-0.1$\,pc. We varied the enforced minimum number of timesteps between allowed refinement or de-refinement operations per cell from $\sim3-3000$ (see \paperone); ran tests in which the temperatures and/or ionization states were arbitrarily reset to $\sim 10^{4}$\,K and/or fully-ionized (sacrificing energy conservation) after a refinement step (to see how quickly the chemistry reverted to standard behavior); forced the code to revert to a first-order reconstruction and solver for any cell which experienced (de)refinement in the last 10 timesteps (see \citealt{hopkins:gizmo}); applied an extra 10-iteration Poisson-type divergence cleaning layer following \citet{hopkins:cg.mhd.gizmo} after each refinement; adopted the alternative method from \citet{su:2021.agn.jet.params.vs.quenching} to assign ${\bf B}$ after refinement to maintain divergence control; and varied the separation distance and axis of separation of the child cells within each parent cell split \citep{grudic:starforge.methods}. So long as the refinement was sufficiently rapid that the target mass $\Delta m$ could be reached before gas accreted through our boundary, and sufficiently slow so as to not create memory errors or prevent simulation progress (by e.g.\ creating billions of new cells in a short number of timesteps), we found none of these choices has a systemic effect on our results interior to the high-resolution region.

{\bf Resolution variations:} Our default refinement pushes to the boundary of what is technically possible, so we cannot simply increase the resolution further. However, we have considered resolution tests in \paperone, and we experiment here with refining further to $\sim 30 \times$ improved mass resolution in a single narrow radial annulus. Even that is challenging, as material moves in and out rapidly but making the annulus much larger increases computational expense excessively, so we evolve for just $\sim 1$ orbital time, but see no qualitative effect on disk structure in this experiment. More easily, we run an experiment where we degrade the target mass resolution everywhere by an order of magnitude. As expected we see less small-scale disk structure and turbulence, but the zeroth-order properties like accretion rates and disk thickness remain similar everywhere (at the very innermost radii, this does make the vertical disk thickness somewhat poorly resolved, however). The refinement variations above also provide resolution tests of different sorts, as several (such as fixed spatial resolution, or disabling de-refinement in winds or cold/short thermal-scale-length gas) provide enhanced resolution in different gas phases (at the cost of more poor resolution in other phases). We describe some additional resolution tests in \S~\ref{sec:resolution.thermal.scale}.

{\bf Inner boundary variations:} We have varied the inner boundary/accretion radius in shorter restarts from $\sim 3-300\,R_{g}$. The larger radii approach our original simulation accretion radius, so are consistent with those simulations as expected. But the results {\em at a given radius} do not appear strongly sensitive to this choice. The main limitation of going to smaller radii is the much smaller dynamical and light-crossing times (and therefore numerical timesteps) involved, limiting the duration of the experiments. Below we show the structure that appears in our experiments with smaller inner radii. 

{\bf Initial conditions:} We restarted our simulation beginning at three different times (snapshots) randomly sampled from the parent simulation after it had reached quasi-steady state (see \papertwo). We also consider snapshots from both the ``fast'' and ``slow'' protostellar disk accretion sub-grid model variant simulations in \citet{hopkins:superzoom.imf} -- this choice only changes the sink-particle description of accretion for stars forming in the accretion disk, so unsurprisingly has no effects here. We also experimented with altering our outer ``frozen'' radius, making it larger or smaller by a factor of $\sim 3$. We ran tests in which we reset all initial temperature and/or ionization states of the snapshot before restart (making the ICs uniformly $T=10^{4}$\,K and/or fully ionized). We also considered a run in which the inner boundary gradually shrunk (decreasing by a factor of $\sim 2$ every two orbital times at each radius) as opposed to simply setting it to $\sim 30\,R_{g}$ initially. Again none of these had a significant impact on our results. As emphasized in the text, because our simulations are trying to predict uncertain quantities like the inner disk accretion rates, luminosities, and outflow properties, our inner boundary is a simple sink, but we have also considered simulations where the inner boundary features an outgoing steady radiation flux with a luminosity $=3\times10^{45}\,{\rm erg\,s^{-1}}$ directed with a $|\cos{\theta}|$ intensity pattern outwards at the inner boundary; this can have some significant effects on the inner boundaries in simulations where the inner boundary is large ($\sim 300\,R_{g}$ or larger), but once we refine to small radii, we find that starting from this modified IC makes no difference to the qualitative disk structure that forms (it is similar to restarting from a different snapshot in our default setup). 

{\bf Star formation and stellar physics:} As noted below, we experimented with removing any star particles, turning off new star formation and stellar physics entirely, or considering the different fast/slow sub-grid circumstellar accretion models from \paperthree. None of these has any effect at radii $\sim 3-300\,R_{g}$ where we focus here, as there are basically no stars and no star formation in any of the cases considered.

{\bf Gravitational potentials:} We experimented with switching from a Paczy{\'{n}}ski-Wiita (PW) potential for the BH back to Newtonian, replacing the gas cells potential with PW (though not self-consistent, this is irrelevant because their individual cell $R_{g}$ is vastly smaller than our resolution), or disabling entirely the external potential from matter exterior to the ``frozen'' larger radii $\gtrsim$\,pc. None of these had any major effect at $\gtrsim 30\,R_{g}$, and even removing the PW potential at $\sim 3-30\,R_{g}$ only produces modest quantitative (not qualitative) effects on the disk properties.

{\bf Radiation-hydrodynamics:} We have considered removing all radiation transport, either assuming radiation is locked to gas (so contributes to the pressure but alway has zero cell-to-cell flux, aking to an adiabatic-like $\gamma=4/3$ equation-of-state assumption) or simply deleting all radiation immediately (assuming optically-thin escape of all cooling radiation), which both have significant effects as described in the main text. Less radically, we can also turn on and off different radiation bands of our multi-group solver in a modular fashion. In the inner disk, we can remove the EUV, NUV, and OIR bands and reduce to a single photo-ionizing band and see very little effect, since at these wavelengths narrow-band effects like Lyman-Werner radiative dissociation or photo-electric heating are not the dominant sources of radiation or opacity. Removing any treatment of ionizing photons (e.g.\ removing all ionizing bands {\em and} not calculating ionization from the adaptive-wavelength band) produces an artificial result that the disk remains atomic at all radii, which in turn means the opacities are orders-of-magnitude lower than they should be, and so the effective optical depth $\tau^{\ast} \sim \sqrt{\tau_{a} (\tau_{a} + \tau_{s})}$ remains below unity and the disk never thermalizes or effectively traps radiation as it should. But so long as this is accounted for, we can capture the most important physics with our single adaptive band as this still allows for a flexible scaling of $T_{\rm rad}$ which is very different across the disk. We have modified various numerical slope and flux-limiters used to stabilize the operator-splitting of the thermochemistry and radiation transport (see \citealt{hopkins:gizmo.diffusion}), and experimented with the alternative operator-splitting approach used in GIZMO radiation-hydrodynamic TDE simulations in \citet{bonnerot:2021.gizmo.rhd.tde.sims}, and switched from our default comoving formulation of the RHD equations (since the code is Lagrangian) to a boosted-Eulerian mixed-frame RHD formulation per \citet{mihalas:1984oup..book.....M}. None of these technical numerical choices produces significant qualitative effects.

{\bf Cooling physics \&\ Opacities:} More details of the cooling physics and opacities included are given in \S~\ref{sec:opacities}. We have experimented with a number of the salient choices there. For example, turning off the self-absorption correction for continuum radiation, relativistic Compton cooling (just using non-relativistic expressions) and Comptonization, including or excluding Klein-Nishina corrections to free electron opacity, excluding our approximate treatment of two-temperature plasmas (relevant for extremely hot gas), removing the iron line-blanketing opacities, changing the dust sublimation temperature (and rapidity of sublimation above $T_{\rm sub}$) between $\sim 1000-2000$\,K, or adding all cooling radiation to our adaptive continuum band versus spreading it between bins according to the estimated emission wavelengths. Most of these have at most modest (order-unity) quantitative effects on gas temperatures in some specific regions of parameter space, but do not significantly alter the mean (mass-weighted) gas temperatures or ionization structure or radiation temperatures, let alone the global disk dynamics. The same is true if we replace our default cooling network with just our grey-band opacities or conversely calculate the effective opacities not directly but simply from the cooling subroutine emissivities, as described in \S~\ref{sec:opacities} -- these can produce significant inaccuracy in the outer, optically-thin atomic disk (more ISM-like conditions) but for the innermost radiation-pressure dominated disk more simple approximations appear to behave reasonably.

\section{On the ``Minimum Resolvable'' Scale Heights/Lengths}
\label{sec:spatial.resolution}

In Eulerian (e.g.\ fixed-grid) MHD methods, the minimum spatial scale $\Delta x_{\rm min} = \Delta x$ which can be represented is trivially defined by the cell size, but there is no unique definition of the minimum mass scale $\Delta m_{\rm min} =  \rho_{\rm min} \Delta x^{3}$ (in terms of some ``minimum possible density'' $\rho_{\rm min}$). Conversely, with a Lagrangian methods (like ours), the minimum mass scale $\Delta m_{\rm min} = \Delta m$ is fixed, but there is no unique definition of the minimum resolvable spatial scale $\Delta x_{\rm min} = (\Delta m/\rho_{\rm max})^{1/3}$ (noting the 1D cubic-cell equivalent cell size is $(m_{i}/\rho_{i})^{1/3}$, and now in terms of a ``maximum possible density'' $\rho_{\rm max}$). Some examples of the latter salient for different physical applications for our {\small GIZMO} code (e.g.\ Jeans fragmentation) are discussed in \citet{hopkins:lagrangian.pressure.sph,hopkins:gizmo,hopkins:mhd.gizmo,hopkins:fire2.methods}. 

A well-posed and salient question here is what the minimum resolvable disk thickness $H_{\rm min}$ would be. If we started, for example, from an arbitrarily tall gas cylinder with $H\rightarrow \infty$, in a stratified potential, with only thermal pressure support, and allowed it to cool to zero temperature, it would collapse vertically, with the instantaneous minimum cell size ${\rm MIN}(m_{i}/\rho_{i}[t])^{1/3}$ shrinking as the disk collapses (so this instantaneous value is clearly not the minimum resolvable). However, at some point, if the cell centers-of-mass (mesh-generating points, moving with the fluid) collapse to a razor-thin configuration, then the separation in the horizontal ($xy$ or $R\phi$) direction $\Delta x$ between cells will become much larger than their vertical separation ($\Delta z$). At this point, quantities like vertical gradients can no longer be defined (or more formally in {\small GIZMO}, the condition matrix for the face and gradient reconstruction becomes ill-conditioned; see \citealt{hopkins:gizmo} Appendix~C). This is what it means for a structure to be ``one cell thick'' in such a method. The vertical thickness at which this occurs is easily calculated as $\Delta z_{\rm min} \approx ({\rm d} N/{\rm d}A)^{-1/2} = (\Delta m/\Sigma_{\rm gas})^{1/2}$, in terms of the number of cells per unit area (vertically integrating through the disk scale height) ${\rm d}N/{\rm d}A$. Direct numerical tests also validate this scaling for practical questions \citep{hopkins:rad.pressure.sf.fb,hopkins:fb.ism.prop,hopkins:gizmo,few:2016.disk.spiral.arm.sims.gizmo.vs.ramses.vs.sph,deng:gravito.turb.frag.convergence.gizmo.methods,deng:2021.magnetic.disk.frag.in.gizmo.small.planetesimals}. This defines the resolution limit for $H/R$ plotted in Fig.~\ref{fig:profiles.general}. Note that this is, by approximate design of our refinement criterion, order-of-magnitude similar to the thermal scale length $c_{s}/v_{\rm K}$ of the midplane gas at most radii.

However, this does not mean that local structures with even smaller sizes ($\Delta x \ll \Delta z_{\rm min}$) cannot be represented. Indeed we see some dense structures in \paperone\ and here with sizes below this limit. The key is that these are small-scale structures ($\Delta x \ll H$) -- a ``clump'' of gas cells could, for example, collapse under self-gravity in principle in this numerical method to arbitrarily small scales (i.e.\ form a star). In practice this occurs at large radii in the disk and is truncated only by the sink particle prescription for star formation (see detailed discussion in \citealt{hopkins:superzoom.imf}, but in brief we see at least some structures with sizes $\sim 10\,$au at $R\gtrsim 5\,$pc, or $\Delta x/R \lesssim 10^{-5}$). More relevant in the inner disk, strong radiative shocks can generate denser structures, for which $\Delta x$ can drops to smaller values.

\section{Tests with Enforced-Resolved Thermal Scale-Lengths}
\label{sec:resolution.thermal.scale}

\begin{figure*}
	\centering
	\includegraphics[width=0.97\textwidth]{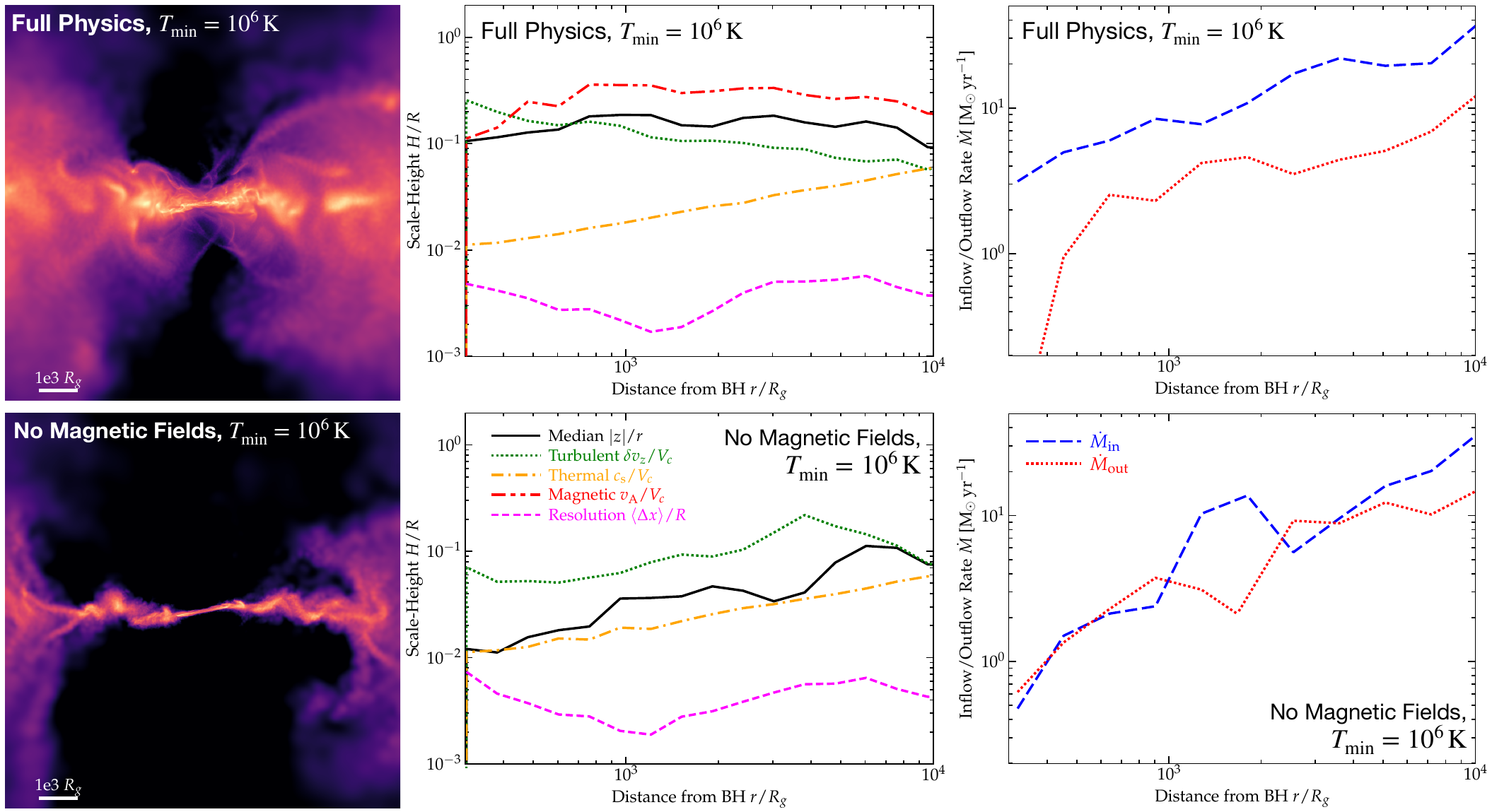} 
	\caption{Tests from \S~\ref{sec:resolution.thermal.scale} where we enforce a well-resolved minimum thermal scale-length in the disk. We re-run a portion of the disk from our simulation snapshots (with inner boundary at $300\,R_{g}$), run for time $\sim 300 \Omega^{-1}$ ($\sim 1\Omega^{-1}$) at radius ($10^{4}\,R_{g}$), adding an arbitrary temperature floor $T > T_{\rm min} \equiv 10^{6}\,$K in the thermochemistry to ensure the thermal scale-length $H_{\rm therm} \sim c_{s}/\Omega$ is well-resolved at all radii shown.
	We compare full-physics otherwise ({\em top}) and a restart where we remove magnetic fields (radiation-hydrodynamics only, {\em bottom}). 
	{\em Left:} Edge-on density projection in a slice through the angular momentum axis as Fig.~\ref{fig:zoomies.cyl}. 
	{\em Middle:} Density scale height ($|z|/r$) and predicted thermal, turbulent, magnetic scale-heights and resolution in annuli, versus radius $r$. 
	{\em Right:} Mass inflow $\dot{M}_{\rm in}$ and outflow $\dot{M}_{\rm out}$ rates through each annulus. 
	With full physics, the un-ambiguously resolved $H_{\rm therm} \gg \Delta x$ from this test changes some of the small-scale gas clumping, but none of the global properties of the disk: notably the \Alf, turbulent, and density scale-heights, and accretion rates, remain similar to our default case. Removing magnetic fields however leads to immediate collapse of the disk to the thermal scale-height, with order-of-magnitude lower inflow rates (and even those are mostly radial ``sloshing''/spiral/eccentric motions, hence $\dot{M}_{\rm in} \approx \dot{M}_{\rm out}$, with the vertical motions $\delta v_{z}$ dominated by the prominent warps/breaks seen in the disk). 
	\label{fig:tmin.compare}}
\end{figure*}

In our simulations, spatial scales like the turbulent energy range ($\sim |\delta v|/\Omega \sim H$), density scale-height ($H$), \Alf\ scale ($\sim v_{A}/\Omega \sim H$), mean-field and density gradient scale-lengths, as well as dynamical and light-crossing times and mass scales of the disk are well-resolved (Figs.~\ref{fig:resolution}, \ref{fig:profiles.general}). Indeed our typical $\Delta x/H \sim 0.01$ is comparable to the highest resolution often achieved even in historical shearing-box type simulations \citep[see discussion in][]{zier:2022.mri.dynamo.arepo,dhang:2024.mri.dynamo.shearing.box.sims}. But it is more challenging to even uniquely define (let alone resolve) the thermal scale-length $H_{\rm therm} \sim c_{s}/\Omega$. In Fig.~\ref{fig:profiles.general} we see our resolution is comparable to this ($\Delta x \sim H_{\rm therm}$) defined for a weighted mean $c_{s}$ in the midplane. But recall the disks are multi-phase with a wide range of $T_{\rm gas}$ and $c_{s}$ (Fig.~\ref{fig:profiles.general}), so $H_{\rm therm}$ of the hot gas is quite well-resolved, while (especially in the outer disk) some gas cools to molecular temperatures $\lesssim 100\,$K giving local $H_{\rm therm}/R$ as small as $\sim 10^{-5}$. Resolving this is not possible even in shearing-box simulations. But this gas is precisely that in which thermal pressure is least significant (they have magnetic, turbulent, and radiation pressures orders-of-magnitude larger than thermal), so we expect finite resolution not to matter for global disk properties (though it could be important for microphysics like the sizes of the smallest ``cloudlets'' in the BLR), as commonly seen in other simulations of multi-phase disks in galaxy and star formation \citep{hopkins:rad.pressure.sf.fb,hopkins:fb.ism.prop,hopkins:fire2.methods,2015MNRAS.454..238W,benincasa:2016.selfreg.sf.disk.modes,benincasa:2020.gmc.lifetimes.fire,guszejnov:universal.scalings,grudic:sfe.gmcs.vs.obs,grudic:2022.sf.fullstarforge.imf,orr:non.eqm.sf.model}. 

That said, in \citet{squire:2024.mri.shearing.box.strongly.magnetized.different.beta.states} and Guo et al., in prep., we have studied the highly-idealized problems in magnetically-dominated disks, specifically shearing boxes and homogeneous collapse, respectively, with greatly simplified physics (isothermal, ideal MHD, in analytic Keplerian potentials, idealized geometries, and no other physics) and different code (ATHENA). While some of the conclusions for $\beta_{\rm therm} \ll 1$ (though still $\beta_{\rm therm} \sim 0.01$, much higher than the outer disks here) are similar to our ``full-physics'' simulations here, in these idealized experiments there is a hint of some non-linear resolution dependence where magnetic flux-loss becomes more efficient when $\Delta x$ becomes smaller than $H_{\rm therm}$ (or sufficiently small relative to $v_{A}/\Omega$), before becoming less efficient again at smaller $\Delta x$, which can increase $\beta_{\rm therm}$ to $\sim 0.1-1$ in the midplane. Repeating some of these experiments in GIZMO (Tomar et al., in prep.) shows qualitatively similar results, despite the very different method and meaning of resolution (\S~\ref{sec:spatial.resolution}). 

In our ``full physics'' simulations, however, we do not see any obvious resolution dependence of the mean field or disk height $H/R$ (\S~\ref{sec:methods:variants}), varying $\Delta x$ by order-of-magnitude. But again, it is impossible to reach $\Delta x \ll H_{\rm therm}$ for all cold gas. Therefore here we consider a simple experiment, restarting our fiducial simulation enforcing a temperature floor in the cooling function. We consider both $T_{\rm min} = 10^{6}\,$K (so $H_{\rm therm}/R \propto R^{1/2}$) and $T_{\rm min} \propto R$ (chosen so $H_{\rm therm}/R = 0.1$), evolving the outer disk (inner boundary moved out to $\sim 300\,R_{g}$) for $\sim 20$ orbital times ($>100\,\Omega^{-1}$) at these radii. These values of $T_{\rm min}$ are chosen so that the disk retains $\beta_{\rm therm} \ll 1$, but with a well-resolved hierarchy $v_{A}/\Omega \gg  c_{s}/\Omega \gg \Delta x$. We also add the refinement criterion that no cell can have individual instantaneous size $\Delta x_{i} \equiv m_{i}/\rho_{i}$ larger than $0.01\,r$.\footnote{Note, in our tests in \S~\ref{sec:resolution.thermal.scale}, the thermal scale-height is well-resolved with $c_{s}/\Omega \gg \Delta x$ whether we use the definition of the minimum-resolvable $\Delta x$ from \S~\ref{sec:spatial.resolution} as appropriate for a Lagrangian code, or simply take the average $\langle \Delta x_{i} \rangle \equiv \langle m_{i}/\rho_{i} \rangle$ of cells in the disk.} The resulting $H/R$ and vertical profiles are shown in Fig.~\ref{fig:tmin.compare}. We see that there is no major loss of toroidal magnetic flux, or change in either the magnetic ($v_{A}/\Omega$) or turbulent ($ \delta v_{z}/\Omega$) or density ($H$) scale-heights of the disk. Indeed, considering essentially all the plots from the main text, the only noticeable change is some suppression of small scale structure with sizes $\ll H$ in the disk (e.g.\ some of the narrow shocks and dense gas clumps), which is completely expected given the much larger thermal scale length and lower (by construction) sonic Mach number. 

Thus there is no major qualitative change in global properties of these disks owing to explicitly resolving the thermal scale-length (consistent with our initial speculation), nor is there a change with resolution in the range $\Delta x \sim 0.01-1\,H$. This suggests that there is some important difference in our global, galaxy-scale, full-physics simulations compared to the idealized tests in \citet{squire:2024.mri.shearing.box.strongly.magnetized.different.beta.states} and Guo et al., in prep. This may owe to the effects of finite radiation pressure/live radiation transport, more realistic thermochemistry and multi-phase structure, boundary/initial conditions, pre-existing turbulence and strong fields, or other effects not captured in those test problems.

If, however, we remove magnetic fields from this experiment in a single timestep (keeping the radiation and turbulent and thermal support fixed), shown in Fig.~\ref{fig:tmin.compare}, we can confirm the expected behavior, that the disks collapse rapidly to the input thermal scale-height, without any of its thick vertical structure. But we also see the behaviors noted in \papertwo\ at larger radii if we begin our refinement from a galaxy-scale setup without magnetic fields: the accretion disk actually becomes a {\em decretion} disk on average, with a widening central hole featuring only occasional very small accretion of gas clumps so that the time-averaged gas accretion rate supported drops by an order of magnitude in a short time (and is continuing to drop at an increasing rate at the final times to which we run the test). In these simulations, magnetic fields play a critical role regardless of how well one resolves the thermal scale-length.

\end{appendix}

\end{document}